\documentclass[12pt]{article}\usepackage{amsmath,amssymb,multirow,jheppub,graphicx}
\usepackage{centernot}
\usepackage{mciteplus}
\usepackage{color}
\allowdisplaybreaks[1]
\usepackage{tikz}
 \tikzset{node distance=2cm, auto}

\renewcommand{\Re}{\text{Re}\;}
\renewcommand{\Im}{\text{Im}\;}

\def\W{ {\cal W}}

\def\Im{\text{Im}\;}
\def\Re{\text{Re}\;}

\def\tfrac#1#2{{\textstyle{\frac{#1}{#2}}}}

\def\hat{\widehat}
\def\bar{\overline}

\def\x{{\mathfrak x}}

\def\p{{\mathfrak p}}

\def\O{{\Omega}}

\def\hpp{{\hat \psi}_{+}}
\def\hpm{{\hat \psi}_{-}}
\def\I{{\cal I}}
\def\be{\begin{equation}}
\def\ee{\end{equation}}
\def\Z{{\mathbb Z}}
\def\R{{\mathbb R}}
\def\coeff#1#2{{\textstyle {\frac {#1}{#2}}}}

\def\half{\coeff 12}
\def\N{{\cal N}}
\usepackage{verbatim}   
\usepackage[all]{xy}
\usepackage{bm}  
\def\Dslash{{\rlap{\raise 1pt \hbox{$\>/$}}D}}
\def\Pslash{{\rlap{\raise  1pt \hbox{$\>/$}}\,\partial}}

\title{Toward Picard-Lefschetz Theory of  Path Integrals, Complex Saddles and Resurgence}
\author[a,c]{Alireza Behtash,}
\author[b]{Gerald  V. Dunne,}
\author[a]{Thomas Sch\"afer,}
\author[a]{Tin Sulejmanpasic,}     
\author[a,c]{\qquad Mithat \"Unsal}
\affiliation[a]{Department of Physics, North Carolina State University, 
Raleigh, NC 27695, USA}
\affiliation[b]{Department of Physics, University of Connecticut, Storrs, 
CT 06269-3046, USA}
\affiliation[c]{Department of Mathematics,  Harvard University, Cambridge, MA, 02138, USA}
\abstract{
We show that the semi-classical analysis of generic Euclidean path integrals  
necessarily requires  complexification of the action and measure, and 
consideration  of complex saddle solutions. We  demonstrate  that complex 
saddle points have a natural interpretation in terms of the Picard-Lefschetz 
theory. Motivated in part by the semi-classical expansion of QCD with 
adjoint matter on $\R^3\times S^1$, we study quantum-mechanical systems 
with bosonic and fermionic (Grassmann) degrees of freedom  with harmonic 
degenerate minima, as well as (related) purely bosonic systems  with 
harmonic non-degenerate minima.
We find exact finite action non-BPS bounce and bion solutions to the
holomorphic Newton equations. We find not only real solutions, but 
also complex solution with non-trivial monodromy, and finally complex
multi-valued and singular solutions. 
Complex bions are necessary for obtaining the correct non-perturbative 
structure of these models. 
In the supersymmetric limit the complex solutions govern the ground 
state properties, and their contribution to the semiclassical expansion 
is necessary to obtain consistency with the supersymmetry algebra. 
The multi-valuedness of the action is either related to the hidden 
topological angle or to the resurgent cancellation of ambiguities. 
We  also show that  in the approximate  multi-instanton  description
the integration over the complex quasi-zero mode thimble produces 
the most salient features of the exact solutions. While exact complex 
saddles are more difficult to construct in quantum field theory, the 
relation to the approximate thimble construction suggests that such 
solutions may be underlying some remarkable features of approximate 
bion saddles  in quantum field theories. }

\begin{document}
\maketitle

\newpage
\section{Introduction and  physical motivation from QFT}
\label{sec:intro}

 In this paper we pursue two goals, one related to the proper formalism  
of path integration in semi-classical analysis, and the other related
to novel phenomena in quantum mechanics and quantum field theory 
associated with new semi-classical contributions.
Our first aim is   
to argue that  the correct framework for studying the semi-classical 
representation of a Euclidean path integral necessarily involves 
complexifying  the configuration space,   measure and action. 
This obviously  implies complexification of classical mechanics arising from the Euclidean action.
Typically,  once this is done, there are  complex saddle configurations,  
solutions to the complexified equations of motion.  
Our second goal is to  elucidate 
the  physical role of complex saddle  configurations in the path integral 
formulation of quantum mechanics and quantum field theory.   The examples  
reveal surprising new phenomena, with implications that force us  to 
re-consider our  intuition about  the semi-classical approach to path 
integrals.

The motivation for the present study comes from an intuition that developed in  
supersymmetric and non-supersymmetric quantum field theories, in particular
semi-classical studies of QCD with $N_f$ adjoint fermion on $\R^3\times S^1$  
\cite{Unsal:2007jx,Poppitz:2011wy,Anber:2011de,Misumi:2014raa}. 
The case $N_f=1$ is ${\cal N}=1$  supersymmetric gauge field theory. 
In this context, monopole-instantons are solutions to the BPS (self-duality) 
equations, and magnetic and neutral bions, correlated 2-events, are manifestly 
non-self-dual, and not a solution to  first order BPS equations of motion. 
However, bions have an interesting property that suggests that they may in 
fact be associated with exact saddle points of  the path integral. Bions 
have a calculable characteristic size, $r_b$,   parametrically larger 
than the monopole-instanton radius $r_m$ ($r_m$ is dictated by the scale of 
the Higgs phenomenon of the Wilson line). The scale $r_b \sim r_m/(g^2N_f)$ 
determines the critical point of the quasi-zero mode integration.  Neither 
$g^2$ nor $N_f$ enters  the second order  Yang-Mills equations of motion, 
and hence,  bions are not solutions of the second order Euclidean equation.

A critical point on the quasi-zero mode integration contour at a {\it finite} 
separation (either real or complex) is an unusual property. In the standard 
textbook treatment of instantons in bosonic systems, it is shown that for 
the quasi-zero mode associated with the instanton anti-instanton separation   
the  critical point  is at {\it infinite} separation, reflecting the fact 
that the superposition of an instanton with an anti-instanton  is  not a 
solution at finite separation. This is related to the non-linearity  of 
the underlying instanton equations, see the discussion in textbook
\cite{Coleman:1978ae,ZinnJustin:2002ru,Rubakov:2002fi,Wen:2004ym, Altland:2006si,Weinberg:2012pjx} 
or reviews \cite{Schafer:1996wv,Dorey:2002ik,Vandoren:2008xg} on the subject. 
This raises the question of what the meaning of a critical point at finite  
(real or complex) quasi-zero mode separation is. In particular, is it 
possible that these configurations are exact saddle points in some 
suitable formalism, and not just approximate descriptions?   
What is the distinction between a critical point at real separation 
and at complex ``separation"?  Since the quasi-zero mode direction is 
a particular direction in  field space, real separation is a real field 
configuration and complex direction is a complexified  field configuration. 
What then is the natural habitat for these saddle points?  What is the 
role of complex fields in the semi-classical treatment of real path 
integrals?  

 At this point in time we cannot provide definitive answers to all these 
questions in quantum field theory.  However, attacking the quantum field 
theory problem in the more tractable context of  quantum mechanics, we 
study quantum  systems  with  one  bosonic and $N_f$ fermionic (Grassmannian)  
fields, corresponding to a particle with internal spin $(\half)^{N_f}$. 
$N_f=1$ corresponds to the ${\cal N}=1$  supersymmetric quantum theory. In 
each case we demonstrate  that both real and complex non-self dual saddle 
configurations  exist, and that they solve complexified equations of motion 
which can be obtained by integrating out fermions exactly.\footnote{Although 
we are mostly motivated by theories with fermions, it is important to keep 
in mind that these systems can be represented in terms of purely bosonic 
systems after quantizing fermions, and projecting on fermion number 
eigenstates, and that \emph{all} the conclusions of this work are equally 
valid for  purely bosonic theories with generic  (non-symmetric) bosonic 
potentials.}
We show that  they  must be included in a consistent semi-classical 
expansion of the quantum path integrals. In certain limits the exact 
solution has an approximate description in terms of instanton-anti-instanton 
correlated 2-events with real and complex quasi-zero mode separation. This 
Picard-Lefschetz thimble interpretation generalized more directly to 
quantum field theory, and suggests that even in quantum field theories 
exact complex saddle points may be underlying the known bion analysis 
\cite{Unsal:2007jx,Poppitz:2011wy,Anber:2011de,Argyres:2012ka,Argyres:2012vv,Dunne:2012zk,Dunne:2012ae,Cherman:2013yfa,Misumi:2014raa}.

\subsection{Complex Saddle Points in Complexified Euclidean Path Integrals}
\label{sec:complex}

The basic question is the following: Given an ordinary  path integral 
over {\it real fields} in a general Euclidean version of QFT or quantum 
mechanics, 
\begin{align}
Z_{\rm bos} =   \int Dx(t) \;   e^{ -	\frac{1}{\hbar} S[x(t)]   }  
 = \int Dx(t) \;   
    e^{ - \frac{1}{\hbar}  \int dt  \left( \half \dot x^2 + V(x) \right)}\, , 
\label{R-1}
\end{align}
how does one perform the semi-classical expansion? The same question 
also applies to theories with both bosonic and fermionic degrees of freedom.
 
This question is usually answered by studying real saddle points, single-valued
and smooth configurations, such as  instantons or multi-instantons 
\cite{Coleman:1978ae,Wen:2004ym,ZinnJustin:2002ru,Schafer:1996wv,Rubakov:2002fi,Weinberg:2012pjx,Dorey:2002ik,Altland:2006si,Vandoren:2008xg}. 
However, we show in this work that important physics is missed in the 
standard approach which involves only real-valued ``instanton'' solutions.
These solutions satisfy Newton's equation in the inverted potential $-V(x)$: 
\begin{align}
 \frac{ \delta {S} }{ \delta x}=0 \Longrightarrow \frac{d^2 x}{dt^2}  
 = + \frac{\partial V}{\partial { x}} \, . 
\label{R-2}
\end{align}	
Instantons and anti-instantons are solutions of the second order 
equations of motion by virtue of the real first-order equations:
\begin{align}
\frac{d x}{dt}  = \pm  \sqrt{2 V(x)} \, . 
\label{INS-1}
\end{align}
The two  main new observations made in this work are: 
\begin{itemize} 
\item[\bf 1)] In order to perform  the semiclassical expansion in a 
Euclidean  quantum mechanical path integral, the action and measure  
must be complexified. The partition function is 
\begin{align}
Z_{\rm bos}&=  \int_\Gamma  Dz \;   e^{ -	\frac{1}{\hbar} {\cal S}[z(t)]   } 
=    \int_\Gamma Dz (t) \; e^{ -\frac{1}{\hbar}  \int dt \left( \half \dot z^2  
               +   V(z)  \right)} \, , 
\label{C-1}
\end{align}
where $\Gamma$ is an integration cycle to be determined. Despite the 
fact that the original path integral is a sum over real configurations,  
the semi-classical expansion may receive physically important contributions 
from complex configurations.  
The critical points of the complexified path integral are found  
by solving the {\it holomorphic Newton equations} in the 
inverted potential $-V(z)$:
\begin{align}
 \frac{ \delta {\cal S} }{ \delta z}=0 \Longrightarrow   \frac{d^2 z}{dt^2}  
 = + \frac{\partial V}{\partial { z}}  \, . 
 \label{C-2}
\end{align}
Clearly this provides a larger basis of classical configurations for the 
semi-classical expansion, and entails a physical interpretation in terms 
of Picard-Lefschetz theory  \cite{Fedoryuk,Kaminski,Pham,Berry657,Howls,delabaere2002,Witten:2010cx}.
We demonstrate that for generic potentials, physical properties of the ground
state are determined by complex rather than real saddle points.   We stress 
that this occurs in systems for which the physical couplings are manifestly 
real, so that the theories have a precise Hilbert space interpretation, which 
can be directly matched to the semi-classical analysis.

\item[\bf 2)] The generic saddle point configurations which contribute  
to the ground state properties of a general quantum mechanical system 
are {\it complex} solutions of the holomorphic Newton equations in 
the inverted potential, and they may even be {\it multi-valued} and 
{\it singular}.  Despite this, the action of these saddles is finite, 
but typically with multi-valued  imaginary part.  The multi-valuedness 
and complexity of the action (recall that the path integral for  real 
coupling must be real),  may appear to be a serious problem, but 
it is indeed necessary. This phenomenon is related to either {\it i)}  
the resurgent cancellations with the (lateral) Borel resummation of 
perturbation theory, or {\it ii)} the hidden topological angle phenomenon 
\cite{Behtash:2015kna, Behtash:2015kva}.
The former  provides  a rigorous version   of the Bogomolny/Zinn-Justin (BZJ) 
prescription  for controlling the ambiguities inherent in the analysis of 
instanton/anti-instanton correlated two-event amplitudes  in the instanton 
gas picture in bosonic theories \cite{Bogomolny:1980ur,ZinnJustin:1981dx}.
\end{itemize}

In finite dimensional systems the analysis of exponential type integrals 
via the complex generalization of Morse theory, Picard-Lefschetz theory, 
and via  resurgent trans-series are parallel constructions. Phenomena like 
Borel resummation, and the associated ambiguities from asymptotic analysis, 
find a geometric  realization, respectively, as integration over Lefschetz 
thimbles and Stokes jumps of the thimbles.   A discussion of multi-dimensional 
integrals, which defines a steepest surface using a complex gradient flow 
system, can be found in \cite{Fedoryuk,Kaminski}.  
In \cite{Pham,Howls,Berry657,delabaere2002}, this  construction  is  
implicit, and the saddle-point method is already based on the assumption of 
the appropriate Lefschetz thimble decomposition. The  {\it all-orders} 
steepest descents analysis, which includes all contributions beyond the 
usual Gaussian approximation, reveals a rich and intricate structure 
of relations between the contributions from different saddles 
\cite{Berry657,Howls,delabaere2002}. 
An application to infinite dimensional path integrals appeared in 
\cite{Witten:2010cx} in the context of Chern-Simons theory. In this
case, due to the elliptic nature of the gradient flow system, the nice 
properties of the finite dimensional system carry over to the infinite
dimensional case.
Nevertheless, a surprising amount of this resurgent structure is also 
inherited by configuration space path integrals \cite{Basar:2013eka}, for 
which the gradient flow system is parabolic. 
A number of recent talks by Kontsevich also emphasize the precise relation  
between resurgence and Lefschetz thimbles in finite dimensions, and discuss 
the extent to which these generalize to infinite dimensional problems 
\cite{Kontsevich-1,Kontsevich-2, Kontsevich-3}.  The point of view  he 
presents is close to our perspective,  viewing the perturbative expansion 
around perturbative saddle points  as a constructive approach to defining 
the path integral, even though the geometrization of the infinite dimensional 
path integral is more complicated and richer than the finite dimensional 
case.  Kontsevich further claims  that  this framework could provide a 
rigorous mathematical  replacement for path integrals, which is also an 
optimism  that we share.  We are also motivated by the well-known 
importance of complexified dynamics in the theory of algebraically 
integrable quantum systems \cite{adler,Sklyanin:1991ss,Nekrasov:2009rc,Nekrasov:2009ui,Kozlowski:2010tv}.

From a complementary  perspective, the uniform WKB approach  provides  a 
constructive and explicit realization of the relations between 
saddle points \cite{Dunne:2013ada,Dunne:2014bca,Misumi:2015dua}, similar 
to the finite dimensional relations. One finds that fluctuations about 
different non-perturbative multi-instanton sectors are related in a precise 
quantitative way, as has been confirmed recently by explicit quantum field 
theoretic multi-loop computations \cite{Escobar-Ruiz:2015nsa,Escobar-Ruiz:2015rfa}. 
  
The perspective emanating from resurgence and finite dimensional examples 
of Lefschetz thimbles indicates that whenever we consider the semi-classical 
analysis of the path integral (either QFT or QM), we are required to  start  
with a complexified/holomorphic  version of it.  In this paper,  we explore 
consequences of this apparently innocuous step, which, in turn, leads to 
many surprises.

\subsection{Motivation from QFT translated to quantum mechanics} 

The physical  basis for the QFT intuition from studies of magnetic and 
neutral bions  in QCD(adj)  
\cite{Unsal:2007jx,Anber:2011de,Misumi:2014raa, Poppitz:2011wy,Argyres:2012ka,Argyres:2012vv}
and  2d non-linear sigma models with fermions  \cite{Dunne:2012zk,Dunne:2012ae,Cherman:2013yfa,Cherman:2014ofa,Misumi:2014jua,Misumi:2014bsa,Nitta:2014vpa}, 
translated to the quantum mechanical context  with $N_f$ Grassmann valued 
(fermion) fields, amounts to  the following.  First, consider an 
instanton/instanton or an  instanton/anti-instanton  
pair in the bosonic theory ($N_f=0$).  Since the BPS equations are non-linear, 
it is clear that  a superposition of two individual solutions is not a solution
at any finite separation $\tau$ between them. In fact, the action of the 
configuration changes with separation as 
\begin{align}
S_\pm(\tau) \equiv  2S_I + {\cal V}_{\pm} (\tau) 
 = 2S_I  \pm  \tfrac{A a^3}{ g} e^{- m_b \tau}  
          \qquad {\rm bosonic \;  models,}
\label{bosonic}
\end{align}
where the $+$ sign is for an $[\I\I]$ configuration, and the $-$ sign 
for an $[\I \bar \I]$ configuration. Thus, the critical point of the 
potential between the two-events is at  {\it infinite separation}, 
$\tau^*=\infty$, which we refer to as a ``critical point at infinity''.

This is the point where one first realizes that something may be special 
about bions in theories with fermions.  For bions in such theories the 
potential between the constituent instantons has a critical point at a 
{\it finite separation}.\footnote{This is a universal phenomenon.  
\eqref{fermions} coincides with the quasi-zero-mode (QZM) integrations 
in QFTs, such as non-linear sigma models in 2d and QCD(adj) 
\cite{Unsal:2007jx,Argyres:2012ka,Dunne:2012zk,Dunne:2012ae,Cherman:2013yfa,Cherman:2014ofa}, 
and $\N=1$ SYM in 4d by a simple change of variables.}  
For example,  the potential between two instantons in QM is of the form:
\begin{align}
\label{fermions}
S_\pm(\tau) \equiv  2S_I + {\cal V}_{\pm} (\tau) 
= 2S_I  \pm  \tfrac{Aa^3}{ g} e^{- m_b \tau}  +  N_f\,  m_b\, \tau  
   \qquad {\rm theories \;  with \;  fermions,} 
\end{align}
and there exists a critical point at finite-separation $\tau^*< \infty$:
\begin{subequations}
 \begin{align}
&\tfrac{d {\cal S}_{+}(\tau) }{ d \tau}  = 0 
   \qquad \Rightarrow\quad   m_b\,\tau^* =  
           \log \left( \tfrac {Aa^3}{g\, N_f} \right) 
   \qquad  \qquad {\rm for} \;\;  [\I\I]\, , \label{1.9a} \\
& \tfrac{d {\cal S}_{-}(\tau) }{d \tau}  = 0 
   \qquad  \Rightarrow\quad    m_b\,\tau^* =  
           \log \left( \tfrac {A a^3}{g\, N_f} \right) \pm i \pi  
\qquad {\rm for} \;\;  [\I \bar \I]\, .  \label{1.9b}
\end{align}
\end{subequations}
The corresponding critical amplitude is
\begin{subequations}
\begin{align}
&\exp\left[-\frac{1}{g} S_+\right]
      =\left(\frac{g\, N_f}{Aa^3}\right)^{N_f}\, e^{-2S_I}
       \quad {\rm for} \;\;  [\I\I]\, ,  \label{1.10a}\\
&\exp\left[-\frac{1}{g} S_-\right]
      =e^{\pm i\pi N_f} \left(\frac{g\,  N_f}{Aa^3}\right)^{N_f}\, e^{-2S_I}
       \quad {\rm for} \;\;  [\I\bar \I]\, .  \label{1.10b}
\end{align}
\end{subequations}
The first case suggests that the  $[\I\I]$-two-event may actually be 
an approximate form of an exact real  solution, while the latter case 
suggests that the $[\I\bar \I]$-two-event    may be an approximate form of 
 a complex (possibly multi-valued)   exact solution.  
The factor $e^{\pm i\pi N_f}$, where $N_f$ can also be continued to non-integer 
values, associated with the $ [\I\bar \I]$ plays a crucial role in the 
behavior of supersymmetry in these fermion/boson systems. We show 
explicitly in Sections \ref{sec:DoubleWell} and \ref{sec:periodic} that 
there are  exact non-BPS saddle point solutions with size $\Re({\tau^*})$ in 
these two cases.  See Fig.\ref{tau-plane}.

 Based on these results we  demonstrate in quantum mechanics a precise 
relation between the nature of the integration over the  complex quasi-zero 
mode (QZM) thimble   and  the existence of exact solutions of the full path 
integral: 
\begin{itemize} 
{\item [{\bf i)}] $\tau^*=\infty \Longleftrightarrow$ 
                   Approximate quasi-solution. } 
{\item [{\bf ii)}] $\tau^* \in \mathbb \R^{+}, \;  \tau^* <\infty  
         \Longleftrightarrow$  Exact real solution.}
{\item [{\bf iii)}] $\tau^* \in \mathbb{C}\backslash { \mathbb \R^{+}}, \;   
   \Re\tau^* < \infty   \Longleftrightarrow$  Exact complex solution.}  
\end{itemize}
We show that the integral over the QZM-thimble reproduces the most 
important features of the exact non-BPS solutions, such as the critical 
size, the ambiguous imaginary part of their action, and the hidden 
topological angle(s) associated with the exact solutions. Based on these, 
we conjecture that this deduction is also valid in general QFTs.

\subsection{Necessity of complexification: Simple pictures}
\label{sec:necessity}

\begin{figure}[t]
\begin{center}
\includegraphics[angle=0, width=0.50\textwidth]{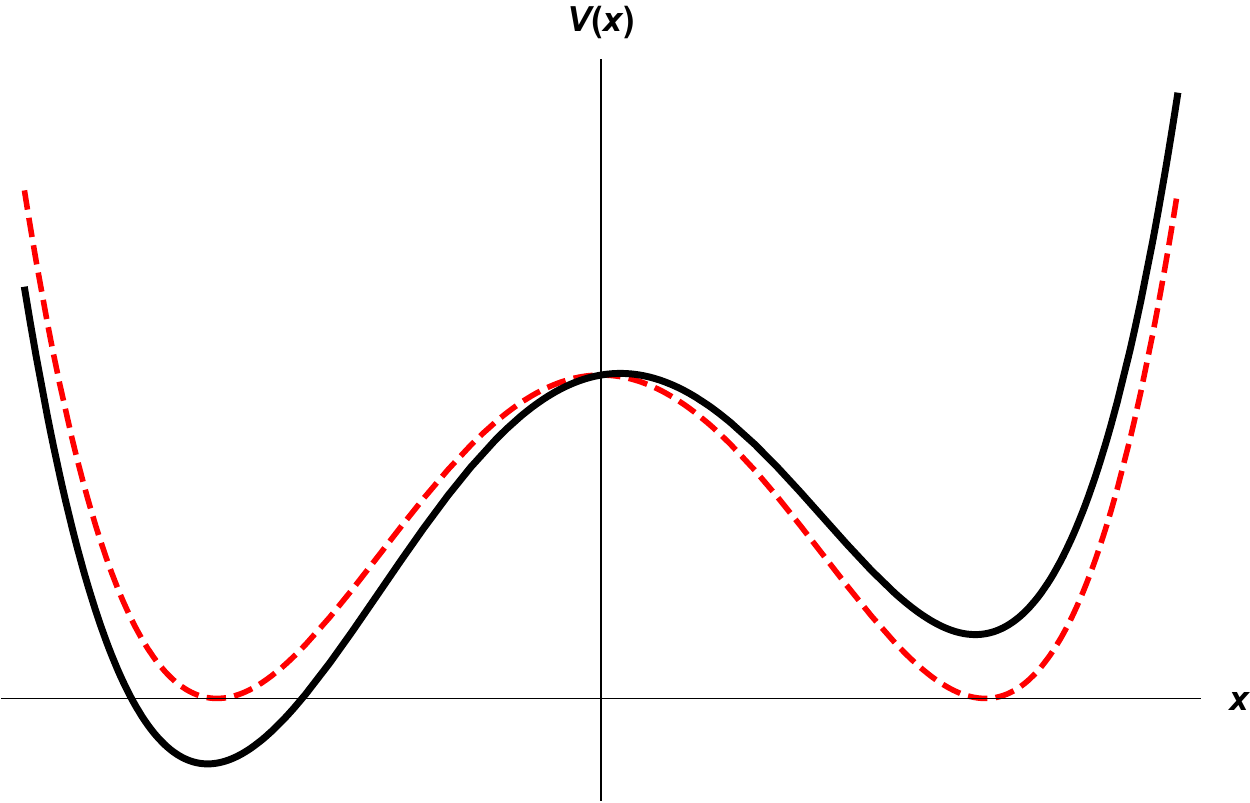} 
\caption{Bosonic potential $V_{\rm bos}(x)$ (red dashed) vs. quantum modified 
potential (black solid) $V(x)$ for double-well system. The quantum modified 
potential (\ref{eq:dw-mod}) is obtained upon quantizing the fermions and 
projecting  to a fermion number (or spin) eigenstate.  }
\label{pot-quantize}
\end{center}
\end{figure}

\begin{figure}[t]
\begin{center}
\includegraphics[angle=0, width=0.60\textwidth]{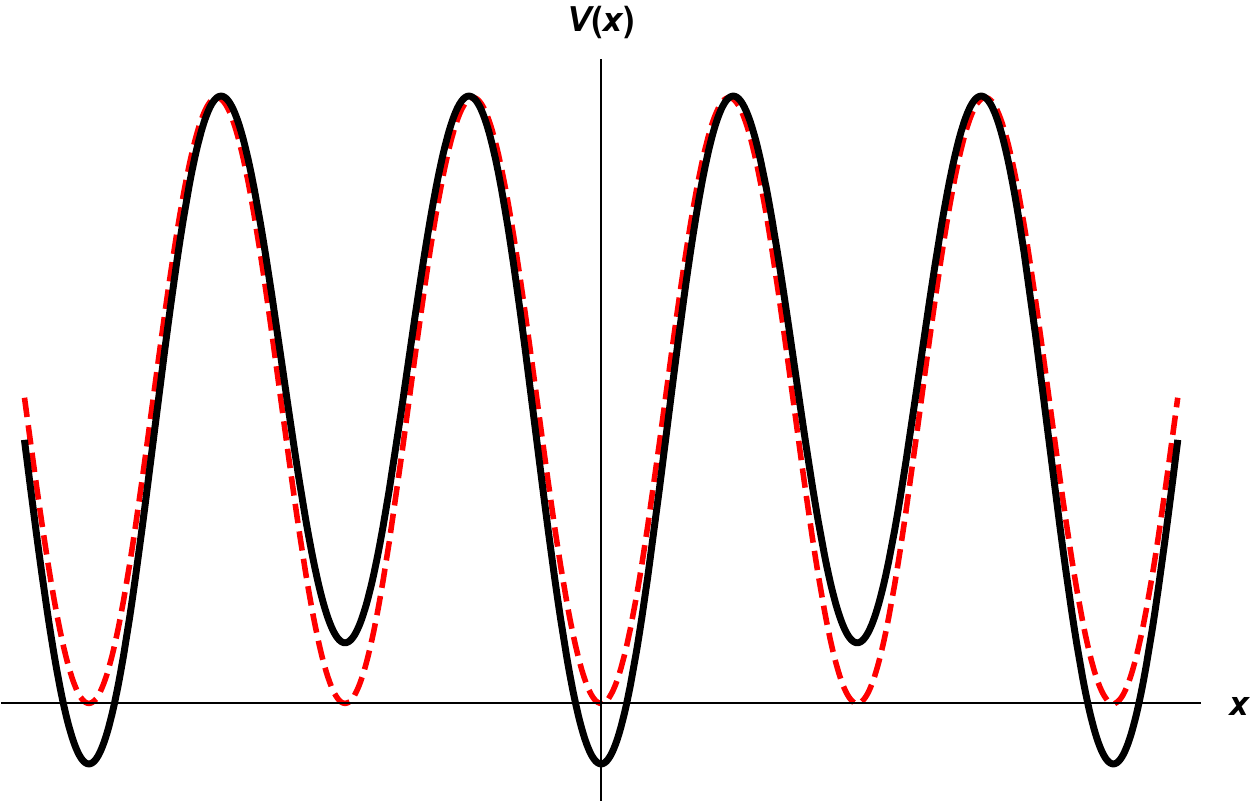}
\caption{Bosonic potential $V_{\rm bos}(x)$ (red dashed) vs. quantum modified 
potential (black solid) $V(x)$ for the periodic potential. The quantum 
modified potential (\ref{eq:sg-mod}) is obtained upon quantizing the 
fermions and projecting  to a fermion number (or spin) eigenstate. }
\label{pot-quantize-2}
\end{center}
\end{figure}

To demonstrate the necessity of considering complex solutions, we first 
provide some simple illustrations. We will show that the following quantum 
mechanical problems are closely related:
\begin{itemize}
\item{Systems with degenerate  harmonic  minima coupled to $N_f$ Grassmann 
valued fields.}
\item{Bosonic systems with non-degenerate  harmonic minima.}
\end{itemize}
In order to be specific, we consider two different bosonic potentials, a 
double-well potential and a periodic potential, see Fig~\ref{pot-quantize} 
and \ref{pot-quantize-2}. We couple Grassmann fields to these potentials
and project on fermion number eigenstates. This leads to effective bosonic
potentials in the sector with fermion number $p$
\begin{align} 
V(x)=   \underbrace{ \half (W'(x))^2 }_{V_{\rm bos}(x)}  
         + \tfrac{pg}{2}  W''(x), \quad p=2k-N_f, \label{Gen-NP-3}  
\end{align}
which we refer to as the {\it quantum modified potential}.  The fermion
number $p$ is given by $p=2k-N_f$ with $k=0, \ldots, N_f$.  In a SUSY 
theory, with $N_f=1$, the function $W(x)$ is the superpotential 
\cite{Witten:1981nf,Witten:1982df}. For $N_f =2,3 \ldots $,  we refer 
to $W(x)$ as an {\it auxiliary potential}.

 In this sense, the first class of systems (with fermions) that we study  
can be expressed as a collection of systems of the second type. 
``Integrating out" fermions lifts the degeneracy of the harmonic minima. 
It is also useful to consider $p$ as an arbitrary parameter, 
which can even be taken to be complex.  
 
The bosonic double-well (DW) potential, with $W(x)=\frac{1}{3}x^3-a^2x$, 
leads to a quantum modified tilted-double-well  potential 
\begin{equation}
V_\pm =\frac{1}{2}(x^2-a^2)^2\pm pg\, x \, . 
\label{eq:dw-mod}
\end{equation}
Similarly, the  bosonic periodic Sine-Gordon (SG) potential, with $W(x)=
4a^3\cos\left(\frac{x}{2a}\right)$, leads to a quantum modified 
double-Sine-Gordon potential
\begin{equation}
V_\pm =
2a^4 \sin^2\left(\frac{x}{2a}\right)
      \mp \frac{p g a }{2} \cos\left(\frac{x}{2a}\right)\, . 
      \label{eq:sg-mod}
\end{equation}
It is well known that there are instanton solutions for the {\it untilted} 
bosonic potentials, with $p=0$. These are real solutions to Newton's equation 
in the inverted potential, $-V(x)$, given by solutions to the first-order 
equation \eqref{INS-1}. By connecting degenerate harmonic vacua,  this  
results in the familiar non-perturbative  level splitting 
\cite{Coleman:1978ae,ZinnJustin:2002ru,Rubakov:2002fi,Wen:2004ym, Altland:2006si,Weinberg:2012pjx}. 
Instantons are associated with the separatrix (see \cite{arnold1989}) in 
the classical phase space of the inverted $p=0$ potential. 

\begin{figure}[t]
\begin{center}
\includegraphics[angle=0, width=0.6\textwidth]{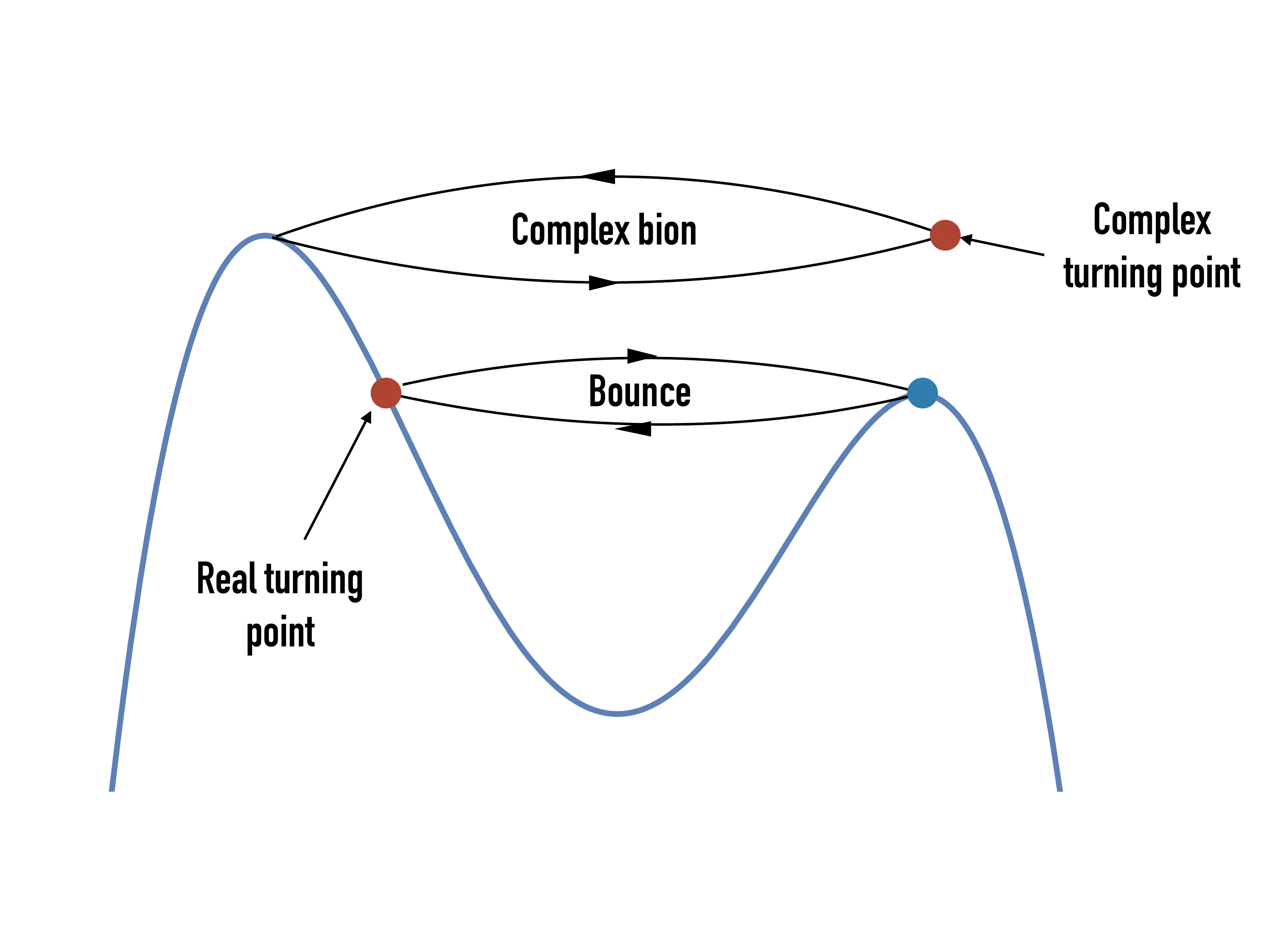}
\caption{Inverted quantum modified potential in a theory with a symmetric 
bosonic double-well potential. There are no real finite action saddle 
configurations contributing to ground state properties, but there are exact 
complex saddle solutions. The Euclidean description of the vacuum is a 
dilute gas of these complex saddle points, which we call complex  bions. 
The complex bion is necessary in order to explain the non-perturbatively 
induced  vacuum energy of the corresponding $N_f=1$ SUSY QM model. }
\label{Sep-pot}
\end{center}
\end{figure}
\begin{figure}[t]                
\centering
\includegraphics[width=10cm]{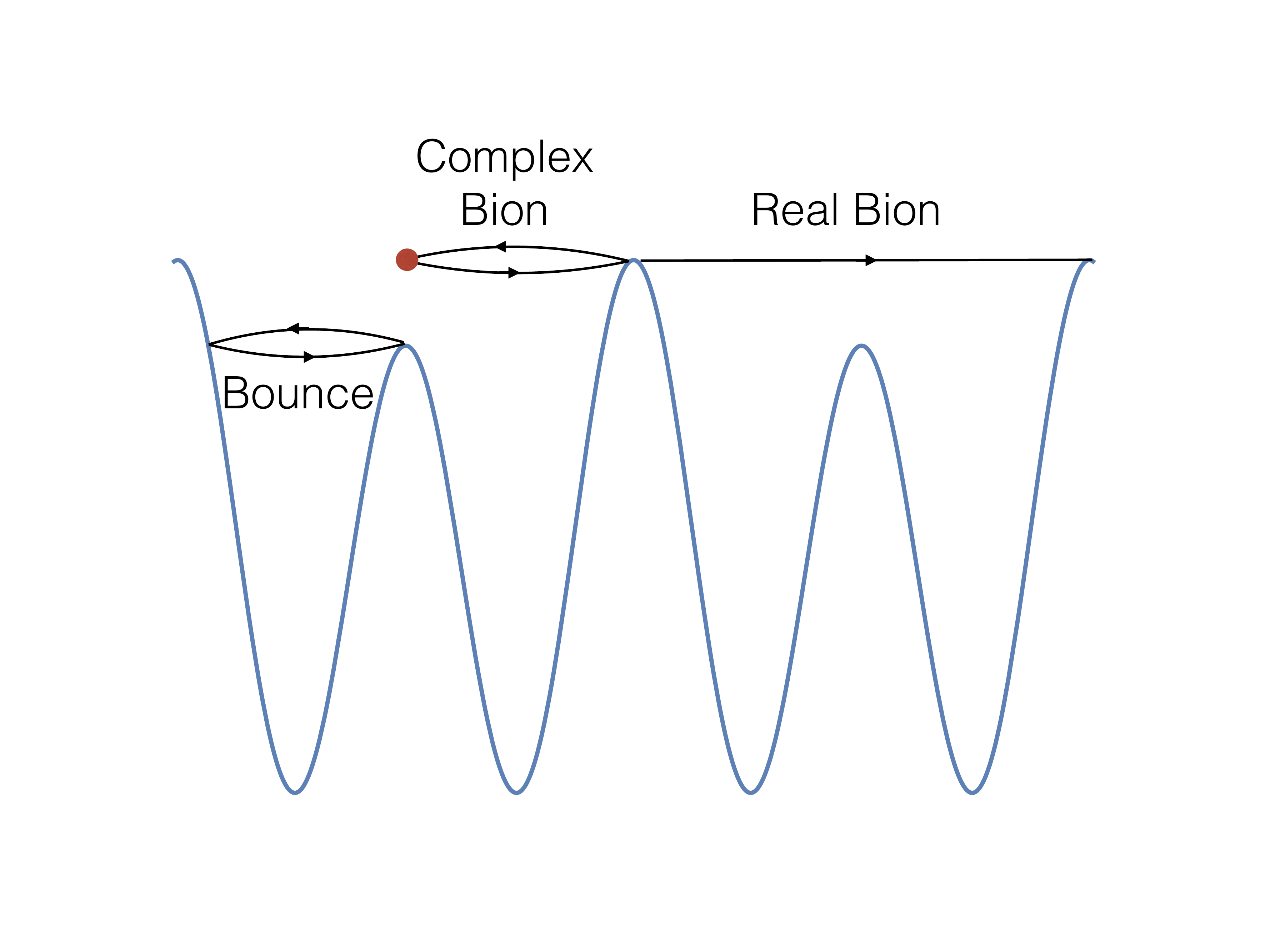}
\vspace{-1cm}
\caption{Inverted quantum modified potential in a theory with a periodic  
bosonic  potential.   There  are exact real and complex saddle solutions in the 
quantum modified Sine-Gordon model.  Both 
the real and complex bions are necessary in order to explain the 
non-perturbative vanishing of the vacuum energy of the corresponding 
$N_f=1$ SUSY QM model. The Euclidean description of the vacuum is a dilute 
gas of real and complex bions.   		 }
\label{three_bions}
\end{figure}

For $p\ne 0$, the standard lore of semi-classics instructs us to search 
for finite action classical solutions of the equations of motion for the 
Euclidean action, which correspond to classical motions of a particle in 
the inverted potential. But this leads to two immediate problems:
\begin{itemize}
\item Puzzle 1: For $N_f=1$ SUSY QM with a bosonic tilted double-well 
potential, it is well-known that the perturbatively vanishing ground 
state energy is  lifted non-perturbatively,  and supersymmetry is 
spontaneously broken \cite{Witten:1981nf,Witten:1982df}.  However, 
there is no real classical finite action solution contributing to the 
ground state energy.

\item Puzzle 2: For $N_f=1$ SUSY QM with a bosonic Sine-Gordon potential, 
it is known that the perturbatively vanishing ground state energy remains 
zero  non-perturbatively \cite{Witten:1981nf}. However, there is only 
one real finite action solution  which tunnels from one global maximum 
of the inverted potential to the next. We call this solution the \emph{real 
bion} solution because its profile looks like two consecutive instantons. 
This solution makes a negative contribution to the ground state energy. As 
it stands, semi-classics would be in conflict with  the SUSY algebra,  
and positive-definiteness of the SUSY Hamiltonian. 
\end{itemize}

We will show that the semi-classical configuration responsible for SUSY 
breaking in the former case is a complex bion solution which tunnels 
from the global maximum of the inverted potential to one of two complex 
turning points, as shown in Figure \ref{Sep-pot}. Moreover, the contribution 
to ground state energy is  $-e^{-S_{\rm cb}}=-e^{-2S_I\pm i\pi}$,  admitting an 
interpretation in the two-instanton sector of the formulation with fermions, 
and the $\pm i\pi$  imaginary part of the complex bion action, referred to 
as the hidden topological angle (HTA) in  \cite{Behtash:2015kna}, is crucial 
for the positive definiteness of the non-perturbative vacuum energy. In the 
absence of $\pm i \pi$, one would end up with a negative ground state energy,  
and semi-classics would be in conflict  with the constraints of  the 
supersymmetry algebra \cite{Witten:1982df}.

In the Sine-Gordon  case we will show that in addition to the real bion 
solution there is a second solution, the complex bion, which tunnels from 
one global maximum of the inverted potential to one of two complex turning 
points, as shown in Figure \ref{three_bions}. Moreover, the real part
of the action is exactly equal to the real bion action, and roughly equal 
to the twice the instanton action $S_I$.
The net contribution to the ground state energy is $-e^{-S_{\rm rb}}
-e^{-S_{\rm cb}}=-e^{-2S_I}-e^{-2S_I\pm i\pi}=0$. Notice that the 
imaginary part of the complex bion action or  HTA,   $\pm i\pi$,   is 
necessary for the non-perturbative 
vanishing of the vacuum energy \cite{Behtash:2015kna}.  One dramatic aspect 
in this case is that the complex bion solution is both multi-valued, as well 
as singular, but has finite action with smooth real part. Similar multi-valued 
solutions also appeared in Ref.\cite{Harlow:2011ny}, which we discuss in 
Section \ref{sec:HMW}.) Since the hidden topological angle is defined modulo 
$2\pi$,   the naively multi-valued action is in fact single valued, and no 
pathologies arise for the supersymmetric theory. For non-supersymmetric 
theories for which the complex bion term comes out as $e^{\pm i p \pi}$, the 
semiclassical analysis  involves both the HTA and resurgent cancellations.

 There is a particularly simple way to obtain the complex bion solutions 
that we will employ in this work, and one can check that the resulting 
solutions satisfy the holomorphic Newton equations.  In Figs.~\ref{Sep-pot} 
and \ref{three_bions}, there is an obvious exact bounce solution. Starting 
with the bounce solution,  consider an analytic continuation in $p$ 
\begin{align}
{p \in \mathbb R} \rightarrow {p e^{i \theta}  \in \mathbb C}  \, . 
\label{analytic-cont-p}
\end{align}   
As $\theta $ changes, the path integral moves into an  {\it unphysical 
region of parameter space}, similar to the discussion of \cite{Harlow:2011ny}, 
where the theory ceases to have a physical Hilbert space interpretation. But 
the end-points of this unphysical region, $\theta=0$ and $\theta=\pi$, 
correspond to  $V_{+}(\x)$ and $V_{-}(\x)$, which are physical theories. 
However, in the cases studied in this work, the two  potentials are  
related either by parity  (mirror images   of one another) or a simple 
shift,  and the path integral representation and  the set of 
non-perturbative saddle points associated with them are identical.

 Turning on $ \theta$ gives the analytic continuation of the bounce into 
a complex saddle.   These complex saddles are plotted in Fig.~\ref{dw-1} and  
Fig.~\ref{ac-1}. The continuation to $\theta=\pi$ results in complex smooth 
saddles for the tilted DW case and complex singular saddle for the double-SG 
example. 
One interesting aspect of the analytic continuation for the DW system  is 
that one can show that the monodromies associated with the solutions are 
non-trivial. In fact, as $p$ changes its  phase by $2\pi$,  the potential 
$V_{+}(\x)$ turns back to itself, but the two complex bion solutions are 
interchanged.  Thus, the solutions has a monodromy of order 2, reflecting 
the two-fold ambiguity in the choice of the exact solutions. 
 
\subsection{What is  surprising (and what is not)?} 

 The necessity of  complexification is not surprising from the point of 
view of the steepest descent method for ordinary integration.   Since the 
path integral is a particular form of infinitely many ordinary  integrals,  
complexification is in fact a {\it natural} step. What is interesting and 
surprising  is the   important new effects  that appear in {\it functional} 
integrals. 

 As is well known, complexification is both a necessary and  sufficient step 
to capture a complete steepest descent cycle decomposition for ordinary 
integration. Let $f(x)$ be a real function, and   consider an exponential type 
integral  $I(\hbar)= \int_{-\infty}^{\infty} dx \;  e^{-\frac{1}{\hbar} f(x)} $ which 
exists for $\hbar >0$. (We will also consider the continuation $\hbar 
\rightarrow \hbar e^{i \theta}$.) To tackle the integration via the steepest 
descent method, the first step is to complexify: 
\begin{align}
(f(x), \R) \longrightarrow (f(z), \Gamma \in {\mathbb C})\, . 
\end{align}
Since ${\mathbb C}$ has twice the real dimension of $ \R$ the integration 
is restricted to  a certain middle-dimensional cycle $\Gamma$ in ${\mathbb C}$. 
The standard procedure is:
\begin{align}
I(\hbar)= \int_{-\infty}^{\infty} dx \;  e^{-\frac{1}{\hbar} f(x)}  
   \underbrace{\longrightarrow}_{\rm  steepest \; descent \;  method} 
\sum_\sigma n_\sigma  \int_{\cal J_\sigma} dz \; e^{- \frac{1}{\hbar} f(z)} \, , 
\label{eq:cycle}
\end{align}
where ${\cal J}_\sigma$ is the steepest descent cycle attached to the critical 
point $z_{\sigma}$ of $f(z)$, i.e., $f'(z_\sigma)=0$  and the interval 
$\int_{[-\infty, \infty]}=  \sum_\sigma n_\sigma \int_{\cal J_\sigma} = \int_\Gamma $ 
is a sum over the homology cycle decomposition of the pair $(f(z), \mathbb C)$ 
despite the fact that the original integration is over $\R$.  ${\cal J_\sigma} 
(\theta)$ cycle is found by solving a complex version of the gradient flow 
equation (also called Picard-Lefschetz equation)
\begin{align}
 \frac{dz}{du} = e^{i \theta} \frac{\partial \bar f}{\partial \bar z}\, , 
 \label{cgf-inst}
 \end{align} 
where $u$ is gradient flow time. For $\hbar >0$, we take $\theta=0$. 
Note that the vanishing of the right hand side determines the critical 
point set $z_\sigma$ for the one dimensional integral 
given in \eqref{eq:cycle}.

If $\hbar$ is analytically continued to $\hbar e^{i \theta}$, complex saddles 
$z_\sigma$ contribute once their multipliers  $n_\sigma\neq 0$.  In  generic 
situations this is the case. However, consider the case where $\hbar >0$, 
positive and real. Then, the cycle is $\R$ and the decomposition is 
saturated by the real saddles on $\R$. Namely, despite the fact that 
complex saddles are present, their multiplier  $n_\sigma$ is zero.   
Contributing cycles  ${\cal J}_\sigma$ still live in  $\mathbb C$, but 
in the linear combination  $ \sum_\sigma  n_\sigma {\cal J_\sigma}$, the 
segments that move into the complex domain,  always back-track by some 
other contribution   ${\cal J}_{\sigma'}$, so the full  cycle is just $\R$. 
In other words, complex saddles in this case do  not contribute to real 
integration.

 
  Consider now a real Euclidean path integral. Based on the fact that the 
partition function is real, and guided by the arguments given for ordinary 
integrals above, complex saddles (which would typically give complex 
contributions) were usually deemed irrelevant, and did not receive 
sufficient attention. There are two potential ways around the problem
that the partition functions must be real and, interestingly, path 
integrals take advantage of both:
\begin{itemize} 
\item If the action of a complex saddle point is  $S_{\rm r} + i \pi$ 
(mod $2\pi i$), the reality of the partition function will be preserved.  
Indeed, path integrals of many  supersymmetric theories are of this type. 
\item If the action of a complex saddle point is  $S_{\rm r} + i p \pi$, where 
$p\in \R$ arbitrary, this  will (naively) render the partition function 
complex. But this issue  is resolved via  resurgence, and renders the 
trans-series sum over saddles real. An explicit example of this phenomenon  
is discussed in Section \ref{sec:csr}.
\end{itemize}

Both of these mechanisms do {\bf not} take place in one-dimensional real 
exponential integrals, as we demonstrate in Appendix \ref{app:hta}. In this 
sense, the one dimensional  exponential integral cannot be used as a guide 
on this issue. 

In fact, the way that the second item is resolved is  interesting.   The 
action of saddles contributing to the path integral is often multi-valued 
(even for a supersymmetric theory). In particular, the action  is of the 
form  $S_{\rm r} \pm i p \pi$.   In the past, multi-valued saddles were 
viewed as ``disturbing", since one is trying to calculate a physical and 
unambiguous quantity. A recent serious deliberation on this issue (which 
at the end remains undecided) can be found in the recent work 
\cite{Harlow:2011ny},  which we will comment on in detail in 
Section \ref{sec:HMW}.

The main surprise is the following: In order to get a physical real result 
for the path integral,  multi-valued saddle solutions and multi-valued 
actions are required.   The reason is that Borel resummation  of perturbation 
theory is also generically multi-fold ambiguous, with ambiguities related to 
the action of the  multi-valued saddles.   The only possible way that the 
resulting path integral can be real is by virtue of the exact cancellation 
between these two types of ambiguities. 

\subsection{Holomorphic Newton equations}

The holomorphic classical Euclidean equations \eqref{C-2} describe coupled 
motion of the real and imaginary parts of $z(t)=x(t)+i y(t)$. Expressing 
the holomorphic potential  in terms of real and imaginary parts of the 
potential,  $V(z) = V_{\rm r}(x,y) +i V_{\rm i}(x,y)$, the holomorphic 
Newton equation can be re-expressed as:
\begin{eqnarray}\label{eq:eom1}
\frac{d^2 z}{dt^2}=\frac{\partial V}{\partial z}  
      \qquad {\rm or \; equivalently}  \qquad   
 \frac{d^2 x}{dt^2}  = + \frac{\partial { V_{\rm r}} }{\partial x} \, , \cr  \cr
\qquad \qquad\qquad\qquad\qquad 
 \frac{d^2 y}{dt^2}  = - \frac{\partial { V_{\rm r}} }{\partial y}   \, ,
\end{eqnarray}
where  the Cauchy-Riemann equations,  $\partial_x V_{\rm r} = \partial_y 
V_{\rm i}$ and $\partial_y V_{\rm r} = -\partial_x V_{\rm i}$, are used to  
obtain the second form. \eqref{eq:eom1}  implies  that  the force in the 
$x$-direction is  $\nabla_x V_{\rm r}(x,y)$, and 
the force in the $y$-direction is  
$-\nabla_y V_{\rm r}(x,y)$.  In this sense, these are different from the 
classical motion in the inverted real potential $-V_{\rm r}(x, y)$ in 
two space dimensions. This aspect underlies some of the exotic features 
of our exact solutions.

To make this idea more concrete, as it will provide much intuition for 
the exact solutions, consider  the Euclidean Lagrangian $ {\cal L}^E= 
\half \dot x ^2 + \half x^2  $, and  energy conservation $ E= \half \dot x ^2 
- \half x^2  $.  Obviously,  a particle which starts at rest at a typical 
point in the close vicinity of $x=0$    will simply roll down.    
   Making the Lagrangian holomorphic, and  taking advantage of the 
Cauchy-Riemann equations,  we may write 
\begin{align}
  {\cal L}^E = \half \dot z^2 + \half z^2, \qquad \Re {\cal L}^E 
 = \half ( \dot {x}^2 - \dot y^2) + \half    (x^2-y^2 )\, ,
\label{hol-1}
\end{align}
which means  $-V_{\rm r}(x,y) = -x^2 +y^2$.  A particle which starts at rest 
at a typical point in the  the close vicinity of $(x,y)=(0,0)$ will roll 
down in $x$, but  will {\it roll up} in the $y$ direction! This is 
counter-intuitive if we just think in terms of potential  $-x^2 +y^2$, but 
it makes sense because of the relative sign in the  kinetic term 
$ \dot{x}^2 - \dot y^2$.  For example, a particle that starts at $(x, y)=
(0, y(0))$ will evolve in time as $x(t)=0, y(t)= y(0) \cosh(t)$, rolling 
up in  the $y$-direction to infinity.   

Our goal is to study the path integral  corresponding to the holomorphic 
Lagrangian in \eqref{C-1}, and find saddles in the system where we 
analytically continue $p$ as given in \eqref{analytic-cont-p}. 
The reason 
we do so is that it will be fairly easy to find single valued 
solutions for general $\theta$, but we will in fact observe non-trivial 
effects exactly at the physical theories $\theta=0,\pi$. The saddle point
contributing to ground state properties is in general complex, sometimes 
even multi-valued and singular, i.e, either  the $\theta=0$ or $\theta= 
\pi$ direction may correspond to a branch-cut, and we will obtain solutions 
in the cut-plane. The multi-valued solutions are either related to 
resurgent cancellations or to a hidden topological angle.  Turning 
on $\theta$ provides a useful  regularization of the singular solutions. 

 The coupled Euclidean equations of motion  for the tilted-double-well 
system are 
\begin{align}
\frac{d^2 x}{dt^2}  &= +2 x^3-6 xy^2 -2a^2x +pg \cos \theta,  
          \cr\qquad   \qquad       
\frac{d^2 y}{dt^2}  &= -2 y^3 + 6  x^2 y  -2a^2y  + pg \sin \theta,
\label{complex-3}  
\end{align}
and for the double-Sine-Gordon system they are given by 
\begin{align}
&\frac{d^2 x}{dt^2}  =    a^3\cosh (\tfrac{y}{a}) \sin (\tfrac{x}{a})   
 + \frac{pg}{4} \Big(  \cos \theta \cosh (\tfrac{y}{2a}) \sin(\tfrac{x}{2a}) 
  - \sin \theta  \cos(\tfrac{x}{2a})  \sinh(\tfrac{y}{2a})   
   \Big)    \, ,
       \qquad \qquad \qquad \qquad   \cr \cr
&\frac{d^2 y}{dt^2}  = a^3\cos(\tfrac{x}{a}) \sinh(\tfrac{y}{a})   
 + \frac{pg}{4}  \Big( \sin \theta  \cosh( \tfrac{y}{2a})  \sin(\tfrac{x}{2a})  
 +  \cos \theta  \cos (\tfrac{x}{2a})  \sinh(\tfrac{y}{2a})  
 \Big) \, . 
\label{complex-4}   
\end{align}
Note how these differ  from the corresponding real Newton equations even 
at $\theta=0$ or $\pi$, at which \eqref{complex-3}  and \eqref{complex-4}   
still remain  a coupled set of equations, while the equations in the space 
of real paths are given by
\begin{align}
&\frac{d^2 x}{dt^2}  = +2 x^3  -2a^2x +p g \, , \qquad   \qquad    
\label{real-3}  \\
&\frac{d^2 x}{dt^2}  =  a^3\sin\left(\frac{x}{a}\right) 
   + \frac{p g }{4} \sin\left(\frac{x}{2a}\right) \, . 
\label{real-4}  
\end{align}
In terms of real paths, \eqref{real-3}  and \eqref{real-4} do not have 
finite action non-trivial  solutions which start arbitrarily close to the 
global  maximum   of the inverted potential.  
For example, for the tilted-double-well case,  any {\it real} 
solution starting at  the global maximum of the inverted tilted-double-well 
potential will go off to infinity and have infinite action.  
For the periodic 
double-Sine-Gordon potential, there is a real solution connecting neighboring 
global maxima of the inverted potential; 
this is the real bion solution.

On the other hand, \eqref{complex-3} and \eqref{complex-4} also admit  
{\it complex finite action} solutions, with non-vanishing real and 
imaginary parts. We will see that these exact solutions are crucial for 
the semiclassical analysis. The existence of the complex bion solutions 
follows from consideration of the energy conservation equation  for a 
particle starting at the global maximum of the inverted potential of 
Figure \ref{Sep-pot} and Figure \ref{three_bions}:
\begin{equation}
\frac{1}{2}\dot z^2-V(z)=E\, . 
\end{equation}
The turning point equation $V(z)=-E$ has 4 complex solutions. This 
is clear for the tilted double well, which is described by a fourth
order polynomial, and is also true for the 
double-Sine-Gordon, which can be described by a fourth order polynomial
after a suitable change of variables. Two of the turning points are 
degenerate roots at the global maximum of the inverted potential, 
while the other points form a complex conjugate pair near the local 
maximum of the inverted potential (the point corresponding to the false 
vacuum).

There  indeed exist exact complex solutions 
\footnote{The approximate form of the  exact  complex bion solutions 
was noted long ago by Balitsky and Yung \cite{Balitsky:1985in} in the 
case of tilted double well potential, see Section \ref{sec:comments}.} 
for which a particle follows a complex trajectory that reaches one of 
the complex turning points and backtracks to the initial position. We call 
such solutions {\it complex bions}, shown in Figure \ref{DWcomplexbion} 
for the double-well system. Its real part is reminiscent of an 
instanton-anti-instanton $[{\cal I} \bar {\cal I}]$ 
configuration, and in fact $[{\cal I} \bar {\cal I}]$ superposition at a 
particular \emph{complex separation} (see Section \ref{correlated} for 
details) is a systematic  approximation  to the exact complex bion solution. 	                
  
Normally, an $\mathcal{O}(g)$ effect to the bosonic potential would be 
negligible  with no dramatic effects. However, whenever $V_{\rm bos} (z)$ has 
degenerate harmonic  minima, (this is the class of examples we consider), 
the instanton is described as the  separatrix of the inverted-potential.   
The ``small" term coming from integrating out fermions and projecting
onto a definite sector of the Fock space leads to a  perturbative 
splitting of the  separatrix  by an $\mathcal{O}(g)$ amount. This has 
dramatic effects as it causes the instanton to cease being an exact solution.  
A single instanton has infinite action compared to the vacuum, as one of 
its ends is in the true vacuum, while the other is in the false vacuum. 
However what used to be an instanton--anti-instanton configuration, for 
a specific value of a finite  complex separation,   becomes an approximation 
to an exact solution.  This exact solution is the \emph{complex bion}. 

\subsection{Relation to complex gradient flow  and Picard-Lefschetz  theory}

Attached to each critical point there exists a cycle $\cal J_{\sigma}$, 
as in \eqref{eq:cycle}, which can be found by solving a complex gradient 
flow equation, or  the Picard-Lefschetz theory, to describe the 
Lefschetz-thimbles.\footnote{Various applications of the Lefschetz thimbles  
to simple systems, and  QFT at finite density can be found in 
\cite{ Witten:2010cx, Cristoforetti:2012su, Cristoforetti:2013wha, Fujii:2013sra, Aarts:2014nxa, Kanazawa:2014qma,Tanizaki:2014tua,  Cherman:2014ofa}. 
Complex saddles appear naturally in QCD at finite chemical potential, 
e.g, \cite{Nishimura:2014rxa, Nishimura:2014kla}, and in real time Feynman 
path integrals \cite {Tanizaki:2014xba,Cherman:2014sba}.}
In a theory with a field $z(t)$ and action functional ${\cal S}(z)$, this 
amounts to 
\begin{align}
&\frac{\partial z  (t, u)}{\partial u} 
  =  +  \frac{ \delta \bar {\cal S} }{ \delta \bar z}    
  =  + \left(  \frac{d^2 \bar z}{dt^2}  
            -  \frac{\partial\bar V}{\partial {\bar z}}   \right)     
      \, , \qquad  \cr
&\frac{\partial  \bar z (t, u) }{\partial u} 
   =  +  \frac{ \delta  \cal S }{ \delta z}  
   =  + \left(  \frac{d^2 z}{dt^2}  
  -  \frac{\partial V}{\partial { z}}   \right)   \, ,
\label{PLW-3}
\end{align}
where $u$ is the gradient flow time. 
Note that vanishing of the right 
hand side of the gradient system  \eqref{PLW-3}  is the complexification of 
the equations of motion,  (the holomorphic Newton equations), as it should be. 
The limit $u=-\infty$ is 
the critical point of the action, for which  the right hand side is 
zero and gives just the complexified equations of motions \eqref{eq:eom1}.

The  partition function \eqref{C-1} can also be expressed  as
\begin{align}
Z &=  \int_\Gamma  D(x + i y) \;   
   e^{ - \frac{1}{\hbar} \left( \Re {\cal S}  + i   \Im {\cal S}  \right)	 } 
  =  \int_\Gamma  D(x + i y) \;   
   e^{ - \frac{1}{\hbar} \int \half (\dot x^2 -\dot y^2) + i \dot x \dot y + \ldots }\, . 
\label{complex-path}
\end{align}
The ``kinetic term" of $y(t)$ is not bounded 
from below, and one may worry about the convergence of the path integral. 
However, this is  an illusion.  The thimble construction 
guarantees that the integral is convergent.  It is an easy exercise to prove, for example for the simple harmonic oscillator, 
that  the middle-dimensional cycle $\Gamma$ reduces to the standard path integral.

\noindent{\bf Functional Cauchy-Riemann equation:} In terms of  real and 
imaginary parts of the complexified field  $z(t)=x(t)+iy(t)$, the  
action  satisfies a functional  (infinite dimensional) version  
of the Cauchy-Riemann equation:
\begin{align}
{\rm Cauchy-Riemann:} \qquad \frac{\delta{\cal S}}{\delta \bar z} =0, 
       \qquad  \Longleftrightarrow  \qquad 
& \frac{\delta{ \Re \cal S}}{\delta x} = +\frac{\delta{ \Im \cal S}}{\delta y} 
      \, , \qquad   \cr 
& \frac{\delta{ \Re \cal S}}{\delta y} = - \frac{\delta{ \Im \cal S}}{\delta x} 
      \, , \qquad   
\label{CR-eq}
\end{align}
which means that action is a holomorphic functional of  $x(t) + i y(t)$. 
Now, it is also worthwhile to rewrite the complexified equations of motions: 
\begin{align}
{\rm Equation \;of\; motion:} \qquad  \frac{\delta{\cal S}}{\delta  z} =0, 
        \qquad \Longleftrightarrow  \qquad 
& \frac{\delta{ \Re \cal S}}{\delta x} = -\frac{\delta{ \Im \cal S}}{\delta y} 
       \, , \qquad   \cr 
& \frac{\delta{ \Re \cal S}}{\delta y} = + \frac{\delta{ \Im \cal S}}{\delta x} 
       \, .  \qquad   
\label{C-EOM}
\end{align}
The combination of \eqref{C-EOM} with the holomorphy condition \eqref{CR-eq} 
has a simple consequence. The critical points  (saddles) of the real part of 
a holomorphic functional, $\Re{\cal S}$, which may be viewed as a Morse 
functional, are the same as the critical points of  ${\cal S}$. In particular, 
using  \eqref{C-EOM} and  \eqref{CR-eq},   the  complexified equations of 
motion can be re-written as 
\begin{equation}
 \frac{\delta{ \Re \cal S}}{\delta x}   =0 \, , \qquad\qquad
 \frac{\delta{ \Re \cal S}}{\delta y}   =0 \, .     
\label{C-EOM-2}
\end{equation}
Consider the saddle $z_\sigma(t) $ and the Lefschetz thimble $\cal J_{\sigma}$  
attached to it. The imaginary part of the action remains invariant under 
the gradient flow time on the thimble. Using  \eqref{PLW-3} and the chain rule, 
\begin{align}
\frac{\partial  {\rm Im} [{\cal S}] }{\partial u}  = 0\, ,
\label{PLW-4}
\end{align}
meaning that  
\begin{align}
{\rm Im} [{\cal S}(z)] ={\rm Im} [{\cal S}(z_\sigma)]
\label{inv}
\end{align}
is {\it invariant} under the flow.   

  Given a set of saddles in a quantum mechanical path integral, in the 
weak coupling regime, there is always a representation where the Euclidean 
vacuum of the theory may be viewed as a proliferation of these configurations, 
corresponding to a ``dilute gas".    Since  ${\rm Im} [{\cal S}(z_\sigma)]$ 
is an invariant associated with these saddles and thimbles, it will crucially 
enter into the Euclidean vacuum description, and play important physical 
roles. This is the origin of the hidden topological angle (HTA) discussed
in \cite{Behtash:2015kna}.  We will provide concrete examples of HTAs. 
An interesting observation is that the HTA associated with a critical saddle 
field \eqref{inv} is identical to the one obtained by studying the  much 
simpler QZM-thimble \eqref{sp}.

We may consider the real part of the complexified action as a Morse functional, or more precisely a Morse-Bott functional, $h =  \Re{\cal S}$ which obeys
\begin{align}
\frac{\partial  {\rm Re} [-{\cal S}] }{\partial u}   \leq 0 \, ,
\label{PLW-5}
\end{align}
indicating downward flow nature of the flow. A Morse function on a manifold 
$\mathcal M$ is a  function which has no degenerate critical points on the 
manifold. Technically the real part of an action with any kind of symmetry 
is not a Morse function, because there are degeneracies associated with the 
symmetry of the system, and therefore there is a critical manifold, i.e. a 
manifold of critical points related by a symmetry. If the Hessian of all 
directions orthogonal to the critical manifold is non-singular, then such a 
function is referred to as a \emph{Morse-Bott function}. The critical points 
of the real part of a holomorphic functional $h=\Re{\cal S}$   
are the same as the critical points of  ${\cal S}$, found by solving 
\eqref{C-EOM-2}. In this paper we  study these problems in two prototypical 
examples, and make these ideas more concrete. 

\section{Graded Hilbert spaces}

{\scriptsize{\bf Outline:} In this section, we  generalize Witten's 
construction of graded Hilbert spaces for supersymmetric quantum 
mechanics \cite{Witten:1982df}  ($N_f=1$ or spin-$\half$ particle) 
to multi-flavor non-supersymmetric theories ($N_f >1$ or internal 
spin $(\half)^{N_f}$ particle.)  This provides a calculable version 
of the path integral based on a graded representation for the fermionic 
determinant. The graded system provides the staring point for obtaining 
exact non-BPS solutions. The second  (and more important) step 
is the idea of complexification.  }

\subsection{Supersymmetric QM and its non-supersymmetric generalization}
\label{sec:general}

Consider the Minkowski space action of supersymmetric quantum mechanics 
with superpotential $\W(\x)$ \cite{Witten:1982df} in canonical 
normalization:\footnote{{\bf Canonical vs. non-perturbative normalization:} 
We use $\x$ to denote the bosonic coordinate  and  $\W$ to denote auxiliary 
potential (superpotential)  with a canonical normalization,  $ \frac{1}{2}
\dot \x(t)^2 +  \half (\W')^2 + \ldots$. For non-perturbative normalization, 
we use $x(t)$ and  $W(x)$, and Lagrangian reads  $ \frac{1}{g}[ \frac{1}{2}
\dot x(t)^2 +  \half (W')^2 + \ldots ]  $ where  $W(x)$ is $g$ independent.  
The relation between the two is 
\begin{equation}
x(t)=\sqrt{g}\x(t), \qquad 
\W(\x)= \frac{1}{g} W(\sqrt{g}\x(t)) \, . 
\end{equation}
The latter is more suitable for the non-perturbative treatment of instantons, 
and the determination of other non-self-dual saddle points. 
\label{fn:norm}}
\begin{equation}
S_M= \int dt \left( \half \dot \x^2  -  \half (\W')^2  
   + \half i (\bar \psi \dot \psi-   \dot {\bar \psi}   \psi ) 
   + \half  \W'' [\bar \psi, \psi] \right)\, . 
 \qquad \qquad \qquad \qquad \qquad 
\label{lagsusy}
\end{equation}
We will consider a manifestly non-supersymmetric  multi-flavor generalization 
of this class of quantum systems. We introduce $N_f$ flavors, coupled as 
\begin{equation}
S_M= \int dt \left( \half \dot \x^2  -  \half (\W')^2 
   + \half i (\bar \psi_i \dot \psi_i -   \dot {\bar \psi}_i \psi_i) 
   + \half \W'' [\bar \psi_i, \psi_i ]  \right)\, ,  
   \qquad i=1, \ldots, N_f \, . 
\label{lag}
\end{equation}
For the non-supersymmetric theory, we refer to   $\W(\x)$ as the ``auxiliary 
potential". It plays a role similar to the superpotential in the $N_f=1$ 
theory.    The study of  $N_f=1$ case is old, and 
$N_f \geq 2$ generalization and its study  is new.\footnote{This type of generalization  is motivated by the quantum field theory studies of QCD(adj) and 
${\cal N}=1$ SYM on  $\R^3 \times S^1$, where the former is the multi-flavor 
non-supersymmetric generalization of the latter, and despite the 
absence of supersymmetry,  it admits  a semi-classical calculable regime 
\cite{Unsal:2007jx}.  }

 We can canonically quantize this Lagrangian.  Introduce bosonic position  
$\hat \x$ and momentum $\hat \p$ operators obeying canonical commutation 
relation, as well as fermionic creation $\hpp^i$ and annihilation $\hpm$ 
operators obeying anti-commutation relations:
\begin{eqnarray}
  && [\hat \p, \hat \x]= -i \, ,  \qquad  
     [\hat \p, \hat \p]=
     [\hat \x, \hat \x]= 0\, ,   \cr  \cr
  && \{\hpp^i, \hpm^j \} = \delta^{ij}\, ,  \qquad  
     \{\hpp^i, \hpp^j \} =\{\hpm^i, \hpm^j \}=0\, .   
\end{eqnarray}
The Hamiltonian of  the multi-flavor theory reads  
\begin{align} 
\hat H & = \underbrace{ \half \hat \p^2 + \half (\W')^2}_{\hat H^b}  
   + \half  [\hpp^i, \hpm^i ] \W''  \cr 
       & =  \hat H^b + (  \hat N - \half N_f)  \W''\, , 
\label{Ham-nf}
\end{align} 
where summation over repeated indices is assumed. In bringing the 
Hamiltonian into the latter form, we used $\sum_{i=1}^{N_f} [\hpp^i, \hpm^i ]
= \sum_{i=1}^{N_f} (2 \hpp^i\hpm^i  -1) \equiv 2 \hat N -N_f$, where $ \hat N$ 
is the ``fermion number" (or equivalently spin) operator. $\hat H^b$ denotes  
the Hamiltonian of the $N_f=0$ (purely bosonic) theory.

 The Lagrangian is invariant under a $U(1) \times SU(N_f)$ global flavor 
symmetry: 
\begin{equation}
 \psi_i  \rightarrow  e^{ i \alpha} U_{ij} \psi_j, \; \;  \; e^{i \alpha } 
        \in U(1), \; \;U \in SU(N_f) \; .
\end{equation}
This means that the  Hilbert space is  decomposable into irreducible 
representations of $ U(1) \times SU(N_f)$.  The charges associated with 
these global symmetries are
\begin{align} 
\hat N =  \sum_{i=1}^{N_f}  \hpp^i\hpm^i \, , \qquad 
\hat S^a =   \hpp^i (t^a)_{ij} \hpm^j   \, ,
\end{align}
where $t^a$,  $a=1, \ldots N_f^2-1$ are the generators of the defining 
representation of the $\mathfrak{su}(N_f)$ algebra. Clearly, these 
generators commute with the Hamiltonian, 
\begin{align} 
[ \hat N,  \hat H ] = 0, \qquad  [ \hat S^a, \hat H ]= 0 \, . 
\label{g-sym}
\end{align}
$\hat N $ allow us to express the full Hilbert space as a direct sum 
of graded Hilbert spaces according to fermion number, and graded Hilbert 
spaces naturally fall into irreducible anti-symmetric representation of 
the $SU(N_f)$ global symmetry. 

\subsection{Fermionic Fock space}  
\label{ffs}

 The fermionic operators can be used to build a finite dimensional 
fermionic  Hilbert space ${\cal F}$. Let $| \O\rangle$ denote the Fock 
vacuum, singlet under the $SU(N_f)$ flavor symmetry. It obeys 
\begin{equation}
\hpm^i | \O\rangle =0 \qquad  \forall  \; i \in [1, N_f] \, . 
\end{equation}
There is one more singlet under the $SU(N_f)$ flavor symmetry. It is the fully 
occupied state, defined as 
\begin{equation}
 | \widetilde \O\rangle =   \hpp^{1}  \hpp^{2} \ldots   \hpp^{{N_f}}   
    | \O\rangle 
\label{empty}
\end{equation}
and by the Pauli exclusion principle, it obeys 
\begin{equation}
\hpp^i   | \widetilde  \O\rangle =0 \qquad   
	\forall  \; i \in [1, N_f] \, .
\label{filled}
\end{equation}
We can now construct the fermionic Hilbert space, which furnishes  $k$-index  
anti-symmetric irreps of $SU(N_f)$, where $k=0, \ldots, N_f$ denotes the
{\it fermion number sector}, where $k$ is eigenvalue of the  
 fermion number operator  $\hat N$. 
There are  $N_f+1$  levels, and  $2^{N_f}$  states:  
\begin{eqnarray} 
{\cal F}=  \bigoplus_{k=0}^{N_f} {\cal F}^{(k)} \, , 
\end{eqnarray}
where 
\begin{align} 
& {\cal F}^{(0)}  :=  \left\{  | \O\rangle \right\},  \cr	 
& {\cal F}^{(1)}  :=  \left\{ \;  \hpp^i | \O\rangle \right\},  \cr
& {\cal F}^{(2)}  :=  \left\{ \;  \hpp^i \hpp^j | \O\rangle \right\},  \cr
& \vdots \cr 
& {\cal F}^{(N_f-2)} :=\left\{   \hpm^{i} \hpm^j    | \widetilde \O\rangle  
\right\}, \cr
& {\cal F}^{(N_f-1)} :=\left\{   \hpm^{i}   | \widetilde \O\rangle  \right\}, \cr
& {\cal F}^{(N_f)}   :=\left\{  | \widetilde \O\rangle \right\},
\end{align}
with degeneracies  at level $k$ given by ${\rm deg}_k={N_f \choose k}$ given 
by the dimension of $k$-index anti-symmetric irreps of $SU(N_f)$. The total 
number of states is then
\begin{equation}
\sum_{k=0}^{N_f}  {\rm deg}_k  = \sum_{k=0}^{N_f}   {N_f \choose k}  
   = \textstyle 1+ N_f + \frac{N_f (N_f-1)}{2} + \ldots +  N_f +1  
   = 2^{N_f}\, .
\end{equation}
Both the unoccupied state and the fully occupied state are singlets 
under $SU(N_f)$.  Since the ground states of quantum mechanics must be 
singlets under bosonic global symmetries, we should expect the 
$|\O\rangle$  and  $| \widetilde \O\rangle$ states  to play significant 
roles in the ground state properties.

\subsection{Graded Hilbert space: Fermion number representation} 
\label{GHS}

 The Hamiltonian of the multi-flavor theory is diagonal in the fermion 
number operator.  This means that the decomposition of the fermionic 
Fock space translates in a simple way to the decomposition of the full 
Hilbert space.  Any eigenstate of the full Hamiltonian can be written 
as $| n \rangle_B \otimes  |\chi \rangle_F$, where $|\chi \rangle_F \in 
{\cal F}$. If  $|\chi \rangle_F \in {\cal F}^{(k)}$, then,  $(2 \hat N -N_f)  
|\chi \rangle =  (2 k -N_f)|\chi \rangle$, and  the  action of the 
level $k$ Hamiltonian on the  state $| n \rangle_B$ is 
\begin{equation} 
{\hat H}_{(N_f, k)}  = \frac{\p^2}{2}+  \half (\W'(\x))^2 
         +    \tfrac{p}{2}  \W''(\x),   \qquad p
     = 2k-N_f, \qquad \qquad  k=0,  \ldots, N_f \, . 
\label{Hamk}
\end{equation} 
This permits us to decompose the Hilbert space  
${\cal H}$ of $N_f$ flavor theory  into sub-Hilbert spaces 
${\cal H}_{(N_f,k)}$  graded  under the ``fermion number" $k$.  The 
Hamiltonian  ${\hat H}$  and the Hilbert space  ${\cal H}$ of the 
$N_f$ flavor theory can be decomposed as: 
\begin{align} 
{\hat H} & = \bigoplus_{k=0}^{N_f}  {\rm deg} ( {\cal H}_{(N_f, k)})   
                  {\hat H}_{(N_f,k)}, \qquad    
{\rm deg} ( {\cal H}_{(N_f, k)}) = \textstyle{ {N_f \choose k}} \cr
{\cal H}&= \bigoplus_{k=0}^{N_f}  {\rm deg} ( {\cal H}_{(N_f, k)})   
            {\cal H}_{(N_f,k)},  
\label{directsum-0}
\end{align}
where  ${\rm deg} ( {\cal H}_{(N_f, k)}) $ is the degeneracy of a 
given sector. This construction  generalizes  Witten's discussion in 
supersymmetric QM to non-supersymmetric theories \cite{Witten:1981nf}.  We obtain a triangle 
of graded Hilbert spaces, where black ($k$-even) are bosonic,  and red 
($k$-odd) are fermionic spaces:
\begin{eqnarray}
&{    1{\cal H}_{(0,0)}} & \nonumber \\ \cr
 {   1{\cal H}_{(1,0)}}
&& 
{\color{red}  1{\cal H}_{(1,1)}}  \nonumber\\ \cr
{  1{\cal H}_{(2,0)}}\qquad \qquad & 
{ \color{red} {2 \cal H}_{(2,1)}}&
\qquad \qquad
{1 {\cal H}_{(2,2)}} \\   \cr
{  1{\cal H}_{(3,0)}}\qquad \qquad
{  \color{red} {3 \cal H}_{(3,1)}}
&& 
{  3{\cal H}_{(3,2)}} \qquad \qquad
{  \color{red} {1 \cal H}_{(3,3)}} 
\cr\cr
{  1{\cal H}_{(4,0)}}
\qquad \qquad
{  \color{red} {4 \cal H}_{(4,1)}} \qquad \qquad 
&{   6{\cal H}_{(4,2)}}
  & \qquad \qquad 
{   \color{red} {4 \cal H}_{(4,3)}}\qquad \qquad
{  1{\cal H}_{(4,4)}}   \nonumber
\label{triangle}
\end{eqnarray}
The $(N_f, k)=(0,0)$ cell is the bosonic Hilbert space, ${\cal H}_{(0,0)}$.  
The $N_f=1$ row contains the paired Hilbert spaces of supersymmetric quantum 
mechanics \cite{Witten:1981nf}.  The study of $N_f \geq 2$ is new and has some 
surprising elements.

 It is also useful to further grade the Hilbert spaces under fermion number 
modulo two, $(-1)^F$. For example, if the state $|0 \rangle_B \otimes  
|\Omega \rangle$ is a ground state,  then,  obviously,  ${\cal H}_0$ is 
bosonic,  $ {\cal H}_1$ is fermionic,  $ {\cal H}_2$ is bosonic, etc. If 
$N_f$  is odd, then $ {\cal H}_{N_f}$ is fermionic, while if $N_f$  is 
even, then $ {\cal H}_{N_f}$ is bosonic.

The advantage of the graded formulation is the following. Given the twisted 
partition function of the theory with multiple fermions, we can now convert 
it into the sum of bosonic partition functions over the graded sectors:
\begin{align}
\label{eq:tpf}
\tilde {Z}(L) \equiv& {\rm Tr}_{\cal H}  (-1)^F e^{-L H}   \cr
  =& \sum_{k=0}^{N_f} (-1)^{N_f -k}  {\rm deg}(  {\cal H}_k)            
       {\rm Tr}_{{\cal H}_k} e^{-L H_k}    \cr
  = & \sum_{k=0}^{N_f} (-1)^{N_f -k}  {\rm deg}(  {\cal H}_k)  Z_k \, . 
\end{align}

%

\subsection{Derivation from the fermionic determinant} 

We can also derive \eqref{eq:tpf} by using a functional integral 
formulation.    Consider 
\begin{align}
\tilde {Z}(L)^{\rm original}  
  &= \int  D\x \prod_{i=1}^{N_f}  D\psi_i D\psi_i   \; 
        e^{ -\int dt \left( \half \dot \x^2  +  \half (\W')^2 
          +   \bar \psi_i (\partial_t + \W'') \psi_i  \right)}  \cr
  &= \int D\x   \; e^{ -\int dt {\cal L}_{\rm bos} }  \;  
          \Big[{\rm det}_{\pm} (\partial_t + \W'')\Big]^{N_f}   \, .
\label{useful}
\end{align} 
The fermionic determinant can be calculated exactly \cite{Gozzi:1984jx}. 
The result for periodic ($+$) and anti-periodic ($-$) boundary conditions 
on the fermions is given by
\begin{align}	
[{\rm det}_{\pm} (\partial_t + \W'')]^{N_f} = \left\{ 
\begin{array} {ll}
  \Big[ 2 \sinh \left( \half \int \W'' dt \right)\Big]^{N_f}  
          & \qquad {\rm for }    \;\;   \psi_i(L) = + \psi_i(0)\, ,    \\ \\
   \Big[ 2 \cosh \left( \half \int \W'' dt \right)\Big]^{N_f}  
          & \qquad {\rm for }    \;\;   \psi_i(\beta) =  - \psi_i(0)\, .  
\end{array} \right.
\end{align}
This implies that we can express the twisted partition function as 
\begin{align}
\label{eq:tpf-5}
\tilde{Z}(L)   &  = \int D\x \;   
  e^{ -\int dt {\cal L}_{\rm bos}  - N_f  \log \left( 2 \sinh \half \int dt  \W''\right)}\, .
\end{align}
This strategy is commonly used in QFT applications of the fermionic 
determinant.  We obtain a closed form result, but it is manifestly 
non-local\footnote{
Notice the fact that the determinant makes the sign problem manifest if 
$\W(\x)$ is an odd function of $\x$.  Under $\x(t) \rightarrow -\x(t)$, the 
second term in the action picks up an extra phase  $e^{i \pi N_f}$, leading to 
destructive interference between the Euclidean paths $\x(t)$ and $-\x(t)$ 
for $N_f$ odd, and to constructive interference for $N_f$ even. This is a 
manifestation of the sign problem, $\tilde{Z}(L)\sim (1+ e^{i \pi N_f})$ in 
the odd-$N_f$ theory, see for example, \cite{Baumgartner:2014nka} for 
the $N_f=1$ case.  The sign problem in these examples is related to exact 
spectral cancellations.}. 
In the present case, however, there is a more useful representation. 
Instead of exponentiating the fermionic determinant, if we just expand 
it into a binomial expansion, we obtain the link between the fermionic 
determinant and the graded Hilbert spaces. We find
\begin{align}	
\tilde {Z}(L)^{\rm original} 
 &= \int D\x \;   e^{ -	 \int dt   {\cal L}_{\rm bos} }   \; 
       \left[ 2 \sinh \half    \int dt  \W''  \right]^{N_f}  \cr
 &= \int D\x \;   e^{ -	 \int dt   {\cal L}_{\rm bos} }  
      \sum_{k=0}^{N_f} {N_f  \choose k} (-1)^{N_f -k}   
          \left[e^{ \int dt  \frac{\W''}{2} } \right]^k   
         \left[ e^{ -\int dt  \frac{\W''}{2} } \right]^{N_f-k}   \cr
 &=  \sum_{k=0}^{N_f} {N_f  \choose k} (-1)^{N_f -k}   
    \int D\x \;   e^{ -\int dt \left( \half \dot \x^2  +  \half (\W')^2    
     + (2k-N_f) \frac{\W''}{2}  \right)} \cr 
 &= \sum_{k=0}^{N_f} {N_f  \choose k} (-1)^{N_f -k}  Z_k 
  = {Z}(L)^{\rm graded}\, . 
\end{align}	
This simple equality has great utility.  In particular, the graded 
formulation will allow us to find new exact saddles in the problem. 
In the original formulation where we keep 
the fermions, it is more difficult to demonstrate the existence of exact 
solutions.  The 
beauty of the  graded formulation is that we can show that these 
approximate solutions are actually approximations to the exact solutions. 


Consider taking the compact radius $L$ to be infinitely large, and restrict  to solutions starting and ending in a particular harmonic minimum, which we label by $\x_i$. These are the solutions which are interesting if we wish to quantize around the classical minimum at $\x=\x_i$. Then $\W''(x(t))\approx \W''(x_i)=\omega_i$, the natural local harmonic frequency,  almost everywhere on this solution. The determinant in the infinite volume limit $L\rightarrow \infty$ will then become a simple exponential
\begin{align}	
[{\rm det}_{\pm} (\partial_t + \W'')]^{N_f}\underbrace{=}_{L\rightarrow \infty} \left\{ 
\begin{array} {ll}
  \exp \left(\frac{\text{sign}(\omega_i) N_f}{2} \int \W'' dt \right)
          & \qquad {\rm for }    \;\;   \psi_i(L) = + \psi_i(0)\, ,    \\ \\
    (\text{sign}(\omega_i))^{N_f}\exp\left( \frac{\text{sign}(\omega_i) N_f}{2} \int \W'' dt \right) 
          & \qquad {\rm for }    \;\;   \psi_i(\beta) =  - \psi_i(0)\, .  
\end{array} \right.
\end{align}
The above form yields a perfectly local action which corresponds to the  projection to the $k=0$ (empty) or $k=N_f$(fully occupied) fermion number eigenstates. This corresponds to 
\emph{maximal spin projection}, i.e. $S_z=\pm N_f/2$. The reason this is the case is that the ground state(s) lives in the  maximal spin state, and sending $L\rightarrow \infty$  
projects onto the  ground-state(s) only. 

\subsection{Graded Hilbert space: Spin representation}	
\label{sec-spin}
 For $N_f=1$, the full Hamiltonian can be interpreted as describing
the motion of a spin-$\half$ particle in a spin-independent potential, 
$(\W')^2$, and a ``magnetic field", $\W^{''}$, which couples to the spin.   
This is the case with exact supersymmetry in the interpretation of 
Ref.~\cite{Witten:1982df}. For  
$N_f=2$, we have a  spin  $\half \otimes \half= 1\oplus 0$ particle,  
a spin-1, and a spin-0 particle in a magnetic field. For few low-$N_f$, 
we have
\begin{align} 
 &N_f=0, \qquad 0 \, , \cr
 &N_f=1, \qquad  \half    \, , \cr 
 &N_f=2, \qquad  \half \otimes \half = 1(1) \oplus 1(0) \, , \cr
 &N_f=3, \qquad  \half \otimes \half \otimes  \half   
   =  1(\textstyle {\frac{3}{2}})  \oplus 2 (\half)  \, , \cr
 &N_f=4, \qquad  \half \otimes \half \otimes \half \otimes  \half  
   = 1(2)+ 3( 1)  \oplus 2(0)  \, , \cr
 &N_f=5, \qquad  \half \otimes \half \otimes \half \otimes  
        \half \otimes  \half  
   =1( \textstyle {\frac{5}{2}}  )+ 4(  \textstyle {\frac{3}{2}}  )  
        \oplus 5(  \textstyle {\frac{1}{2}} ) \, . 
\end{align} 
In general,  decomposing    
\begin{align}
 (\half)^{N_f} =  \bigoplus_{S=S_{\rm min}}^{S_{\rm max}}  
   {\rm mult}(S) \; S \, ,
\label{spin}
\end{align} 
where  
\begin{align}  
S_{\rm max} = \frac{N_f}{2}, \qquad S_{\rm min} 
          =  \left\{ \begin{array} {ll} 0  \qquad  
 &  N_f{\rm \; even}\, ,  \cr
\half  \qquad  &  N_f{\rm \;odd}\, ,  \cr 
\end{array}\right. 
\end{align} 	 
the  multiplicity of the spin-$S$ sector is given by  
 \begin{align}
S = \frac{N_f}{2} -k, \qquad 	{\rm mult}(S) 
  =  \left\{ \begin{array} {ll}
    1  \qquad  &  k =0 \, , \cr
   {N_f \choose k} - {N_f \choose k-1}  \qquad  &  1 \leq k \leq  
                \lfloor \frac{N_f}{2} \rfloor \, . \cr 
    \end{array}\right.
\end{align} 
Therefore, the  fermionic  Fock space admits a representation both in 
terms of fermion number sectors as well as spin sectors. It is easy to 
show that the dimension of the Fock space agree:
\begin{align}
 2^{N_f} = \sum_{k=0}^{N_f} \deg({\cal H}_k)  
        = \sum_{S=S_{\rm min}}^{S_{\rm max}}    {\rm mult}(S) (2S+1).
\end{align} 
\begin{figure}[t]
\begin{centering}
\includegraphics[angle=0, width=0.60\textwidth]{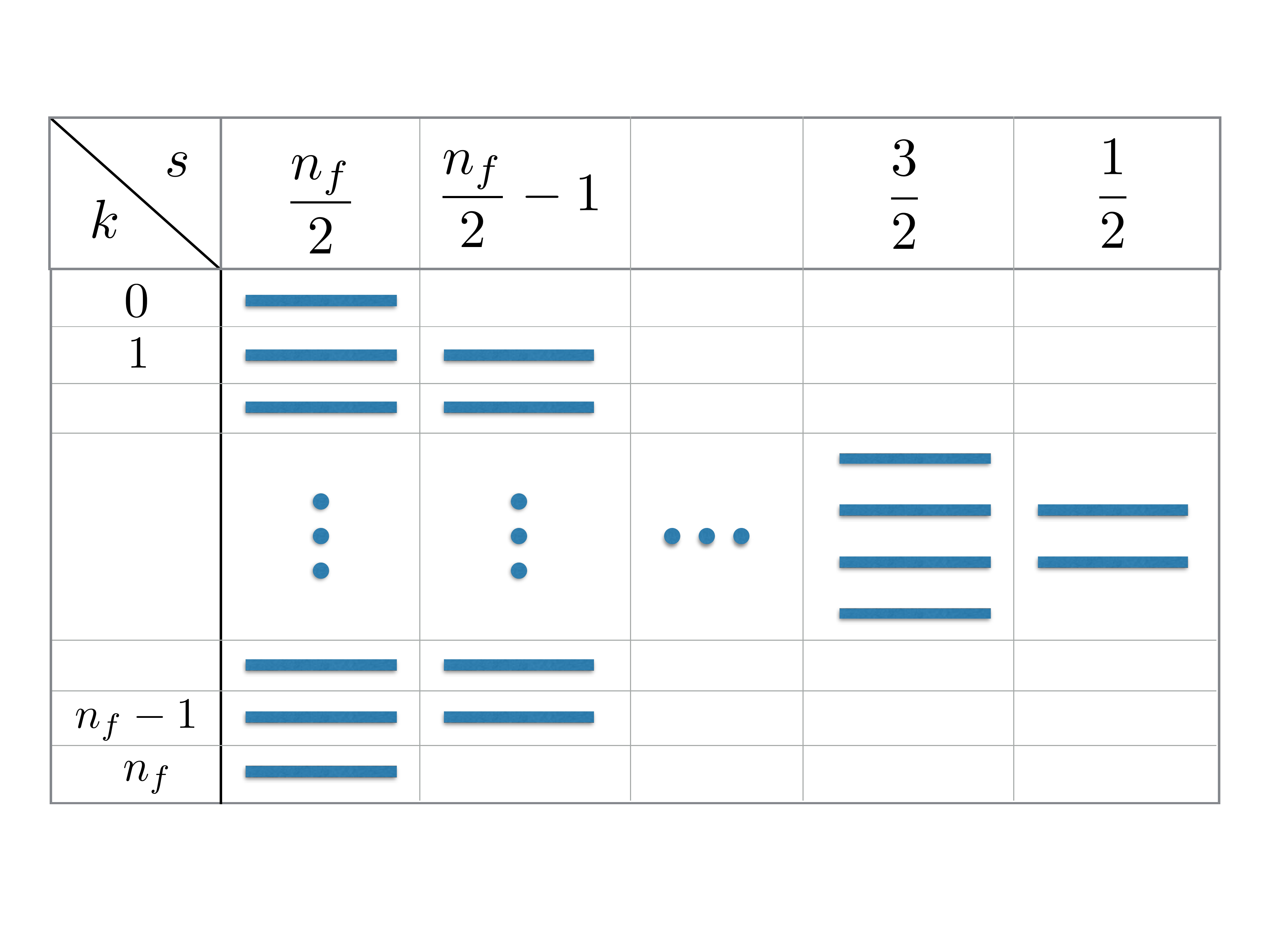} 
\vspace{-1cm}
\caption{
Fermion number  vs. spin representation for the multi-flavor theory 
with Lagrangian  \eqref{lag}.    $N_f$ is chosen to be odd.   The system 
describes a particle with internal $(\half)^{N_f}$-spin. }
\label{spin_pyramid}
\end{centering}
\end{figure}	
 
              	
One  can re-interpret the Hamiltonian of the ``multi-flavor" theory  
as a  direct sum over allowed spin quantum numbers, appearing on the 
right hand side of  \eqref{spin}: 
\begin{align}  
\label{spin-1}
{\hat H} = \bigoplus_{S=S_{\rm min}}^{S_{\rm max}}  {\rm mult}(S) { \hat H}^{(S)} \, .  
\end{align} 
The Hamiltonian in the spin-$S$ sector is given by: 
\begin{align}  
\label{spin-2}
&\hat H^{(S)} = \half \hat \p^2 + \half (\W')^2  + S_z   \W'', 	 
        \qquad  {\rm where}  \;\; 
S_z^{} ={\rm Diag} \left( s, s-1, \ldots , -s+1, -s \right), \;\;
\end{align} 
is the diagonal spin matrix. This provides a second representation for 
the thermal partition function: 	
\begin{align}
\label{eq:spin-rep}
{Z}(\beta) \equiv \sum_{S=S_{\rm min}}^{S_{\rm max}}    {\rm mult}(S) Z_S\, , 
\end{align}
where $Z_S$ is the partition function in the spin-$S$ sector. Note  
that odd $N_f$ is associated with a sum over half-integer spins, and 
even $N_f$ is associated with a sum over integer spins. 

  
 The partition function for the $S= \half$ case (or $N_f=1$ supersymmetric 
QM) is  given by  (see \cite{Wen:2004ym, Altland:2006si} for  discussions concerning the path integral representation of spin)
 \begin{align}
\label{eq:spf}
{Z}(\beta)  &= \int  D\x D(\cos\theta) D \phi  \; 
   e^{ -\int dt \left( \half \dot \x^2  +  \half (\W')^2 + S_B(\theta, \phi)  
          + S_{WZ}(\theta, \phi)  \right)}\, ,  \cr
\qquad  S_B(\theta, \phi) &= \int  d t \half \W''  \cos \theta  \, , \qquad 
  S_{WZ}(\theta, \phi) = i \half   \int  d t  (1- \cos \theta ) 
    \partial_t \phi \, , 
\end{align}
where  $(\theta, \phi) \in {\bf S}^2$ parameterize the Bloch sphere.
$S_B(\theta, \phi)$ denotes the spin-``magnetic field"  interaction, 
and $ S_{WZ}(\theta, \phi) $ is the abelian Berry phase (or Wess-Zumino 
term).  For a spin-$S$ sector, the  generalization of this expression is\footnote{The result we give here is for the thermal partition function. For the twisted partition function an insertion $(-1)^{N_f-k}$ needs to be inserted. Since $S_z=k-N_f/2$, with $k$ being the fermion number, and since $S_z=S\cos\theta$, to implement this twist we can insert a term $i\pi\frac{1}{L}\int_0^L dt\; (S\cos\theta-N_f/2)$ into the action.} 
 \begin{align}
\label{eq:spf-2}
{Z}_S(\beta) = \int  D\x D(\cos \theta) D \phi  \; 
   e^{ -\int dt \left( \half \dot \x^2  +  \half (\W')^2 
    +  S  (  \W''  \cos \theta   +  i (1- \cos \theta ) \partial_t \phi
   \right)} \;. 
\end{align}
In this work we will not take full advantage of the spin-representation. 
However, it is useful to keep in mind that the most physically appropriate 
interpretation of the problem is actually that of a  particle with internal 
spin $(\half)^{N_f}$.  In fact, the equivalence of 
\eqref{eq:spf-2} and \eqref{eq:tpf} is straightforward.
The relation between the  partition function graded according to fermion 
number and spin representations is given by 
\begin{align}
\label{eq:spin-ghs}
Z(\beta) \equiv   \underbrace{ \sum_{k=0}^{N_f} {\rm deg}(  {\cal H}_k)  
        Z_k  }_{\text {sum over columns  in Figure \ref{spin_pyramid}}}
 = \underbrace{\sum_{S=S_{\rm min}}^{S_{\rm max}}    
    {\rm mult}(S) Z_S }_{\text  {sum over rows in Figure \ref{spin_pyramid}}}\, . 
\end{align}

\subsection{Twisted partition function vs. supersymmetric Witten index}
\label{sec:tpf}

The twisted partition function \eqref{eq:tpf} $\tilde{Z}(L)$ is identical 
to the supersymmetric Witten index  $I_W$ \cite{Witten:1982df}  for the 
$N_f=1$ theory. For non-supersymmetric theories with $N_f >1$ the twisted 
partition function has a  set of remarkable properties, which are still 
connected to spectral cancellations (under the conditions stated below)  
in  non-supersymmetric theories. In particular, we will show that for 
certain choices of the auxiliary potential $\W(\x)$, the twisted partition 
function \eqref{eq:tpf} vanishes for any odd $N_f$. 

  
 First note the following three properties regarding the structure of 
the graded Hilbert space \eqref{triangle}:
\begin{itemize}
\item[\bf i)] The degeneracies of states in ${\cal H}_k$ and ${\cal H}_{N_f -k}$ 
are the same: $ {\rm deg}( {\cal H}_k) = {\rm deg}({\cal H} _{N_f -k})  = 
{N_f \choose k}$ as dictated by the dimensions of corresponding irreps 
of $SU(N_f)$.  

\item[\bf ii)] For $N_f$ even,  ${\cal H}_k$ and ${\cal H} _{N_f -k} $ are  
either Bose-Bose or Fermi-Fermi  pairs. For $N_f$ odd, ${\cal H}_k$ 
and ${\cal H} _{N_f -k}$ form Bose-Fermi pairs.
  
\item[\bf iii)]  If the auxiliary potential is an odd function of $\x$,   
$\W(-\x) =-\W(\x) $, or a periodic function, 
then ${\cal H}_k$ and ${\cal H} _{N_f -k} $ exhibit spectral 
degeneracy, $ {\rm spec}( \hat H_k) = {\rm spec}(\hat H_{N_f -k}) $. 
\end{itemize}

If these conditions hold, the twisted partition function  for odd-$N_f$ 
theories is ``trivial" and vanishes:
\begin{align}
\label{eq:trivial}
\tilde{Z}(L) \equiv {\rm Tr}_{\cal H} (-1)^F e^{-L H} 
  =& \sum_{k=0}^{N_f} {\rm deg}(  {\cal H}_k)   (-1)^{N_f -k}   
       {\rm Tr}_{{\cal H}_k} e^{-L H_k}  = 0\, .   
\end{align}
Readers familiar with supersymmetric quantum mechanics will immediately 
realize   that with the choice of $\W(\x)$ quoted in {\bf iii)}, the Witten 
index is identically zero, $I_W=0$.  As a reminder,  we note that $I_W=0$ 
means one of two things: 
{\bf a)} Supersymmetry is spontaneously broken,  and there are degenerate 
{\it positive energy} grounds states.
{\bf b)} Supersymmetry is not  broken,  and there are degenerate  {\it zero  
energy} grounds states.  
In {\it both cases}, regardless of spontaneous breaking of supersymmetry, 
there is an exact spectral cancellation between bosonic and fermionic Hilbert 
spaces. Remarkably enough, the spectral  cancellation generalizes to 
multi-flavor non-supersymmetric quantum mechanics for this general class 
of auxiliary potentials, and we will take advantage of that.   

  
\subsection{Isospectral pairs} 
\label{isospec}

 Let $\W(\x)$ be an odd-polynomial, which means $\W'(x)$ is even and $\W''(x)$ 
is odd.  Consider
\begin{align} 
\hat H_k =\hat H_{\rm bos}  +   ( k - \frac{N_f}{2})  \W'',  \qquad 	
\hat H_{N_f-k} =\hat H_{\rm bos}  -  ( k - \frac{N_f}{2})  \W''\, . 
\label{Hamk-deg}
\end{align} 
Let $\psi_{k,n}(x) $ be an eigenstate of ${\hat H}_k$ with eigenvalue  
$E_{k,n}$: $\hat H_k  \psi_{k,n}(x) =  E_{k,n}  \psi_{k,n}(x)$.  Acting with  
parity operator from the left, and realizing  simple identity\footnote{This 
identity fails if $\W(\x)$ is an even-polynomial. In that case,  $P \hat H_k P = \hat H_k$ with no non-trivial implications. }:
\begin{align}
P \hat H_k P  =  \hat H_{N_f-k},
\label{PHP}
\end{align}
we reach to the conclusion that $P \psi_{k,n}(x) = \psi_{k,n}(-x) $  is an 
eigenstate of $\hat H_{N_f-k}$ with the same eigenvalue. Thus, we identify  
$\psi_{k,n}(-x) = \psi_{N_f-k,n}(x)$, and $E_{k,n}= E_{N_f-k, n}$. This 
demonstrates isospectrality for odd-polynomial auxiliary potentials. 
 
  
\begin{figure}[t]               
\centering
\includegraphics[angle=0,width=.9\textwidth]{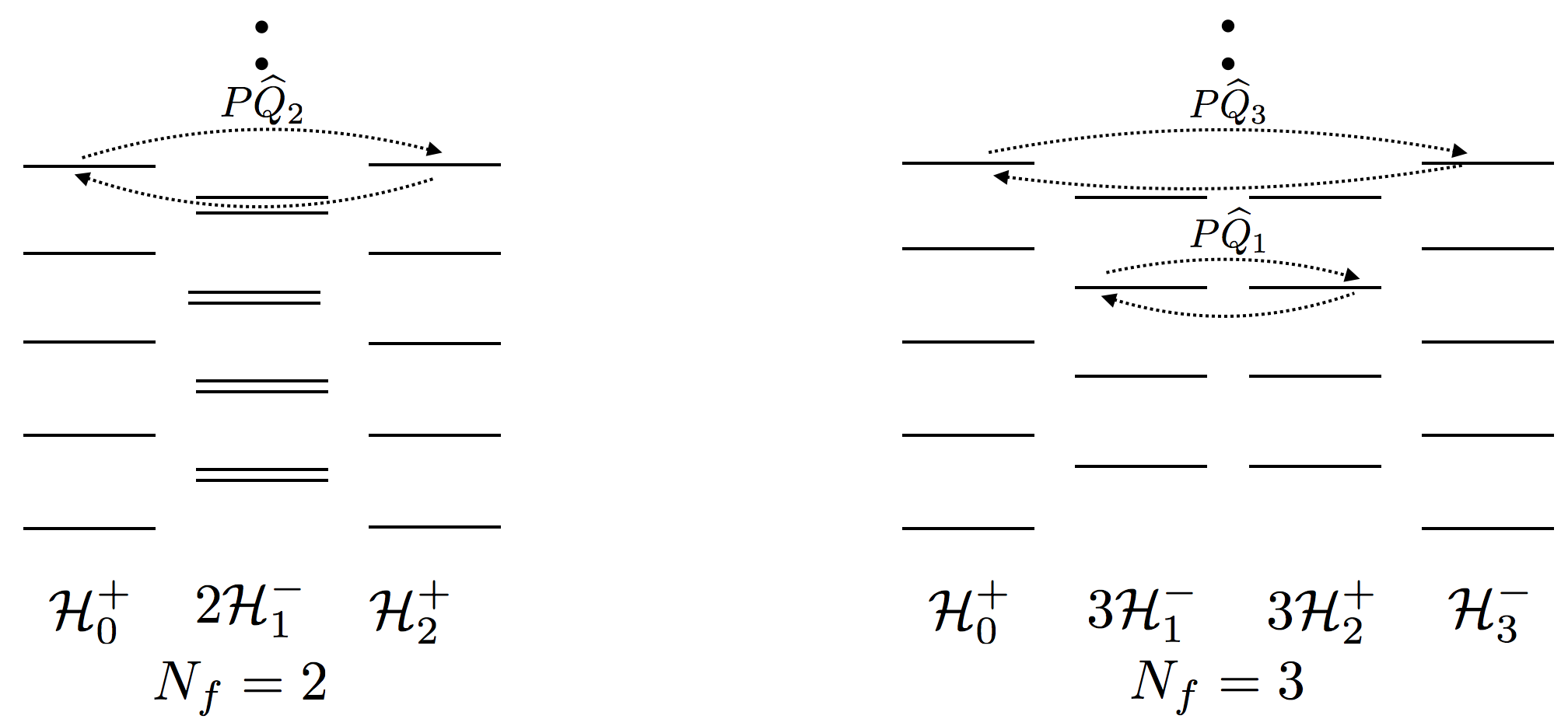}
\caption{  Hilbert-space of the $N_f=2$ (prototype for even $N_f$)  
and $N_f=3$  (prototype for odd $N_f$) flavor theories with odd auxiliary 
potential $W(x)$.   The Hilbert space decompose according to fermion number.  
We use the notation $({\rm deg}_k) {\cal H}_k^{\pm}$ to label a graded 
Hilbert space, where  $\pm$  is the value under $(-1)^F$. For odd $N_f$, 
all sectors are paired via nilpotent operators, $PQ$, for even $N_f$, all 
but the middle one are paired. In both cases, there are two ground states.}
\label{Spec23}
\end{figure}

 For periodic potentials, such as $\W(\x)=\frac{4a^3}{g}\cos\frac{\x\sqrt{g}}
{2a}$,  the demonstration is almost the same, with a minor difference. The 
potential appearing in  $\hat H_k$ is of the form $\half (\W')^2 + 
(k - \frac{N_f}{2})\W''$. This implies that the period of $(\W')^2$ is 
$\frac{\x\sqrt{g}}{2a} \sim \frac{\x\sqrt{g}}{2a} +   \pi$, while the period 
of $\W''$ is $2 \pi$. Based on this observation
\begin{align}
 T_\pi \hat H_k T^{\dagger}_\pi  =  \hat H_{N_f-k} ,
\label{THT}
\end{align}
where  $T_\pi$ is shift operator by $\pi$. This also implies, for periodic 
potentials, $\hat H_k$  and  $\hat H_{N_f-k}$ are isospectral. 
 
   
 Although this is sufficient to demonstrate isospectrality, it is also 
useful to construct this mapping in terms of operators acting in the full 
Hilbert space ${\cal H}$, i.e, restore the fermion operators in the above 
discussion. To this end, we note that we can define the following  
classes of {\it nilpotent} operators as maps from  $ {\cal F}^{(k)}$ to 
${\cal F}^{(N_f-k)} $, namely, 
\begin{align} 
&  \hat Q_{N_f} =    \Big\{  \hpp^{1}  \hpp^{2} \ldots   \hpp^{{N_f}} \Big\} 
  \; \;   \qquad  \qquad  \qquad    \qquad \qquad :  
    {\cal F}^{(0)}    \longrightarrow     {\cal F}^{(N_f)}   \cr 
&  \hat Q_{N_f-2}=    \Big\{  \hpp^{1}  \hpp^{2} \ldots  \hpp^{i-1} \hpm^{i} 
         \hpp^{i+1}  \ldots   \hpp^{{N_f}} \Big\} 
   \; \;    \qquad   \; : {\cal F}^{(1)}    \longrightarrow    
        {\cal F}^{(N_f-1)}, \qquad     etc.
\end{align} 
Clearly, these operators do not commute with the Hamiltonian, for example, 
\begin{align}
 [\hat N, \hat Q_{N_f} ] = N_f \hat Q_{N_f},  \qquad  
 [\hat H, \hat Q_{N_f} ] =  N_f \W''  \hat Q_{N_f},
\end{align}
On  the other hand, the operators  $ P \hat Q_{N_f}, P \hat Q_{N_f-2},\dots$ 
are  nilpotent and they commute with the Hamiltonian, 
\begin{align}
(P \hat Q_{N_f})^2 =  ( P \hat Q_{N_f-2} )^2 = \ldots=0 
          \qquad {\rm and}  \;\; \;   
[\hat H,P \hat Q_{N_f} ] | \Psi \rangle  = 0 
          \qquad \forall  \; | \Psi \rangle \in {\cal H}\, . 
\end{align}
Note that, for odd (even) $N_f$,  $P \hat Q_{N_f},  P \hat Q_{N_f-2},  
\ldots $ pair up bosonic states with fermionic  (bosonic) states. 
See Figure \ref{Spec23} for a demonstration in the cases $N_f=2$ 
and $N_f=3$.

{\bf Use of analytic continuation in conjunction with isospectrality}
The isospectrality will be useful in path integration as well. The partition 
functions of the isospectral  pairs are obviously identical. In the path integral picture, 
there is a trivial change of variable which takes one path integral to the 
other, by $\x(t) \rightarrow -\x(t)$ for the double-well, or  a shift 
$\frac{\x\sqrt{g}}{2a} \sim \frac{\x\sqrt{g}}{2a} +   \pi$ for the periodic potential.  

Alternatively, one can also 
view these paired systems as  follows: Consider the analytic continuation of 
the potential in sector $k$, \eqref{directsum-0}, into $V(x)= \half (\W')^2 +   
( k - \frac{N_f}{2}) e^{i \theta} g W'' $, where we inserted $e^{i \theta}$ into the fermion induced term.  
Clearly, interpolating in $\theta$ 
from $\theta = 0$ to $\theta =\pi$ interchanges the two potentials, 
Hamiltonians, and Hilbert spaces: 
\begin{align}
(\hat H_k, {\cal H}_k) (\theta =\pi) 
  = (\hat H_{N_f-k},  {\cal H}_{N_f, k}) (\theta =0) \, . 
\end{align}
When we discuss exact saddles, we will often find one  simple bounce solution 
in one of  the mirror pairs (it does not matter which) which is related to the 
 excited state,  and by analytic continuation in $\theta$ all the way to $\pi$, we will  
land on the more exotic  complex bion saddles on the mirror which determines 
the properties of the ground state!  This is discussed in depth in 
section \ref{sec:DoubleWell}  and \ref{sec:periodic}.

\section{Original formulation and local-global relations}

{\scriptsize {\bf Outline:} In this section we provide an analysis of the 
problem using conventional methods: the BPS-equations, instantons, local 
harmonic analysis, and index theorems.   The non-perturbative contribution 
to the  ground state energy in multi-flavor theories $N_f \geq 1$ is always 
a two-instanton effect (unlike the symmetric DW, where it is a one-instanton 
effect). We also build up a Hilbert space interpretation for the ground 
states in terms of local harmonic states, which is again different from 
the symmetric DW.}

\subsection{Auxiliary Potential}

 We first consider a general auxiliary potential $\W(\x)$. For the $N_f=1$,
supersymmetric, theory it coincides with the superpotential. Our goal is 
to give an intuitive explanation of certain local vs. global relations, 
which have close relations with real Morse theory, but are fairly easy to 
explain in physical terms. In the case of a $\W(\x)$ for which the bosonic 
potential $V_{\rm bos}= \half (\W')^2$  has local harmonic minima,  there are 
deep connections between certain local and global properties of the system. 
\begin{itemize}
\item{Two consecutive ground states in harmonic approximation always 
alternate, $|0\rangle \otimes |\Omega\rangle $ vs.  $|0\rangle \otimes 
| \widetilde\Omega\rangle $,  where  $|\Omega\rangle$ is unoccupied  
and $| \widetilde \Omega \rangle$ is the fully occupied state.} 
   
\item{The index for Dirac operator for  two consecutive instantons\footnote{
We take the definition of instantons as right tunneling events, and 
anti-instanton as left tunneling. See Section \ref{sec:conventions} 
for details.} 
{\it always} alternate in sign.}

\item{These two statements are related to the finiteness of the fermionic 
Fock space,  the absence of a fermion number anomaly, and the absence of 
a Dirac sea in quantum mechanics.}
\end{itemize}


 It may appear surprising that the perturbative vacuum structure in the 
harmonic approximation ``knows'' about the index theorem and its 
implications.  This, too,  is related to the information encoded in 
$\W'(\x)$ and to real Morse theory.  We will not dwell on this subject 
here. Let us first start with local harmonic analysis. 

\subsection{Local harmonic analysis,  states, and auxiliary Morse function}

\begin{figure}[t] 
\centering
\includegraphics[angle=0, width=0.90\textwidth]{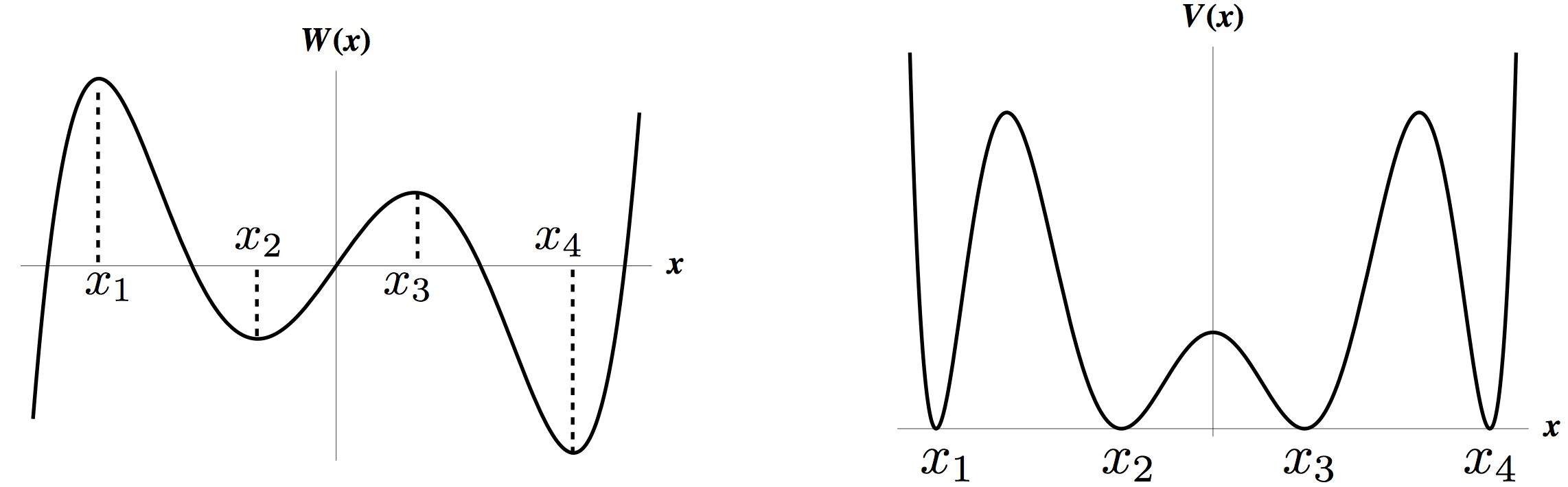}
\caption{A typical auxiliary potential $\W(\x)$ and associated real potential 
$V_{\rm bos}(\x)= \frac{1}{2}(\W'(\x))^2$.  The auxiliary potential  (which may also be viewed 
as a Morse function) is useful in understanding local harmonic analysis of 
Hilbert spaces at each well, and in the study of the index for  the  Dirac 
operator in the presence of an instanton. These two  (naively) different 
type data must be related because of the absence of Dirac sea and absence of fermion number anomaly 
in quantum mechanics. }
\label{fig:QMpot7}
\end{figure}

 In order to understand the connection between the local harmonic analysis 
of the Hilbert space at each minimum and the index of the Dirac operator
it is useful to realize that the auxiliary potential $\W(\x)$ is a Morse 
function, a real-valued smooth function of Euclidean time.  Let us  further 
choose the potential to be a {\it square-free} polynomial:
\begin{align}
V_{\rm bos}(\x) =  \half (\W')^2, \qquad   \W'(\x)=(\x-\x_1) \ldots (\x-\x_n),   
  \qquad  \x_1 < \x_2 < \ldots < \x_n
\label{square-free}
\end{align}   
A typical potential $V_{\rm bos}(\x)$ is shown in Figure \ref{fig:QMpot7}. At 
first glance, all the  minima $\x_i$  of the potential $V(\x)$  may be 
viewed  on the same footing, and in fact, in a purely bosonic theory, this 
is the case. But even at the level of the harmonic approximation, in 
theories with fermions, there are differences between consecutive minima.  
This difference is best described if one plots the auxiliary potential  
$\W(\x)$, which we  view as a real Morse function.  The extrema (minima 
and maxima) of  $\W(\x)$  are the zeros of the potential $V_{\rm bos}(\x)$, and 
there are differences in the description of states depending on $\x_i$ 
being a local minimum or maximum of the auxiliary function $\W(\x)$.  A 
typical auxiliary potential and its  bosonic  potential  are shown in 
Figure \ref{fig:QMpot7}. 
	
Evaluating  the Hamiltonian at the $i^{\rm th}$ harmonic minimum to quadratic 
order, we find that 
\begin{align} 
\hat H_i  &= \half \p^2 + \half  (\W''(\x_i))^2   (\delta \x)^2    
   +  ( {\rm sign}  \W''(\x_{i}))   |\W''(\x_{i})|    
     \sum_{i=1}^{N_f}(\hpp^i \hpm^i  - \half)    \cr
 &=   \omega_i  \left [ a_b^\dagger a_b + \half    
        + (-1)^{\mu_i}   (  \hat N_f - \half N_f)    \right], \qquad   
        \omega_i=  |\W''(x_{i})|  \cr 
 &=   \omega_i \left \{ 
\begin{array} {ll} 
      [\hat N_b -  \hat N_f +  \textstyle{\frac{N_f+1}{2}} ]   
              & \qquad  { \rm if } \; \mu_i = 1,\cr  \cr
     [\hat N_b +   \hat N_f +  \textstyle{\frac{1-N_f}{2}} ]  
              &  \qquad { \rm if } \;  \mu_i= 0,
\end{array} \right. 
\label{pert-spec-general}
\end{align} 
where  $\omega_i$ is the natural frequency of the harmonic oscillator at    
$i^{\rm th}$ minimum and $\mu_i$ is the Morse index (number of downward flow  
directions),  which is one if $\x_i$ is a maximum, and zero if it is a  
minimum.  The crucial point is the appearance of $(-1)^{\mu_i} = {\rm sign}  
\W''(\x_{i}))$ in the local quadratic Hamiltonian.  For a real polynomial 
$\W(\x)$, two consecutive extrema are of opposite nature,  thus,  ${\rm sign}  
(\W''(\x_{i})) $ alternates in its sign. This implies that the ground states 
in two consecutive vacua are also alternating:
\begin{align} 
\text{local harmonic ground states}  
  =  \left \{ \begin{array} {ll} 
  |0\rangle \otimes |\widetilde \Omega\rangle   
   =   |0\rangle \otimes  | \uparrow \ldots \uparrow    \rangle 
       & \qquad  { \rm for } \; \mu_i = 1 \,  ,\cr  \cr
  |0\rangle \otimes |\Omega\rangle    
   =   |0\rangle \otimes  | \downarrow \ldots \downarrow    \rangle      
       &  \qquad { \rm for } \;  \mu_i= 0 \, . 
   \end{array} \right.  \qquad \qquad \qquad \qquad 	
\label{pert-ground-states}
\end{align}

\subsection{Instantons and index theorem for Dirac operator} 
\label{sec:conventions}

 Instantons are tunneling events that involve transitions between 
two critical points, say, $x_i$ to   $x_{i+1}$, of 
$W(x)$.  
We use physics conventions for instantons, i.e., we refer to { \it right 
tunneling events} along which $x(t)$ increases  in the  $-\infty < t < 
\infty$   as ``instantons". Consequently, our instanton obeys: 
\begin{align}
\label{instanton-1}
 \dot x =  -   W'(x) \qquad  {\rm if} \qquad    W'(x) <0  \;\;\;  
                         {\rm  for}   \;\;\;  x_i < x < x_{i+1}, \cr
 \dot x =  +   W'(x) \qquad  {\rm if} \qquad    W'(x) >0  \;\;\;  
                         {\rm  for}   \;\;\;  x_i < x < x_{i+1},
\end{align}
so that the right hand side is always positive and hence, $x(t)$ increases  
with $t$.  Our anti-instantons are left tunneling events, with opposite 
signs.\footnote{ 
According to (real) Morse theory, if the flow from  the critical point  $x_i$ 
to $x_{i+1}$ is down-ward flow, the  flow from  the critical point  $x_{i+1}$ 
to $x_{i+2}$ must be an upward flow. This is the reason for the sign in our 
instanton equation. Otherwise, if we take a uniform sign convention for 
instantons, {\it two consecutive right tunneling} events would be called instanton
and anti-instanton.}
The action of these instanton events is given by  
\begin{align}  
g S_{I, i}=   \left|  \int_{x_i}^{x_{i+1}} dW \right| 
  =  | W(x_{i+1}) - W(x_{i}) |\, , 
\end{align}
which is just the difference of the magnitude of the auxiliary Morse function  
$W(x)$ at two consecutive points. 


  In this work we consider a cubic auxiliary potential $W(x)$ for  
which the bosonic potential $V_{\rm bos}(x)$ is the standard double-well 
potential, as well as a cosine potential $W(x)$ which leads to the 
Sine-Gordon bosonic potential.  Using non-perturbative normalization, see   Footnote.~\ref{fn:norm}, 
\begin{eqnarray}
W(x) &=&\frac{ {x}^3}{3} - {a^2 x} \, , \\
W(x) &=&  4a^3 \cos \left(\frac{x}{2a}   \right)\, .  
\label{canonical}
\end{eqnarray}
In non-perturbative normalization,  the first order BPS equation and its solution are  independent of the
coupling $g$. The smooth instanton solutions and their actions are given by 
\begin{subequations}
\begin{align} 
\label{instanton-2}
& {\rm DW}: \qquad 
 x_{\I }(t) =  a \tanh \frac{m_b}{2} (t-t_c), 
       \qquad \;\;\;\;\;\;\;\; \;\;\;\; 
 m_b = 2a, \qquad 
 S_I =  \frac{4a^3}{3g} \, ,   \\
& {\rm SG} \; \; : \qquad  
 x_{\I }(t) =  4a\arctan(\exp[ m_b(t-t_{c})]), 
      \quad \;\;   
 m_b = a, \; \qquad 
 S_I =  \frac{8a^3}{g}  \, . 
\end{align}
\end{subequations}
$t_c \in \R $ is the position modulus, a  bosonic zero mode of the instanton 
solution. The parameter $m_b$ is the mass parameter, which is the natural 
frequency of small oscillations around the harmonic minima. 


The instanton amplitude is given by 
\begin{eqnarray}
\I = J_{t_c}  \; e^{-S_\I} \;  \left[\frac{ \det' M }{ \det M_0}\right]^{- {1 \over2}} =  \left\{ 
\begin{array}{ll}     \sqrt \frac{  6S}{  \pi}   e^{-S}   
                       & \qquad ({\rm DW}),  \cr 
  \sqrt \frac{  2S}{  \pi}   e^{-S}   
                       &  \qquad ({\rm SG}),  
\end{array}  \right.
 \label{instanton-amp}
\end{eqnarray} 
  where   $J_{t_c} =  \sqrt \frac{  S}{ 2 \pi}  $ is the Jacobian associated with the bosonic zero mode,  
$  {\bf M}=-\frac{d^2}{d t^2}+V''(x)|_{x=x_{\I}(t) } $ 
  is the  quadratic fluctuation operator  in the background of  the instanton and   prime in  ${\det' {\bf M} }$ denotes removal of the  zero mode, 
 and  $\det M_0$ is  for normalization.   
 The determinant can be evaluated in    multiple different ways \cite{ZinnJustin:2002ru, Marino:2012zq}, for example, via the Gelfand-Yaglom method   \cite{Dunne:2007rt}, and gives 
\begin{align} 
 \frac{ \det' M }{ \det M_0}  =  \left\{ 
\begin{array}{cc}   \frac{1}{12}  & \qquad   ({\rm DW}),  \cr 
  \frac{1}{4}  &  \qquad ({\rm SG}),  
\end{array}  \right.
 \end{align}
 resulting in the instanton amplitudes given in  \eqref{instanton-amp}.


\noindent{\bf Fermionic zero modes and index theorem:}  In the theory with fermions, an instanton has a certain number of fermion 
zero modes, as a result of an index theorem for the one dimensional Dirac 
operator. Defining 
\begin{eqnarray}
D= +d_t + W''(x_I),  \qquad  D^{\dagger} = - d_t + W''(x_I),
\end{eqnarray}
the index is equal to the spectral asymmetry of the Dirac operator. For 
each flavor of fermions, 
\begin{eqnarray} 
\label{index}
n_\psi- n_{\bar \psi}  
 &=& {\rm Index}
     = {\rm dim \;  ker} (D)  - {\rm dim \;ker} (D^{\dagger} )  \cr \cr
 &=& \frac{1}{2} \left[ {\rm sign}  \{W''(x_I(t=\infty))\}  - {\rm sign}  
           W''(x_I(t=-\infty))  \right]  \cr \cr
 &=&  \frac{1}{2} \left[  (-1)^{\mu_i}  - (-1)^{\mu_{i+1}}  \right]  \cr
 &=& \left \{ \begin{array} {ll}  
	       1 & \qquad  { \rm for } \; \mu_i = 1,\cr  
             - 1 & \qquad  { \rm for } \; \mu_i = 0.
              \end{array} \right.
\end{eqnarray}
Thus, the ``instanton amplitudes"  differ for an instanton interpolating 
from $x_i$ to $x_{i+1}$ relative to the consecutive instanton interpolating 
from $x_{i+1}$ to $x_{i+2}$. In particular, one has 
\begin{eqnarray}
\I_{i, i+1} &&  \sim  \sqrt { \tfrac{S_I}{2 \pi}}   e^{-S_I}   
``   \psi_{1}  \psi_{2} \ldots   \psi_{{N_f}}  " \, , 
  \cr 	
{\I}_{i+1, i+2}&&\sim  \sqrt { \tfrac{S_I}{2 \pi}}  e^{-S_I}   
   ``\bar \psi_{1}  \bar \psi_{2} \ldots  \bar  \psi_{{N_f}} " \, .   
\label{amplitude}
\end{eqnarray}	
The anti-instanton amplitudes are given by the conjugates of these.  We put 
the annihilation and creation operators $\psi_i,\bar\psi_i$  in quotes is 
to remind the reader that this is a symbolic notation, inspired by the 
index theorem. But the meaning is different from QFTs, where instantons 
induce an anomaly. In quantum mechanics, of course, there is no anomaly, 
and the instanton amplitudes do not violate fermion number symmetry (or spin),
see Section \ref{sec:anomaly}.


\subsection{Difference between QFT and QM: Hilbert's hotel  or not} 
\label{sec:anomaly}

 There is a significant difference between the physical implications of 
the index theorem for the Dirac operator in 1d quantum mechanics compared
to QCD or other QFTs. In  QCD, in the background of an instanton, there 
is chiral charge non-conservation, i.e. the $U(1)_A$ is anomalous. A left 
handed particle can annihilate and create a right handed one, violating 
$N_R-N_L$ by two units. In the Euclidean formulation, the density of these 
events is finite, thus, in an infinite volume, this process may happen 
infinitely many times. This does not lead to any pathology in QFT because 
there is an (infinite) Dirac sea reservoir for both chiralities, i.e, the 
Dirac sea is a ``Hilbert hotel". 

  In quantum mechanics, there is no anomaly because of the finiteness of 
the number of degrees of freedom and ultimately, due to the absence of the
Dirac sea. In particular, there is no fermion number anomaly.  Despite 
the fact that the instanton amplitude looks like it violates fermion 
number by $N_f$, this does not imply an anomaly. 
  
 Consider QCD with one-flavor Dirac fermion. The instanton reduces the
$U(1)_A$ symmetry down to ${\mathbb Z}_2$ because of the anomaly.  In
contrast, consider $N_f=2$ QM. The instanton amplitude is formally the 
same,  and one may think that this also reduces $U(1)_F$ down to 
${\mathbb Z}_2$. However, this is incorrect, because of the absence of 
a fermion number anomaly: 
\begin{align}
&\I_{4d} \sim e^{-S_I} \psi_L \psi_R   \implies  
      U(1)_A \longrightarrow   {\mathbb Z}_2  \qquad  {\rm for \; QCD,} \cr
&\I_{1d} \sim e^{-S_I} \psi_1 \psi_2  \;  \centernot \implies  
      U(1)_F \longrightarrow   {\mathbb Z}_2  \qquad     {\rm for \; QM.} 
\end{align}
However while in QCD the instanton vertex $\psi_L\psi_R$ is a genuine operator 
accompanying an instanton in the low energy effective theory, in QM writing 
$\psi_1\psi_2$ is symbolic and it implies the existence of two zero-modes. 
The meaning of $\psi_1\psi_2$ is that the instanton \emph{saturates} a matrix 
element   connecting states whose fermion number differs by two units.

	
Further, let $|{\mathsf G_{i}}\rangle$ denote a ground state of either theory. 
Then the difference between QFT and QM is that in QCD, for the present example 
of 1-flavor QCD, the ground state is unique, and $U(1)_A$ breaking operators 
have a non-zero expectation value in the ground state. In QM, on the other 
hand, there is more than one ground state. The operator charged under 
$U(1)_F$ has vanishing matrix elements in each of the ground states. Instead,
the operator $ \hat\psi_1 \hat\psi_2$ has non-zero matrix elements
between states whose fermion number differs by two units.  
\begin{align}
& {\rm for \; QCD:}   \qquad 
   \langle {\mathsf G_i} |  \hat\psi_L \hat\psi_R| {\mathsf G_i} \rangle  
           \neq 0 \, ,   \qquad  \cr \cr
& {\rm for \; QM:}     \qquad  
   \left \{  \begin{array}{l}
   \langle  {\mathsf G_i} |\hat\psi_1\hat\psi_2  \;  | {\mathsf G_i}\rangle  
      = 0 \, ,   \cr \cr
   \langle  {\mathsf G_j} |\hat\psi_1\hat\psi_2  \;  | {\mathsf G_i}\rangle  
     \neq  0 \, .    \qquad  \exists (i, j), \;\;\;   (i\neq j).
    \end{array} \right.
\end{align}
Indeed, in the semi-classical approximation,  two consecutive harmonic 
vacua are, respectively, full and empty.  The alternating vacuum structure 
and alternating index theorem conspire to give an alternating harmonic 
vacuum  chain among the perturbative vacuum states:  
\begin{align}
 \underbrace{\longrightarrow}_{\I \sim \bar\psi_1 \ldots \bar\psi_{N_f} }|0 \rangle_i  
      \otimes  |  \widetilde \O \rangle_i   
\underbrace{\longrightarrow}_{\I_i  \sim  \psi_1 \ldots  \psi_{N_f}  }  |0 \rangle_{i+1}
      \otimes  |  \O \rangle_{i+1}    
\underbrace{\longrightarrow}_{\I_{i+1}\sim\bar\psi_1 \ldots\bar\psi_{N_f} }|0\rangle_{i+2}
      \otimes  |  \widetilde \O \rangle_{i+2}   
\underbrace{\longrightarrow}_{\I\sim  \psi_1 \ldots  \psi_{N_f} }
\end{align}
Starting with a fully occupied harmonic state an instanton event in QM is 
associated with a transition matrix element with the insertion of operator  
$\hat \psi_1 \ldots  \hat \psi_{N_f}$ for which the fermion number changes  
by $N_f$. Thus, the subsequent state is empty. The next instanton event is 
associated with a fermion number increasing matrix element, and  refills 
all $N_f$ unoccupied states, and so and so forth. One cannot have two 
consecutive instanton events where all are reducing the fermion number. 
This clashes with finiteness of the Fermion Fock space.  Unlike the chiral 
charge in QCD which does not commute with the QCD Hamiltonian,  the 
fermion number  operator  (i.e. z-component of Spin) {\it commutes} with 
the Hamiltonian \eqref{Ham-nf}. As a result, there are a multitude of 
interesting and somehow unconventional instanton effects in this class 
of quantum mechanical theories, discussed in the next section. 
 
	
 To summarize, we show that the following three concepts are intertwined: 
\begin{itemize}
\item{Finiteness of Fock space, or  absence of Dirac sea.}
\item{Alternating index  for instantons, absence of fermion number anomaly.}
\item{Alternating  fermion numbers in perturbative vacua in harmonic 
approximation.}
\end{itemize}

\subsection{Mixing of harmonic states in QM  with fermions: New instanton 
effects}

For simplicity, we discuss a particle with internal spin $(\half)^{N_f}$ in 
a DW potential. This has almost all  interesting features  that also takes 
place for more general potentials. 


 {\bf $\bm {N_f=0}$ (reminder):}   The theory is bosonic and the states on 
either well are harmonic oscillator eigenstates, $|L, n\rangle$ and $|R, n
\rangle$.  These states are degenerate to all orders in perturbation theory.  
Non-perturbatively, the degeneracy between  $|L,0\rangle$ and  $|R,0\rangle$ 
is lifted due to tunneling/instanton events. The ground state is the parity 
even combination  and the first excited state is the parity odd combination 
of the two lowest lying modes. These are  
\begin{eqnarray}
 |\Psi_\epsilon  \rangle 
    = \frac{1}{\sqrt 2} ( |L,0\rangle + \epsilon| R,0 \rangle),  \qquad 
 {\bf P}    | \Psi_\epsilon  \rangle 
    = \epsilon  | \Psi_\epsilon  \rangle,  \qquad \epsilon= \pm, 
\label{GS-DW}
\end{eqnarray}
where  ${\bf P}$ is the parity operator.  The tunneling amplitude is  equal 
to the instanton amplitude  given in  \eqref{instanton-amp}:
\begin{eqnarray}
 \langle -a  | e^{-TH} |a \rangle = {\cal N}\,\I=  
   {\cal N}\sqrt  \frac{6S_I}{\pi}  e^{-S_I}\, ,
 \label{tunnel-b}
\end{eqnarray}  
where ${\cal N}=\frac{1}{2}\langle R,0|a\rangle^2 (\omega T)e^{-\omega T}$.
Consequently, the  {\it non-perturbative  splitting} or lifting of the  
perturbative two-fold degeneracy  is an instanton effect  $(E_{-}-E_{+}) 
/\omega= \sqrt \frac{6S_I}{\pi} e^{-S_I}$ at leading non-perturbative order.   
The reason for restating this well-known result is that instantons do 
not lead to the level splitting effects in 
theories with $N_f \geq 1$, as described below.


{\bf $\bm {N_f=1}$ (supersymmetric) theory:} 	 
In the harmonic approximation the states in both the left or the right well
are described by the Hilbert space of supersymmetric harmonic oscillator. 	The lowest lying states in  ${\cal H}_L$ and ${\cal H}_R$ as well as the 
corresponding eigen-energies  are:
\begin{eqnarray}
&& E_3=3   \qquad   \overbrace{  |L, 2\rangle | 0  \rangle }^{\cal B} 
  \longleftrightarrow   
        \overbrace{|L,  3 \rangle  |1\rangle  }^{ \cal F} 
  \qquad \qquad  
        \overbrace{|R, 3 \rangle  |0\rangle }^{\cal B}  
  \longleftrightarrow   
        \overbrace{|R, 2 \rangle  |1\rangle  }^{\cal F} \cr
&& E_2=2   \qquad   |L, 1\rangle  |0\rangle 
  \longleftrightarrow 
                   |L,  2 \rangle |1  \rangle  
	  \qquad   \qquad   
                   |R, 2 \rangle  |0  \rangle   
  \longleftrightarrow 
                   |R, 1 \rangle  |1  \rangle 
	\cr 
&& E_1=1   \qquad  |L, 0\rangle |0  \rangle 
  \longleftrightarrow 
                  |L,  1 \rangle  |1 \rangle 
  \qquad   \qquad 
                  |R, 1 \rangle  | 0  \rangle  
  \longleftrightarrow 
                  |R, 0 \rangle   | 1  \rangle  
	  \cr 
&& E_0=0 \qquad   \underbrace{   \quad \qquad  \qquad    \;\;\;  
                  |L,0 \rangle |1  \rangle}_{{\cal H}_L}   
  \qquad   \;\;\;\;\;\;
  \underbrace{  
                  |R, 0 \rangle   | 0  \rangle 
  \qquad \qquad \qquad   \quad }_{  {\cal H}_R}  
\end{eqnarray}  
In the supersymmetric theory, the two perturbative ground states 
$ |L, 0\rangle |1\rangle$ and  $ |R, 0 \rangle |0 \rangle $ cannot mix 
because of the conservation of  fermion number or spin. As emphasized in  
Sec.\ref{sec:anomaly}, there is no fermion number anomaly,  i.e., 
fermion number operator commutes with the Hamiltonian: 
\begin{align}
\label{N-H-com}
\big [H,  {\hat N_f}  \big] =\big [H,  \sigma_3 \big] = 0 .
\end{align}
Thus, the  matrix element of  $e^{-TH}$ connecting left harmonic ground state to right one vanishes: 
\begin{align}
\label{tunnel-f1}
  \langle -a  | \langle 1 |   e^{-TH} |a \rangle  | 0\rangle  =  0, 
\end{align}
in contrast to the bosonic case \eqref{tunnel-b}, which implies that 
there is no  splitting of the two lowest lying states.  Instead, the 
energy of both states gets simultaneously lifted, and because of 
supersymmetry the shift is exactly the same for both states.  To leading 
order in $\hbar$, the lifting is due to mixing between the harmonic {\it  
ground state} of ${\cal H}_L$, and the {\it  first excited state} of 
${\cal H}_R$. For the other ground state the role of $L$ and $R$ is
exchanged.


 Therefore, in terms of the harmonic states of the left and right well,
the two non-perturbative ground state wave functions are given by
\begin{eqnarray}
&& | \Psi_0\rangle | 1 \rangle 
 = a_1 \underbrace{|L, 0\rangle |1\rangle}_{\rm Ground \;  state  \; in \; {\cal H}_L}
 + a_2 \underbrace{|R, 0\rangle |1\rangle}_{\rm First-excited \;  state  \; in \; 
        {\cal H}_R}     +  \ldots \cr \cr
&& | \Psi_0\rangle | 0  \rangle 
 = a_1 \underbrace{|R, 0\rangle |0\rangle}_{\rm Ground \;  state  \; in \; {\cal H}_R} 
 + a_2 \underbrace{|L, 0\rangle |0\rangle}_{\rm First-excited \;  state  \; in \; 
        {\cal H}_L} +  \ldots\, . 
\end{eqnarray} 
Unlike the bosonic system in which the ground state must be unique, in the 
theory with fermions, the ground state is {\it two-fold degenerate}. The 
ground state energy can be found by calculating the expectation value of 
the Hamiltonian in either one of the two true ground states. We have 
\begin{eqnarray}
E_0&=& \langle  \Psi_0| \langle 0 | \hat H |0\rangle | \Psi_0  \rangle 
    = \half\langle  \Psi_0| \langle 0 |  \{Q, \overline Q\}   
                 |0\rangle | \Psi_0  \rangle 
    = \half\langle  \Psi_0| \langle 0 |  Q \overline Q   
                 |0\rangle | \Psi_0  \rangle =   \cr 
   &=&  \half  \langle  \Psi_0| \langle 0    | Q   
          \left( \sum_{n, \sigma=0,1}  | \sigma \rangle |\Psi_n\rangle  \langle 
          \Psi_n|  \langle \sigma | \right)    \overline Q  
                          |0\rangle | \Psi_0  \rangle 
    = \half   \sum_{n}   | \langle  \Psi_n| \langle 1    |  \overline Q  
                   |0\rangle | \Psi_0  \rangle|^2 \cr
   &\approx&  \half    | \langle  \Psi_0| \langle 1    |  \overline Q  
          |0\rangle | \Psi_0  \rangle |^2, 
\end{eqnarray} 
where $\bar{Q}=(p+iW') \hpp$ and $Q=(p-iW') \hpm$ are the supercharges, 
and in the final line we inserted a complete set of states. Note that since  
$\overline Q$ commutes with Hamiltonian,  but anti-commutes commute with 
$(-1)^F$, the sum over $\sigma$ reduces to $\sigma=1$. For the final
estimate, we kept only the dominant $n=0$ state in the sum. 
	
	
 The ground-state energy is therefore the square of a matrix element, 
$E_0= \half |\epsilon|^2$, where $\epsilon= \langle  \Psi_0| \langle 1    
|\overline Q  |0\rangle | \Psi_0  \rangle  $. This is the quantity 
saturated by an instanton. Instanton processes are associated with the 	
matrix element $\langle  \Psi_0 | \langle 1 | \overline Q | 0\rangle  
|\Psi_0 \rangle$ between the two ground states,  which have different 
fermion numbers (or spin). In the functional integral formulation, this  
result can be written as:
\begin{eqnarray}
 \langle  \Psi_0 | \langle 1 | \overline Q | 0\rangle  |\Psi_0 \rangle     
\sim \langle -a | \langle 1 |  \bar{Q}   | +a \rangle | 0  \rangle    
\sim \int_{x( -\infty)= -a }^{x( \infty)= +a } Dx D \psi D\bar \psi  e^{-S} \; 
         (p + i W') \psi \, . 
\end{eqnarray}
The evaluation of the integral gives  $\epsilon \sim e^{-S_I}$.  See 
\cite{Salomonson:1981ug} for details. The shift in the ground state 
energy is  
\begin{eqnarray}
E_0 \sim e^{-2S_I}\, . 
\label{gse}
\end{eqnarray} 
This is  a second order effect in semi-classical expansion, and it can be 
interpreted in terms of a dilute gas of correlated  instanton-anti-instanton 
$[{\cal I} \bar {\cal I}]$ events. However, this formulation does not tell 
us if the  $[{\cal I} \bar {\cal I}]$  configuration is an exact solution or not.  In fact, there is 
reason to think otherwise.  Since the BPS equations are non-linear, the superposition of the two configuration 
is not a solution to that BPS equation, i.e, the configuration is  non-BPS.  

 In the graded formulation,  
we will prove that an {\it exact non-BPS complex saddle} is responsible 
for the ground state energy \eqref{gse}, and that $[{\cal I} \bar {\cal I}]$ (evaluated on appropriate complex quasi-zero mode thimble,  not over the naive ``separation" cycle)  is an approximation to the  exact solution.  We will also show that the exact saddles  are solutions to holomorphic and quantum modified second order 
Euclidean equations of motions. 


\begin{figure}[t]
\includegraphics[angle=0, width=0.9\textwidth]{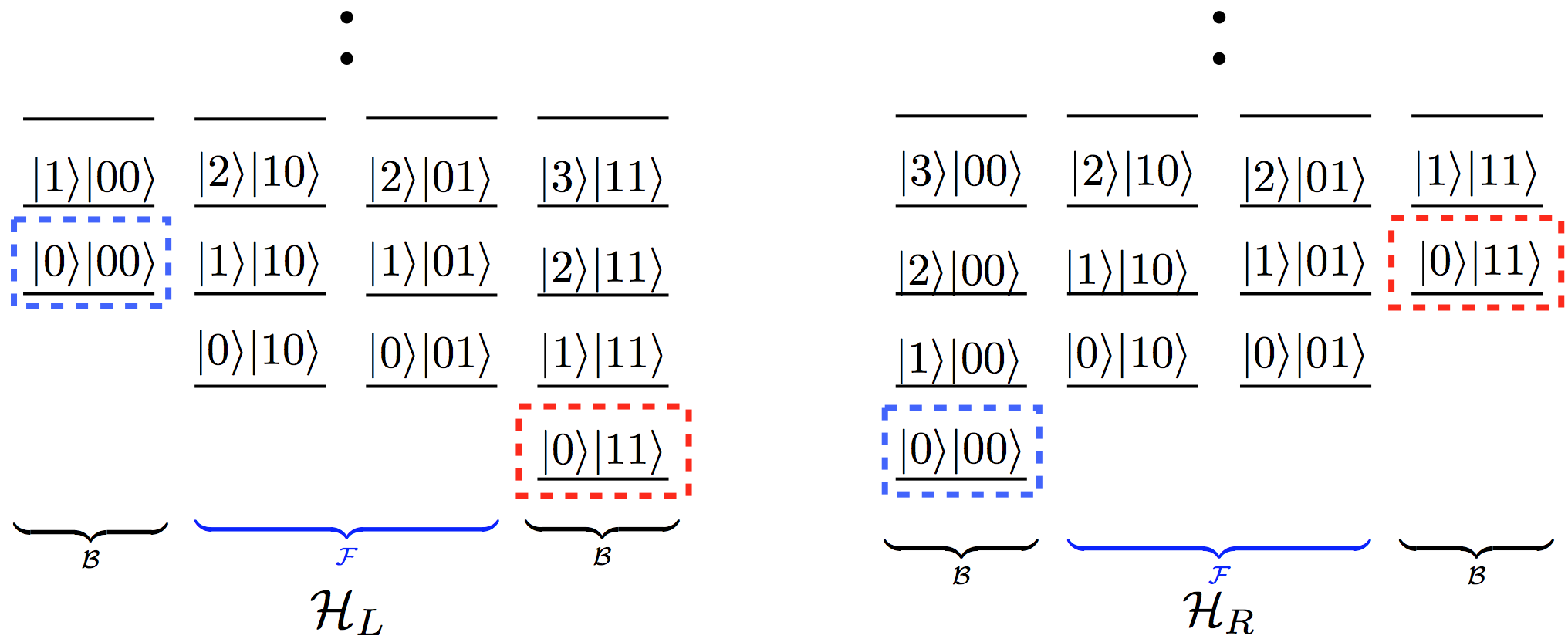}
\caption{ 
Hilbert-space on the left and right well in the $N_f=2$ (multi-flavor) theory 
in harmonic approximation. Perturbative ground state on the left well is 
fully occupied, while the one on the right is empty.  These two cannot mix (form linear combinations) due 
to conservation of  fermion number or spin.
Non-perturbatively, 
the theory has two ground states.  One is (approximately)  a linear 
superposition of ground state on the left and {\it second} excited state 
on the right (with the same fermion occupation number, the ones in red 
boxes),  and the other is the linear superposition of the states in 
blue boxes. \label{fig:LR-Hilbert}}
\end{figure}
 	
{\bf $\bm {N_f \geq 2}$ (non-supersymmetric, multi-flavor) theory:} 
 In harmonic approximation the states in the left and right well are 
described by a natural generalization of the Hilbert space of the 
supersymmetric harmonic oscillator. Our construction  follows  
\cite{Basar:2013sza}. For the free theory, which is relevant in 
the case of the harmonic approximation, one can define $N_f$  
nilpotent conserved fermionic charges: 
\begin{align}
 Q_i = a^{\dag}_b  \hpm^i, \qquad Q_i^2 = 0.
\end{align}
Let $|\psi \rangle= |n\rangle |\Omega\rangle $ be an eigenstate of the 
Hamiltonian $\hat H_R$ with energy $E_{n}$ annihilated by  $Q_i,  \forall 
i = 1, \ldots, N_f$.  It is possible to generate all states in the right 
harmonic  Hilbert space ${\cal H}_R$ by acting with the $Q^{\dagger}_i$'s on  
$|\psi\rangle$. Consider the set:
\begin{align}
 S = \{  |\psi\rangle, \; Q_i^{\dag} |\psi\rangle, \;  
         Q_i^{\dag} Q_j^{\dag} |\psi\rangle, \; \ldots,  \; 
         Q_i^{\dag} \ldots Q_{i_{N_f}}^{\dag} |\psi\rangle\}\, .
\end{align}
If $n \geq N_f$, this procedure, which goes parallel to our construction 
of the fermionic Fock space in Section \ref{ffs}, creates $2^{N_f}$ 
degenerate states at each level:  If  $n < N_f$, some combination of the
operators $Q_{i_1}^{\dag}\ldots Q_{i_k}^{\dag} $ annihilates $|\psi \rangle$.  
For example, in the case $N_f=2$, the lowest lying states in ${\cal H}_L$ 
and their eigen-energies are given by:
\begin{eqnarray}
&&  E_3={\textstyle + \frac{5}{2} }     \qquad   
           |L, 1 \rangle | 00 \rangle 
    \longleftrightarrow 
           |L, 2 \rangle  |10    \rangle 
    \longleftrightarrow 
           |L, 2 \rangle  |01  \rangle 
    \longleftrightarrow  
           |L, 3 \rangle  |11 \rangle   \cr 
&&  E_2={\textstyle + \frac{3}{2} }     \qquad   
           |L, 0 \rangle | 00 \rangle 
    \longleftrightarrow 
           |L, 1 \rangle  |10   \rangle 
    \longleftrightarrow 
           |L, 1 \rangle  |01  \rangle 
    \longleftrightarrow  
           |L, 2 \rangle  |11 \rangle	   \cr 	 
&&  E_1= + \half  \qquad  \qquad \qquad \qquad  \;
	   |L, 0 \rangle  |10   \rangle 
    \longleftrightarrow 
           |L, 0 \rangle | 01   \rangle  
    \longleftrightarrow 
           |L, 1 \rangle | 11  \rangle    \cr 
&&  E_0=-\half   \qquad  \qquad \qquad \qquad \qquad 
    \qquad \qquad\qquad \qquad \qquad \quad  
           |L, 0 \rangle |  11  \rangle  
\end{eqnarray}
and analogously, the eigen-states in ${\cal H}_R$ and their eigen-energies are
\begin{eqnarray}
&&  E_3={\textstyle  + \frac{5}{2} }     \qquad   
           |R, 3 \rangle | 00  \rangle 
      \longleftrightarrow 
           |R, 2 \rangle  |10  \rangle
      \longleftrightarrow 
           |R, 2 \rangle  | 01  \rangle 
      \longleftrightarrow  
           |R, 1 \rangle  |11 \rangle		 \cr 
&&  E_2={\textstyle + \frac{3}{2} }     \qquad   
           |R, 2 \rangle | 00  \rangle 
      \longleftrightarrow 
           |R, 1 \rangle  |10    \rangle 
      \longleftrightarrow 
           |R, 1 \rangle  |01  \rangle 
      \longleftrightarrow  
           |R, 0 \rangle  |11 \rangle	   \cr 
&&  E_1= + \half  \qquad  
           |R, 1 \rangle  |00   \rangle 
      \longleftrightarrow 
           |R, 0 \rangle |10 \rangle  
      \longleftrightarrow 
           |R, 0 \rangle |  01  \rangle    \cr 
&&    E_0=-\half  \qquad   
           |R, 0 \rangle | 00  \rangle  \, . 
\end{eqnarray}
As in the $N_f=1$  supersymmetric theory, for $N_f >1$ as well,  the two 
perturbative ground states $ |L, 0\rangle | \widetilde  \Omega \rangle$ and  
$|R, 0 \rangle | \Omega  \rangle $ cannot mix because of the conservation 
of  fermion number \eqref{N-H-com}. Therefore, the transition matrix element vanishes:
\begin{align}
\label{tunnel-f2}
 \langle -a  | \langle \widetilde  \Omega |e^{-TH} |a \rangle | \Omega\rangle  
 =  0,  \qquad \forall \;  N_f  \; \geq 1 \, . 
 \end{align}
Similar to the supersymmetric theory, and again unlike the bosonic theory, 
this implies that there will be  no splitting of the two lowest lying states. 
Instead, the energy of both states gets simultaneously lifted, and they do 
get lifted exactly by the same amount. 
	 	 
		 
 To leading order in $\hbar$, the lifting is due to mixing between the 
ground state of ${\cal H}_L$,  $|L, 0 \rangle  | \widetilde \Omega\rangle$ 
and the  $N_f^{\rm th}$ excited state of ${\cal H}_R$,  $|R,0 \rangle |\widetilde
\Omega \rangle $, and vice versa for the other ground state.\footnote{The 
mixing between the lowest state on one side  with the $N_f^{\rm th}$ excited 
state on the other side is also present for the $N_f=0$ (bosonic) and 
$N_f=1$ (supersymmetric) examples. For general $N_f$, this is a new result.}
Therefore, the two non-perturbative ground state wave functions are 
\begin{eqnarray}
&& | \Psi_0\rangle | \widetilde \Omega    \rangle 
 = a_1  \underbrace{|L, 0 \rangle |\widetilde\Omega
         \rangle}_{\rm Ground \;  state  \; in \; {\cal H}_L}  
 + a_2  \underbrace{ |R,  0 \rangle | \widetilde \Omega    
         \rangle}_{\rm N_f^{\rm th}-excited \;    state  \; in \; {\cal H}_R}     
  +  \ldots \cr \cr
&& |  \Psi_0\rangle |  \Omega   \rangle 
 = a_1 \underbrace{ |R, 0\rangle |  \Omega  
         \rangle}_{\rm Ground \;  state  \; in \; {\cal H}_R} 
 + a_2  \underbrace{ |L,  0\rangle |  \Omega  
         \rangle}_{\rm N_f^{\rm th}-excited\;  state  \; in \; {\cal H}_L} +  \ldots
\label{g-s-2}
\end{eqnarray} 
Unlike the bosonic system in which the ground state must be unique, for the 
theory with fermions, the ground state is {\it two-fold degenerate}. The 
ground state energy can be found by calculating the expectation value of the 
Hamiltonian in either one of the two true ground states.  
 
 
 In contrast to the supersymmetric theory, the shift in the ground state
energy has perturbative and non-perturbative contributions. We have 
\begin{eqnarray}
E_0= \langle  \Psi_0| \langle \Omega |\hat H |\Omega \rangle | \Psi_0 \rangle 
   = E_0^{\rm pert}(g) + E_0^{\rm non-pert} (g) \, , 
\end{eqnarray} 
where the non-perturbative part is a second order effect in the semi-classical 
expansion,
\begin{eqnarray}
 E_0^{\rm non-pert} (g)  
 \sim  | \langle  \Psi_0| \langle  \widetilde \Omega |  \bar Q_{N_f}  
           | \Omega \rangle | \Psi_0  \rangle |^2 \, , 
\end{eqnarray} 
given by a square of a matrix element,   $\epsilon=\langle\Psi_0| \langle  
\widetilde \Omega | \overline Q_{N_f}    | \Omega \rangle | \Psi_0  \rangle$. Because 
the instanton contributes to a fermion number changing matrix element, and 
because fermion number commutes with Hamiltonian, the non-perturbative 
contribution to ground state energy  can only start at second order 
in semi-classical expansion. In terms of a functional integral we can write
\begin{eqnarray}
\epsilon &=& \langle  \Psi_0| \langle  \widetilde \Omega |  
          \hat Q_{N_f}  | \Omega \rangle | \Psi_0  \rangle 
 \sim  \langle  -a | \langle  \widetilde \Omega    |  
          \hat Q_{N_f}  | \Omega \rangle | +a  \rangle  \cr
 &=&
 \int_{x( -\infty)= -a }^{x( \infty)= +a } Dx  
  \prod_{i=1}^{N_f} D \psi_i D\bar \psi_i   \; 
      e^{-S} \; \psi_1\psi_2 \ldots \psi_{N_f} \, . 
\end{eqnarray}
The evaluation of this integral gives  $\epsilon \sim e^{-S_I}$,  because the  $\psi_1\psi_2 \ldots \psi_{N_f} $ inversion  gets soaked up 
to the unpaired zero modes in the measure $D \psi_i (t) D\bar \psi_i(t)$.  
The non-perturbative shift in the ground state energy is 
\begin{eqnarray}
E_0^{\rm non-pert} (g) \sim e^{-2S_I}\, . 
\label{gse-2}
\end{eqnarray} 
Similar to  the supersymmetric case, this can be interpreted in terms of 
a dilute gas of  correlated instanton-anti-instanton $[{\cal I}\bar{\cal I}]$ 
events. Similar to the supersymmetric theory,  in the graded formulation,  
we  prove that an {\it exact non-BPS complex saddle} is responsible 
for the  non-perturbative lifting of the ground state energy,  and that $[{\cal I} \bar {\cal I}]$   is an approximation to the  exact solution.

\section{Approximate versions of exact solutions from  quasi-zero mode thimbles}
\label{correlated}

{\scriptsize{\bf Outline:} 
We discuss  correlated two-instanton events  on the  complexified quasi-zero 
mode Lefschetz thimbles. The treatment of the QZM integration over the  
Lefschetz thimbles represents a reduced version of the  complexification of  
the full  field space. The conventional wisdom is that instanton-anti-instanton 
events can at best be approximate solutions, due to the non-linearity of the 
underlying BPS-equations. This intuition turns out to be incorrect, as we 
later find exact saddles.  Remarkably, the Lefschetz thimble treatment of 
the multi-instantons  provides a systematic approximation to the exact result 
and  reproduces the important features of the exact solutions discussed
in Section \ref{sec:graded-dw}  and \ref{sec:periodic}. }

\subsection{Multi-instantons and  boson and fermion induced interactions}

 In the previous section we saw that for $N_f \geq 1$  the non-perturbative 
contribution to the ground state energy is of the form $e^{-2S_I}$, rather 
than $e^{-S_I}$. In the graded theory, we will in fact find an {\it exact 
non-BPS  bion saddle}  which is responsible for this ground state energy 
\eqref{gse}. However, in order to understand the connection between the 
exact saddle and the instantons in the original description, we first sketch 
the calculation of the contribution  to correlated 2-events $[{\cal I}  
{\cal I}]$ or $[{\cal I}\bar {\cal I}]$ in the  original formulation.   
Note that in the double-well potential we only have $[{\cal I}\bar {\cal I}]$, 
while in the periodic potential, we have both.  

	
 The instanton amplitude is given in \eqref{amplitude}. As discussed earlier, 
an instanton has an exact position modulus. If we consider a two instanton 
configuration then the center position of the two is still an exact zero mode, 
but the relative separation $\tau$ is a quasi-zero mode.  The reason for this 
fact is the interaction (in the euclidean sense) between the two instantons. 
The bosonic interaction can be described as due to the exchange of bosonic 
fluctuations, derived in detail in \cite{ZinnJustin:2002ru}. 
The fermion induced interaction can be deduced either 
by studying the fermionic determinant in the background of an 
instanton-(anti)instanton pair, or equivalently, by looking to the connected 
correlator of the fermion zero mode structure $\langle \psi_1(0) \ldots 
\psi_{N_f}(0) \bar \psi_1(\tau) \ldots \bar \psi_{N_f}(\tau) \rangle 
\sim e^{-N_f m_b \tau}$ \cite{Behtash:2015kva,  Balitsky:1985in}.
Here, instead of these technical arguments, we provide a more intuitive physical description. 
 
An instanton will interact with an anti-instanton due to the spin interaction with the ``magnetic field'' $W''(x)$.  The  ``magnetic field'' at the two classical vacua are $W''(\pm a)=\pm 2a=\pm \omega$, where $\omega$ is the natural frequency.   The spin-dependent part of the Hamiltonian  is $-S_z W''(x)$, where $S_z \in[-N_f/2,N_f/2]$ is the spin. Therefore the ground state in the left well will have $S_z=N_f/2$ (i.e. all spins up) while the ground state on the right will have $s=-N_f/2$ (i.e. all spins down) (see Fig. \ref{fig:LR-Hilbert} ). Now let us consider an instanton--anti-instanton pair interpolating from vacuum at $x=-a$ to the vacuum at $x=+a$ and back, and let the time that this configuration spends at $x=+a$ be $\tau$. Since the configuration spends an infinite amount of time in the vacuum $x=-a$, the spin $s$ must be $s=+N_f/2$, and it cannot change along the instanton path, as spin is   conserved. Then, the spin induced action cost is the penalty for the Euclidean time spent in the "false vacuum".  
 Then the action cost (compared to the vacuum) will  be a product of $2\omega$ (the change in the magnetic field from left well to the right well), the spin $s=N_f/2$ and the time spent in the right well $\tau$. Therefore $S_{int}= N_f \omega \tau$.

 We denote the interaction potential between two  
instantons by ${\cal V}_{+}(z)$,  and the one between an instanton and 
anti-instanton by ${\cal V}_{-}(z)$. 
\begin{align}
\label{interaction}
{\cal V}_{\pm}(\tau) =  \pm \frac {Aa^3}{g} e^{-m_b \tau} + N_f m_b \tau \qquad  
A=  \left \{ \begin{array} {ll} 16  \qquad & {\rm for \; DW,} \cr
32 \qquad  & {\rm for \; SG.} 
\end{array}  \right.
\end{align}
In both cases the fermion zero mode induced interaction is attractive. 
For  ${\cal V}_{+}(z)$ the boson induced interaction is repulsive, and 
for  ${\cal V}_{-}(z)$ the bosonic part is attractive. In the space of 
fields (or paths), the quasi-zero mode direction is non-Gaussian, and 
the integral over the QZM needs to be treated exactly in order to obtain 
the correlated 2-event amplitudes.  

\begin{figure}[t] 
\centering
\includegraphics[width=14cm]{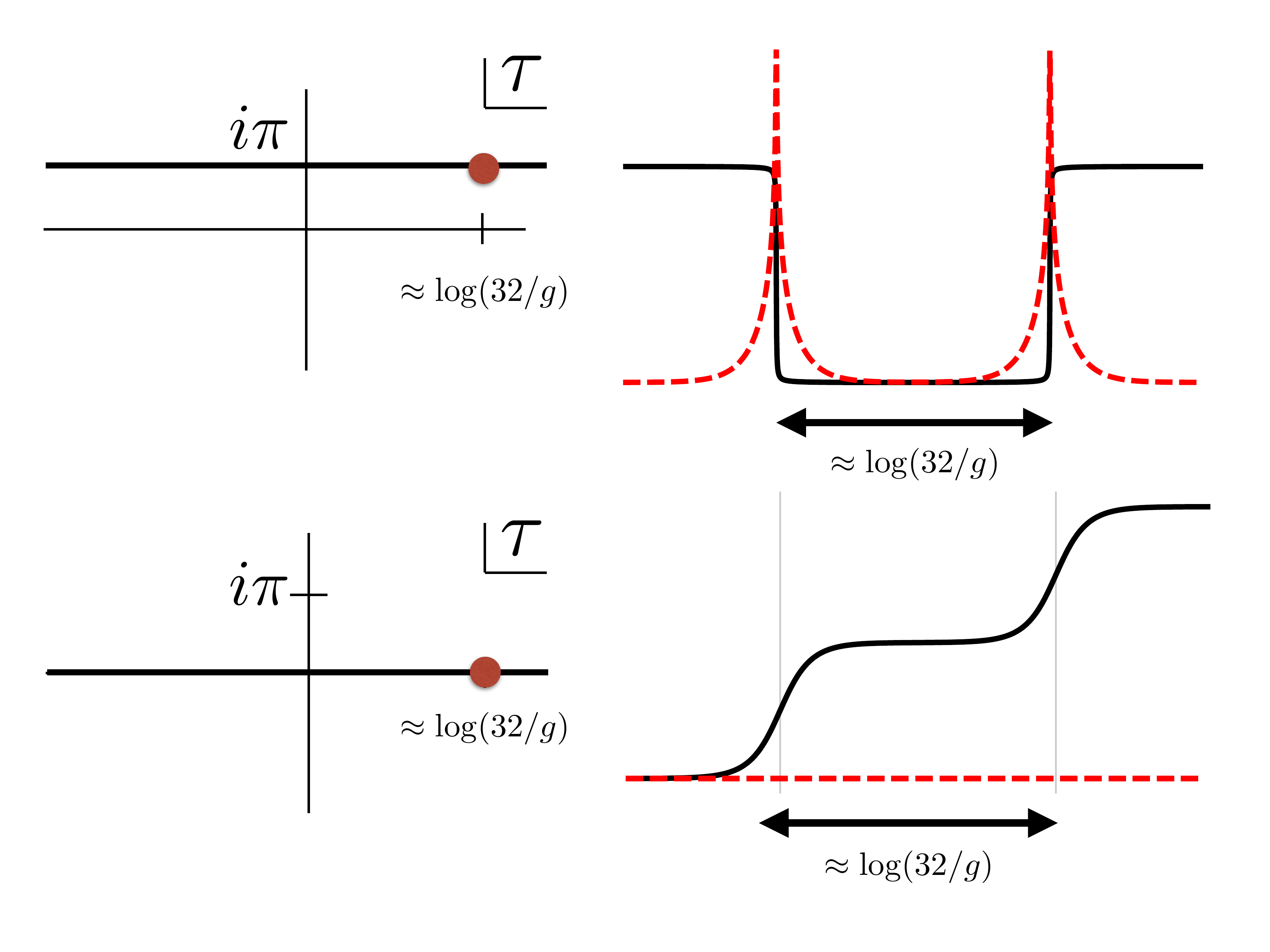}  
\caption{The quasi-zero mode (QZM) integral  between two instanton events 
has a critical point at a finite (real) separation. The QZM integration 
for the instanton-anti-instanton has a a critical point at a finite 
(complex) separation. It is usually believed that two-events can at best 
be quasi-solutions due to non-linearity  of the BPS equations. In this 
work, we prove that the figures on the right are actually exact non-BPS 
solutions to holomorphic Newton equations.  We call these solutions  
complex (top) and real (bottom)  bions. The figures are  for $N_f=1$ 
and periodic potential.}
\label{tau-plane}
\end{figure}

\subsection{$[{\cal I}{\cal I}]$ thimble  integration and approximate form of real bion }
The QZM integration for the  $[{\cal I}{\cal I}]$ correlated event is 
 \begin{equation}
 I_{+}(N_f, g)=\int_{\Gamma^{\rm qzm}_{+}}  d(m_b\tau)  e^{-  {\cal V}_{+}(\tau)}
=  \int_{\Gamma^{\rm qzm}_{+}}   d(m_b\tau)   
       e^{- \left( \frac {Aa^3}{g} e^{-m_b \tau} + N_f m_b \tau \right)}  \; ,
\end{equation}
where ${\Gamma^{\rm qzm}_{+}}$ is the QZM-cycle for the QZM-integration, shown 
in Fig. \ref{tau-plane}, lower figures. 
The critical point of the integration is located at 
\begin{align}
\label{critical-1}
\frac{d {\cal V}_{+}(\tau) }{d \tau}  = 0\, ,  \qquad  
\tau^{*} =  m^{-1}_b \ln \left( \frac {Aa^3}{gN_f} \right)\, .  
\end{align}
The descent manifold (Lefschetz thimble) associated with this critical 
point is the real axis, ${\Gamma^{\rm qzm}_{+}} = \mathbb R$, as can be 
deduced by solving the stationary phase condition 
\begin{align}
\Im[ V(\tau) - V(\tau^*)]=0\, . 
\label{sp}
\end{align}   
Substituting  $u=e^{-m_b\tau}$, we map the integral to the standard 
representation of the Gamma-function:
\begin{align}
 I_{+}(N_f, g) =     \int_0^{\infty} du \;  u^{N_f-1} \;  e^{-   \frac{Aa^3}{g} u  }   
=    \left(\frac{g}{Aa^3}\right)^{N_f} \Gamma(N_f) .
 \end{align}
The $[\I\I]$  event, obtained by integrating over the QZM amplitude, has 
the form 
\begin{align}
[\I\I] & =  I_{+}(N_f, g) \times  [\I]^2   
=   \left(\frac{g}{Aa^3}\right)^{N_f} \Gamma(N_f) \times \tfrac{S_I}{2 \pi}  \textstyle{ \left[\frac{ \det' M }{ \det M_0}\right]^{-1}  } e^{-2S_I} \, . 
\end{align}

\noindent 
{\bf Meaning of the critical point at finite real separation and a puzzle:} 
It is interesting to note that the critical point of the interaction  
between two instantons \eqref{interaction}, unlike the bosonic case $N_f=0$ 
where it is located at infinity $m_b\tau^*=\infty$, is located at a {\it finite} 
real separation \eqref{critical-1}. This phenomena appeared first in field 
theory, in QCD(adj) for magnetic bions  \cite{Unsal:2007jx}, and makes one 
suspect that there may be something ``special" about such correlated 
two-events. The idea is that perhaps these correlated two-events are 
related to some exact solution with size $ \tau^{*} = m^{-1}_b \ln \left( 
\frac {Aa^3}{gN_f} \right)$. There is an apparent problem with this idea,
however. Neither in the BPS-equation, nor in Newton's equation for the 
inverted bosonic potential, does the coupling constant $g$ or the number 
of fermions $N_f$  appear. But $ m_b\tau^{*} $, the characteristic size 
of the correlated-two event depends on both  $g$ and $N_f$. So, it seems impossible for 
these configurations to be solutions to the standard equations of motion 
in the inverted potential.   We will solve this problem in this work, 
and show that the two-instanton correlated event is actually an 
approximation to an exact solution,  that we refer to as a real bion, 
see Section \ref{sec:sg-real-bion}, 
 which solves the equation for  a quantum modified potential.

\subsection{$[{\cal I}\bar {\cal I}]$ thimble integration and approximate 
form of complex  bion}

The QZM integration for the  $[{\cal I}  \bar {\cal I}]$ correlated event is  
\begin{equation}
\label{qzm-mm}
 I_{-}(N_f, g)=\int_{\Gamma^{\rm qzm}_{-}}  d(m_b\tau)   e^{-  {\cal V}_{-}(\tau)}
     =  \int_{\Gamma^{\rm qzm}_{-}}   d(m_b\tau) 
       e^{- \left( - \frac {A a^3}{g} e^{-m_b \tau} + N_f m_b \tau \right)}  \; ,
\end{equation}
where ${\Gamma_{-}^{\rm qzm}}$ is the QZM-cycle for the $[{{\cal I}\bar{\cal I}}]$
QZM-integration. The critical point of the integration is located at a complex 
point: 
\begin{align}
\label{critical-2}
\frac{d {\cal V}_{-}(\tau) }{d \tau}  = 0 \qquad   
\tau^{*} =m^{-1}_b \left[ \ln \left( \frac {Aa^3}{gN_f} \right) \pm i \pi  \right]  = \left\{ 
\begin{array}{ll} 
 \frac{1}{2a}  \left[ \ln \left( \frac {16 a^3}{gN_f} \right) \pm i \pi  \right]  & \qquad   ({\rm DW}),  \cr \cr
 \frac{1}{a}  \; \left[ \ln \left( \frac {32 a^3}{gN_f} \right) \pm i \pi  \right]  &  \qquad ({\rm SG}).  
\end{array}  \right.
\end{align}
The descent manifold associated with this critical point is  in the 
complex plane,  ${\Gamma^{\rm qzm}_{-}} = \mathbb R \pm i \pi$, as can be 
deduced by solving the stationary phase condition \eqref{sp}: see Fig. \ref{tau-plane}. Substituting  
$u=e^{-m\tau}$, we map the integral to the standard representation of 
Gamma-function:
\begin{equation}
 I_{-}(N_f, g)
   = e^{\pm i \pi N_f}    \int_0^{\infty} du \;  u^{N_f-1} \;  e^{-\frac{Aa^3}{g} u  }  
   =  e^{\pm i \pi N_f}  \left(\frac{g}{Aa^3}\right)^{N_f} \Gamma(N_f)\, . 
\end{equation}
The $[\I\bar \I]$  event, obtained by integrating over the QZM-thimble, 
has the form 
\begin{align}
[\I \bar  \I]_{\pm}  
   =  I_{-}(N_f, g) \times  [\I]^2 
   =  e^{\pm i \pi N_f}  \left(\frac{g}{Aa^3}\right)^{N_f} \Gamma(N_f)   \times 
            \tfrac{S_I}{2 \pi}   \textstyle{ \left[\frac{ \det' M }{ \det M_0}\right]^{-1} }  e^{-2S_I} \, . 
 \end{align}
The integration over ${\Gamma_{-}^{\rm qzm}}$-thimbles is a rigorous 
version of the BZJ-prescription, and generalized to the multi-flavor theory.

\noindent
{\bf Meaning of the critical point at finite complex separation, and another 
puzzle:} 
For the instanton-anti-instanton, there is a critical point at a finite complex 
separation.  All the concerns stated in the previous item stand.  On top 
of that, apparently, the quasi-zero mode, a part of field space, is 
necessarily complexified in order to make sense of $[{{\cal I}\bar{\cal I}}]$. 
If we plot, for example, an instanton-anti-instanton separated by $\tau^*$, 
we see that the combination is actually complex, see the top part of
Figure \ref{tau-plane}.  Evidently, this cannot be a solution to the real 
Newton equation for the inverted bosonic potential. We will also solve 
this problem, and show that the  instanton-anti-instanton event is an 
approximation to an exact  complex solution,  that we call complex bion, discussed in Sections 
\ref{sec:dw-complex-bion} and \ref{sec:sg-complex-bion}, 
which solves the holomorphic Newton  equation. 
  
\subsection{Other aspects of thimble integration}

{\bf Meaning of poles at $\bm {N_f=0, -1, -2, \ldots}$} 
The $\Gamma(N_f)$ function is meromorphic in the complexified $N_f$ plane, 
with  poles at $N_f=0,-1, -2, \ldots $ if one analytically continues $N_f  
\in \mathbb N$  to complex numbers. Recall that a Hilbert space interpretation 
requires $N_f \in \mathbb N$, but as a statistical field theory, we are 
free to move to complex $N_f$ and view $N_f$ as a coupling constant 
(parameter)  in the functional integral.  

 
 Thus, the correlated $[\I\I]$-amplitude is meromorphic in the whole 
complex $N_f$ plane except for the poles, which requires an interpretation. 
For $N_f\to 0^+$, the convergence factor due to fermion zero modes that 
regulates the large $\tau$ behavior disappears. The physical explanation
of this phenomenon is that these events are uncorrelated, and are already 
included in the leading order semi-classical result. Expanding the result 
around $\epsilon=0$ we obtain 
 \begin{align}
 I_{+}(\epsilon, g) 
&=  \left(\frac{g}{Aa^3}\right)^{\epsilon} \Gamma(\epsilon) 
 =  \frac{1}{\epsilon}   - \underbrace{ \left( \ln  \left(\frac{Aa^3}{g}\right) 
+ \gamma \right) }_{Q_1(N_f=0, g)}+ O(\epsilon) \, . 
 \end{align}
Subtracting off the $\frac{1}{\epsilon}$-pole in order to get rid of double 
counting of the uncorrelated two instanton events, we obtain the correlated 
two-event amplitude in the $N_f=0$ theory.  The $N_f=0$ result is derived  in   \cite{CDU}
by using Lefschetz  thimble of the bosonic theory. 
A similar subtraction is also 
needed for $N_f=-1, -2, \ldots$ as well. The result for the correlated 
2-events  for general $N_f$ is 
\begin{eqnarray}
[\I\I]&=&    \left(\frac{g}{Aa^3}\right)^{N_f} \Gamma(N_f)   
          \tfrac{S_I}{2 \pi}  \textstyle{ \left[\frac{ \det' M }{ \det M_0}\right]^{-1} }   e^{-2S_\I},    
           \qquad  \qquad \qquad \;\;  
          N_f \in \mathbb C\backslash  \{0, -1, -2, \ldots\},  \cr  \cr
[\I\I]&=&       Q_1(N_f, g)   \tfrac{S_I}{2 \pi}   \textstyle{ \left[\frac{ \det' M }{ \det M_0}\right]^{-1} }     e^{-2S_\I},    
         \qquad \qquad   \qquad   \qquad \;\;\; \; 
          N_f =0, -1, -2, \ldots \, .  
\label{amplitude-3}
\end{eqnarray}
For $N_f=0$, this results agrees with Bogomolny \cite{Bogomolny:1980ur}  
and Zinn-Justin's bosonic result \cite{ZinnJustin:1981dx} and can also 
be obtained also via the WKB approximation. For $N_f=1$ this agrees with 
the result obtained with the use of supersymmetry 
\cite{Salomonson:1981ug,Balitsky:1985in}. For  small values of $N_f$ 
the polynomials $Q_1(N_f, g)$ are given by 
\begin{eqnarray} 	
 Q_1(N_f=0, g) \;\;\;&=&  - \gamma  -  \ln [Aa^3/g]  \, ,  \cr	
 Q_1(N_f=-1, g) &=& -1 + \gamma + \ln[Aa^3/g]  \, , \cr	
 Q_1(N_f=-2, g) &=& \tfrac{1}{4} (3 - 2 \gamma - 2 \ln[Aa^3/g])\, ,    \cr	 
 Q_1(N_f=-3, g) &=&  \tfrac{1}{36}   (-11 + 6 \gamma + 6 \ln[Aa^3/g])\, . 	 	
 \label{amplitude-4}	
\end{eqnarray}


\noindent
{\bf Differentiating ambiguities from hidden topological angle:} 
The  $[\I\bar \I]$ amplitude can be written as
\begin{eqnarray}
[\I\bar \I]_{\pm}  &=&   e^{ \pm i \pi N_f}   
 \left(\frac{g}{A}\right)^{N_f} \Gamma(N_f) \tfrac{S_I}{2 \pi}   
   \textstyle{ \left[\frac{ \det' M }{ \det M_0}\right]^{-1} }    e^{-2S_\I},    
          \qquad  \qquad \qquad \;\;  
    N_f \in \mathbb C \backslash \{0, -1, -2, \ldots\} \, , \cr  \cr
[\I \bar \I]_{\pm}&=&   Q_1(N_f, e^{ \pm i \pi} g)   \tfrac{S_I}{2 \pi}   
  \textstyle{ \left[\frac{ \det' M }{ \det M_0}\right]^{-1} }  e^{-2S_\I},  
   \; \; \qquad \qquad   \qquad   \qquad \;\;\; \; 
    N_f =0, -1, -2, \ldots  \, ,
\label{amplitude-5}
\end{eqnarray}
where $Q_1(N_f, g)$ are the polynomials given in \eqref{amplitude-4}. This 
amplitude is  divergent for $N_f=0,-1,-2, \ldots$. This  divergence is 
the same as in the $[{{\cal I}{\cal I}}]$ case, and is subtracted off 
to prevent double-counting.  


The ambiguous imaginary part is related to resurgence. However, the ambiguous 
imaginary part disappears for  $N_f  \in  {\mathbb  N}^{+} $. This may imply
the disappearance of the first would-be singularity in the Borel plane, 
located at $2S_I$. Singularities at $4S_I, 6S_I, 
\ldots$ etc are still expected.  This is to a certain degree similar to the disappearance of 
the first renormalon singularity in Borel plane in $\N=1$  SYM theory 
\cite{Dunne:2015eoa}.


 For  $N_f\in   {\mathbb  N}^{+} $, there is a more subtle effect. 
$\Re[\I\bar \I]  \propto e^{\pm i \pi N_f} $ and this means that the 
contribution to the ground state energy due to  $[\I\bar \I]$ is 
proportional to $(-1)^{N_f} \in \Z_2$, alternating in sign depending 
on $N_f$ being even or odd.  Recall that $N_f$ even  correspond to the 
integer spin, and  $N_f$ odd  correspond to the  half integer spin.  
When we discuss exact solutions, we will provide an interpretation for 
this hidden topological angle in terms of topology in complexified 
field space.

\section{Graded  formulation: Double-well potential } 
\label{sec:DoubleWell}
\label{sec:graded-dw}

{\scriptsize {\bf Outline:}
In this section we construct new exact saddle solutions to the holomorphic 
Newton equations.  In addition to the familiar real bounce solution, there 
is also a complex bion solution. The proliferation of the complex bions  
provides the Euclidean description of the  ground state.  }

\subsection{Quantum modified potential  and exact non-BPS solutions}

\begin{figure}[t]
\begin{center}
\includegraphics[width=0.8\textwidth]{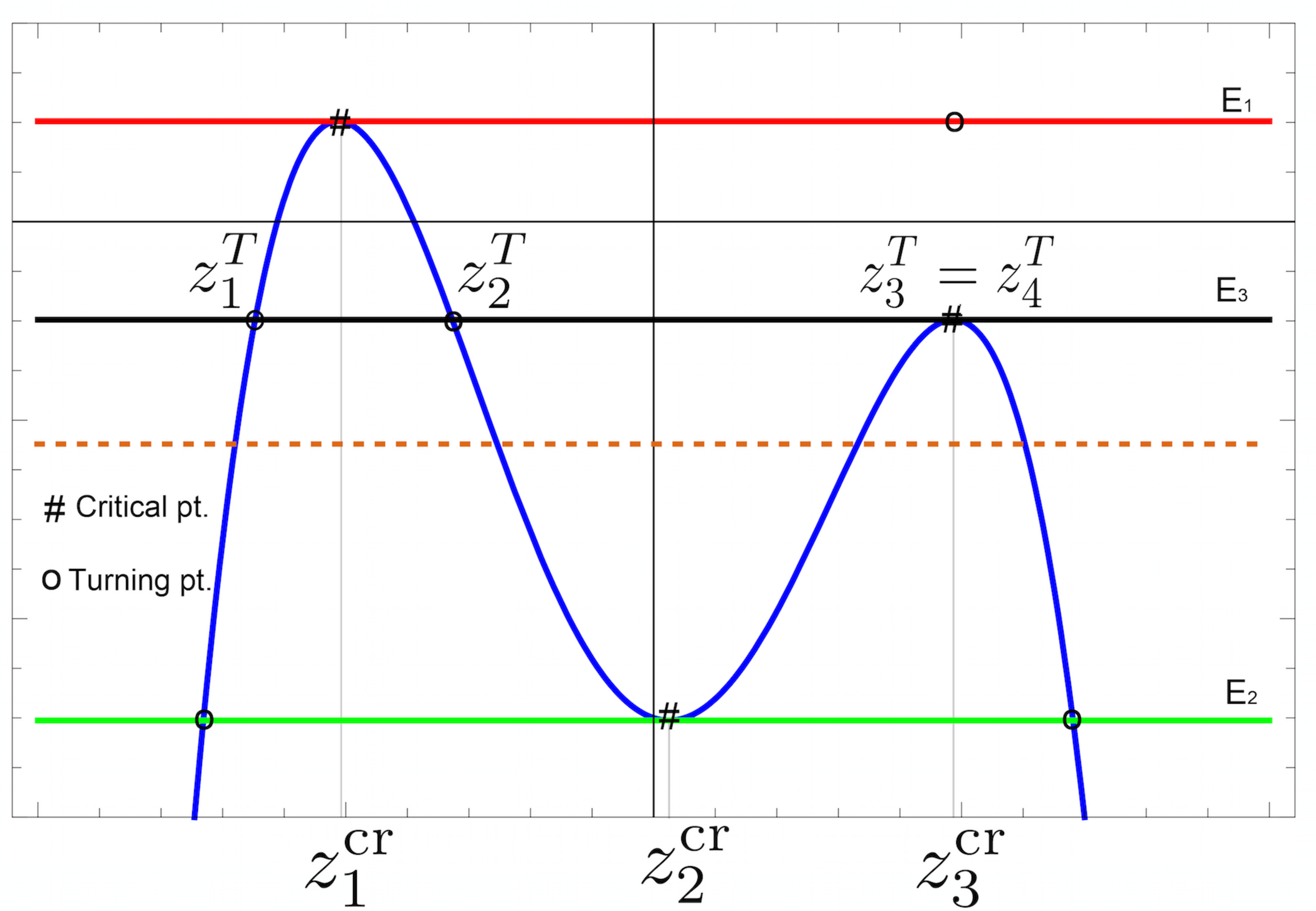}
\caption{Inverted tilted DW potential. The three  critical points are denoted $z_a^{\rm cr}$, $a=1, 2, 3$, and  $E_a$, $a=1, 2, 3$ labels the Euclidean energy at these critical points.}
\label{QMpot1-tilt}
\end{center}
\end{figure}

 As described in Section \ref{GHS}, we grade the system with fermions 
according to fermion number, resulting in a sequence of quantum modified 
bosonic models, which we then complexify: 
\begin{align} 
{\cal L} (\dot x, x, \psi) 
     \longrightarrow 
 \bigoplus_{k=0}^{N_f}  {\rm deg} ( {\cal H}_k)   {\cal L}_{k} (\dot x, x)   
     \longrightarrow   
 \bigoplus_{k=0}^{N_f}  {\rm deg} ( {\cal H}_k)   {\cal L}_{k} (\dot z, z) \, . 
\label{Main-steps}  
\end{align}
The  quantum modified (or graded) bosonic potentials are of the form 
\begin{align}
\label{GenDW-3} 
V_k(z)=     \underbrace{ \frac{1}{2}  (W' (z))^2 }_{V_{\rm bos}(z)} 
   +  \underbrace{ \frac{p g}{2}  W'' (z)}_{\rm fermion-induced},  
      \quad p\equiv (2k-N_f),  \quad k=0, \ldots, N_f\, . 
\end{align}
For the case of the double-well we have $W^\prime(z)=z^2-a^2$, $W^{\prime\prime}
=2z$, so the fermion-induced potential introduces a tilting, breaking the 
degeneracy of the ground state of the bosonic potential. The non-perturbative 
normalization of the action  makes it manifest that the fermion induced 
potential is $\mathcal{O}(g)$ with respect to bosonic potential $V_{\rm bos}
(z)$.  Since  $V_{\rm bos}$ has exactly degenerate minima, an order 
$\mathcal{O}(g)$ tilting effect lifts the degeneracy of the classical 
potential, see Figure \ref{QMpot1-tilt}, and produces significant physical 
effects.


 Recall that instantons satisfy a first order equation, $\dot{z}=W^\prime(z)$, 
already given in \eqref{instanton-1}. Differentiating  once with respect to 
time,  the instanton also  satisfies  the second order equations of motion 
for a classical particle moving  in the inverted potential, $-V_{\rm bos} = 
-\half(W')^2$, namely 
\begin{align} 
\ddot z_{\rm instanton} = \dot z_{\rm instanton}\,  
  W'' =   W' W'' = +\frac{ \partial V_{\rm bos}}{\partial z}  \, . 
\label{2nd-order}
\end{align}
Obviously, \eqref{instanton-1} and  \eqref{2nd-order}  do not involve $g$, 
the coupling constant, and neither does the solution. So, the correlated 
two-event discussed in Section \ref{correlated}, which has characteristic 
size $\tau^{*} = m^{-1}_b \ln \left( \frac {A a^3 }{g\, N_f} \right) $ cannot be 
the solution to  \eqref{instanton-1} and  \eqref{2nd-order}.  

 However,  the saddle equations that one needs to solve are  actually not 
these equations. Instead, we need to solve the complexified Newton equations 
for the inverted quantum modified potential:
\begin{align} 
 \ddot z_{\rm saddle} 
   =  W'  W'' + \frac{gp }{2} W''' 
   = +\frac{ \partial V_{\rm bos}}{\partial z}   +  \frac{g\, p}{2} W''' \, . 
\label{bion-eq-2}
\end{align}
This equation involves both $g$ and $p$ (i.e. spin projection), due to the 
fermion induced term, 
and we now show that it gives new exact solutions with characteristic size 
$\tau^{*} = m^{-1}_b \ln \left( \frac {A a^3 }{g\, N_f} \right) $. 
	
 Since the quantum modified potentials are quartic polynomials, the saddles 
are given in terms of standard elliptic functions, as follows. The saddles 
are found by using conservation of energy for a classical particle in the 
inverted potential. 
Consider a classical particle with Euclidean energy 
$E$ in potential $ -V(z)$:
\begin{align}
{\cal L}&= \half \dot z^2  + V(z) \, , \qquad  \cr
E&= \half \dot z^2   -V(z) \, . 
\label{first-order}
\end{align}
For  $p g/a^3 \in \R$ and $p g\leq \frac{4a^3}{3\sqrt{3}}$, the potential has 
three real critical points, $z_\ell^{\rm cr}$, $\ell=1, 2, 3$, whose expressions 
involve complicated cube roots. 
These can be usefully parametrized as follows:
\begin{eqnarray}
z_\ell^{\rm cr} = - \frac{2a}{\sqrt{3}} 
  \cos \left(\frac{1}{3} {\rm arccos}\left(\frac{3\sqrt{3}}{4a^3} p g\right) 
    -\frac{2(\ell-1) \pi }{3}\right)\quad, \quad \ell=1, 2, 3
\label{eq:dw-critical} 
\end{eqnarray}
For small $p g /a^3 \ll 1$, the tilting is small and $z_1^{\rm cr}\approx -a$, 
$z_2^{\rm cr}\approx 0$, and $z_3^{\rm cr}\approx +a$:
\begin{eqnarray}
z_1^{\rm cr} &\approx & -a-\frac{p g}{4a^2} + \frac{3 (p g)^2}{32a^5} 
      - O((p g)^3) \, , \nonumber\\
z_2^{\rm cr} &\approx & \frac{p g}{2a^2} + \frac{(p g)^3}{8a^8} 
      + O((p g)^5) \, ,  \nonumber\\
z_3^{\rm cr} &\approx & a-\frac{p g}{4a^2} - \frac{3 (p g)^2}{32a^5} 
      - O((p g)^3) \, . 
\label{eq:dw-critical-approx}
\end{eqnarray}
The corresponding Euclidean energies, $E_k\equiv -V(z_k^{\rm cr})$, also play 
an important role in the description of the exact saddle solutions. As shown 
in Figure \ref{QMpot1-tilt}, these Euclidean energies mark the boundaries at 
which turning points  coalesce. These are points where two roots of the quartic $V(z)+E=0$ coalesce. 
For example,  as $E$ approaches $E_3$ from below, the right-hand pair of real turning points 
coalesce, while for $E> E_3$, the turning points $z_3^T$ and $z_4^T$ become complex (they 
form a complex conjugate pair for $p g/a^3 \in \R$). Similarly at $E=E_1$ the 
turning points $z_1^T$ and $z_2^T$ coalesce; they are real for $E_2\leq E \leq 
E_1$, but are complex for $E> E_1$.


It is useful to view the energy conservation relation as the  projection of 
a complex algebraic curve to real momentum and position. Consider the complex 
algebraic curve: 
\begin{align}
{\cal P}^2 = 2V(z)+2 E = z^4 -2 z^2 a^2 + 2gp z + a^4+ 2E  \, . 
\label{eq:curve1}
\end{align}
This ``energy conservation" \eqref{eq:curve1} can be integrated by 
quadratures  to give:
\begin{align} 
\int_{t_0=0}^{t} dt =  \int_{z_T}^{z}  \frac{dz}{ {\cal P}   }   
  =  \int_{z_T}^{z}  \frac{dz}{\sqrt {2(E+V(z))}}  \, . 
\end{align}
This integral can be computed exactly in terms of the  Weierstrass 
$\wp$-function, and takes a particularly simple form when $z_T$ is a 
turning point; i.e., if $z_T$ is any root of the equation $E+V(z)= 0$ 
(see Appendix \ref{app:elliptic}).  The solution is 
\begin{equation}
 z(t) = z_{T} + 
  \frac{ \tfrac{1}{2}V'({z_T})}{\wp(t; g_2, g_3)-\tfrac{1}{12}V''(z_T)}. 
\label{Master}
\end{equation}
Here, the algebraic invariants $g_2(E)$ and $g_3(E)$ are given by 
\begin{eqnarray}
g_2(E)&=& 2E + \frac{4 a^4}{3} \, , \cr
g_3(E)&=&-\frac{8a^6}{27} - \frac{2 E a^2}{3} - \frac{p^2 g^2}{4}\, .
\label{ALI-1}
\end{eqnarray}	
The exact solution \eqref{Master} has a center position modulus, associated 
with translation symmetry in $t$,  which can be restored by setting  
$t \rightarrow t- t_{\rm c}$.  Without loss of generality, we set   
$t_{\rm c}=0$ from now on. 	


Since the classical saddle bounce and bion solutions begin and end on 
turning points (either real or complex), they can all be expressed in 
closed form using the expression \eqref{Master}. All that needs to be 
done is to match to the appropriate boundary conditions. 


\begin{figure}[t] 
\centering
\includegraphics[width=13cm]{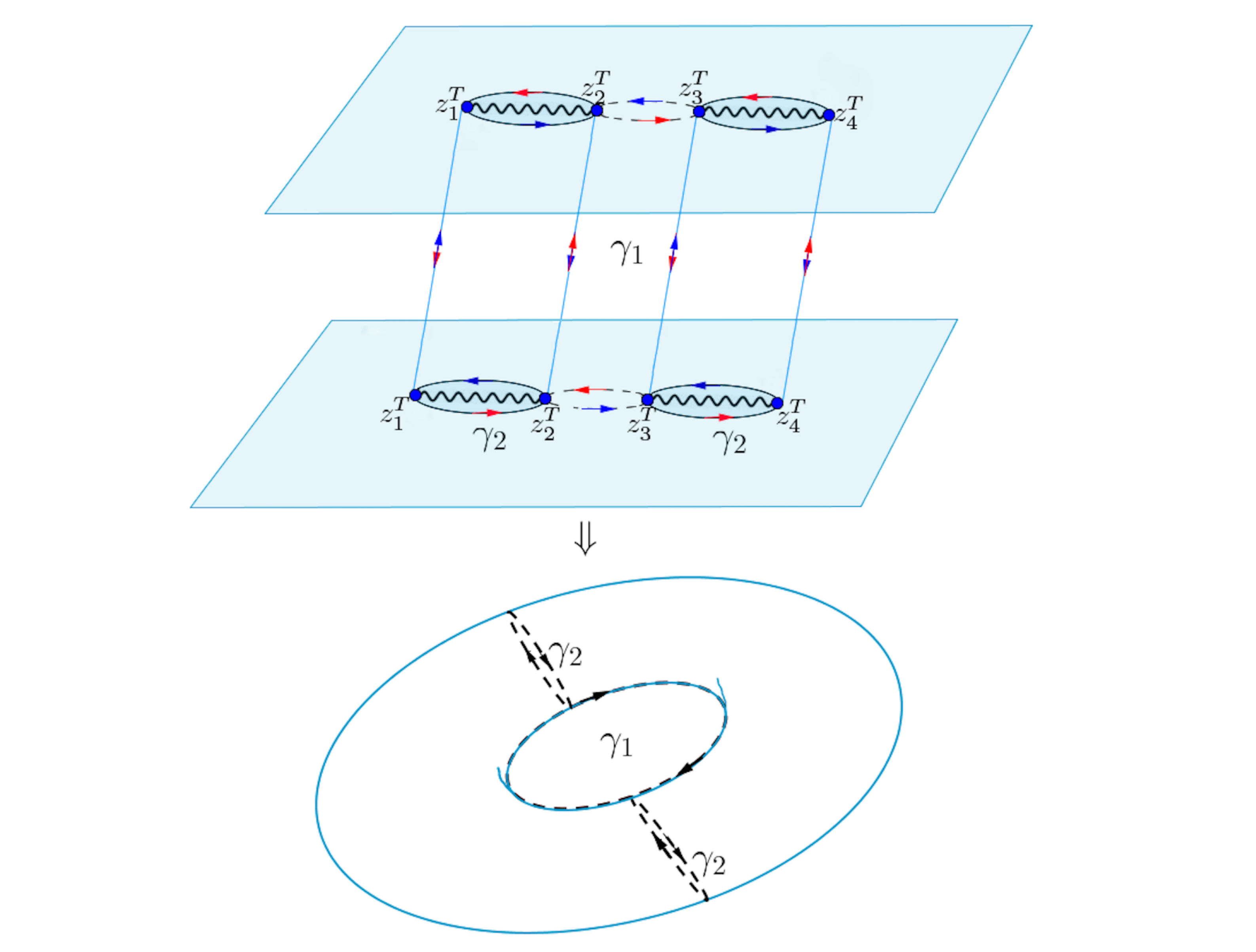}
\caption{Doubly periodic structure at a typical energy $E$.   At critical energies   shown in Fig.\ref{QMpot1-tilt}, two of the turning points degenerates.}
\label{fig:Riemann}
\end{figure}

 The Weierstrass $\wp$-function is doubly periodic, with a real and complex (in our case, purely  imaginary) 
period\footnote{Standard mathematical notation for the two  periods are 
$(2\omega_1, 2\omega_2)$, which we call here $(T_1, T_2)$. See, for example,  
Whittaker and Watson \cite{WhWa27}.}, $T_1$ and $T_2$.
There is a simple interpretation of the double periodicity.  The real period  
is the period of  bounded motion in the potential $-V(z)$ with energy $E$. 
To understand the imaginary period, take $t \rightarrow it$  (back to 
Minkowski space for a moment) and observe that this reverses the signs 
of potential and energy in the conservation of energy:  $(-V(z), E) 
\rightarrow  (V(z), -E) $. The magnitude of the  imaginary period is  
the period of classical motion either on the left or right well. Note 
that  the left-well and right-well periods are actually equal 
to each other, for any asymmetric double-well quartic potential!  This remarkable fact helps us to write down 
simpler expressions for exact solutions. 
Explicit expressions for the two periods are (see Figure \ref{fig:Riemann}).
\begin{align}
T_1(E) &=   \int_{\gamma_1}  \frac{dz} {{\cal P}}  
        = 2 \int_{ z_2(E)}^{ z_3(E)}  \frac{dz} {{\cal P}} \, ,   \cr
T_2(E) &=   \int_{\gamma_2}  \frac{dz} {{\cal P}}   
        = 2 \int_{ z_1(E)}^{z_2(E)}    \frac{dz} {{\cal P}}  
        = 2 \int_{ z_3(E)}^{ z_4(E)}  \frac{dz} {{\cal P}} \, . 
        \label{per-rel}
\end{align}	
where $\gamma_{1,2}$ are the cycles defined on a Riemann torus 
	
\subsection{Exact (real) bounce solution} 

The bounce solution is a solution with Euclidean energy $E=E_3$, equal 
to minus the potential at the lower of the maxima of $-V(z)$, see Figure 
\ref{QMpot1-tilt}. At this energy there are three real turning points. It is useful  to 
parametrize all the turning points (which are cumbersome in terms of original parameters of the potential), as well as exact solutions  in terms of 
a  critical  point. This provides relatively simple expressions:
\begin{eqnarray}
\{z_1^{T},z_2^T,z_{3}^T=z_4^T\}=\left\{-z_3^{\rm cr}-\sqrt{\frac{p g}{z_3^{\rm cr}}},
                  -z_3^{\rm cr}+\sqrt{\frac{p g}{z_3^{\rm cr}}}, 
                   z_3^{\rm cr}\right\}\, . 
\label{eq:bounce-turning}
\end{eqnarray}
The ``bounce'' solution  \cite{Coleman:1978ae} bounces off the (real) 
turning point $z_T=-z_3^{\rm cr}+\sqrt{\frac{p g}{z_3^{\rm cr}}}$ at some 
finite time $t=0$, tending as $t\to\pm \infty$ to the lower of the (real) 
maxima of the inverted potential, $z=z_3^{\rm cr}$. At this particular energy, 
$T_1$ diverges and the doubly-periodic Weierstrass function in \eqref{Master} 
degenerates to a singly-periodic function:
\begin{equation}
\wp(t)=\frac{\omega_{\rm bn}^2}{4}
 \left(\frac{1}{\sinh^2\left(\omega_{\rm bn}\,t/2\right)}+\frac{1}{3}\right)\, , 
\label{P-func1}
\end{equation}
where 
\begin{eqnarray}
\omega_{\rm bn}= \sqrt{V^{\prime\prime}(z_3^{\rm cr})}\, , 
\label{eq:T}
\end{eqnarray}
is the curvature of the potential at $z_3^{\rm cr}$.\footnote{The linear frequency on the right (R) well at  $z= z_3^{\rm cr}$ (the value of the frequency as one take $E \rightarrow -E_3$ limit in the original potential)   is equal to the 
{\it non-linear} frequency on the left  (L) well at  
$z=z_1^{\rm cr}$ for the energy level $E = -E_3$! 
\begin{align}
\omega_{\rm bn}= \omega^{\rm R}_{\rm linear} (E=-E_3) =  \omega^{\rm L}_{\rm non-linear}(E=-E_3)\, . 
\end{align}
This  identity is a consequence of the relation \eqref{per-rel} for the periods of the complex algebraic curve.  
Physically, in the singly periodic limit of the Weierstrass function,   $i \omega^{\rm L}_{\rm non-linear}(E=-E_3)$ is the (pure imaginary) frequency which remains finite, while the other real frequency tends to zero. } 
 This corresponds to the 
singly-periodic limit where $\omega_{\rm bn}^2=-18 g_3(E_3)/g_2(E_3)$, in 
terms of the algebraic lattice  invariants $g_2(E)$ and $g_3(E)$ in 
\eqref{ALI-2}.

\begin{figure}[t]
\center
\includegraphics[scale=0.7]{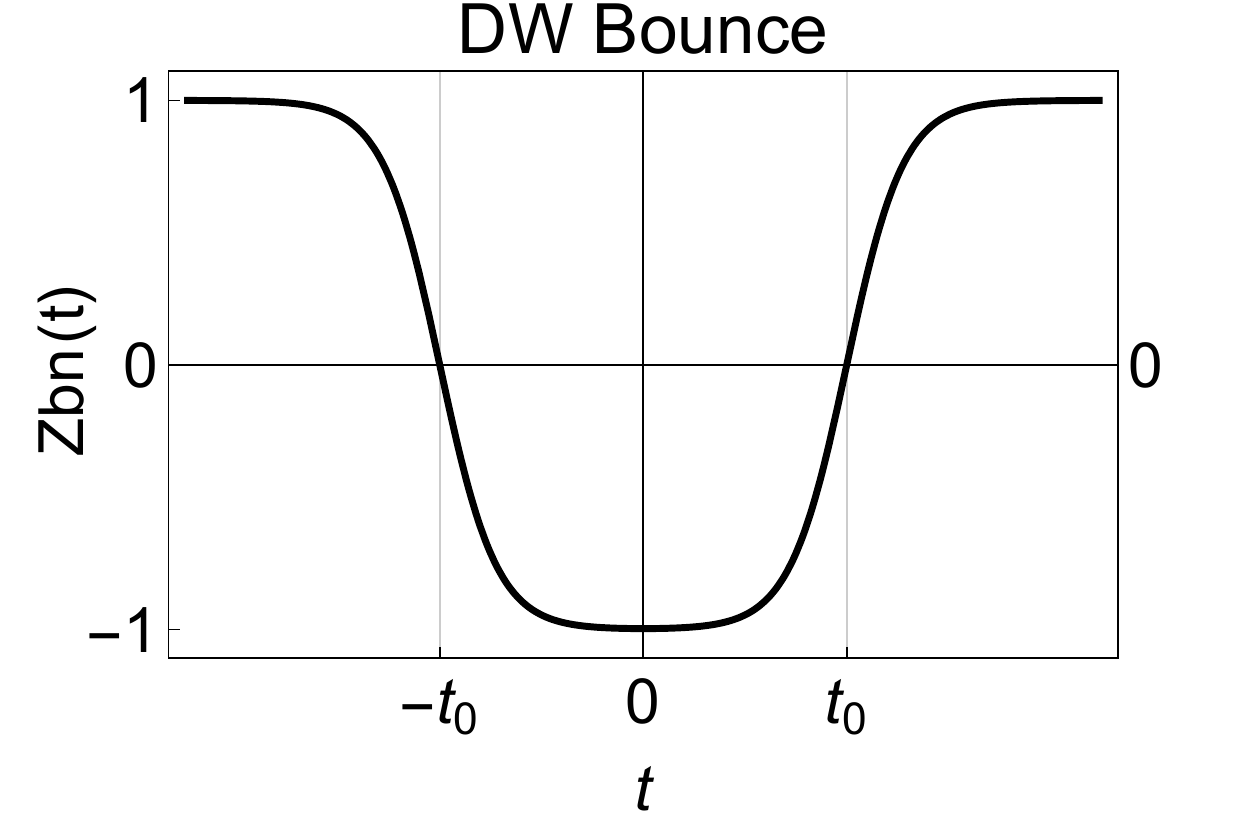}
\caption{Bounce solution in the double well case. Note that the 
separation between instanton-anti-instanton is $ 2t_{0}\approx \frac{1}{2a} 
\ln \left(\frac{16a^3}{p g}\right)$.}
\label{DWbounce}
\end{figure}

Thus, the exact bounce solution can be written as  (here 
$z_T=-z_3^{\rm cr}+\sqrt{\frac{p g}{z_3^{\rm cr}}}$)
\begin{eqnarray}
z_{\text{bn}}(t)
 &=& z_3^{\rm cr}+\left(z_T-z_3^{\rm cr}\right) 
   \frac{\cosh^2\left(\omega_{\rm bn}\, t_0/2\right)}
        {\cosh^2\left(\omega_{\rm bn}\, t/2\right)
              +\sinh^2\left(\omega_{\rm bn}\, t_0/2\right)} \\
 &=& z_T+\left(z_3^{\rm cr}-z_T\right) 
    \frac{\sinh^2\left(\omega_{\rm bn}\, t/2\right)}
        {\sinh^2\left(\omega_{\rm bn}\, t/2\right)
              +\cosh^2\left(\omega_{\rm bn}\, t_0/2\right)}\, ,
\label{Bounce_final_1}
\end{eqnarray} 
where (see Appendix \ref{app:elliptic})
\begin{eqnarray}
t_0=\frac{2}{\omega_{\text{bn}}}\, 
    {\rm arccosh} \left(\sqrt{\frac{3}
              {1-V^{\prime\prime}(z_T)/\omega^{2}_{\text{bn}}}}\right)\, .
\label{eq:dw-t0}
\end{eqnarray}
Comparing with \eqref{eq:id}, we see that this solution satisfies the 
desired boundary conditions:
\begin{eqnarray}
z_{\text{bn}}(\pm \infty)= z_3^{\rm cr}, \qquad    
z_{\text{bn}}(0)  =z_T\equiv -z_3^{\rm cr}+\sqrt{\frac{p g}{z_3^{\rm cr}}}\, . 
\label{eq:bounce-bcs}
\end{eqnarray}
The bounce solution is shown in Figure \ref{DWbounce}.

 It is physically more intuitive to rewrite the exact bounce solution 
\eqref{Bounce_final_1} as
\begin{align}
z_{\text{bn}}(t)
   &=  z_3^{\rm cr}
     -\frac{(z_3^{\rm cr}-z_T)}{2}\coth\left(\omega_{\rm bn}\frac{t_0}{2}\right)
       \left(\tanh\left[\omega_{\rm bn}\frac{(t+t_0)}{2}\right]
            -\tanh\left[\omega_{\rm bn}\frac{(t-t_0)}{2}\right]\right)\, ,
\label{Bounce_final_2}
\end{align} 
where $t_0$ is given by \eqref{eq:dw-t0}. This expression makes it clear 
that the real bounce solution  has a kink/anti-kink form. In fact, the 
exact bounce solution is nothing but the exact version of the approximate 
correlated anti-instanton/instanton $[\bar \I\I]$ event in the language of 
Section \ref{correlated}. The bounce is an $\bar \I$ event taking place at 
$t=-t_0$, followed by an instanton event $ \I$  at $t=+t_0$. The separation 
between these two events is fixed and is given by 
\begin{align}
2t_0  \approx m^{-1}_b  \ln \frac{16 a^3 }{pg}   
   \equiv \tau^* \quad, \qquad {\rm for} \;  gp  \ll a^3 \, . 
\label{separation-2}
\end{align}
This is precisely the scale which dominates the quasi-zero-mode thimble 
integration: $\tau^{*} = m_{\rm b}^{-1}\ln \left( \frac {Aa^3}{g\, N_f} \right)$,  
where in this case $A=16$, $m_b=2a$, and $p=N_f$ in \eqref{critical-1}.  


 An important characteristic property of the solution is the parametrically 
large plateau, i.e, parametrically large time spent around the turning point $z_T$. 
This makes this solution slightly different from the conventional 
bounce solutions (which looks just like a regular bump). 
Therefore a more appropriate name would be a \emph{flat} bounce. The classical particle 
starts at the local maximum $z_3^{\rm cr}$ at $t=-\infty$ (see Fig. \ref{QMpot1-tilt}). After spending an infinitely 
long time there, it rolls down at an {\it instant} (like an instanton), and 
reaches the vicinity of the turning point $z_2^T$  within a time  scale $\sim a^{-1}$, characteristic of instantons.  Since 
$z_2^T$ is  is parametrically close to  the global maximum, 
 the classical particle spends a large time span $2t_0 \sim  m^{-1}_b\ln  
\frac{16a^3}{g p}$ near the turning point $z^T_2$. The Euclidean particle 
gets parametrically close to the global maximum $z_1^{\rm cr}$, $|z_3^{\rm cr} - z^T_2|\sim    
\sqrt{p g/a}$, and returns to the local maximum $z_3^{\rm cr}$. The crucial point is 
that the size of the flat-bounce  solution {\it is not} controlled by the 
inverse natural frequency of the system $\sim (2a)^{-1}\equiv m^{-1}_b$, 
which is the case for conventional instantons as well as bounce solutions. 
Rather, the size of the flat-bounce solution is parametrically larger than 
the instanton size, by a factor $ \ln \frac{16a^3}{g p}$.  
For a harmonic unstable equilibrium point,  a particle with energy $E \rightarrow  E_1$ from below  will spend a time span of 
$\log |E-E_1|$ in the neighborhood of the critical point. This form of divergence is universal around the critical point. In the present case, 
for $E=E_2$, the time span on the plateau is $-\log |E_2-E_1| = \log \frac{16 a^3}{pg} $.   This feature  
leads to some interesting new effects when we discuss the fluctuation 
operators around these exact new saddles in Section \ref{sec:fluc-op}. 

\subsection{Exact complex bion solution} 
\label{sec:dw-complex-bion}

\begin{figure}[t]
\center
\includegraphics[width=0.5\textwidth]{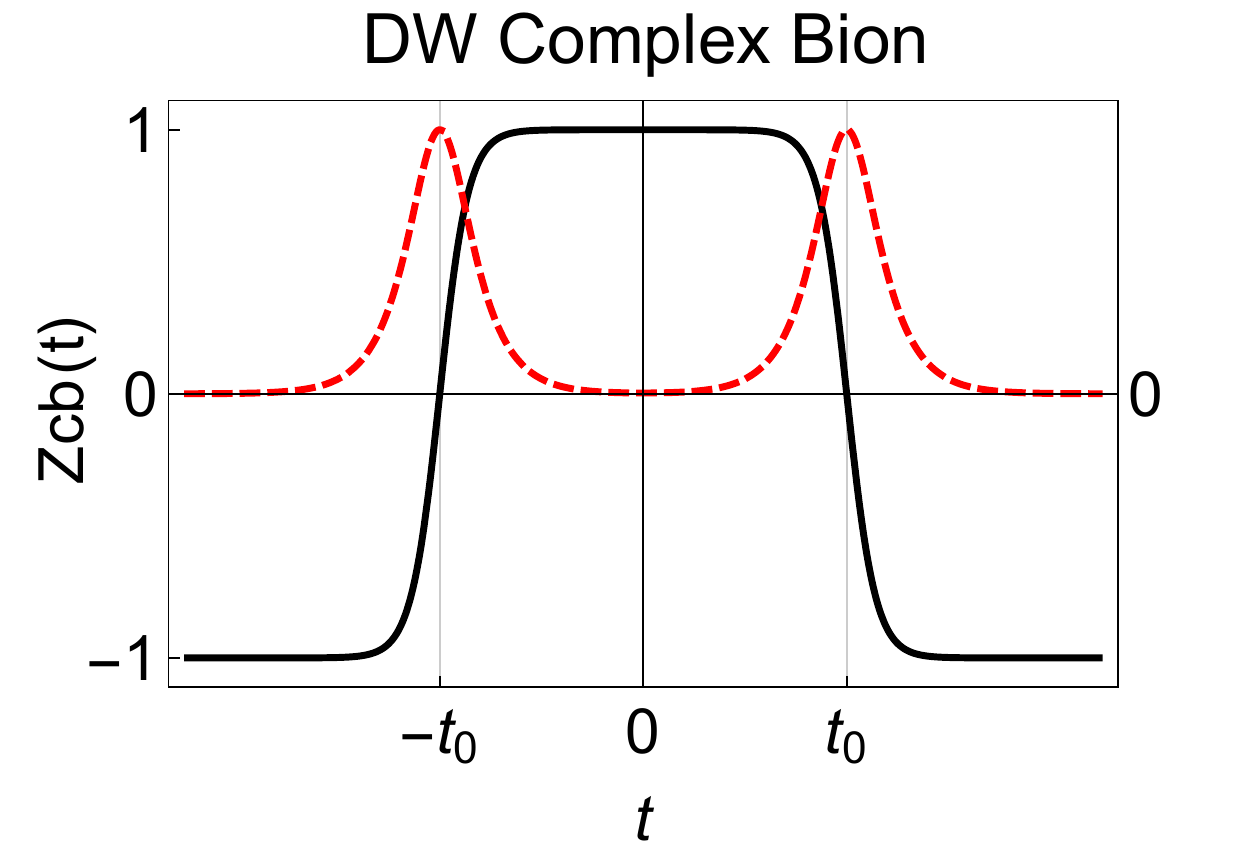}
\caption{Real (black) and imaginary (red) parts of the complex bion for 
the double well case. The  real part of the separation is ${\rm Re} (2t_0)\approx\frac{1}{2a} \ln 
\left(\frac{16a^3}{p g}\right)$. 
}
\label{DWcomplexbion}
\end{figure}

 The complex bion solution is a complex 
classical solution with Euclidean energy $E=E_1$, equal to minus the potential 
evaluated at the higher of the maxima of $-V(z)$. See Figure \ref{QMpot1-tilt}.   
The complex bion bounces off one of the complex turning points 
$z=z_T=-z_1^{\rm cr} + i \sqrt{\frac{p g}{-z_1^{\rm cr}}}$, 
(or its complex conjugate $z=z_T^*$), at some finite time $t=0$, and reaches 
the upper of the (real) maxima of the inverted potential, $z=z_1^{\rm cr}$
as $t\to\pm \infty$, see Figure \ref{QMpot1-tilt}. The doubly-periodic 
Weierstrass function in \eqref{Master} degenerates again to a singly-periodic 
function, and we obtain a solution of the same form as the bounce solution 
in \eqref{Bounce_final_1}, but with different critical point and turning 
point:
\begin{eqnarray}
z_{\text{cb}}(t)
 &=&  z_1^{\rm cr}+\left(z_T-z_1^{\rm cr}\right) 
   \frac{\cosh^2\left(\omega_{\text{cb}}\, t_0/2\right)}
        {\cosh^2\left(\omega_{\text{cb}}\, t/2\right)
              +\sinh^2\left(\omega_{\text{cb}}\, t_0/2\right)} \\
 &=& z_T+\left(z_1^{\rm cr}-z_T\right) 
   \frac{\sinh^2\left(\omega_{\text{cb}}\, t/2\right)}
        {\sinh^2\left(\omega_{\text{cb}}\, t/2\right)
              +\cosh^2\left(\omega_{\text{cb}}\, t_0/2\right)}\, , 
\label{Complex_Bion_final_1}
\end{eqnarray} 
where $z_T=-z_1^{\rm cr} + i \sqrt{\frac{p g}{-z_1^{\rm cr}}}$, and 
\begin{eqnarray}
\omega_{\text{cb}}= \sqrt{V^{\prime\prime}(z_1^{\rm cr})}\, . 
\label{eq:freq-cb-dw}
\end{eqnarray}
The parameter $t_0$ is given by (see Appendix \ref{app:elliptic})
\begin{eqnarray}
t_0 = \frac{2}{\omega_{\text{cb}}}\, {\rm arccosh} 
   \left(\sqrt{\frac{3}{1-V^{\prime\prime}(z_T)/\omega^{2}_{\text{cb}}}}\right)\, . 
\label{eq:dw-t0-bion} 
\end{eqnarray}
Comparing with \eqref{eq:id}, we see that this solution satisfies the desired 
boundary conditions:
\begin{eqnarray}
z_{\text{cb}}(\pm \infty)= z_1^{\rm cr}, \qquad    
z_{\text{cb}}(0)  =z_T=-z_1^{\rm cr} + i\sqrt{\frac{p g}{-z_1^{\rm cr}}}\, . 
\label{eq:cb-bcs}
\end{eqnarray}
We can also write the complex bion as
\begin{align}
z_{\text{cb}}(t)
 = z_1^{\rm cr}-\frac{(z_1^{\rm cr}-z_T)}{2}
            \coth\left(\frac{\omega_{\text{cb}}t_0}{2}\right)
   \left(\tanh\left[\omega_{\text{cb }}\frac{(t+t_0)}{2}\right]
        -\tanh\left[\omega_{\text{cb}}\frac{(t-t_0)}{2}\right]\right) ,
\label{bion_final_2}
\end{align} 
where $t_0$ is given by \eqref{eq:dw-t0-bion}. Notice that $\omega_{\text{cb}}$ 
is real, but $t_0$ is complex, as is $z_T$. Thus, the complex bion has both 
real and imaginary parts. The complex bion is plotted in Figure 
\ref{DWcomplexbion}.


At  weak coupling, we see that an approximate form of the exact solution  
\eqref{bion_final_2} is directly related to the correlated 
instanton-anti-instanton $[\I \bar \I]$ pair discussed in Section \ref{correlated}. To 
see this we can use the fact that, in the weak coupling regime, the complex 
turning point, $z_T$, and the critical point $z_1^{\rm cr}$ have the expansions: 
\begin{align}
\label{qT}
  &z_T     =  a + i \sqrt{\frac{p g}{a}}+ \frac{p g}{4a^2}+\dots \, ,  
&& z_1^{\rm cr}=  -a-\frac{p g}{4a^2} +\dots \, ,\\
  &V''(z_T)=(2a)^2+12 i \sqrt{a pg}+  3pg/a+\dots\;,
&& V''(z_1^{\rm cr})=\omega_{\rm cb}^2=(2a)^2+\frac{3pg}{a}+\dots  \, . 
 \nonumber
\end{align} 
The complex turning points in the bion solutions acquire an elegant 
reinterpretation in terms of  the complex quasi-zero mode in the 
instanton-anti-instanton correlated event. To see this, consider the  
kink/anti-kink form in \eqref{bion_final_2}. The separation between 
the instanton and anti-instanton \eqref{eq:dw-t0-bion}  has  both real and imaginary parts. 
Using the weak coupling expansions given in  \eqref{qT},  the approximate form of the complex separation 
can be written as 
\begin{align}
 2 \, t_{0c}^{\pm} \approx  m_b^{-1} \left[ \ln\left(\frac{16a^3 }{pg}\right) \pm i \pi \right]  \approx  \tau^{*}  \, .   
\end{align} 
	This is precisely the scale which dominates the quasi-zero-mode thimble 
integration: $\tau^{*} = m_{\rm b}^{-1}\ln \left( \frac {Aa^3}{g\, N_f} \right)$,  
where in this case $A=16$, $ \omega_{\rm cb} \approx m_b=2a$, and $p=N_f$ in \eqref{critical-2}.
This proves that the integration over the complexified  quasi-zero mode described in Section \ref{correlated} produces  the most significant physical features of the exact 
complex bion solution, such as the size of the two-event and most importantly, the hidden topological angle.

\subsection{Actions and amplitudes of the (real) bounce and the complex bion}

\begin{figure}[t]                
\includegraphics[width=7cm]{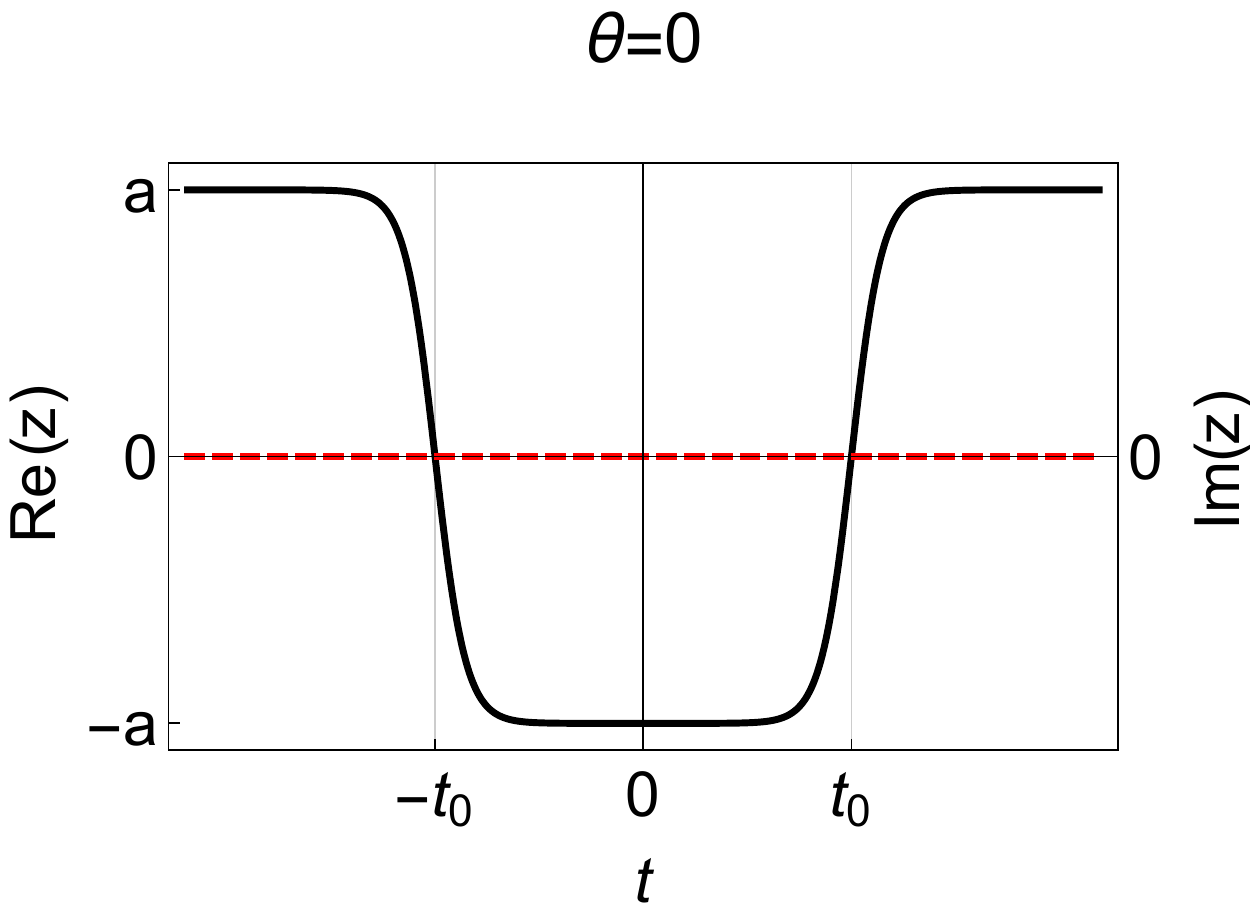}
\includegraphics[width=7cm]{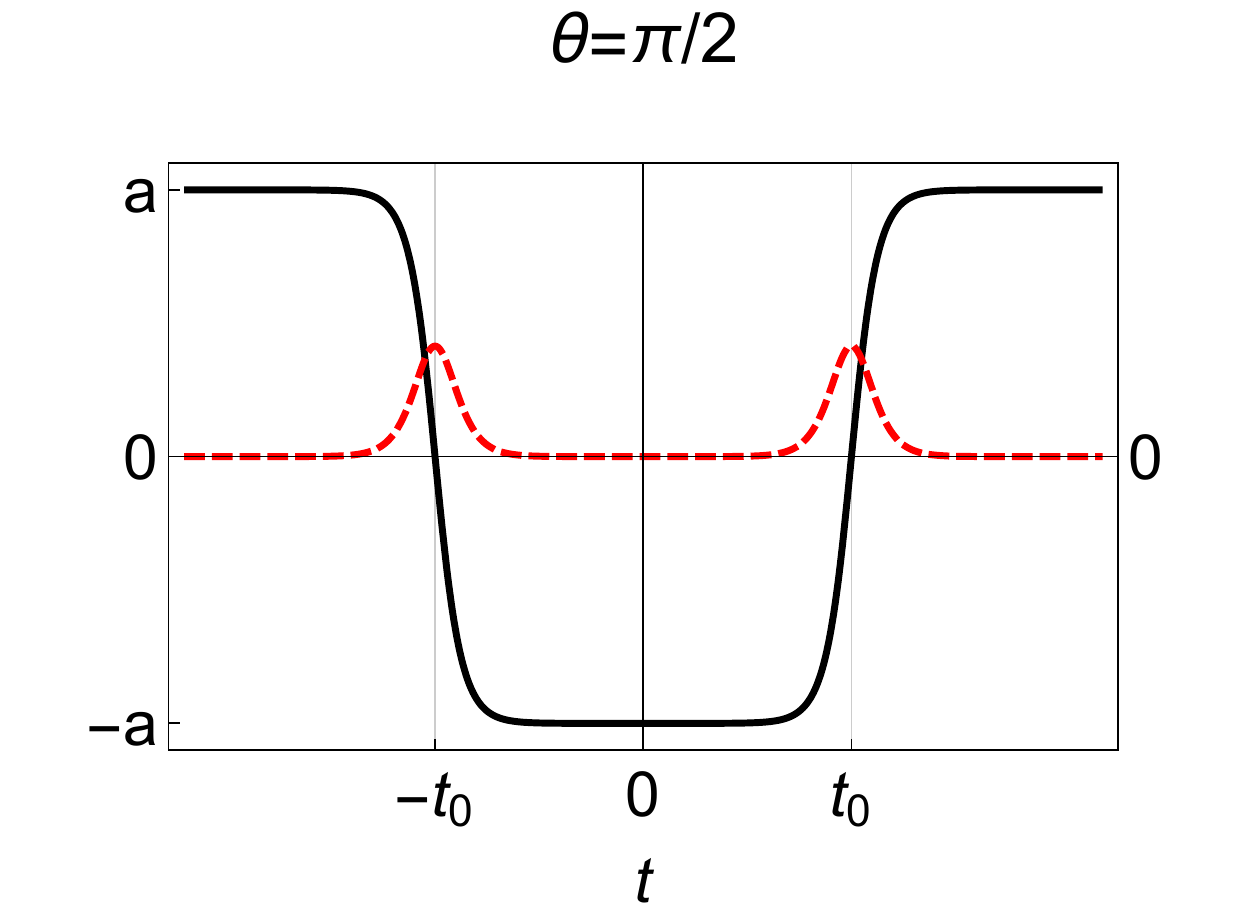}
\includegraphics[width=7cm]{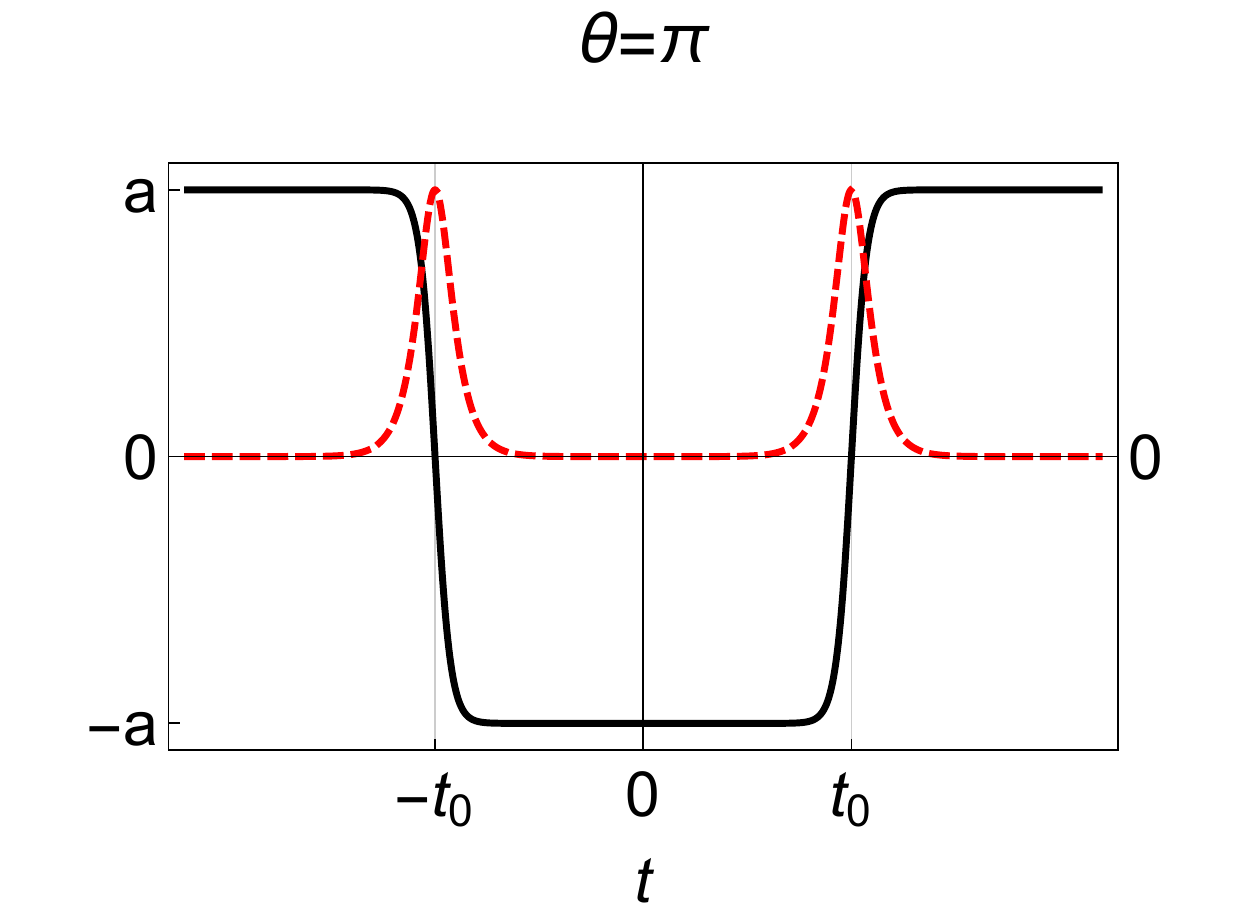} \hspace{1.4cm}
\includegraphics[width=7cm]{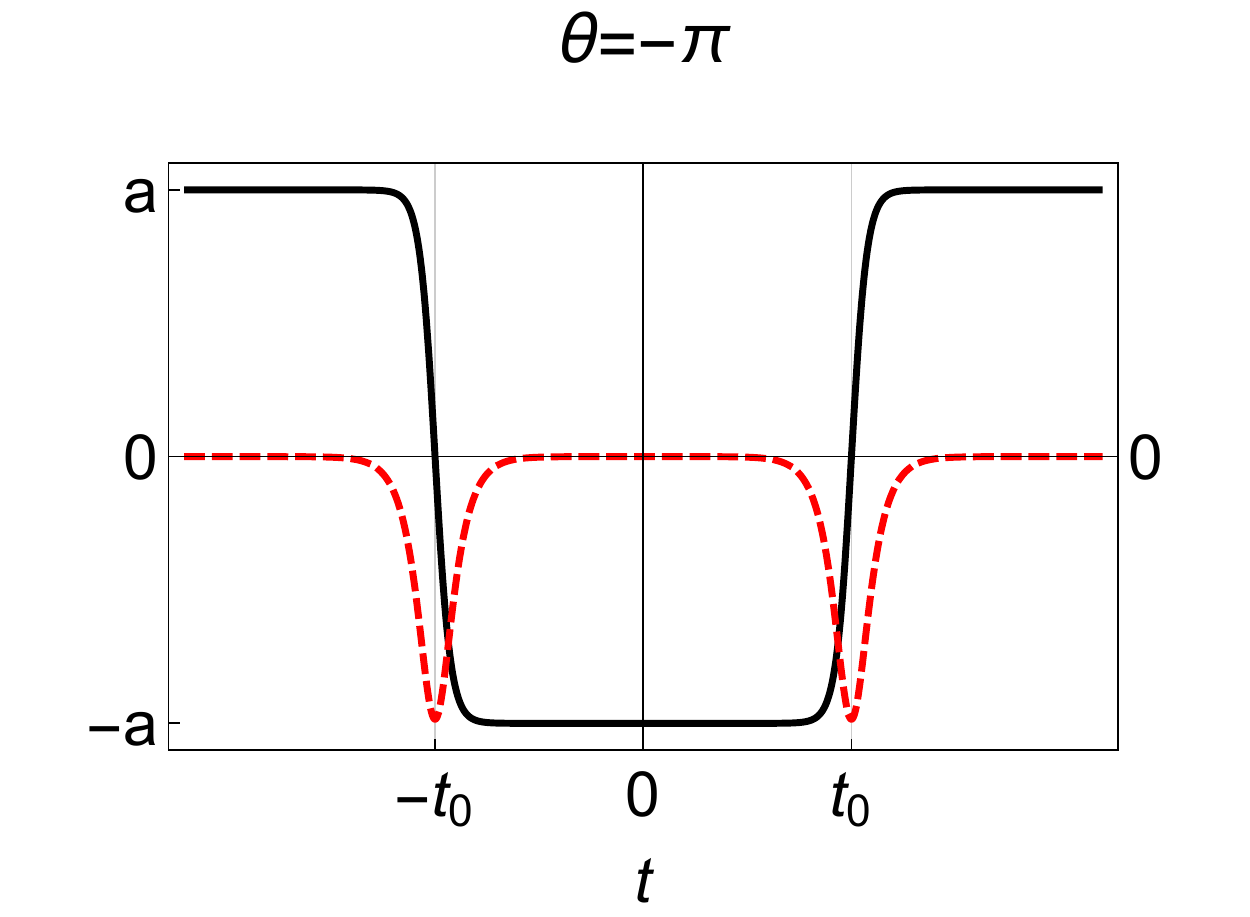}
\caption{This figure shows the real (solid black line) and imaginary 
(dashed red line) parts of the classical solution $z(t)$ in the double 
well potential for $pg/a^3 \ll 1$, and a range of values 
of the $\theta$ parameter.  The 
$\theta=0$ solution corresponds to the real bounce,  $\theta=\pi$ and  $\theta=-\pi$ are 
the complex bions. The solution has monodromy $4\pi$, or order 2 monodromy.  }
\label{dw-1}
\end{figure}

 The bounce and complex bion solutions have finite Euclidean action. To 
obtain the action we simply integrate
\be\label{eq:DWaction1}
 g S=2\int_{z_{\rm cr}}^{z_T} dz\;\sqrt{2E+2V(z)}\;,
\ee
where $z_{\rm cr}$ is the critical point, corresponding to   the initial and final point of the solution, and $z_T$ 
is the turning point.  For the  bounce, the critical point is $z_3^{\rm cr}$ associated with the energy level $E_3$ in Fig.~\ref{QMpot1-tilt}  and for the complex bion, the 
critical point is  $z_1^{\rm cr}$ associated with the energy level $E_1$ in Fig.~\ref{QMpot1-tilt}: 
\begin{subequations}
\begin{align}
\text{bounce (bn)}&: z_{\rm cr}=z_3^{\rm cr}, \qquad 
            z_T=-z_3^{\rm cr}+\sqrt{\frac{p g}{z_3^{\rm cr}}}\, ,  \\
\text{complex bion (cb)}&: z_{\rm cr}=z_1^{\rm cr}, \qquad 
            z_T=-z_1^{\rm cr}+i \sqrt{\frac{p g}{-z_1^{\rm cr}}}\, .
\label{eq:endpts}
\end{align}
\end{subequations}
Note that the turning point $z_T$ is real in the case of the bounce, 
and complex in the case of the complex bion. Both $z_1^{\rm cr}$ and 
$z_3^{\rm cr}$ are real. To evaluate \eqref{eq:DWaction1}, since either $E_1$ level and $E_3$ level are associated with the 
separatrices (critical orbits) where two of the roots of the quadratic polynomial  degenerate, 
we can express the  integrand as
\be
2E+2V(z) = (z-z_{\rm cr})^2(z-z_T)(z-z_T')\, ,
\label{eq:sep}
\ee
where $z_T'$ is the other turning point. Then the action 
integral \eqref{eq:DWaction1} can be performed and it yields
\begin{multline}
g S=\frac{1}{12}\Bigg\{\sqrt{(z_{\rm cr}-z_T)(z_{\rm cr}-z_T')}
          \left[3(z_T-z_T')^2+4(z_{\rm cr}-z_T)(z_{\rm cr}-z_T')\right]\\
    -3(z_T-z_T')^2(2z_{\rm cr}-z_T-z_T')\ln\left(\sqrt\frac{z_{\rm cr}-z_T}{z_T-z_T'}
                    +\sqrt\frac{z_{\rm cr}-z_T'}{z_T-z_T'}\right)\Bigg\}\, . 
\label{eq:dw-action1}
\end{multline}
This expression can be simplified significantly using simple  algebraic 
properties of the critical points and turning points.  Equating \eqref{eq:curve1} with \eqref{eq:sep} at the critical energy levels, 
the vanishing of the cubic term implies:
$2z_{\rm cr}+z_T+z_T'=0.$ 
Comparing the linear   terms in the two expressions, we obtain
$-(z_T+z_T')z_{\rm cr}^2-2z_Tz'_T z_{\rm cr}=2pg$. 
These  two equations allow us to express both $z_T$ and $z_T'$ in 
terms of $z_{\rm cr}$, 
\begin{align}
z_T &=-z_{\rm cr}+\sqrt\frac{pg}{z_{\rm cr}} \, , &
z_T'&=-z_{\rm cr}-\sqrt\frac{pg}{z_{\rm cr}} \, .
\end{align}
which is a useful parametrization. 
Finally, comparing quadratic terms,  one obtains 
\begin{align}
\frac{2z_{\rm cr}^3+pg}{z_{\rm cr}}=2a^2\, . 
\label{crit-solve}
\end{align}
This  equation has three solutions. These are just the three critical 
points in \eqref{eq:dw-critical} and \eqref{eq:dw-critical-approx}. For the 
bounce we choose $z_{\rm cr}=z_3^{\rm cr}$, and for the complex bion we choose 
$z_{\rm cr}=z_1^{\rm cr}$. Using these relations, the action \eqref{eq:dw-action1} 
reduces to
\begin{subequations}
\begin{eqnarray}
S_{\text{saddle}} 
 &=& \frac{4a^2\sqrt{6z_{\rm cr}^2-2a^2}}{3g}
      -4 p \ln\left(\sqrt{\tfrac{1}{2}+\sqrt{z_{\rm cr}^3/pg}}
                   +\sqrt{-\tfrac{1}{2}+\sqrt{z_{\rm cr}^3/pg}}\right) \cr
 &=&\frac{8a^3}{3g}\sqrt{1-\frac{3 p g}{4a^2 z_{\rm cr}}} 
      + 2 p\,  \text{arccosh}\left(2\sqrt{\frac{1}{2}
                        +\frac{z^3_{\rm cr}}{p g}}\right)\, .
                           \label{eq:bounce-2-action}
\end{eqnarray}
\end{subequations}
For the 
bounce we choose $z_{\rm cr}=z_3^{\rm cr}$,  $S_{\text{saddle}}$ is real, 
and for the complex bion we choose 
$z_{\rm cr}=z_1^{\rm cr}$,  $S_{\rm cb}$ has both real and imaginary parts.  In particular, 
it is easy to write that, by simply observing the fact that  $z_1^{\rm cr} <0$, and expressing 
\eqref{eq:bounce-2-action} in terms of real and imaginary parts, 
\begin{eqnarray}
S_{\rm cb}= {\rm Re} [S_{\rm cb}] \pm i p \pi,  \qquad{\rm Im}\left[S_{\text{cb}}\right] =\pm  p \pi.
\label{cb-phase}
\end{eqnarray}
The imaginary part  is the hidden topological angle (HTA) discussed in \cite{Behtash:2015kna} for integer $p$. For non-integer $p$, 
it is related both to HTA and resurgent cancellations with perturbation theory. 
The sign of the HTA depends on the choice of complex turning point, see Figure \ref{dw-1}.
A crucial point is that the imaginary part  is independent of $pg/a^3$ (so long $pg/a^3$ is small enough not to alter the tilted double-well structure.) 


In fact, it is more illuminating to express the actions in the weak coupling regime,  $\frac{pg}{a^3} \ll 1$. 
For the (real) bounce solution, $S_{\text{saddle}}$ is real,  while for the complex bion $S_{\text{saddle}}$ has an imaginary part:
\begin{eqnarray}
S_{\rm bn} &=& \frac{8a^3}{3g} +  
	{p}\ln \left(\frac{pg}{16a^3}\right)+ \mathcal{O} (g^{1/2}), \quad \cr 
S_{\rm cb}  &=& \frac{8a^3}{3g} - 
   {p}\ln \left(\frac{pg}{16a^3}\right)\pm i\pi p+ \mathcal{O} (g^{1/2}) .	
\label{eq:bounce-action}
\end{eqnarray}
Clearly, the real part of the action of the complex  bion is larger than twice the instanton action in the original formulation, which, in turn is 
 larger than the action of the bounce.  


These weak coupling expressions  are in precise agreement with the original formulation in Section \ref{correlated}, 
where we obtained instanton-anti-instanton correlated two-events by integration over the 
quasi-zero mode (QZM) thimble, $\Gamma^{\rm qzm}$.
The corresponding amplitude 
of the complex bion  and real bounce take the form: 
\begin{align}
  I_{\rm bn}  &\sim  \left( {\frac{pg}{16a^3}} \right)^{-p}  e^{-2S_I }, \cr
I_{\rm cb}
   &\sim \left( {\frac{pg}{16a^3}} \right)^p e^{-2S_I  \pm i p \pi},
   \label{amps}
\end{align}
as can be deduced in the original formulation. This one-to-one correspondence between the exact complex and real  solutions and 
the complexification of the  (QZM)-thimbles suggests that  the latter rationale is actually a systematic approximation 
to the saddles in the full quantum theory.

\subsection{Analytic continuation and monodromy:  From  real bounce to complex bion}

\begin{figure}[t]
\center
\includegraphics[width=.4\textwidth]{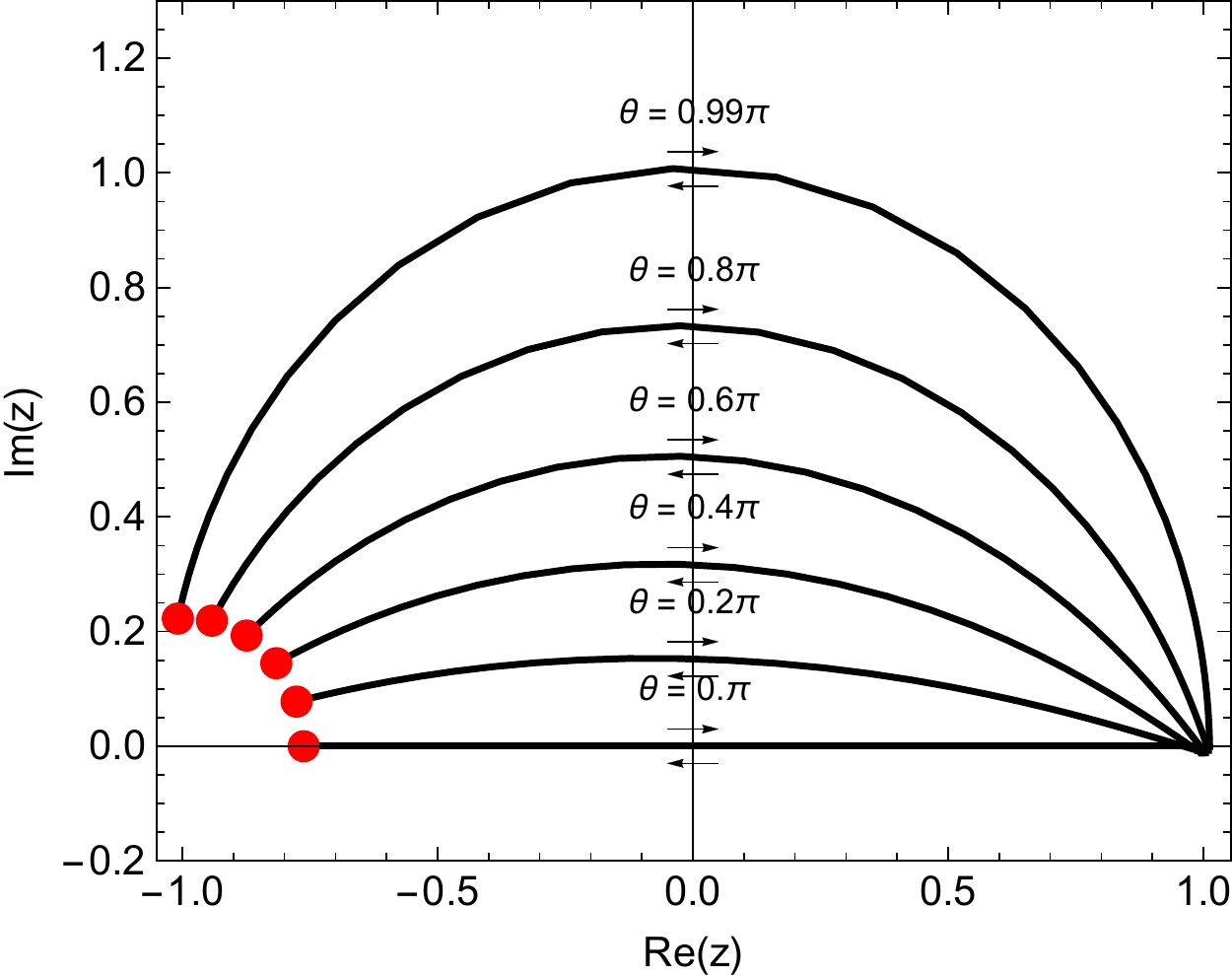}
\caption{
Parametric plot of the real and imaginary parts of the analytic 
continuation of the real bounce solution $z(t)$. We continue $p\rightarrow 
pe^{i \theta}$ and show the solution for a range of $\theta$ parameters.
$\theta=0$ corresponds to the real bounce solution, with a real turning 
point. As one changes $\theta$, the turning point becomes complex. Exactly at 
$\theta=\pi^{-}$, the real bounce turns into a complex bion.}
\label{DWcomplexbion2}
\end{figure}

By the discussion of Section \ref{isospec}, we know that for the tilted double-well potential, 
$P \hat H_{+p} P  =  \hat H_{-p}$,  namely parity maps one Hamiltonian system to an isospectral Hamiltonian system,  by \eqref{PHP}.   
The non-perturbative effects in the paired graded systems  are same, and 
can be obtained from each other by  analytic continuation. In particular,  we can 
obtain the complex bion solution by analytic continuation of 
the real bounce solution, essentially analytically continuing $p\to -p$. 
Consider  analytic 
continuation of the tilted DW theory to the complex $p$ plane, namely:
\begin{align}
p \rightarrow p\, e^{i \theta} \in  \mathbb C
\label{acontinue}
\end{align}
Start with $V_{+}(z) = \half 
(W'(z))^2 +  \tfrac{pg }{2}  W''(z)$. Turn on  $\theta\neq 0$. 
At 
$\theta=\pi$,  we reach  the theory described by the potential 
$V_{-}(z)$, which is  the parity transform  of $V_{+}(z)$.  Upon analytic 
continuation all the way to $\theta=\pi$:
{\bf (i)} The local maximum (of course, restricted to the real axis)  of   $-V_{+}(z)$  becomes the global 
maximum of  $-V_{-}(z)$. 
{\bf (ii)} The global maximum of the $-V_{+}(z)$  becomes the local maximum 
of  $-V_{-}(z)$.
{\bf (iii)} The real turning point of the real bounce solution in the 
$-V_{+}(z)$ system turns into a complex turning point in the $-V_{-}(z)$ 
system.  
At intermediate $\theta$, the complex  solution is always a complex  classical  solution to the holomorphic equations of motions,  and the profiles of 
the real and imaginary parts are shown in Figure \ref{dw-1} and Figure \ref{DWcomplexbion2}  for various values of $\theta$.


\begin{figure}[t]         
\centering
\includegraphics[width=8.5cm]{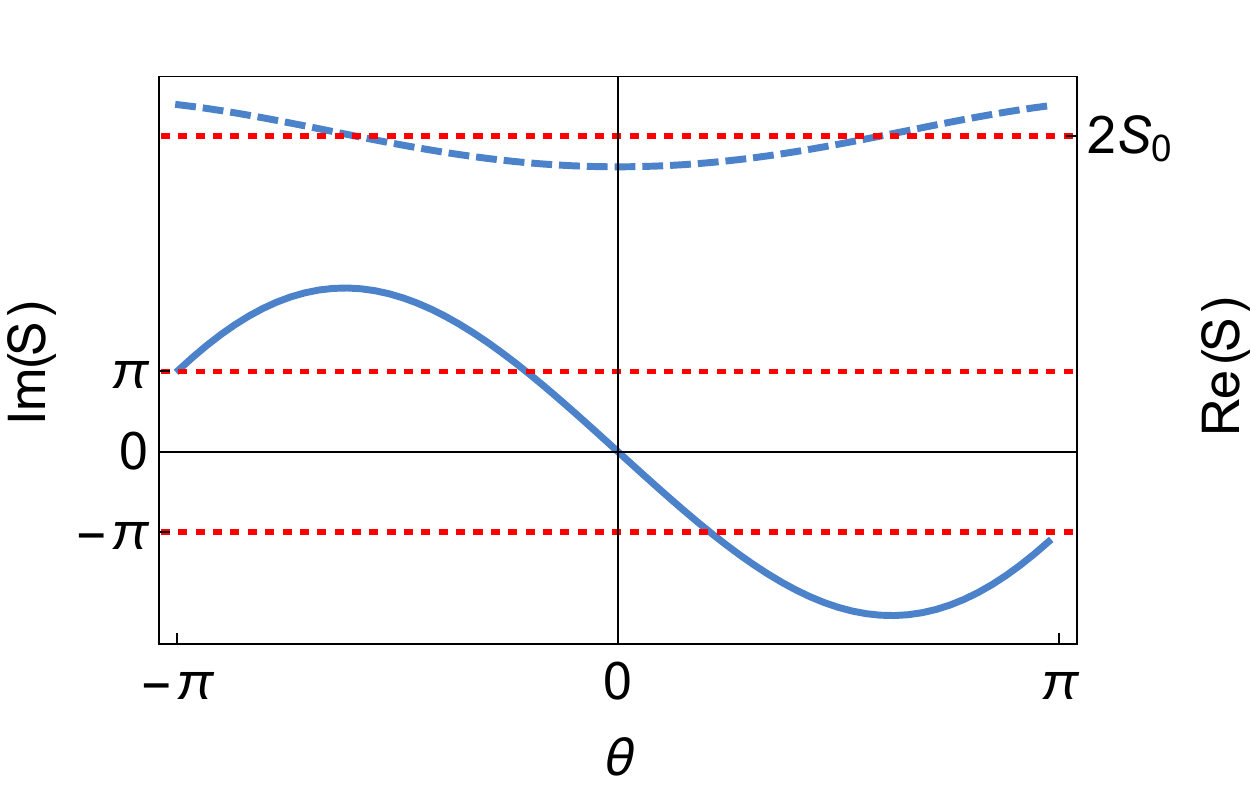}
\caption{Real and imaginary part of the action of the analytically 
continued  real bounce solution for $p=1$. Under $\theta \rightarrow \theta + 2\pi$, a complex solution 
goes to the conjugate, reflecting monodromy 2 nature of the solution.   
}
\label{DWaction}
\end{figure}

Fig.~\ref{DWaction} depicts the real and imaginary part of the action  (${\rm Re}(S)$, ${\rm Im}(S)$)  of  the analytically 
continued  real bounce solution as a function of $\theta$.  
 $\theta=0$ correspond to the real bounce solution, which is purely real.  
 Note that the action of the real bounce is less than two-times instanton action, $2S_I$, in terms of the instanton action of the original formulation. 
 The reason for this is that the bounce is associated with the lower separatrix (critical orbit) associated with energy $E_3$ in Fig.~\ref{QMpot1-tilt}. 
 As  $\theta$ is turned on, it interpolates into a smooth saddle of the analytically continued path integral.   
 At  $\theta=\pi^{-}$, the real bounce turns into a complex bion. The real 
part of the action is continuous at $\theta=\pi$, while the imaginary part 
is discontinuous.  
 Note that the real part of the action of the complex bion is larger than twice the instanton action, 
 $2S_I$. 
 The reason for this is that the bion  is associated with the upper  separatrix  associated with energy $E_1$ in Fig.~\ref{QMpot1-tilt}. 
As for the imaginary part,  for $p\not\in  \mathbb{Z}^{+}$, the ambiguity of the action  is related 
to resurgence, and ambiguity cancellation, and for $p\in \mathbb{Z}^{+}$ 
it is related to the hidden topological angle.

 In fact, as $p$ changes its  phase by $2\pi$ in (\ref{acontinue}), the potential  $V_{+}(\x)$ turns back to itself,  but two complex bion solutions are 
interchanged.  On the pair of solutions, $ (z_{\rm cb},  z_{\rm cb}^{*})^T$, the monodromy matrix acts as 
 \begin{align}
 M(2\pi)= \left( \begin{array} {cc}
 0 & 1 \cr
 1& 0 
 \end{array}  \right), \qquad M(4\pi) = M^2=   \left( \begin{array} {cc}
 1 & 0 \cr
 0& 1
 \end{array}  \right)\, . 
 \end{align}
Thus, the bounce or bion solutions have   monodromy of order 2, reflecting the 
two-fold ambiguity in the choice of the exact solutions for the physical 
theory.  We show in Section \ref{sec:csr} that this two-fold ambiguity is 
related to the two-fold ambiguity in Borel resummation of perturbation theory. 


\subsection{Vacuum  as a  complex bion gas} 

{\bf Reminder of the bosonic case, ground state wave function vs. instanton gas:}	
 It is well-known that in the bosonic (symmetric) double-well potential, 
the dilute gas of instantons dictates the ground state properties of the system. In particular, the non-perturbative level splitting between 
the first excited state and the ground state can be shown to be due to 
instantons.  More specifically, the ground state wave function is symmetric 
with respect to parity, \eqref{GS-DW}, and the probability to find the 
particle on the left and right well is equal. The dilute instanton gas 
reflects this property. In fact, in the dilute instanton gas, the classical 
Euclidean particle spends half of its  (Euclidean) time at $+a$ and the other half at 
$-a$, reflecting the symmetry of the ground state wave function. 


\begin{figure}[t]          
\centering
\includegraphics[width=15.6cm]{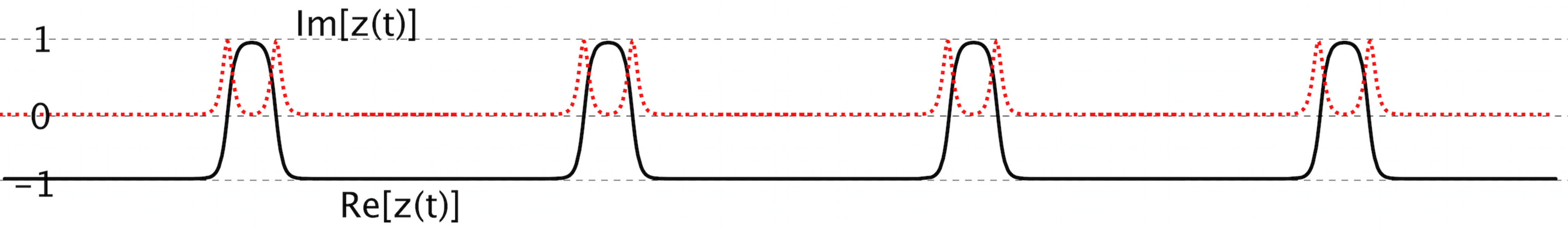}
\caption{The ground state properties  of the theory with fixed number of 
fermions ($N_f \geq 1$) is controlled by a dilute gas of complex exact  bion saddles, 
with density $e^{-2S_0}$. 
Black solid  (red dashed) line is the real (imaginary) part of the complex 
solution. The imaginary part is related to the HTA phenomenon.
}
\label{DW-bions}
\end{figure}

\begin{figure}[t]          
\centering
\includegraphics[width=8cm]{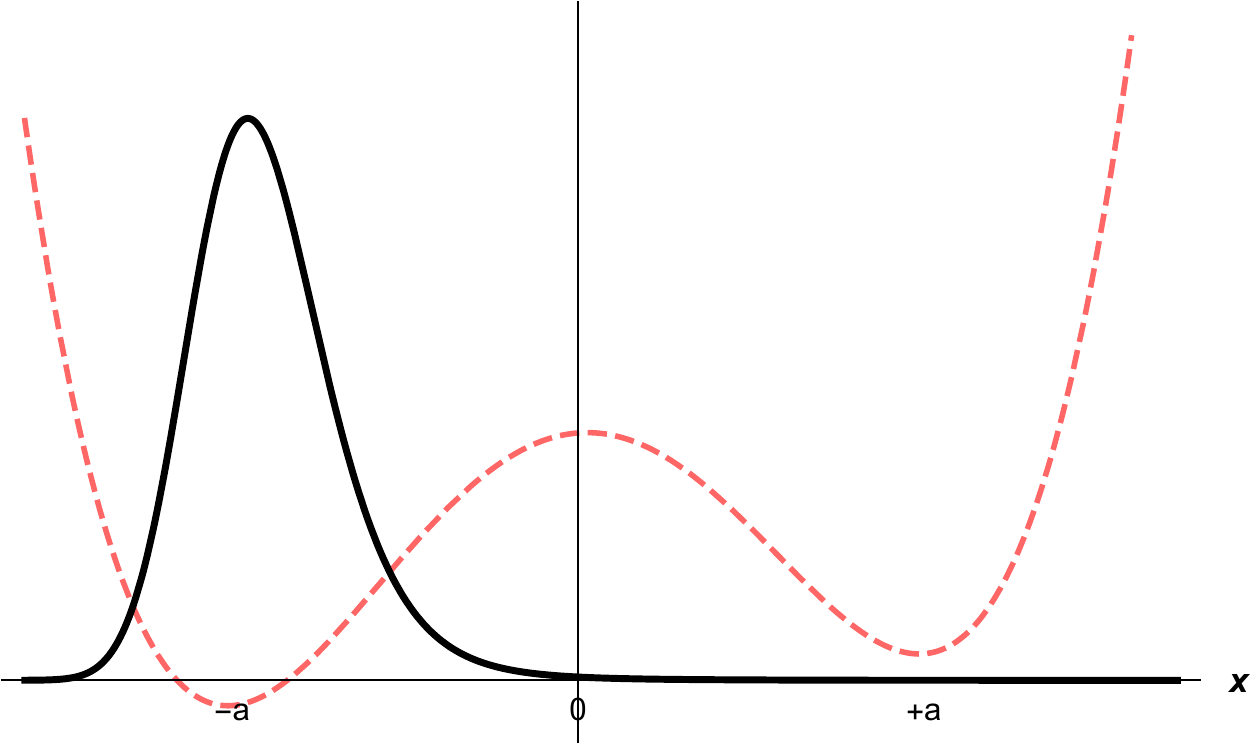}  
\caption{Quantum modified potential (red dashed) for $N_f \geq 1$ theories,  and  one of the two  ground state wave function. The probability to find the particle on an $+a$ well is 
suppressed exponentially, by $e^{-2S_I}$, consistent with the dilute complex bion gas description. As is well-known, in the bosonic   case, $N_f=0$,  the wave function is equally dominant both on the even and odd sites. Note the absence of any  bump in the wave function at $x=+a$. 
}
\label{DW-bions-WF}
\end{figure}

\noindent	
{\bf Ground state wave function vs. bion gas:}	
We have shown that complex bions are exact saddle points of the path integral 
in theories with fermions ($N_f \geq 1$). The Euclidean description of one 
of the vacua of these theories can be viewed as a dilute gas of neutral 
bions, see Figure \ref{DW-bions}. The Euclidean configurations reflects
the fact that the ground state wave function is not symmetric. Fermion
number is conserved, and in any of the two Fock space sectors the particle
spends most of its time in the deeper well. More specifically, we showed
in \eqref{g-s} that one of two ground states is described by
\begin{eqnarray}
 |  \Psi_0\rangle |  \Omega   \rangle 
   =  |\Psi_0\rangle \otimes  | \downarrow \ldots \downarrow    \rangle   
   =  a_1 \underbrace{ |R, 0\rangle | \Omega 
          \rangle}_{\rm Ground \;  state  \; in \; {\cal H}_R} 
    + a_2  \underbrace{ |L,  0\rangle |  \Omega  
          \rangle}_{\rm N_f^{\rm th}-excited\;  state  \; in \; {\cal H}_L} +  \ldots \, .
\label{g-s}
\end{eqnarray} 
and the other ground state is given by $L\leftrightarrow R$. Since the density 
of the complex bions is  $e^{-2S_I}$,  the  probability to find the particle on the shallow well 
 compared to an deep well is    exponentially small.  This can also be shown to be the case by using WKB 
 approximation. 
\begin{align}
 \frac{{\rm {Pr.}}( +a )} {{\rm {Pr.}}(  -a)}  
  \sim e^{-2 \Delta W/g} =  e^{-S_b} \sim e^{-2S_0} 
     \qquad  {\text {for one of the ground states.} }
\end{align}
For the other ground state, odd and even sites are interchanged.
  Also, note that the complex bions {\it increase} the ground state energy for odd-integer $N_f$,  (unlike instantons in bosonic theories), 
 and lower  it for  even-integer $N_f$.


\vspace{0.2cm} 	
\noindent
{\bf What would happen if the complex  bion saddle did not contribute?}	
Consider the $N_f=1$ theory, corresponding to  $\N=1$ supersymmetric QM. In 
this theory supersymmetry is spontaneously broken, and the ground state 
energy is given by $E_{\rm gr} \propto e^{ -2S_I }$. This is an effect due to 
complex  bion configuration. If this configuration is not 
included, one would erroneously conclude that the ground state energy 
remains zero, and supersymmetry is unbroken.  
In the original description involving fermions and  instantons, the ground stage energy 
is  an instanton--anti-instanton $[\I\bar \I]$ effect. 
A  rigorous attempt to compute the $[\I\bar \I]$  contribution  amounts to the  complexification of the quasi-zero mode direction, 
and essentially results in a systematic  approximation to the  complex bion configuration as described in  Section.\ref{correlated}.

\vspace{0.2cm} 
\noindent
{\bf Complexity of the solution and the hidden topological angle (HTA):} 
Again we consider the case $N_f=1$. The ground state energy is 
given by 
\begin{align}
E_{\rm gr} \propto - e^{ -2S_I \pm iN_f \pi}\Big|_{N_f=1} =  e^{-2S_I}  >0\, . 
\end{align}
Although the action is multi-valued and ambiguous for general non-integer 
$N_f$, it becomes unambiguous for even $N_f$. For odd-integer $N_f$, it is 
again unambiguous, because the  imaginary part of the action is defined 
only modulo $2\pi $, and hence, $\pm \pi$ are actually identified.\footnote{
This aspect of the hidden topological angle is the same as the usual 
topological $\Theta$-angle in Yang-Mills  and other theories. Yang-Mills 
theory  is invariant under CP-at $\Theta=0$ (obvious) and at $\pi$, because  
$\pm \pi$ are identified.}
However, the fact that $|{\rm Im}(S)|=\pi$ is extremely important, because 
it determines the sign of the ground state energy. Without the HTA, or if 
the solution were real, we would conclude that the ground state energy must 
be  negative.  Indeed, in some of the older literature on instantons it is 
asserted that non-perturbative saddles always {\it reduce} the ground state 
energy. But this would be inconsistent with the supersymmetry algebra, which 
implies positive-definiteness of the  ground state energy 
\begin{align}
\langle  {\mathsf G_i} | H | {\mathsf G_i} \rangle  
 = \Big| Q   | {\mathsf G_i} \rangle \Big|^2 \geq 0 \, . 
\label{pos-def}
\end{align}
The role of the hidden topological angle (and complexity of the solution)  is to  
restore consistency between supersymmetry algebra and the semi-classical 
approximation.

\section{Graded formulation: Periodic Sine-Gordon potential} 
\label{sec:periodic}

{\scriptsize {\bf Outline:}  In this section we construct new  exact saddle point solutions for the 
system whose bosonic potential is the periodic Sine-Gordon potential. In addition to an exact real bounce solution, 
there is an exact real bion solution, and also an exact complex bion solution. 
The main new phenomenon is that there is a multi-valued bion solution which
is also singular, but has  finite action. We show that the singular bion 
contributes to the path integral, and must be included in the semi-classical 
analysis in order to obtain results consistent with the Witten index, and the SUSY algebra in the $N_f\rightarrow 1$ limit.  This resolves an old  and controversial issue 
regarding the inclusion multi-valued, singular saddles in the evaluation of path integral. }

\subsection{Quantum modified potential  and exact non-BPS solutions}
 The Euclidean Lagrangian for the periodic auxiliary potential in the 
presence of $N_f$ Grassmann-valued fermion fields  is given by a set of  
graded  Lagrangians, as described in \S.\ref{GHS}, with $W^\prime(z)=
-2 a^2\sin(z/2a)$. The graded Euclidean Lagrangian and energy conservation 
relations for the  periodic system are given by 
\begin{subequations}
\begin{align} 
 {\cal L} 
  &= \half \dot z^2 + V(z) 	
   = \half \dot z^2 +2a^4- 2a^4\cos^{2}(z/2a)- \tfrac{pg }{2} a \cos(z/2a), \\
 E&= \half \dot z^2  -V(z) 
   = \half \dot z^2 -2a^4+2a^4\cos^{2}(z/2a)+ \tfrac{p g }{2} a \cos(z/2a), 
\label{p-deformedLagrangian} 
\end{align}
\end{subequations}
for the fermion-number sector $k$, where $p= N_f -2k$, for $k=0, \ldots, N_f$. 
As in the double-well case, the fermion-induced potential is of order $g$.  
  
 The inverted potential is shown in Figure \ref{three_bions}, along with 
the physically important saddles. The bounce solution is straightforward, 
but is not relevant for the ground state properties. The 
real bion is expected, as it exists as a solution of the ordinary (real) 
Newton equation in the inverted potential.  The surprise is the complex bion 
solution to the holomorphic Newton equation, which has no counterpart in 
(real) classical mechanics.  

  	
 To describe the non-perturbative saddles in the problem, we study the 
classical solutions in the inverted potential, $-V(z)$, in Euclidean time. 
The solutions can be found by quadratures:
\begin{align} 
\int_{t_0=0}^{t} dt   
   =  -\int_{z_T}^{z}  \frac{dz}{\sqrt {2(E+V(z))}}   
   =  -\int\frac{dz}{\sqrt{2 E+4a^4-4a^4\cos^{2}(z/2a) -  pg a\cos(z/2a)}}\, . 
\label{eq:integraleom}
\end{align}
Substituting  
\begin{align}
Z=a\cos(z/2a)
\label{mapping-21}
\end{align}
the  integration takes a form familiar from the  Weierstrass elliptic 
functions: 
\begin{align}
t-t_{c}  = \int_{\cos(z_{T}/2)}^{\cos(z/2)}
    \frac{dZ}{\sqrt{Z^{4}+\tfrac{1}{4a^2}pg Z^{3}
       -\tfrac{1}{2a^2}({E}+4a^4)Z^{2}-\tfrac{1}{4}pgZ+a^4+\tfrac{1}{2} E}}\, , 
\label{master-int}
\end{align}
where $t_{c}$ is the integration constant (which becomes the center position 
modulus of the solution), and $z_{T}$ is the turning point. The mapping 
\eqref{mapping-21} maps two physically distinct points, $z$ and $-z$, to 
the same value of $Z$.  The inversion of this mapping is two-valued, and 
requires some care, which  will be easy to understand both on physical 
and mathematical grounds.  There will be another multi-valuedness in the story unrelated 
to this, and tied with the nature of the exact complex  bion solution. 
The integral \eqref{master-int} can be associated 
with a complex algebraic curve:
\begin{align} 
F(Z)= \mathcal{P}^2(Z)
  := Z^{4}+\tfrac{1}{4a^2}pg Z^{3}-\tfrac{1}{2a^2}({E}+4a^4)Z^{2}
       -\tfrac{1}{4}pgZ+a^4+\tfrac{1}{2} E.
\label{PX}
\end{align}
The general solution $z(t)$ for an energy $E$ that corresponds to 
a turning point $Z_T$ (a zero of the polynomial $F(Z)$), is given by
\begin{align}
z(t) = \pm 2a\arccos\left(  
   Z_{T}+ \frac{ \tfrac{1}{4}F'({Z_T})}{\wp(t; g_2(E), g_3(E))
           -\tfrac{1}{24}F''(Z_T)} \right). 
\label{eq:general-sg}
\end{align}
where  $g_2(E), g_3(E)$ are the associated with algebraic  lattice invariants,
given by 
\begin{eqnarray}
g_{2} (E) & = &  \tfrac{4a^4}{3}+\tfrac{2}{3}E +\tfrac{1}{48a^4}E ^{2}
                +\tfrac{1}{64a^2}g^{2}p^{2},\cr 
g_{3} (E) & = & -\tfrac{8a^6}{27}-\tfrac{2a^2}{9}E -\tfrac{5}{144a^2}E ^{2}
                +\tfrac{1}{1728a^6}E^{3}-\tfrac{1}{192a^2}g^{2}p^{2}
                -\tfrac{1}{768a^4}g^{2}p^{2}E.
\end{eqnarray}
Below, we construct the most interesting solutions: First, a simple bounce 
solution, and then the complex and real bion solutions  which determine the 
ground state properties of the quantum system. 

\subsection{Exact (real) bounce solution} 

\begin{figure}[t]
\center
\includegraphics[scale=0.6]{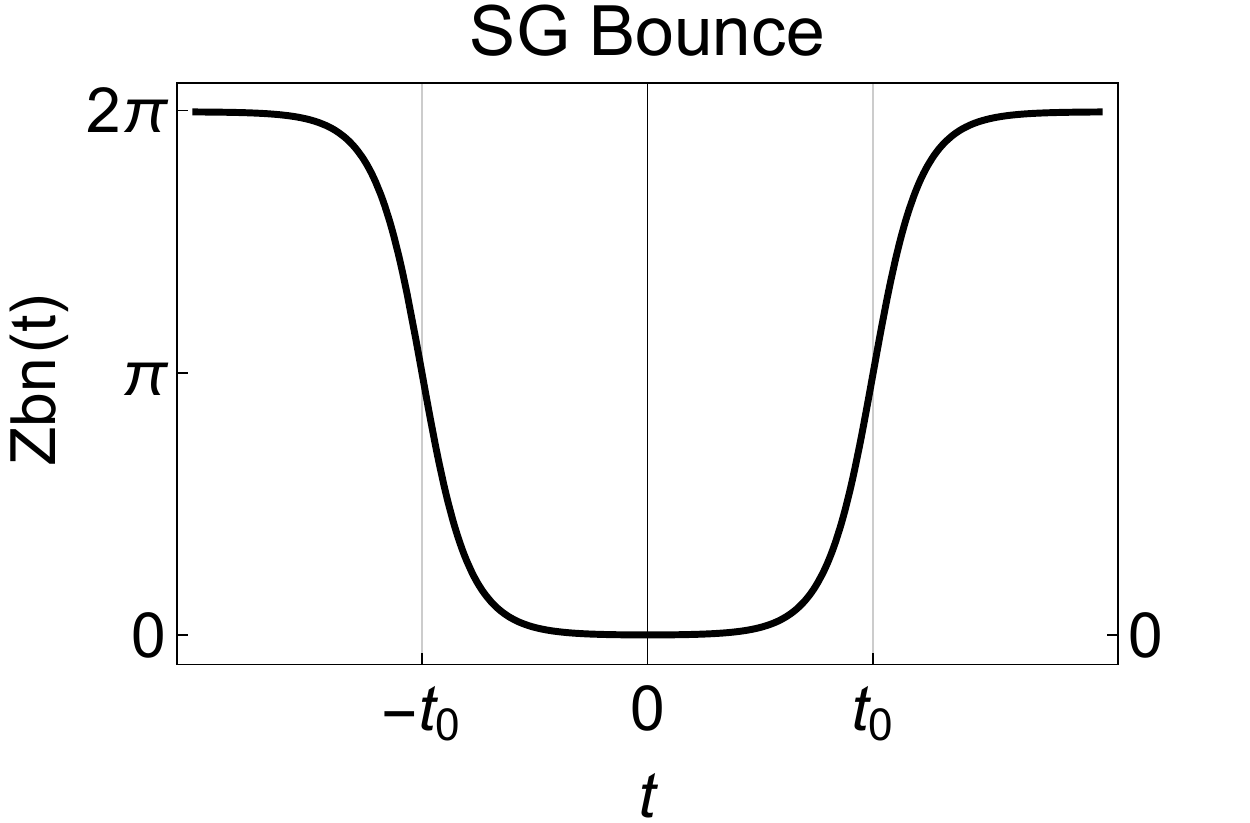}
\caption{Bounce solution for the Sine-Gordon case,  for $a=1$. The size   of 
separation  (or the size of plateau  is $ 2t_{0}\approx a^{-1}  \ln 
\left(\frac{32a^3}{p g}\right)$.}
\label{sg-bounce}
\end{figure}

 The ``bounce'' solution  \cite{Coleman:1978ae,Campbell:1986nu,Mussardo:2004rw} is a classical solution 
with Euclidean energy  
\begin{align}
 {E}=-\tfrac{1}{2}pg a
 \label{lower-sep}
\end{align}
equal to the value of the inverted potential at $z/a=2\pi +  4\pi k$, for 
$k\in \Z$, the lower of the (real) maxima of the inverted potential. 
At this 
energy there  are real turning points at $z=z_T$ where
\begin{align}
 \cos \tfrac {z_{T}}{2a} = 1-\tfrac{gp}{4a^3}, \qquad 
 z_{T}=  \pm 2a\arccos\left(1-\tfrac{gp}{4a^3} \right) 
                  + 4a\pi k , \quad k\in\Z \, .
 \label{turningpoint}
\end{align}
The bounce solution bounces off the (real) turning point $z_T$ at a 
finite time $t=0$, tending as $t\to\pm \infty$ to $z/a=2\pi +  4\pi k$, 
for $k\in \Z$, see Figure \ref{sg-bounce}. 

With these boundary conditions, the general solution \eqref{eq:general-sg} 
simplifies significantly, as the Weierstrass function reduces to a 
singly-periodic function. We find
\begin{subequations}
\begin{align}
z_{{\rm bn}}(t) 
 & =2a\,{\rm arccos}\left(\frac{1-\frac{pg}{8a^{3}-pg}\cosh^{2}
     \left(\omega_{{\rm bn}}t\right)}{1+\frac{pg}{8a^{3}-pg}\cosh^{2}
     \left(\omega_{{\rm bn}}t\right)}\right)\\
 & =4a\,{\rm arctan}\left(\sqrt{\frac{pg}{8a^{3}-pg}}\,
          \cosh\left(\omega_{{\rm bn}}t\right)\right)\, , 
\label{eq:sg-bounce}
\end{align}
\end{subequations}
where $\omega_{\rm bn}$ is the  curvature of the potential at $z= 2\pi a$, 
given by \footnote{The linear frequency on the well at  $z= 2\pi a$ (the value of the frequency as one take $E \rightarrow \tfrac{1}{2}pg a$ limit in the 
original potential)   is equal to the 
{\it non-linear} frequency on the well at  
$z=0$ for the energy level $E = \tfrac{1}{2}pg a$.  
\begin{align}
\omega_{\rm bn}= \omega^{\rm  2\pi a}_{\rm linear} (E= \tfrac{1}{2}pg a ) =  \omega^{\rm 0}_{\rm non-linear}(E=\tfrac{1}{2}pg a  )
\end{align}
This  identity is a consequence of the relation \eqref{per-rel} for the periods of the complex algebraic curve.  
Physically, in the singly periodic limit of the Weierstrass function,   $i  \omega^{\rm 0}_{\rm non-linear}(E=\tfrac{1}{2}pg a  ) $ is the (pure imaginary) frequency which remains finite, 
while the other purely real  frequency tends to zero.} 
\begin{eqnarray}
\omega_{\rm bn}= \sqrt  {V'' (2a\pi ) }  =  a\sqrt{1 -\tfrac{pg}{8a^3}}\, .
\label{tau-2} 
\end{eqnarray}
Alternatively, we can obtain this solution more directly by substituting 
into the energy condition \eqref{p-deformedLagrangian} the ansatz
\begin{eqnarray}
z(t)=4a\ {\rm arctan}\left(f(t)\right) \, , 
\label{eq:ansatz1}
\end{eqnarray}
which produces the following equation for $f(t)$:
\begin{eqnarray}
a^2\dot{f}^2
  = \frac{1}{8}\left(E+\frac{p g}{2} a\right) f^4+\frac{E+4a^4}{4} f^2 
         +\frac{1}{8}\left(E-\frac{p g}{2} a\right)\, . 
\label{eq:ansatz2}
\end{eqnarray}
At the bounce energy, $E=-\frac{p g}{2} a$, the quartic term in $f$ vanishes, 
and we obtain the solution in \eqref{eq:sg-bounce}. The bounce solution 
clearly satisfies the desired boundary conditions (mod $4 a\pi$): 
\begin{equation}
z_{\text{bn}}(\pm \infty)= 2a \pi,  \qquad     
z_{\text{bn}}(0)  =  2a \arccos\left(1 - \tfrac{gp}{4a^3} \right)\, . 
\label{bounce-limits}
\end{equation}
It is physically instructive to rewrite the bounce solution as
\begin{eqnarray}
z_{\rm bn}(t)
 = 4\pi a -4a\Big[ {\rm arctan}\left(\exp\left[ -\omega_{\rm bn}(t-t_0)\right]\right) 
  + {\rm arctan}\left(\exp\left[ \omega_{\rm bn}(t+t_0)\right]\right)\Big]\, ,
\label{eq:sg-bounce2}
\end{eqnarray}
where  we have defined 
\begin{eqnarray}
t_0 \equiv \omega^{-1}_{\rm bn}  
  \ln \left[\sqrt{\frac{8a^3}{p g}}\left(1+\omega_{\rm bn}/a \right)\right]
    \approx \frac{1}{2a} \ln \left(\frac{32a^3}{p g}\right).
\end{eqnarray}
Thus, the real bounce has the form of a Sine-Gordon instanton at $t=-t_0$, 
followed by a anti-instanton at $t=+t_0$, as shown in 
Figure \ref{sg-bounce}. However, unlike the superposition of the instanton-anti-instanton, which is an approximation, 
the real bounce  solution is exact. 

As in the double-well case, the bounce is parametrically flat. The bounce 
spends a parametrically large time in the vicinity of the turning point 
(unlike instantons). The size of the flat-bounce is given by 
\begin{align}
 2t_{0}\approx  m^{-1}_b\ln \left(\frac{32a^3}{p g}\right) \, ,
\label{t0_SG_real_bn}
\end{align}
where $m_b=a$ in the SG case. Note that the size of the bounce obtained 
in the exact solution agrees precisely with the integration over the 
quasi-zero mode thimble ${\Gamma^{\rm qzm}}$.  In particular, the scale \eqref{t0_SG_real_bn}
provides the dominant support of the QZM integral \eqref{critical-1}, and corresponds to the saddle of the QZM-integration. 

 Finally, the action of the bounce solution is given by
\begin{subequations}
\begin{eqnarray}
 S_{\rm bn} 
 &=&  \tfrac{16a^3}{g} \sqrt{1-\tfrac{g p }{8a^3} } 
     - 2 p\,  {\rm arctanh} \left[\sqrt{1-\tfrac{g p}{8a^3}}\right ] \\
  & =&   \tfrac{16a^3}{g} \sqrt{1-\tfrac{g p }{8a^3} }    -  p  \log \left[  \frac{  1+\sqrt{1 - \tfrac{gp }{8a^3}}   } { 1- \sqrt{1- \tfrac{gp }{8a^3}}   }    \right ]  \\
 &=& \frac{16a^3}{g} - p\ln\left(\frac{32a^3}{p g}\right )+\mathcal{O}(g).
\end{eqnarray}
\end{subequations}
 Similar to DW, the action of the real bounce is less than two-times instanton action, $2S_I$. 
 The reason for this is that the bounce is associated with the lower separatrix  associated with energy \eqref{lower-sep}. 

\subsection{Exact real bion solution}
\label{sec:sg-real-bion}
  
\begin{figure}[t]                
\centering
\includegraphics[width=8cm]{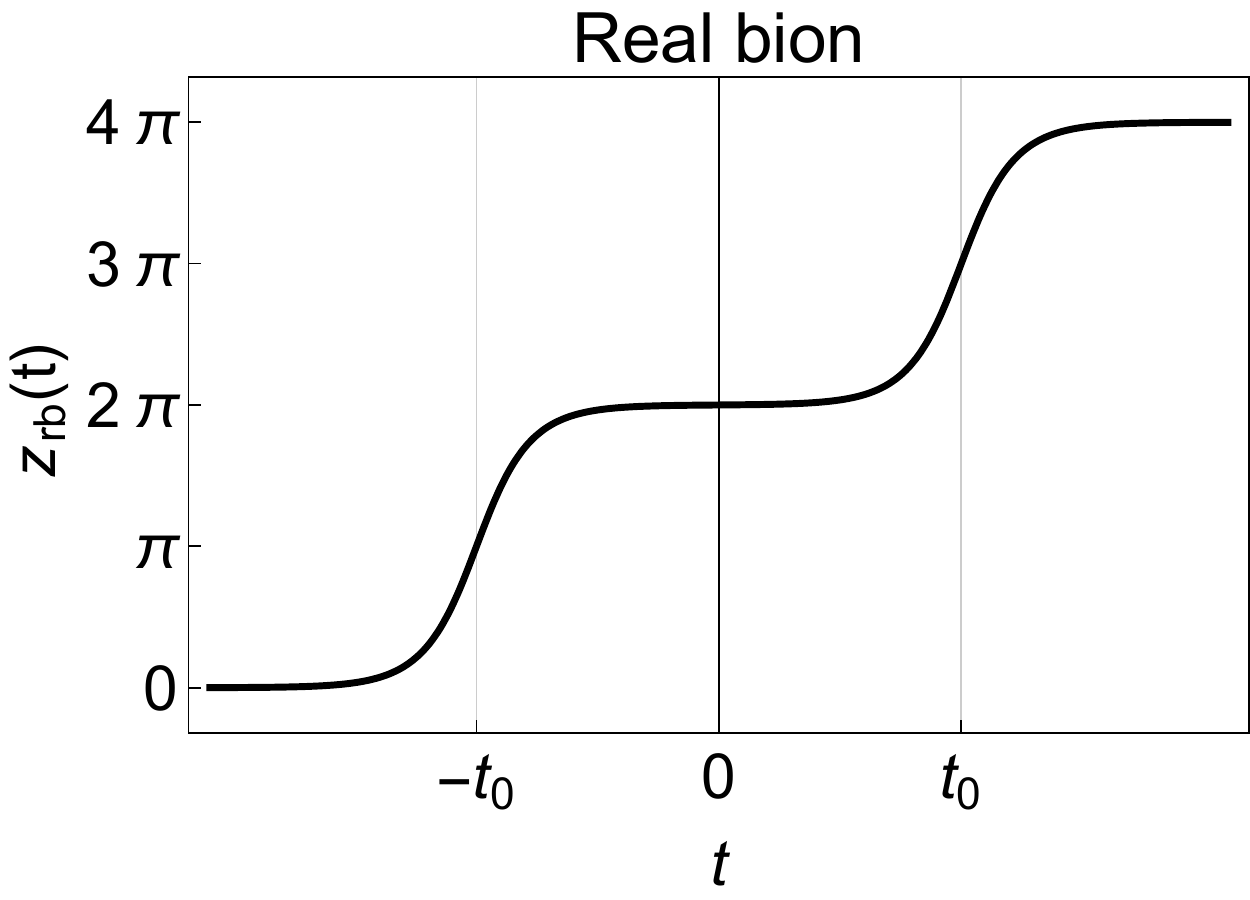}
\caption{Exact real bion solution, interpolating from 0 to $4\pi$, for $a=1$.  
The plateau in the middle is associated with the characteristic size of the 
solution $2t_{0}\approx a^{-1} \ln\left(\frac{32a^3}{pg}\right)$. The ground state 
properties of the quantum theory in leading order semi-classics are described 
in terms of a dilute gas of real bions and complex bions.}
\label{ac-4}
\end{figure}

The real bion solution has Euclidean energy
\begin{align}
 {E}=  \tfrac{1}{2}pg a\, , 
 \label{sep-2}
\end{align}
equal to the inverted potential at its global maxima: $z/a=4\pi k$, for 
$k\in\Z$. 
Qualitatively, the classical particle, after spending an infinite 
amount of time at the global maximum, rolls down at some {\it instant} in 
time (just like the instanton), and then spends a parametrically large but 
finite amount of time in the vicinity of the local maximum near $2\pi a$, and continue its path, reaching to the next equivalent global 
maximum at $4\pi a$ after infinite time. The exact real bion solution is 
shown in Fig. \ref{ac-4}.
	
At the  energy level of upper separatrix \eqref{sep-2},   the general solution \eqref{eq:general-sg} simplifies 
significantly, as the Weierstrass function reduces to a singly-periodic 
function, leading  to 
\begin{eqnarray}
z_{\rm rb}(t)
 &=& 2\pi a+4 a\arctan\Big( \textstyle{\sqrt{\frac{pg}{8a^3+pg}} }
  \sinh(\omega_{\rm rb} t)\Big),
\label{real-bion}
\end{eqnarray}
where $\omega_{\rm bn}$ is the   curvature of the potential at $z= 0$ and is 
 given by
\begin{equation}
\omega_{\rm rb}= \sqrt{V''(0)}  =  a\sqrt{1 +\tfrac{pg}{8a^3}}\, .
\label{tau-rb} 
\end{equation}
This satisfies the boundary conditions:
\begin{eqnarray}
z_{\rm rb}(-\infty) = 0  \quad, \quad 
z_{\rm rb}(0) = 2\pi a    \quad, \quad 
z_{\rm rb}(+\infty) = 4\pi a \, . 
\label{eq:sg-real-bion-bcs}
\end{eqnarray}
Alternatively, applying the transformation $z(t)=2\pi a +4a\arctan f(t)$ 
to \eqref{p-deformedLagrangian}, we obtain 
\begin{eqnarray}
a^2\dot{f}^2 = \frac{1}{8}\left(E-\frac{p g}{2} a\right) f^4
           +\frac{E+4a^4}{4} f^2 +\frac{1}{8}\left(E+\frac{p g}{2} a\right)\, . 
\label{eq:ansatz3}
\end{eqnarray}
At $E=\frac{p g}{2} a$ the quartic term in $f$ vanishes, and we obtain the 
solution in \eqref{eq:real-bion}. It is instructive to rewrite the real bion 
solution as
\begin{equation}
z_{\rm rb}(t) = 2 a\pi +  
  4a\left( {\rm arctan}\left(\exp\left[\omega_{\rm rb}(t+t_0)\right]\right) 
          -{\rm arctan}\left(\exp\left[ -\omega_{\rm rb}(t-t_0)\right]\right)
  \right)\, , 
\label{eq:real-bion}
\end{equation}
where $t_0$ is 
\begin{eqnarray}
t_0 = \omega^{-1}_{\rm rb}
     \ln  \left[\sqrt{\frac{8a^3}{p g}}\left(1+\omega_{\rm rb}/a \right)\right]
\approx \frac{1}{2a} \ln \left(\frac{32a^3}{p g}\right).
\label{eq:sg-t0}
\end{eqnarray}
Hence, one observes that the  real bion solution  is the exact version of the 
correlated instanton-instanton event described in Section \ref{correlated}. 
Despite the fact that the 
$[{\cal II}]$ event is an approximate solution in the original formulation, 
the real bion solution is exact. In the regime $pg \ll a^3$, the relation 
between the two is: 
\begin{eqnarray}
z_{\text{rb}}(t) 
  \approx  x_{{\I}} (t+ t_0) +x_{\I} (t-t_0) + 2 a\pi ,
              \quad\quad\quad
     \text{for \ensuremath{pg\ll a^3.}} 
\label{quasi-zero_exact-2}
\end{eqnarray}
The characteristic size of the real bion solution  \eqref{eq:sg-t0} agrees with  
\eqref{critical-1}, which is the saddle point of the QZM-thimble 
$\Gamma^{\rm qzm}_{+}$  integration in the original formulation, see 
Figure \ref{tau-plane}. 

The action of the real bion solution is given by
\begin{subequations}
\begin{eqnarray}
S_{\rm rb} 
 &=& \tfrac{16a^3}{g} \sqrt{1+\tfrac{g p }{8a^3} } 
   + 2 p\,  {\rm arctanh} \left[ 1/\sqrt{1+\tfrac{g p}{8a^3}}   \right ]\\
 &=& \frac{16a^3}{g} + p \ln\left(\frac{32a^3}{p g}\right )+\mathcal{O}(g).
\label{eq:sg-action-real-bion}
\end{eqnarray}
\end{subequations}
The  action of the real bion is larger than twice the instanton action, 
$2S_I$. The reason for this is that the  real bion  is associated with 
the upper separatrix  with energy \eqref{sep-2}, and the area underneath 
the separatrix in classical phase space is larger than the 
instanton-anti-instanton  separatrix. 

\subsection{Exact complex bion solution}
\label{sec:sg-complex-bion}

\begin{figure}
\center
\includegraphics[width=8cm]{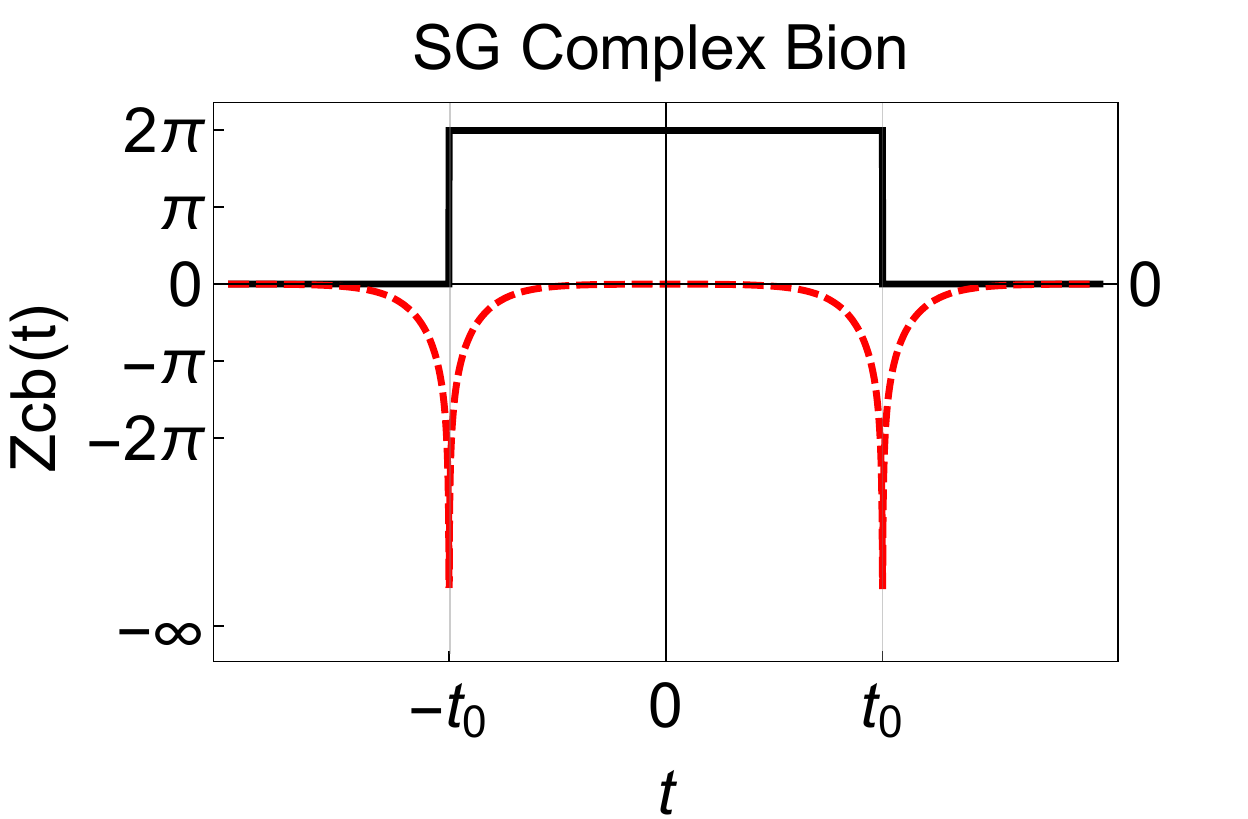}
\caption{Real (black) and imaginary (red) parts of the complex bion for 
the Sine-Gordon case, The separation is $2t_0\approx a^{-1} \ln \left(\frac{32a^3}
{p g}\right)$. The singular solution smooths out upon analytic continuation $p \rightarrow pe^{i \theta}$.}
\label{re_im_cb_sg}
\end{figure}

 To find the exact complex bion solution, we impose the boundary condition 
that at $t=0$ the solution starts at one of the complex turning points  
\begin{eqnarray}
z_T=2 a\pi\pm 2 i a \,{\rm arccosh}\left(1+\frac{p g}{4a^3}\right)\, , 
\label{eq:sg-complex-turning}
\end{eqnarray}
for which the Euclidean energy is
\begin{align}
 {E}=  \tfrac{1}{2}pg a\, .
 \label{sep-3}
\end{align}
Then, from the general solution \eqref{eq:general-sg}, or by substituting 
an appropriate ansatz, we find the two complex conjugate complex bion 
solutions to be  
\begin{eqnarray}
z_{\rm cb} =2 a\pi \pm 4 a \;{\rm arctan}
   \left( i \sqrt{\frac{ p g}{8a^3+p g}}\, \cosh\left(\omega_{\rm cb} t \right)
    \right),
\label{eq:sg-complex-bion}
\end{eqnarray}
where $\omega_{\rm cb}$ is the frequency of the system at  $z=0$ well,  given by:
\begin{equation}
\omega_{\rm cb}= \sqrt  {V'' (0 ) }  =  a\sqrt{1 +\tfrac{pg}{8a^3}}.
\label{tau-cb} 
\end{equation}
This expression is very similar to that of the real bion, which is related to the
fact that both solutions start at the global maximum. The solution is complex, 
and its real and imaginary parts are plotted in Figure \ref{re_im_cb_sg}.

Note that the real part jumps, and the imaginary part is correspondingly 
singular, at a particular $\pm t_0$, where $2t_0$  can be interpreted  in terms of 
the size of this configuration.  
In the past, such configurations would not even be seriously considered as legitimate saddles due to ``disturbing"
discontinuity and singularity.  But the story is more interesting and beautiful, and such saddles  {\it do contribute} to the path integral. 
The physics of this jump behavior is 
discussed below in Section \ref{sec:complexbionsolution}. It is easy to 
show that the complex solution in SG can also be written exactly in terms 
of an instanton-anti-instanton pair as
 \begin{equation}
z_{\rm cb}(t)=2\pi a \pm 4  a \left[
   {\rm arctan}\left(\exp\left(\omega_{\rm cb} (t+t_0)\right)\right)
  +{\rm arctan}\left(\exp\left(-\omega_{\rm cb}(t-t_0)\right)\right)
\right]\, , 
\label{eq:sg-complex-bion-2}
\end{equation}
where
\begin{eqnarray}
t_0= \omega^{-1}_{\rm cb}\ln  \left[\sqrt{ - \frac{8a^3}{p g}}\left(1+\omega_{\rm bn}/a \right)\right]
  \approx \frac{1}{2a } \left [  \ln \left(\frac{32a^3}{p g}\right) \pm i  \pi \right]\, . 
\label{eq:sg-t1}
\end{eqnarray}
The  complex bion solution is the exact form of the $[\I \bar \I]_{\pm}$ 
correlated 2-event found by integrating over the  quasi-zero mode thimble, 
${\Gamma^{\rm qzm}_{-}}$. Again, rather remarkably, the approximation over 
the complexified quasi-zero mode thimble  \eqref{qzm-mm} has all the 
correct features of the complex bion solution, i.e., it is a systematic 
approximation to the exact solution.   In particular,  the complex quasi-zero 
mode  ``separation" saddle  of the  integration  is the same as the size 
of the plateau, $\tau^* \approx 2t_0$,    which is an exact feature of the 
exact solution. 

The action of the complex bion is
\begin{subequations}
\begin{eqnarray}
\label{eq:sg-action-complex-bion}
S_{\rm cb} 
&=& \tfrac{16a^3}{g} \sqrt{1+\tfrac{g p }{8a^3} } 
    + 2 p\,  {\rm arctanh} \left[ \sqrt{1+\tfrac{g p}{8a^3}}   \right ]\\
&=&\frac{16a^3}{g} + p \ln\left(\frac{32a^3}{p g}\right)\pm ip\pi 
    +\mathcal{O}(g).
\end{eqnarray}
\end{subequations} 
The  real part of the action of the complex bion is equal to the action of the real bion, as can  be seen by inspecting  
\eqref{eq:sg-action-complex-bion} and \eqref{eq:sg-action-real-bion}. Furthermore, it is   larger  than two-times instanton action, $2S_I$ since it is associated with upper separatrix. 
They differ in the imaginary part of the action of the complex bion. 
 If  $p=1,2, \ldots $, then, the complex action is unambiguous (because  the imaginary part of the bion action is well-defined modulo $2\pi$), 
   and relates to the hidden topological angle \cite{Behtash:2015kna}.   If $p$ is non-integer, 
then, the action is multi-fold ambiguous,  similar to the action of the  DW complex bion in 
\eqref{cb-phase}   which cancels with the ambiguity associated with the 
non-Borel resummability of the perturbation theory,  as described in more detail in Section \ref{sec:csr}.

\noindent
{\bf Equality of the real bion and  real part of the complex bion actions:}  The equality 
of  the real bion action $S_{\text{rb}}$ with the real part 
of the  complex bion $S_{\text{cb}}$ action can also be shown in a more elegant manner, (without doing computation)
by using contour deformation. 
The complex bion increases the ground state energy 
by an amount $\sim e^{-\Re(S_{\text{cb}})}$, and the real bion lowers it by 
$\sim e^{-S_{\text{rb}}}$. The ground state energy will remain zero (as required by SUSY) if and 
only if $S_{\text{rb}}=\Re(S_{\text{cb}})$.

\begin{figure}[t] 
\centering
\includegraphics[width=.55\textwidth]{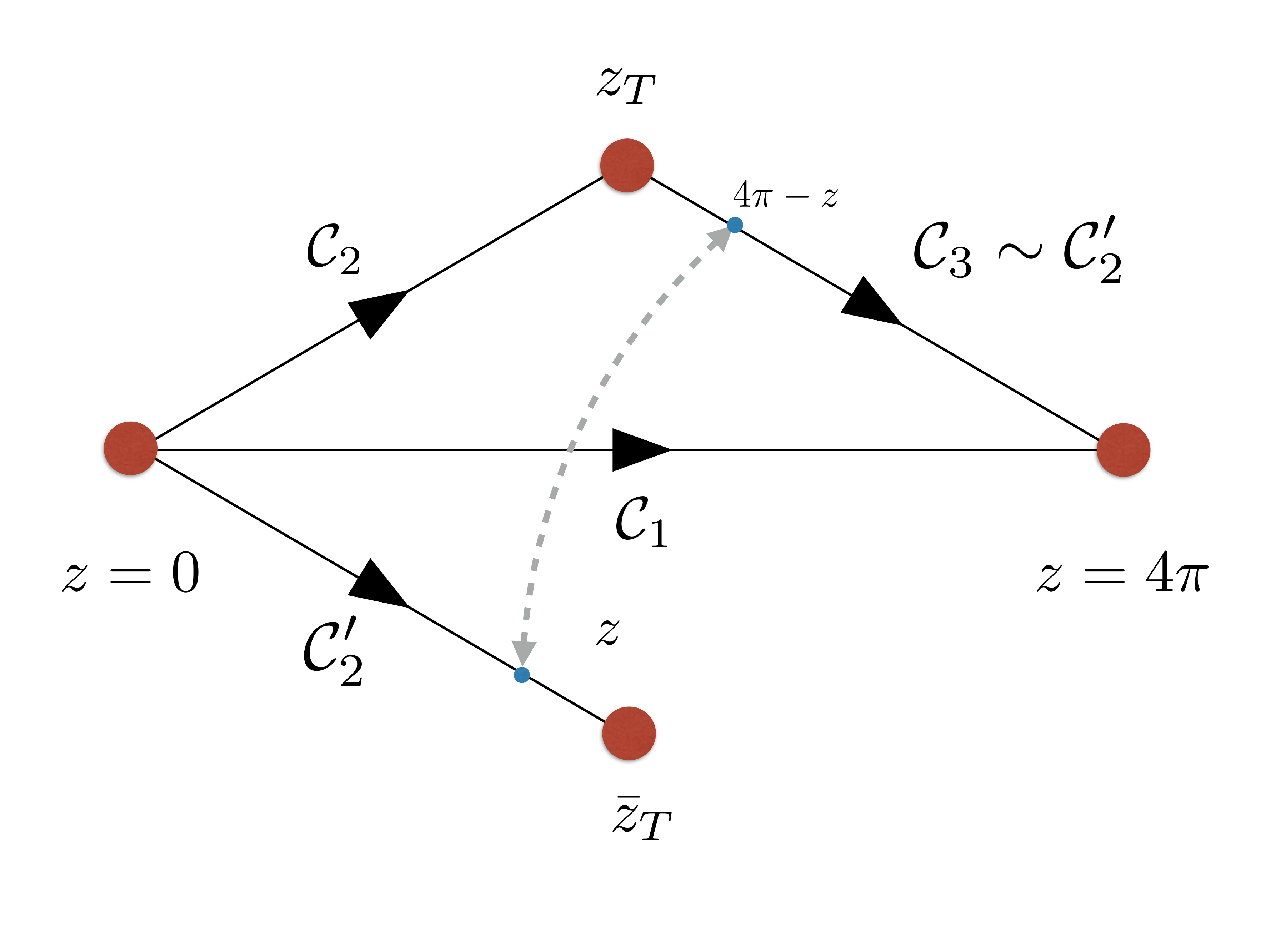} 
\vspace{-1.0cm}
\caption{The complex $z$-plane.  $\mathcal C_{1}$ is associated with real 
bion, and  $\mathcal C_{2}$ is associated with the complex bion with complex 
turning point. Here we have taken $a=1$.}
\label{fig:zplane}
\end{figure}

The actions of the real and complex bion are computed along the contours  
$\mathcal C_{1}$ and $\mathcal C_{2}$ defined in Figure \ref{fig:zplane}:
\begin{align}
&S_{\text {rb}} = \int_0^{4\pi} dz\sqrt{2V(z))}
              = \int_{\mathcal C_1} \sqrt{2V(z)}\;,\\
&S_{\text {cb}} = 2\int_0^{z_T}dz\sqrt{2V(z)}
              = 2\int_{\mathcal C_2} \sqrt{2V(z)}\;,
\end{align}
where $V(z)$ is the quantum modified potential and $z_T$ is the complex 
turning point. On the other hand, $S^*_\text {cb}$ can be written 
as an integral over the contour $\mathcal C_2'$:  $S^*_{\text{cb}}=2
\int_0^{\bar z_T}dz\sqrt{2V(z)}=2\int_{\mathcal C_2'}\sqrt{2V(z)}$. Because
$V(4 a\pi-z)=V(z)$ we can  change variables, $z\rightarrow 4 a\pi-z$, 
and write  $\int_{\mathcal C_2'} = \int_{\mathcal C_3}$, where the contour 
$\mathcal C_3$ is shown in Figure \ref{fig:zplane}. Therefore
\begin{align}
\Re{S_\text {cb}} 
   = \half (S_{\text {cb}}+S_{\text {cb}}^*)
   = \int\limits_{\mathcal C_2\cup \mathcal C_3\equiv \mathcal C_1}dz\sqrt{2V}
   = S_{\text {rb}}\;.
\end{align}
This  proves  the equality of the real parts of the actions  for any positive value of $p=N_f$. 
When  we analytically continue $p\rightarrow p \, e^{i \theta}$, this equality breaks down.

\subsection{Analytic continuation: From  real bounce to  complex bion}
\label{sec:complexbionsolution}

\begin{figure}[t] 
\includegraphics[width=7cm]{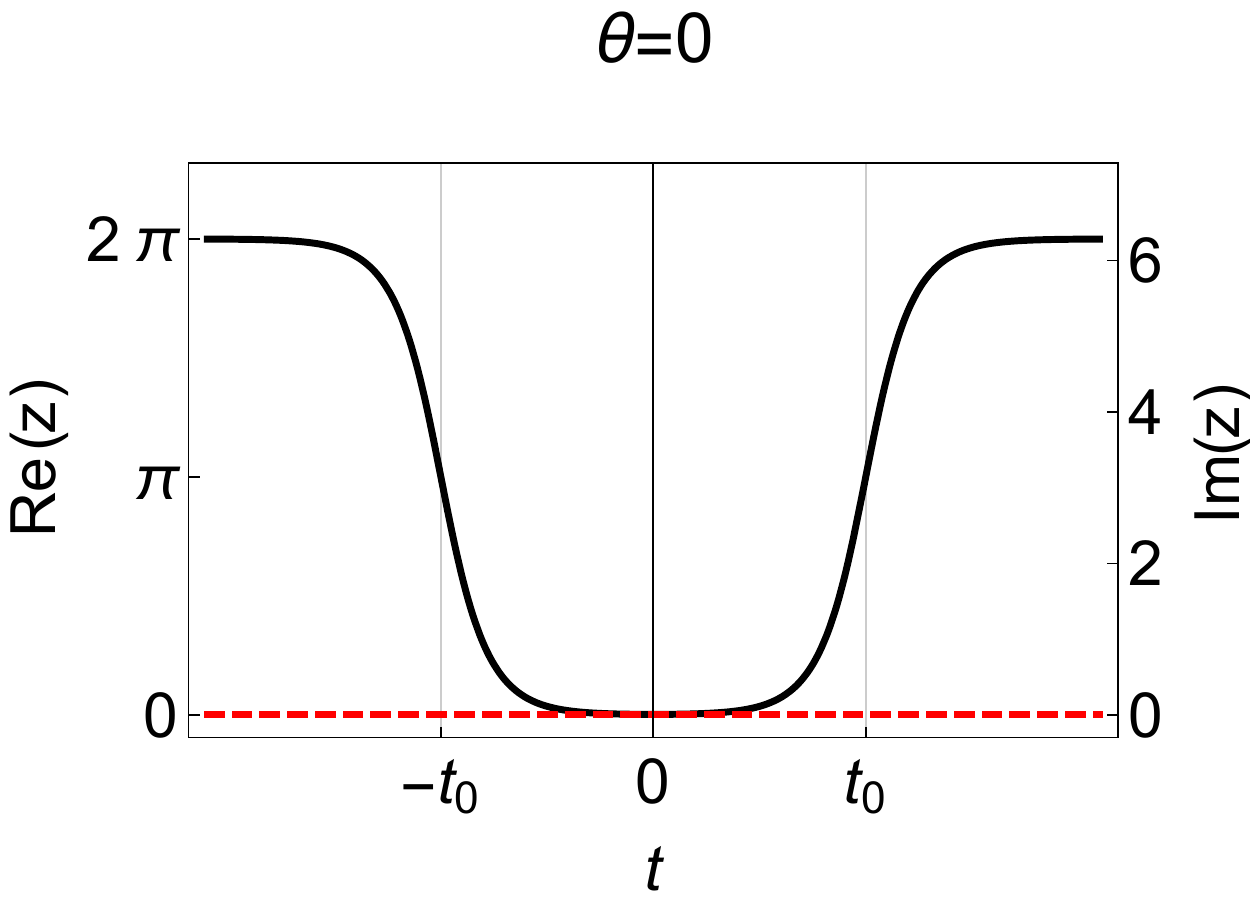}
\includegraphics[width=7cm]{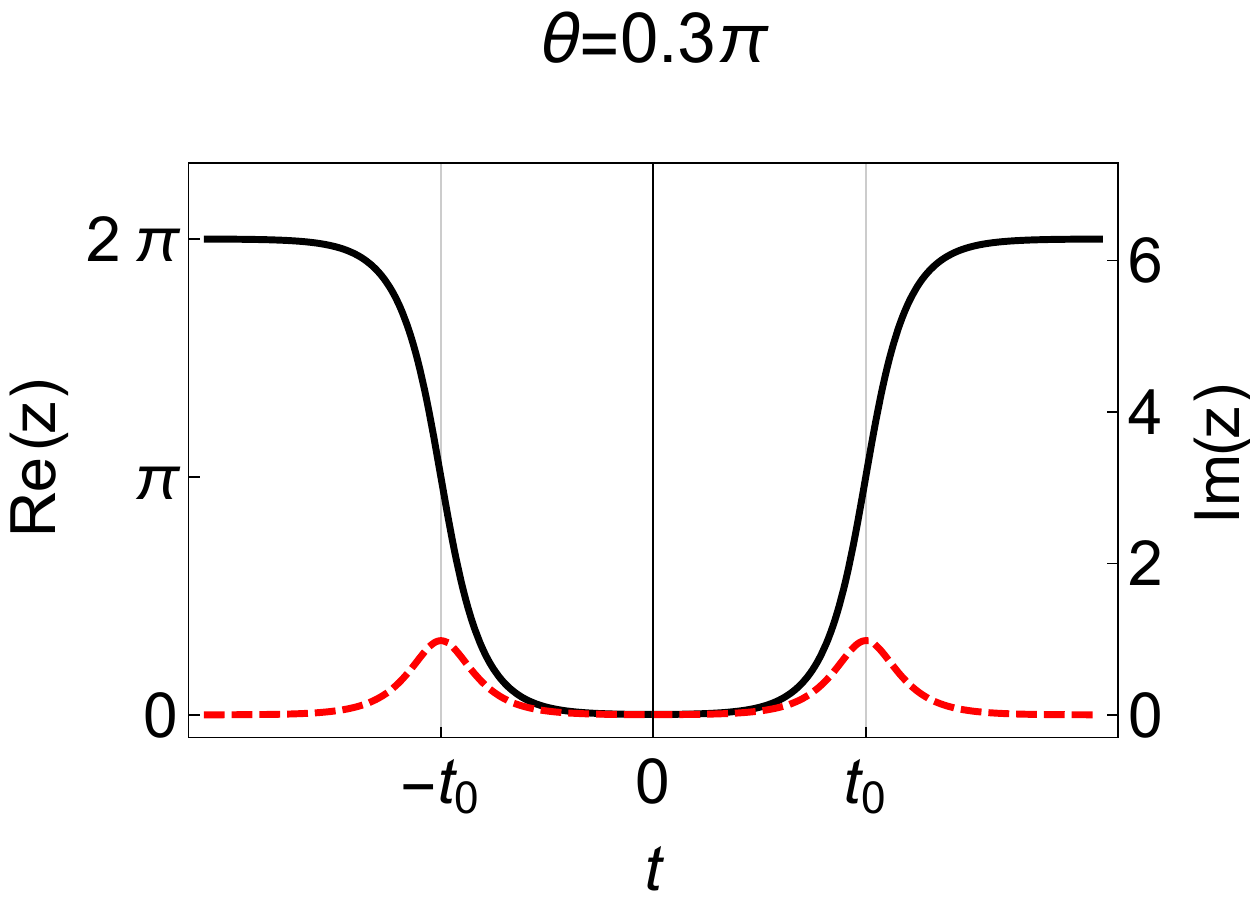}
\includegraphics[width=7cm]{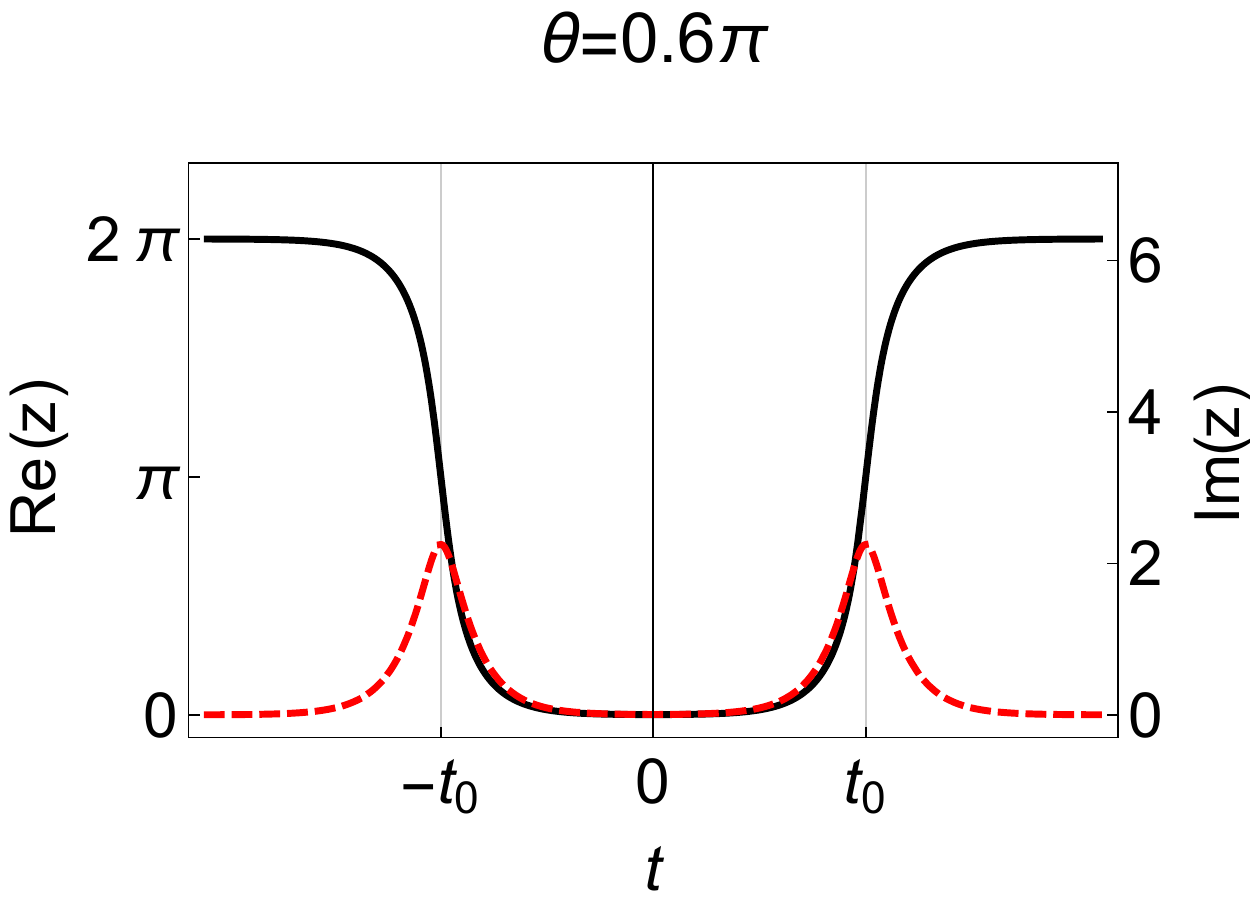}
\includegraphics[width=7cm]{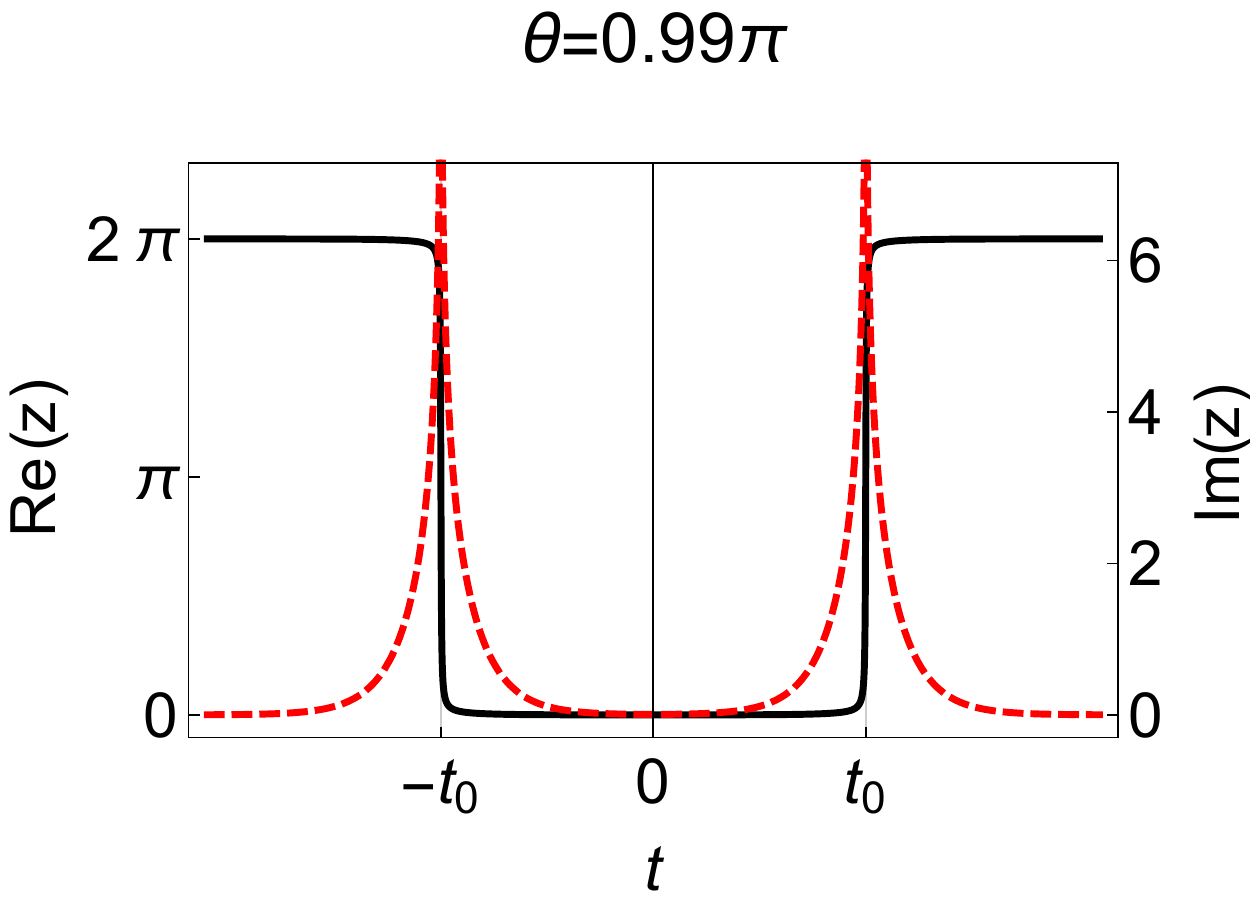}
\includegraphics[width=7.5cm]{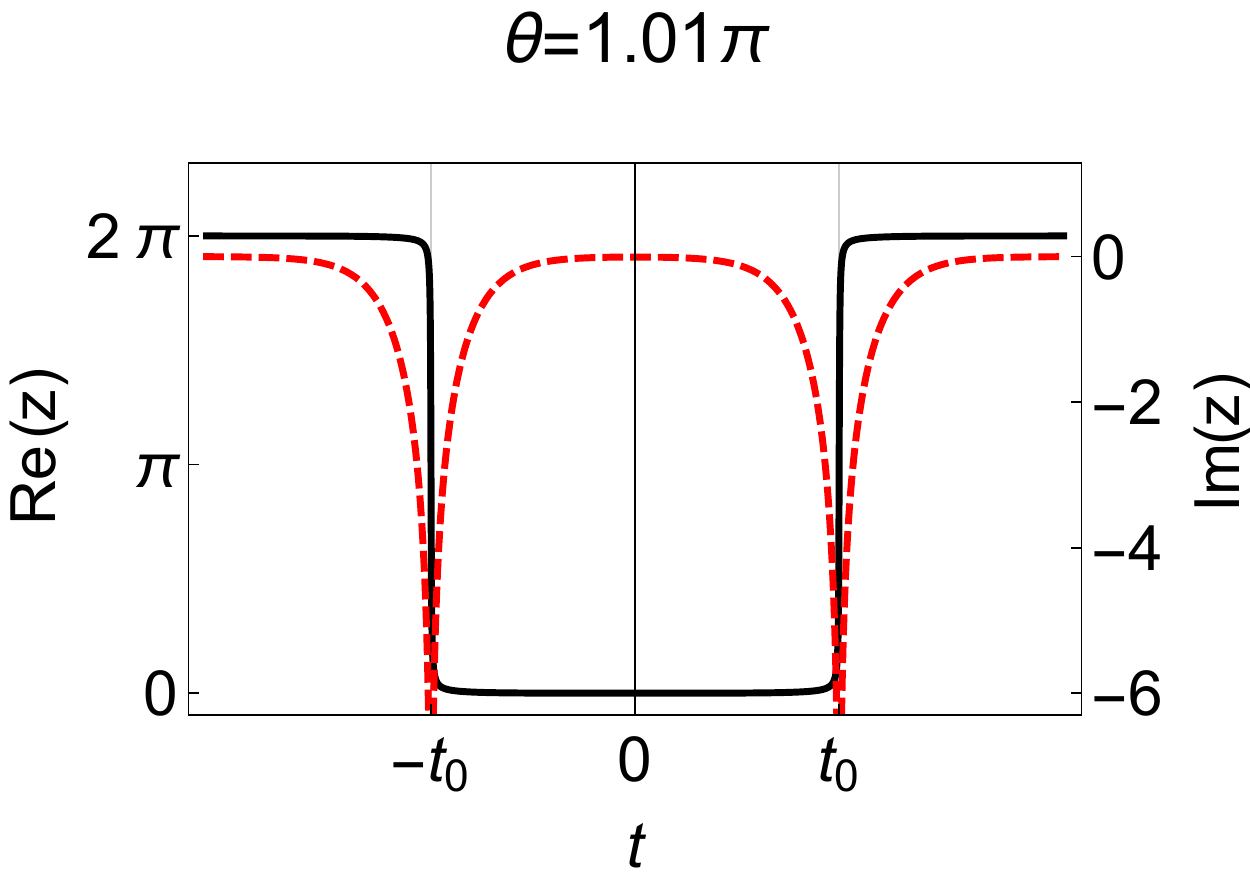}\hspace{1cm}
\includegraphics[width=7.5cm]{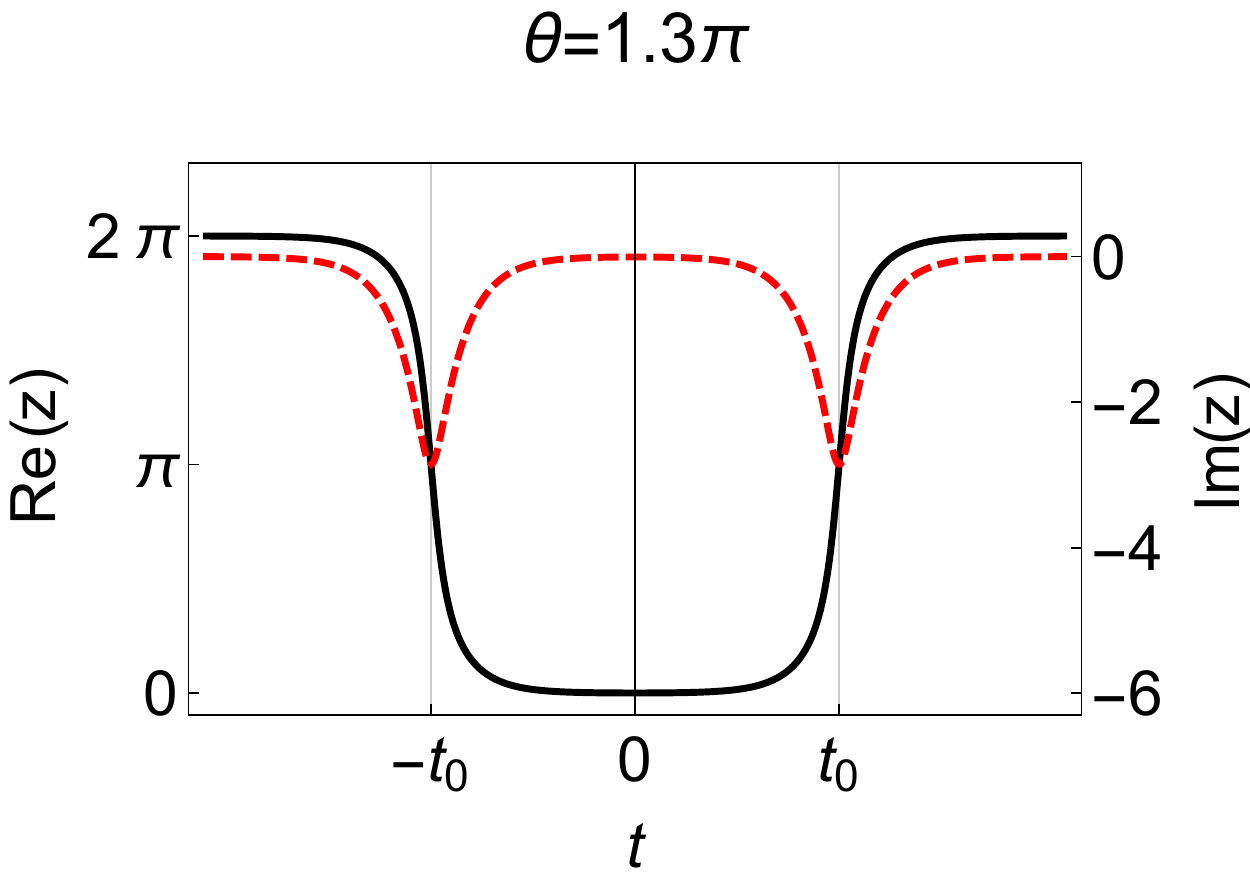}
\caption{Complex saddle solution $z(t,\theta)$.
 Black lines show ${\rm Re}(z)$ and dashed red lines
show ${\rm Im}(z)$. We have chosen $pg/a^3 \ll1$ 
and  set $a=1$. 
$\theta=0$ corresponds to the real bounce, and  $\theta
=\pi^{\mp}$ corresponds to the complex bion. The characteristic 
size of the solution is always $2t_{0}\approx \ln(\frac{32a^3}{pg}) $. }
\label{ac-1}
\end{figure}
	
As in the double-well system, the complex bion can be obtained by analytic 
continuation of the real bounce solution. Let us consider the analytic 
continuation of the Sine-Gordon theory to the complex $p$ plane, namely:
\begin{align}
p \rightarrow p\, e^{i \theta} \in  \mathbb C
\end{align}
The interesting aspect is the  following. Start with $V_{+}(z) = \half 
(W'(z))^2 +  \tfrac{pg }{2}  W''(z)$, and consider $\theta\neq 0$. At 
$\theta=\pi$,  we reach to the theory described by the potential 
$V_{-}(z)$, which is isospectral to $V_{+}(z)$, because the two systems
are related by a translation by $2\pi a$, see  \eqref{THT}. Upon analytic 
continuation all the way to $\theta=\pi$:
{\bf (i)} The local maximum  of the  $-V_{+}(z)$ (along real axis)  becomes the global 
maximum of  $-V_{-}(z)$. 
{\bf (ii)} The global maximum of the $-V_{+}(z)$  becomes local maximum 
of  $-V_{-}(z)$.
{\bf (iii)} The real turning point of the real bounce solution in the 
$-V_{+}(z)$ system turns into a complex turning point in the $-V_{-}(z)$ 
system.  

\begin{figure}[t] 
\centering
\includegraphics[width=6cm]{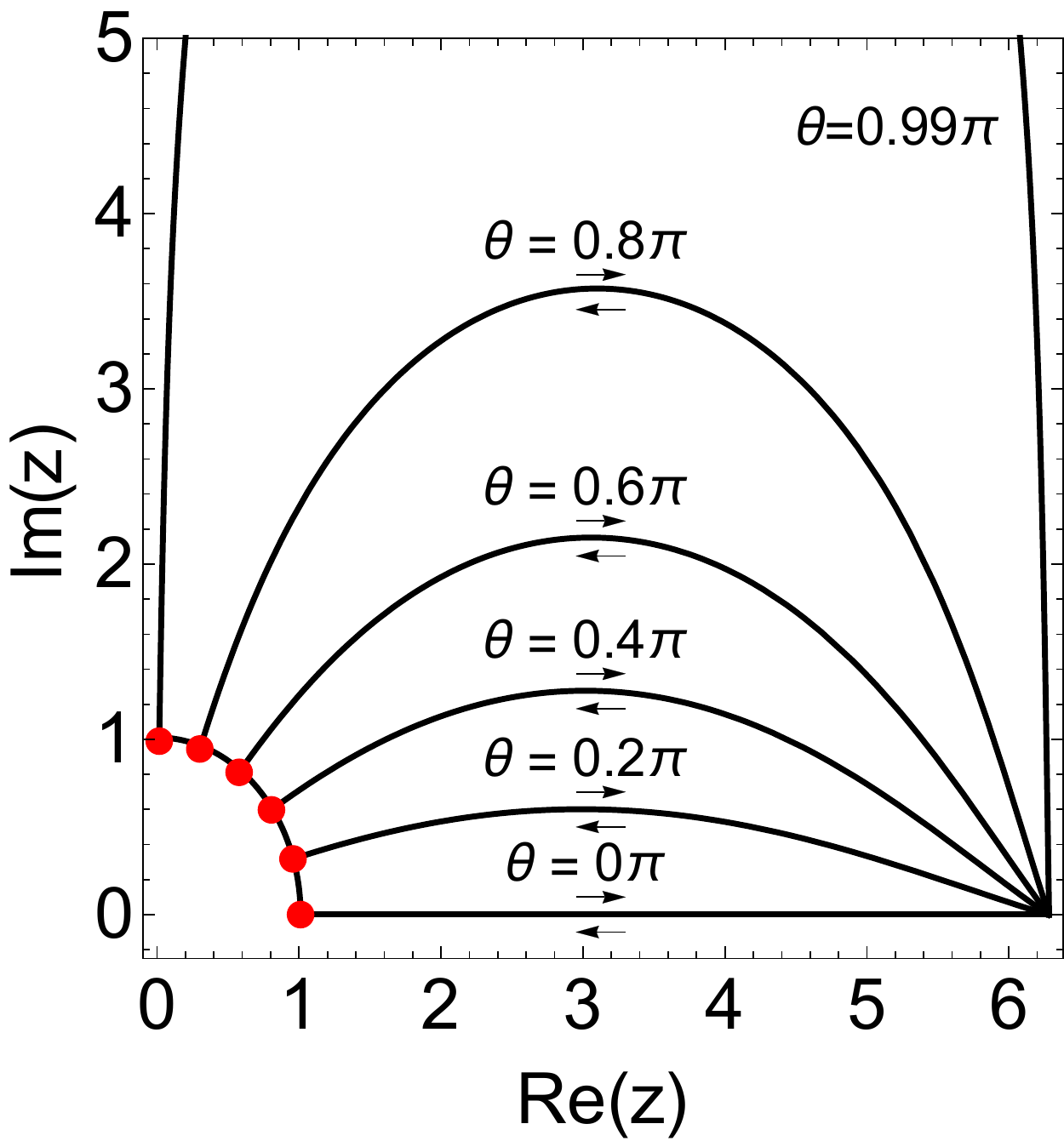}
\caption{Parametric plot of the real and imaginary part of the analytic 
continuation of the real bounce solution, $p\rightarrow p e^{i \theta}$. 
$\theta=0$ correspond to real bounce solution, with a real turning point.   
As one dials theta, the turning point becomes complex. At $\theta=\pi^{-}$, 
the real bounce turns into a complex bion, and the turning points becomes complex. 
}
\label{ac-2}
\end{figure}

\begin{figure}[t]                
\centering
\includegraphics[width=8cm]{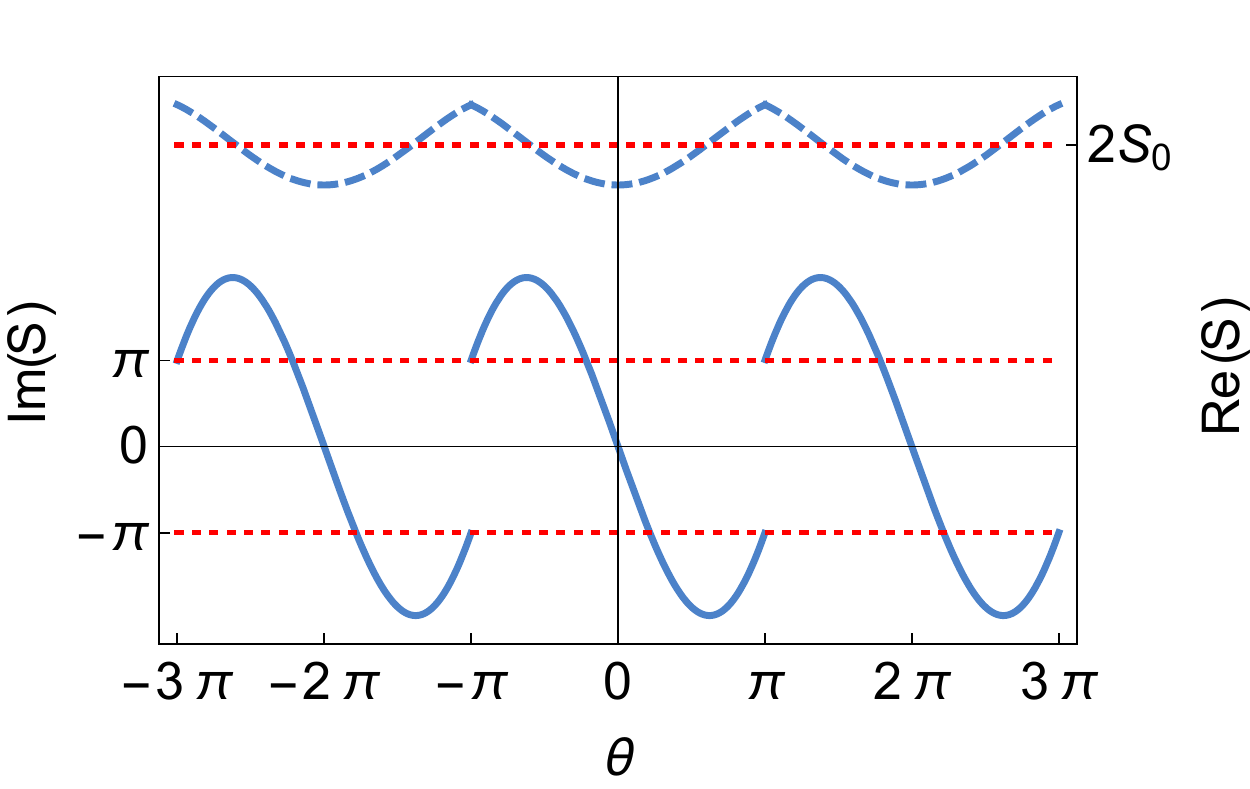}
\caption{ Action of the analytic continuation of the real bounce solution 
under $p\rightarrow pe^{i \theta}$. At $\theta=\pi^{-}$, the real bounce turns 
into a complex bion. The real part of the action is continuous at $\theta
=\pi$, while the imaginary part is discontinuous.  This effect, for 
$p\neq 1,2, \ldots$ is related to  resurgence, and ambiguity cancellation, 
and for $p=1, 2, \ldots$, it is related to the hidden topological angle.}
\label{ac-3}
\end{figure}

The analytic continuation of the bounce solution to arbitrary $\theta$ is 
clearly still an exact solution of the holomorphic Newton equations 
\eqref{complex-4} in the inverted potential. The solutions are obviously 
complex for general $\theta$, with real and imaginary parts plotted in 
Figure \ref{ac-1} for various values of $\theta$. The solution obtained 
via analytic continuation of the bounce is: 
\begin{equation}
z(t, \theta) = 4a\,{\rm arctan}\left(
    \sqrt{\frac{p g\, e^{i \theta}}{8a^{3}-pg\,e^{i \theta}}}\,
    \cosh\left(\sqrt{1 -\tfrac{pg\,e^{i \theta} }{8a^3}}at\right)\right)\, . 
\label{complex_bion_0}
\end{equation}
The  solution is smooth for all $\theta \in [0, \pi)$, but exhibits 
two-valuedness for $\theta=\pi$, see Figure \ref{ac-1},  which will be 
discussed below. It satisfies the correct  boundary conditions:
\begin{equation}
z ( t=\pm \infty, \theta) = 2a \pi,  \qquad     
z (t=0, \theta) =  z_T = 2 a\arccos\left(1 - \tfrac{gp\, e^{i \theta}  }{4a^3} 
  \right) \, , 
\end{equation}
where $z_T$ is the complex turning point, shown in Figure \ref{ac-2}. We 
can calculate the action which is  also multi-valued function at $\theta= 
\pi$. In the $gp/(8a^3)  \lesssim 1 $ regime, the action is is analytic in the cut plane, 
\begin{align}
S(g, \theta) = \tfrac{16a^3}{g} \sqrt{1-\tfrac{gp\, e^{i \theta}  }{8a^3} }  
   - 2 p\, e^{i \theta}  {\rm arctanh} 
  \left[ \sqrt{1-\tfrac{gp\, e^{i \theta}  }{8a^3}}   \right ]\, , 
\label{complex_bion_action}
\end{align}
and two-valued for $\theta=\pi$.  The real and imaginary part of the action 
is plotted in Figure \ref{ac-3}. The real part is continuous, with a cusp 
at $\theta=\pi$. The imaginary part is also continuous except for a 
discontinuity at $\theta=\pi$ (interpreted below). Note that on general 
grounds, the action is well defined only up to $2\pi$ shifts in its 
imaginary part, as  reflected in  Figure \ref{ac-3}. 

\subsection{Physical Role of Singular, Multi-valued  Solutions}

 At $\theta=\pi$,  the potential is again real, and the complex saddles become 
a multi-valued complex bion solution. Let us now discuss 
what happens at $\theta=\pi$ in more detail.  We can consider the behavior 
of the solution at  $\theta= \pi \mp \epsilon$ with  $\epsilon>0$. The 
multi-valued complex bion solution is given by 
\begin{equation}
 z^{\pm} (t)=z (t, \theta= \pi \pm \epsilon)  \, . 
\label{complex_bion}
\end{equation}
This solution is actually not only multi-valued, it is also singular.   (The underlying reason is discussed in the next section.) 


 In the formalism of {\it real} path integration, even the real part of such configurations would 
not contribute to the path integral as saddle points. Clearly, there are 
singular configurations in the real path integral (and these are actually generic), but there cannot exist  
singular saddles with finite action.  The simple reason for this is that 
the kinetic term  $ \int \half \dot x^2 $ blows up for a singular or 
discontinuous configuration. But in complex path integration the situation 
is different.  The kinetic term is now replaced by:
 \begin{align}
   \int \half \dot z^2   = \int  \half (\dot x^2 -  \dot y^2)  
       + 2i  \int \dot x  \dot y \, . 
 \end{align}
Even if the real and imaginary part of a solution $z(t)$ are singular or 
discontinuous, the action may remain finite provided the real and imaginary 
parts lead to cancellations in the action.  For the holomorphic path integral 
this is what is taking place at $\theta=\pi$, see Figure \ref{ac-3}.  


 Remembering that ${\rm arctanh}(z/a)$ is analytic in the cut-plane 
$\mathbb{C}\backslash {\mathfrak {cut}}$, and double-valued along the cut  
$z/a \in (-\infty, -1) \cup (1,\infty)$ we have
\begin{align}
{\mathrm{arctanh}} (z/a)  
=  \left\{  \begin{array}{ll} 
 + \tfrac{1}{2}  \ln \left(\frac{z+a}{z-a}\right) \pm \tfrac{1}{2} \pi i,    
       &\qquad   z/a \in (-\infty, -1) \cup  (1, \infty) \, , \cr \cr
 + \tfrac{1}{2}  \ln \left(\frac{z+a}{z-a}\right),    
       &\qquad   z/a  \in \mathbb{C}\backslash(-\infty,-1)\cup(1,\infty) \, .
\end{array} \right. 
\end{align}
The action of the complex bion acquires an unambiguous real part and a  
discontinuous imaginary part at $\theta=\pi$. Both are shown in Figure 
\ref{ac-3}. The action of the complex bion, in its exact form \eqref{eq:sg-action-complex-bion}, 
 can be rewritten along the cut as 
\begin{subequations}
\begin{align}
S_{\text{cb} }^{\pm} 
  = \tfrac{S(g)}{g} 
& = \tfrac{16a^3}{g}  \sqrt{1+\tfrac{gp   }{8a^3} }  
       + 2 p  \ln \left[  \frac{ \sqrt{1+ \tfrac{gp }{8a^3}}  +1 } 
                               { \sqrt{1+ \tfrac{gp }{8a^3}}  -1 }  \right ]   
            \pm i  p  \pi   \\ 
& =   {\rm Re} [ S_{\text{cb}}]  \pm i  p  \pi, \qquad    \qquad{\rm Im}\left[S_{\text{cb}}\right] =\pm  p \pi \, . 
\label{complex_bion_action-2}
\end{align}
\end{subequations}

Before making a few physical remarks on the implication of the complex saddles in the physical theory, it is useful to state the amplitudes 
associated with the real bounce, real bion and complex bion in the weak coupling regime: 
\begin{align}
  I_{\rm bn}  &\sim  \left( {\frac{pg}{32a^3}} \right)^{-p}  e^{-2S_I }, \cr 
  I_{\rm rb}
   &\sim \left( {\frac{pg}{32a^3}} \right)^{+p} e^{-2S_I  }, \cr
I_{\rm cb}^{\pm}
   &\sim \left( {\frac{pg}{32a^3}} \right)^{+p}  e^{-2S_I  \pm i p \pi},
   \label{amps-2}
\end{align}
This is in precise agreement with the approximate amplitudes of the two-events  obtained in Section \ref{correlated} upon integration over the appropriate QZM-thimbles, $\Gamma^{\rm qzm}$, in the weak coupling regime. A few remarks are in order: 

\begin{itemize}

{\item The exact solution \eqref{complex_bion_0} is discontinuous and 
singular at $\theta=\pi$, but the real part of the action is continuous.   
The imaginary part of the action jumps, making the action, as well as the 
complex bion amplitude $I_{\rm cb}^{\pm}$  two-fold ambiguous  for general 
values of $p$, namely, $p \in \R - \{1,2, \ldots\}$: 
For the consistency of the theory this  discontinuity must be canceled 
by the ambiguity of the (left/right) Borel resummation of perturbation 
theory.\footnote{Discontinuities of this type in saddle amplitudes  have led to some concern in 
the literature, see footnote 33 of \cite{Harlow:2011ny}.  We discuss this important issue and the resolution of the 
problem  in Section \ref{sec:HMW}. } }


\item{For integer values, $p=1,2,\ldots$, where $p=1$ corresponds to 
supersymmetric QM, the ambiguity in the amplitude of complex bion disappears: 
\begin{align}
I_{\rm cb}^{\pm}
   \sim  e^{- S_{cb}^{r}}  e^{ \pm i    \pi} 
    =    \left\{ \begin{array}{ll} 
       +    e^{- S_{cb}^{r}}  &  \qquad {p : 2, 4, \ldots \, , }  \cr
       -    e^{- S_{cb}^{r}}  &  \qquad {p : 1, 3, \ldots \, .}\cr
\end{array} \right.
\end{align} 
In theories for which $p$ is an odd-integer there is a $\Z_2$-worth of hidden  
topological angle (HTA) discussed in our recent work \cite{Behtash:2015kna}. 
For example, for $p=1$, this configuration gives a positive contribution 
to the ground state energy. But there exists another saddle in the problem, 
the real bion discussed in Section \ref{sec:sg-real-bion}, which gives a 
negative contribution to the ground state energy, such that the combination 
of the two cancels exactly.  In this potential, supersymmetry is actually 
unbroken  and the vanishing of the ground state energy relies on the 
existence of this complex  multi-valued, discontinuous saddle in addition 
to the real bion saddle.}
\end{itemize}

\subsection{Physics of the complex bion solution,  multi-valuedness and singularity}

 One may wonder why the complex bion solution exhibits singular behavior 
in the Sine-Gordon system,  in contrast with the complex bion in the  
double well potential discussed in Section \ref{sec:dw-complex-bion}, 
particularly given that the solution was found by mapping the problem 
to the double well potential. 

\begin{figure}[t]
\center
\includegraphics[scale=0.20]{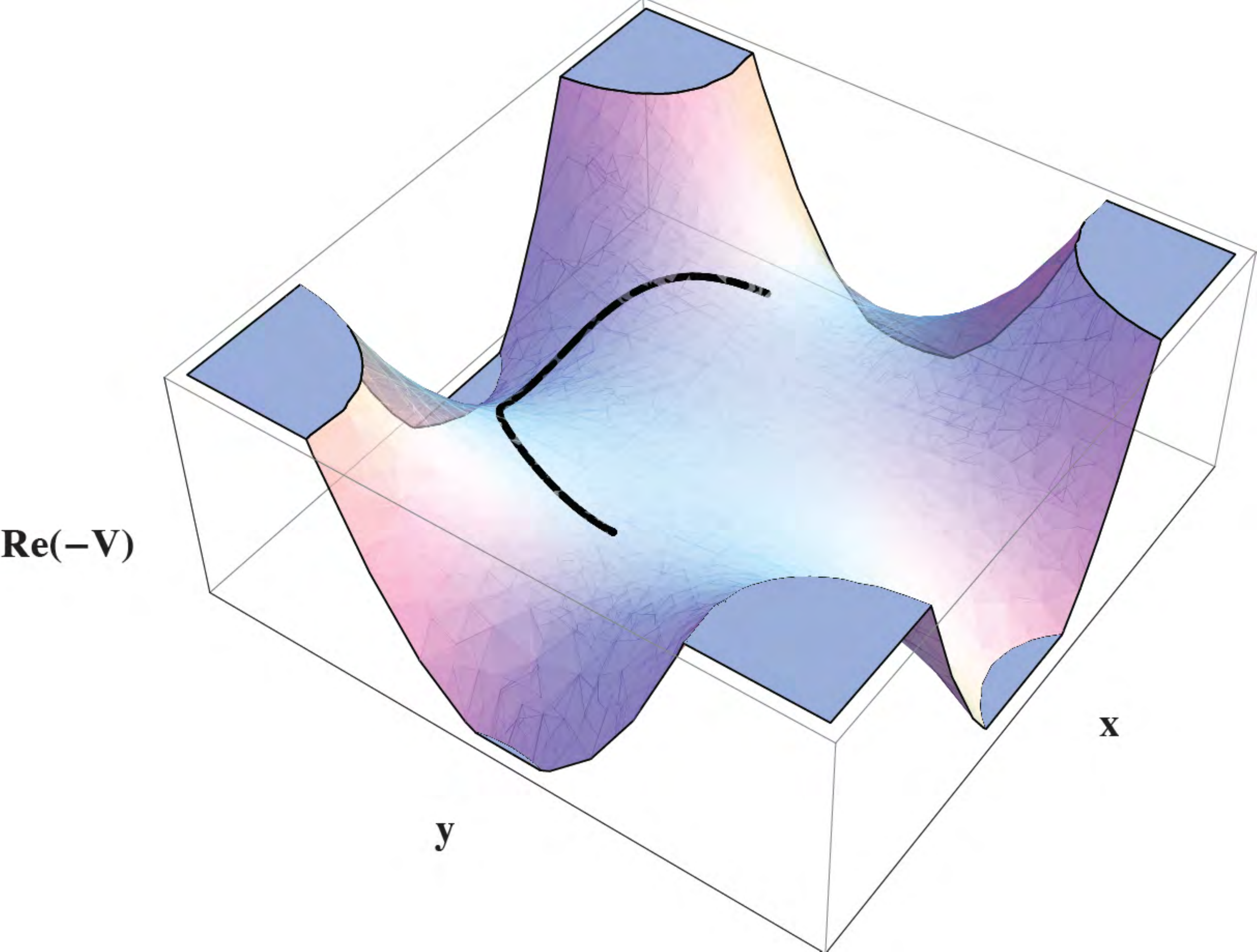}
\caption{Complex bion rolling on the real part 
of the inverted potential for the 
double-well case with $a=1$, and $p g=1/10$. The trajectory begins for $t=0$ at the 
complex turning point and continues to the critical point $z_1^{\rm cr}$ 
as $t\to\pm \infty$. The motion is perfectly smooth.}
\label{3Dcomplex-DW}
\end{figure}

An elementary way to understand the difference between the double well and the Sine-Gordon systems is to consider the problem of integrating the classical equations of motion, starting at 
some complex turning point $z_T$ of the potential $V(z)$ at time $t=0$:
\begin{eqnarray}
z(t) = z_T + t\, \dot{z}(0)+\frac{t^2}{2} \ddot{z}(0)
           + \frac{t^3}{3!} \dddot{z}(0)+ \dots\, . 
\label{eq:numint}
\end{eqnarray}
If the initial point, $z(0)=z_T$ is a turning point, then $\dot{z}(0)=0$, 
so the first step of the evolution is governed by 
\begin{eqnarray}
\ddot{z}(0)=\frac{\partial V}{\partial z}\Bigg |_{t=0}\, . 
\end{eqnarray}
It is easy to see that for the double-well case, $\ddot{z}(0)=2z_T^3-z_Ta^2
+p g$ is complex, with both  non-vanishing  real and imaginary parts.  
Thus the evolution 
moves away from the complex turning point in the complex plane, ending at 
the critical point $z_1^{\rm cr}$. This can be seen pictorially in  Figure 
\ref{3Dcomplex-DW}, which shows the complex bion for the double-well 
``rolling'' on the surface of the real   part of the inverted 
potential.\footnote{
Note that, in contrast to naive intuition, the particle does not fall into  
the infinitely deep well.  The force on a classical particle in the inverted 
potential is not (minus)  the gradient, 
$- \vec \nabla (- V_{\rm r})$.  Instead, 
 $\ddot x  = - \partial_x {(- V_{\rm r})},  \;  
  \ddot y  = + \partial_y {(- V_{\rm r})}   $, 
notice the crucial relative minus sign,  as discussed around \eqref{eq:eom1}.  
The force in the $y$-direction is negated with respect to the usual equations 
of motion due to holomorphy.  As a result, a particle which would, according 
to the regular  Newton equation in 2d,  "roll down" into  a well, may 
instead, ``rolls up".   This is the reason that the classical (holomorphic)  
particle in the inverted potential does not fall into the well in  
Figure \ref{3Dcomplex-DW}. }


\begin{figure}[t]
\center
\includegraphics[scale=0.20]{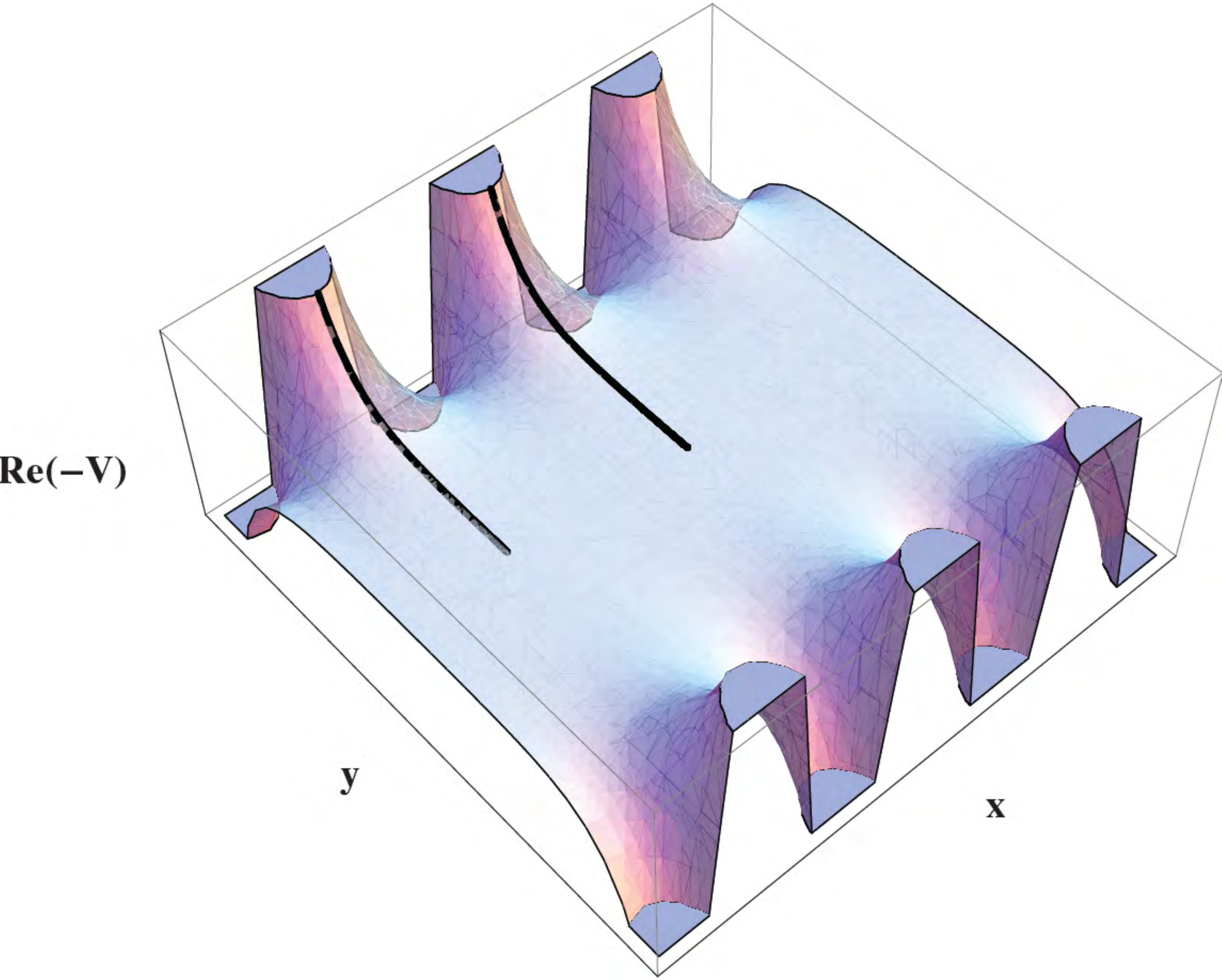}
\includegraphics[scale=0.23]{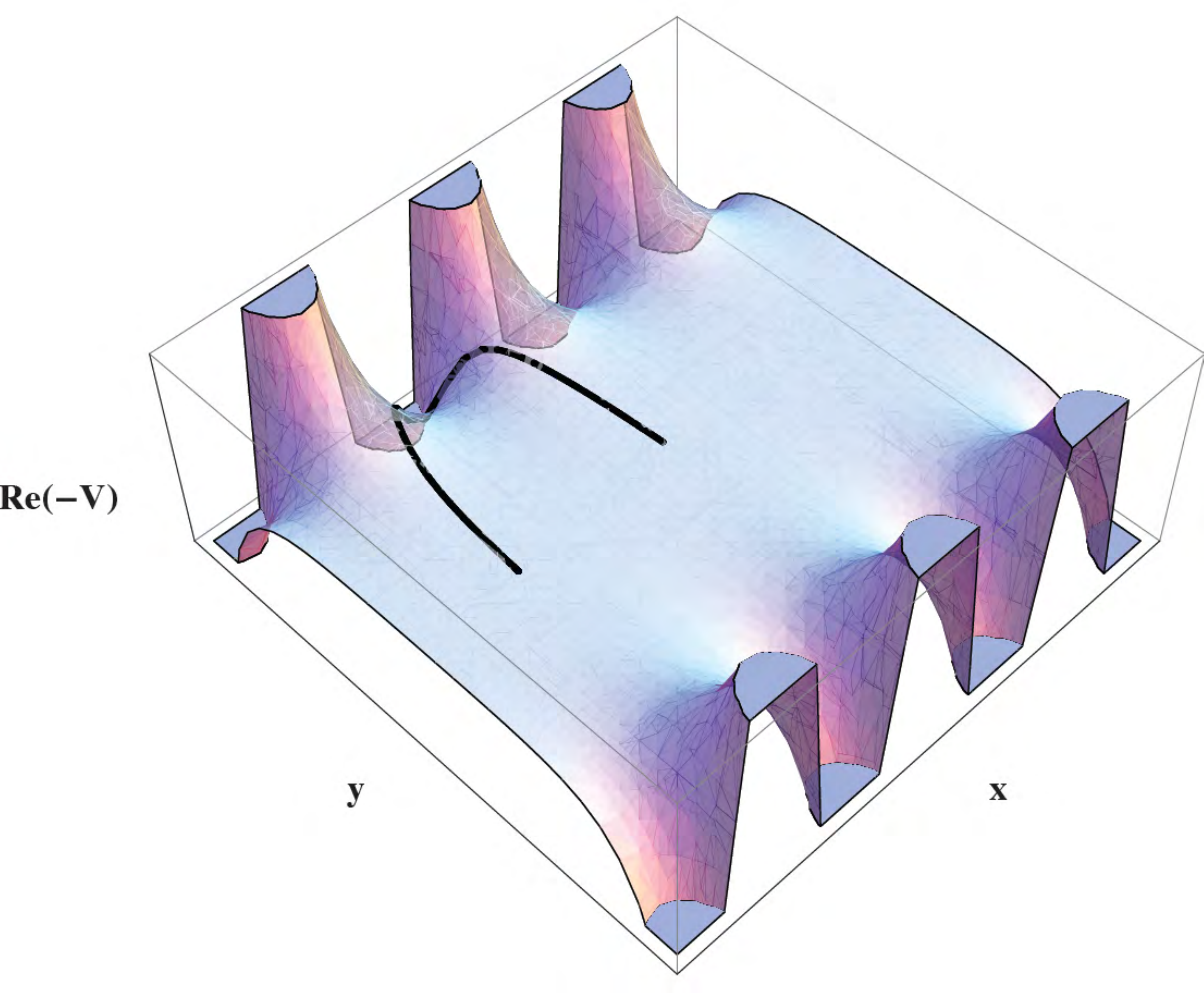}
\caption{First figure: Complex bion rolling on the real part of the inverted potential for the 
Sine-Gordon case with $a=1$ and $p g =1/10$. The particle begins at $t=0$ at the complex 
turning point and rolls out along the ridge at $\Re(z)=2\pi$, reaching 
the point at infinity in finite time, and then rolls back along the ridge 
at $\Re(z)=0$, reaching the critical point at $z=0$ at $t=\pm \infty$. The discontinuous jump occurs at infinity. The second figure shows this curve as well as the real bion with $p\to p\, e^{-i 19\pi/20}$, which effectively smooths out the trajectory so that it passes from one ridge to another in a completely regular way. As $p\to -p$ we obtain the discontinuous complex bion trajectory.}
\label{3Dcomplex-DSG}
\end{figure}

By contrast, for the Sine-Gordon system, starting at the complex turning 
point $z_T/a=2\pi+ 2i\, {\rm arccosh}\left(1+\frac{p g}{4a^3}\right)$, 
we have
\begin{eqnarray}
\ddot{z}(0) 
   =  a^2\sin (z_T/a) +\frac{p}{4}a^2\sin(z_T/2a) 
   = -2ia^2  \sqrt{\frac{p g}{2a^3}}\left(1+\frac{p g}{8a^3}\right)^{3/2}\, , 
\end{eqnarray}
which is pure imaginary. This pattern propagates through the entire series 
in \eqref{eq:numint} and is a consequence of the symmetry \eqref{eq:sym}: all odd derivatives of $z(t)$ vanish at $t=0$, while 
even derivatives are all pure imaginary. Thus, the real part of $z$ must 
remain constant; it cannot change from its initial value of $2 a\pi$, except 
when $\Im(z)$ reaches infinity, as it does in a finite time $t_0$, and at 
this point the real part of $z$ can jump by $2 a\pi$. This can be seen in 
the ``rolling'' of the classical particle shown in Figure \ref{3Dcomplex-DSG}. 
 The particle rolls out along the ridge at $\Re(z)=2 a\pi$, reaching the point 
at infinity in finite time, and then rolls back along the ridge at $\Re(z)=0$, 
reaching the critical point at $z=0$ at $t=\pm \infty$. The sudden jump at 
infinity may be regularized by analytically continuing $p$ in the potential 
to have a phase, as discussed in the previous subsection. In this case, the jump at 
infinity is smoothed out.


\vspace{0.3cm}
\noindent
{\bf Symmetry and  multi-valuedness (or an alternate story for Buridan's donkey):}  There is another more general way to see why the  discontinuity and singularity  occurs. 
First, in classical mechanics   (and its holomorphic version) uniqueness of solutions is guaranteed for any motion given the initial conditions.  This means that  the turning point defined by $-V(z_T)=E$ \emph{uniquely} defines the trajectory of motion in the holomorphic classical mechanics. 
 Since the motion is constrained by constancy of energy, set to be equal to the energy at one of the  turning points (which is on the same critical orbit with a critical point), 
 initializing the motion on the critical orbit to which turning point belongs 
  will necessarily  result in the particle ending its motion at one of the critical points with this energy. 
  

However, if the potential  of motion possesses  an exact symmetry under which turning points remains invariant, then the path of the motion must also remain invariant (since the turning point determines the path uniquely).  On the other hand,  if the critical points
 are \emph{not} invariant under this symmetry,  but instead go into each other,  then the motion cannot connect smoothly to them.  This is similar to the paradox of Buridan's donkey.  
  There are two critical points on the same energy level, equidistant from the turning point, and both are equally attractive. 
 The only option is  for the motion to evolve to infinity. 
If the situation is slightly perturbed (i.e. by changing the potential) so that the symmetry in question is broken even by a very small-amount, the motion will reduce to the generic situation and the Euclidean particle will asymptote to one of the critical points. This is precisely what causes the multivaluedness of the solution, i.e. the trajectory depends on how exactly we perturb the potential. 
The singular nature of the solution also goes hand in hand with multivaluedness, as the only way the particle can move is along a trajectory which obeys the invariance under the symmetry. In the absence of this invariance for critical points, the particle is forced to move off to infinity. This is the singular behavior of the solution.

\vspace{0.3cm}
\noindent {\bf Example:} Let us consider the  double-Sine-Gordon case. (Set $a=1$ momentarily.) The real part of the Lagrangian is: 
\begin{align} 
\Re  {\cal L} &= \half (\dot   
{x}^2 - \dot y^2)  -   2\cos^{2} (\tfrac{x}{2}) \cosh^{2} (\tfrac{y}{2}) +      2\sin^{2} (\tfrac{x}{2}) \sinh^{2}(\tfrac{y}{2})    - \tfrac{pg }{2}  \cos (\tfrac{x}{2}) \cosh(\tfrac{y}{2})    \, , 
\end{align}  
and the holomorphic equations of motion, written in terms of real and imaginary part of $z(t)=x(t) + i y(t)$, take the form:  
\begin{align}
\frac{d^2 x}{dt^2}  &=    \cosh y \sin x   + 
 \frac{pg}{4}  \cosh (y/2) \sin(x/2)   \, ,    \cr       
\frac{d^2 y}{dt^2}  &= \cos x \sinh y   
   + \frac{pg}{4}   \cos (x/2)  \sinh(y/2) \, .  
\label{complex-44}  
\end{align}
The transformation 
\be\label{eq:sym}
x(t) \rightarrow - x(t) +4\pi a\;, \qquad y(t) \rightarrow y(t)
\ee
is a symmetry of the real part of the holomorphic Lagrangian, as well 
as the equations of motion.   For example, the two turning points   located at 
\be
z^T_{\pm}=2\pi a   \pm 2 i a \,{\rm arccosh}\left(1+\frac{p g}{4a^3}\right)
\ee
go back to themselves upon the transformation \eqref{eq:sym}, while the 
critical point at $z^{\rm cr}=0$ goes to   $z^{\rm cr}=4\pi$, and vice versa.

Since only the real part of the potential enters the equations of motion \eqref{complex-44}, if a particle is initialized at the point which obeys the symmetry, then, by uniqueness of the solution, the entire trajectory must obey it. Therefore the Euclidean particle initialized at one of the turning points $z^T_{k,\pm}$ cannot asymptote to one of the critical points $z^{\rm cr} =0 \; {\rm or} \;  4\pi a $, but must fly off to infinity.

However if the system is slightly perturbed, for example by giving $p$ a 
small complex part, $p \rightarrow pe^{ i \theta}$,  the invariance of the 
turning points  $z^T_{\pm}$  and invariance of the Lagrangian and equations 
of motion (see \eqref{complex-4})  is destroyed, and  the motion will 
generically asymptote to one of the critical points. In the limit of real 
$p$, the motion will turn into the singular motion from the turning point 
to one of the critical points, depending how the limit of real $p$ is taken. 
This \emph{is} the multivaluedness of the complex bion.

\subsection{Vacuum  as a  complex and real  bion gas} 

{\bf Reminder of the bosonic case, ground state wave function vs. instanton 
gas:}
In the bosonic periodic potential the Euclidean description of the ground 
state is given  by a dilute gas of instantons (at leading order in semi-classical expansion). The ground state wave function 
is periodic and symmetric. For comparison with the case of a particle with 
spin, consider a the bosonic potential which is periodic with period $2\pi a$, 
but the theory is considered on a circle with circumference $4\pi a$.  Then 
the ground state is symmetric and the first excited state is anti-symmetric 
in the interval $[0, 4\pi]$. This is equivalent to the Bloch wave boundary 
condition $\Psi(x+ 2\pi a) = e^{i \alpha} \Psi(x)$, with $\alpha=0$ for the 
ground state, the lower edge of the band, and $\alpha=\pi$ for the upper 
edge of the band. The non-perturbative splitting between the first excited 
state and the ground state is  due to instantons,  $\Delta E = J_{\tau_0}   \;  \left[\frac{ \det' M }{ \det M_0}\right]^{- {1 \over2}}  e^{-S_\I}$.  
 The probability to find 
the particle either at $x=0$ or $x=2\pi a $ is equal. The dilute instanton 
gas reflects this property.  The classical particle spends half time at 
$x=0$ and the other half at $x=2\pi a$. We see that in the theory of 
a particle with spin (or $N_f \geq 1$) the ground state structure is 
dramatically different. 

\begin{figure}[t]           
\centering
\includegraphics[width=16cm]{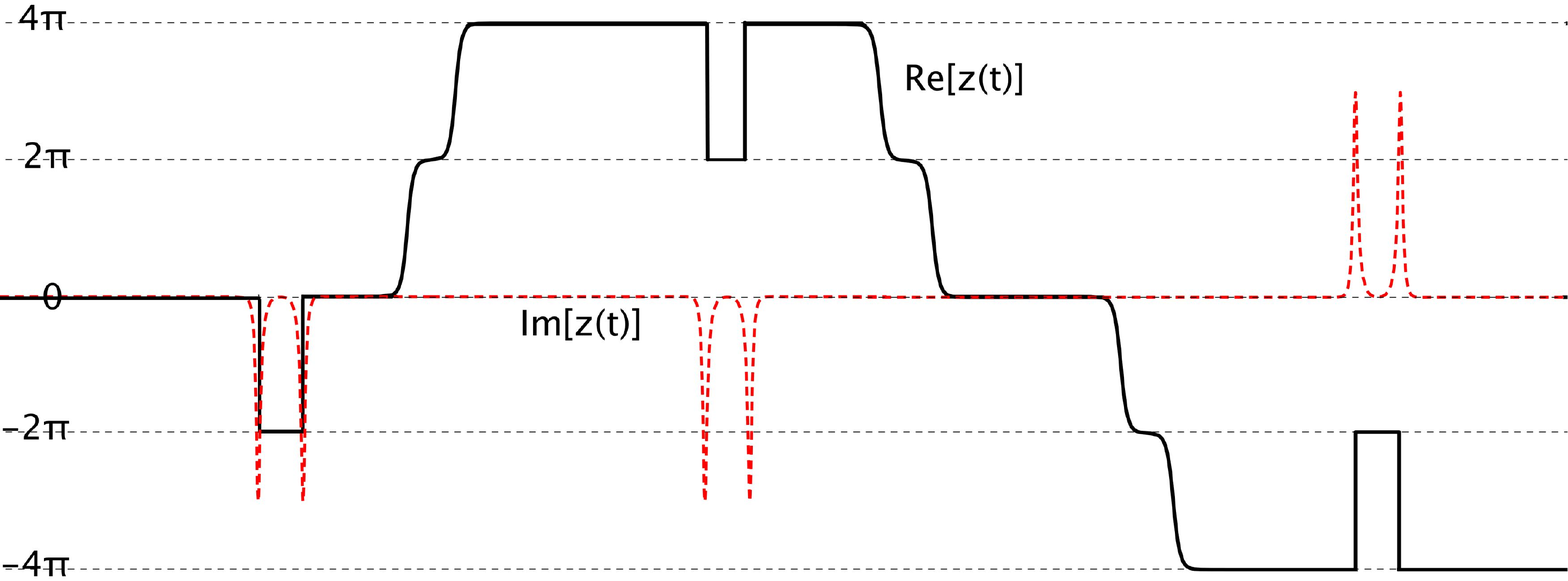}  
\vspace{-3cm}
\caption{The ground state properties  of the theory with fixed number of 
fermions ($N_f \geq 1$) is dictated by a dilute gas of real  (smooth) and complex (singular) 
bion events, both with density $e^{-2S_I}$. Both real and complex bions are exact 
solutions, and  $\Re[2t_0]^{\rm cb}=[2t_0]^{\rm rb} $. Note that the Euclidean time spend on $x=2\pi$ (mod $4\pi$) is exponentially small  compared to the time spent at
$x=0$ (mod $4\pi$) by a factor  $e^{-2S_I}$. 
 }
\label{DGB}
\end{figure}

\noindent
{\bf Ground state wavefunction vs. bion gas:}		
Now consider $N_f\geq 1$. We have shown that real and complex bions are  
exact saddles points of the  path integral for theories with fermions.  
The 
vacuum is two-fold degenerate. The Euclidean description of either one of 
the vacua is given by a dilute gas of real and complex bions, see Figure 
\ref{DGB}. The density of both types of bions is $e^{-S_b} \approx e^{-2S_I}$. 
The semi-classical picture is based on a particle that starts at $0$ (or 
an even site) on the left, and makes occasional excursions to $2\pi$ (odd 
site), where it spends a time $2t_{0}\approx m^{-1}_{b}\ln\frac{32a^3}{pg}  
\equiv \tau^{*}$, and immediately returns back to an even site, either 
back to 0 (complex bion) or forward to $4\pi$. In other words, the Euclidean  particle spends an exponentially
short time  (unlike the bosonic case) on odd sites. In the wave function language, the wave function must be completely 
dominated at even sites. 
 The ratio of the 
probability to find the particle on an odd-site compared to an even 
site is  exponentially small and equal to the bion density: 
\begin{align}
 \frac{{\rm {Pr.}}( 4\pi (k+ \half))} {{\rm {Pr.}}(  4\pi)}  
  \sim  e^{-S_b} \sim e^{-2S_0} 
     \qquad  {\text {for one of the ground states.} }
\end{align}
For the other ground state, odd and even sites are interchanged.

\begin{figure}[t]           
\centering
\includegraphics[width=10cm]{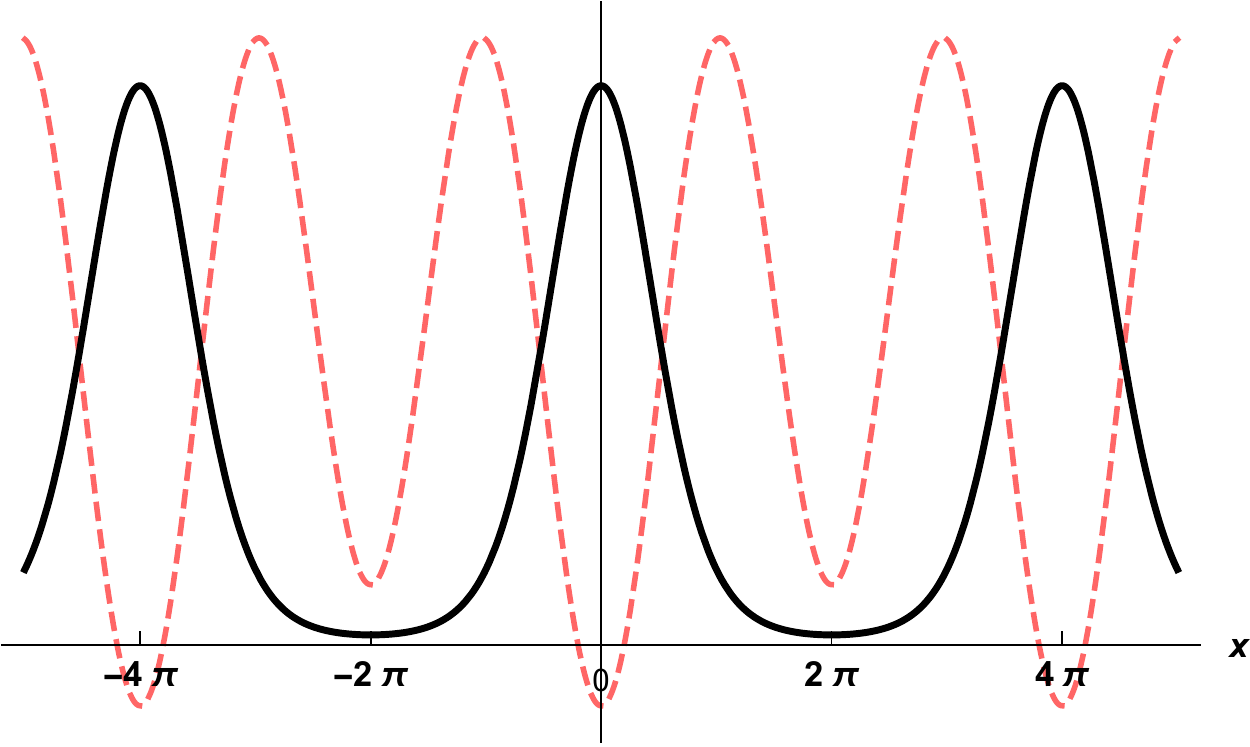}  
\caption{Quantum modified potential (red dashed) for $N_f \geq 1$ theories,  and associated  ground state wave function. The probability to find the particle on an odd-numbered well is 
suppressed exponentially, by $e^{-2S_I}$, consistent with the dilute bion gas description. As is well-known, in the bosonic   case, $N_f=0$,  the wave function is equally dominant both on the even and odd sites. Note the absence of a bump in the wave function at $x=2\pi$. }
\label{DGB-wavefunction}
\end{figure}

  This has a  simple interpretation in the Hamiltonian formulation as well.  
Consider for simplicity the supersymmetric case $N_f=1$. The qualitative 
picture is same for any $N_f \geq 1$. The ground state wave functions are 
given by,  
\begin{align}
 \langle x| {\mathsf G_1} \rangle  
   =  N e^{+ {W(x)}/{g}}  \; | \downarrow   \rangle, \qquad 
          \langle x|  {\mathsf G_2} \rangle  
   =  N e^{- W(x)/g}   | \uparrow   \rangle \, . 
\label{GS-per}
\end{align}
Both  $\pm$ states are renormalizable on the interval $[0, 4\pi]$. Consider 
the $+$ solution. Indeed, this wave function is dominated by even-integer 
wells, and suppressed at the odd-integer wells:
\begin{align}
 \frac{{\rm {Pr.}}( 2\pi) } {{\rm {Pr.}}(  0)}   
   = \Big| \frac{\langle 2\pi | {\mathsf G_1} \rangle} 
                {\langle  0   | {\mathsf G_1} \rangle}  \Big|^2 
  = e^{-2 \Delta W/g}  
  =  e^{-S_b}   = e^{-16a^3/g} \, . 
 \end{align}
The picture  in terms of complex and real bions  shown in Figure \ref{DGB} 
exactly matches the Hamiltonian picture.
     
 This simple system shows that introducing Grassmann valued fields into 
the Lagrangian dramatically alters the ground state structure. This change 
in the ground state structure is the simple quantum mechanics realization 
of the magnetic bion mechanism in QCD(adj) \cite{Unsal:2007jx,Anber:2011de}, 
where introducing the Grasssmann valued fermion fields alter the ground state 
structure in a similar manner  with respect to purely bosonic case of deformed 
Yang-Mills \cite{Unsal:2008ch,Bhoonah:2014gpa}.  

\noindent		
{\bf What would happen if the multi-valued singular saddle did not contribute?}	Consider the $N_f=1$ theory, corresponding to $\N=1$ supersymmetric QM. In 
this example, it is known that supersymmetry is unbroken. The Witten index 
$I_W=0$, but there are Bose-Fermi degenerate two ground states, \eqref{GS-per},
with $E_{\rm gr}=0$. The effect of the real bion is to reduce the ground state 
energy by an amount proportional to $-e^{-2S_I}$. If this was the only 
contribution there would be a contradictions with supersymmetry algebra, which demands positive semi-definiteness of the ground 
state energy.  However,
we showed that there is also  a complex bion solution, which gives a 
contribution of the same magnitude but opposite sign, $E_{\rm gr} \propto 
- e^{ -2S_I \pm  iN_f \pi}\Big|_{N_f=1} =  e^{ -2S_I }>0$. This is the mechanism 
for the non-perturbative vanishing of the ground state energy: 
 \begin{align}
 \label{np-vanish}
E_{\rm gr} \propto   - e^{ -2S_I} - e^{ -2S_I \pm  i   \pi} =0 \, . 
\end{align}
Here, as in the double-well problem,  $e^{i \pi}$  is the $\Z_2$ valued 
hidden topological angle associated with the complexity of the solution \cite{Behtash:2015kna}. 

\subsection{Remark on fluctuation operators}
\label{sec:fluc-op}
In the present work, although we do not examine the fluctuation operators  around the new exact saddle points in detail, 
it is worthwhile to point out some of their interesting  properties. 
Consider the  fluctuation operators 
for the bounce and real bion solution. For the complex bion the analysis  
can either be done by using analytic continuation from the real bounce, 
or by studying the complex fluctuation operator directly. 

\begin{figure}[t]      
\includegraphics[width=7.5cm]{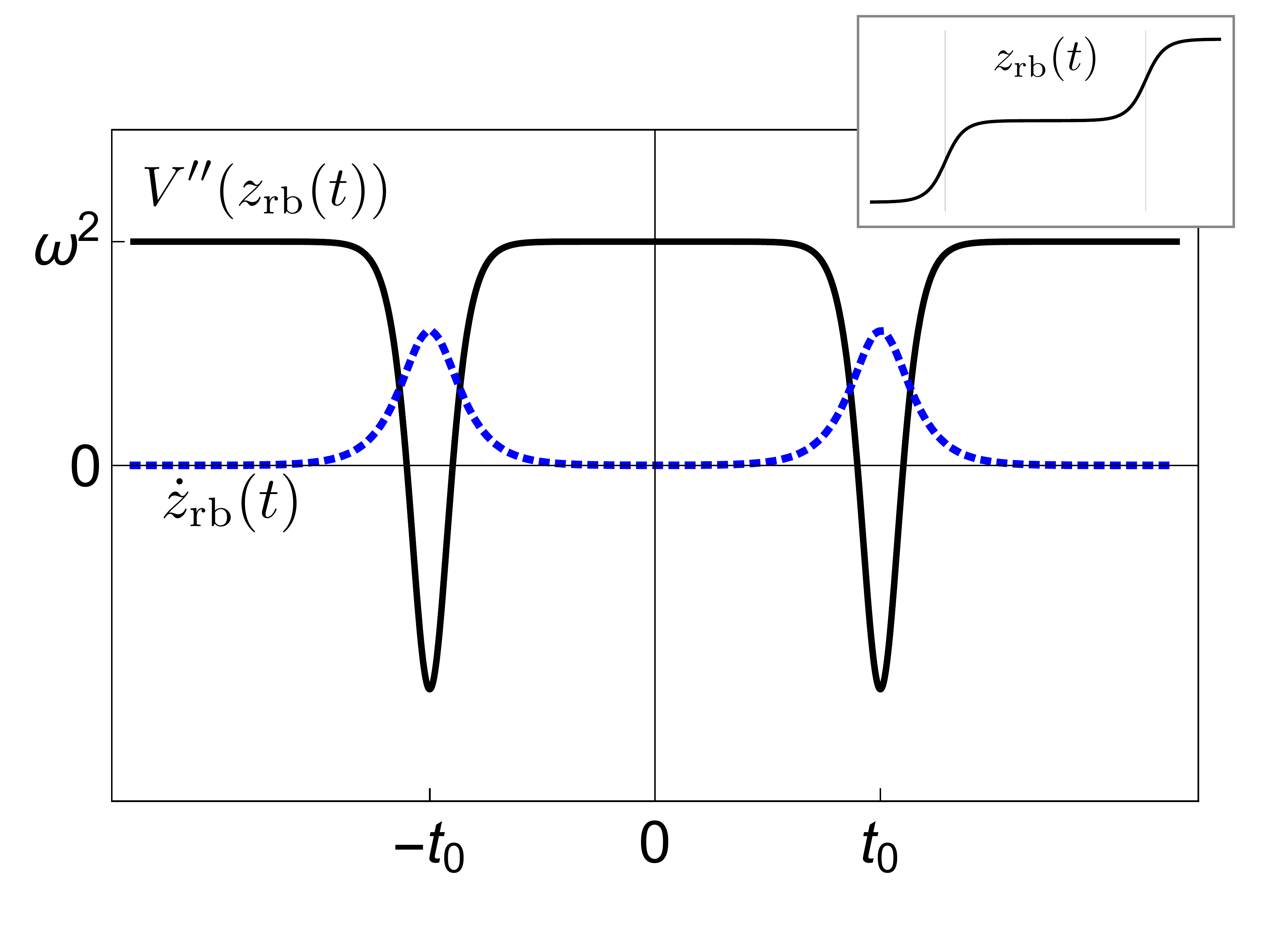}  
\includegraphics[width=7.5cm]{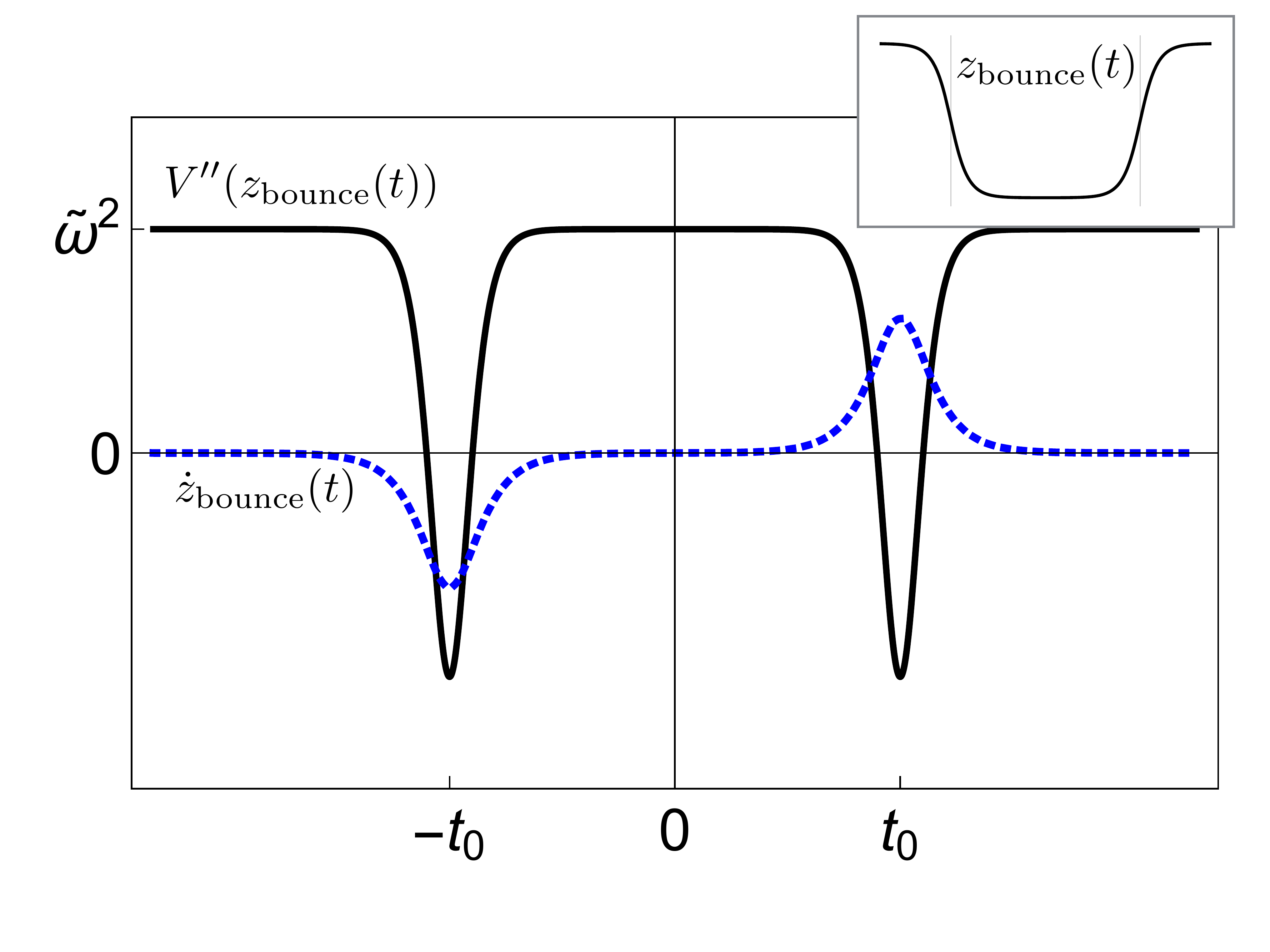}  
\caption{The potential of fluctuation operators for the exact flat-bounce 
solution and real bion solution. }
\label{fig:fluc-op}
\end{figure}

The fluctuation operator around a classical solution  $z_{\text{cl}}$  is  
given by
\begin{equation}
{\bf M}=-\frac{d^2}{d t^2}+V''(z)|_{z=z_{\text{cl}(t) }} \, . 
\label{fluc-op}
\end{equation} 
Figure \ref{fig:fluc-op} depicts the case of the real bion and bounce 
solutions. The characteristic feature in both cases is the plateau and 
the  double-well structure. This implies that the fluctuation operator 
differs from the one in the case of instantons, which is a single-well 
(P\"oschl-Teller type), or the one for a  typical bounce, which is also 
usually single-well. 

\noindent
{\bf Real bion:} For the real bion, the zero mode wave function is 
\begin{align}
 \Psi_0 (t) = \dot z_{\text {rb}}(t) \, . 
\end{align}
Since the fluctuation operator ${\bf M}$ commutes with parity, $[{\bf M},
{\bf P}]= 0$, the ground state must be parity even. Since $\Psi_0 (-t)= 
\Psi_0 (t)$ and it is nodeless, the ground state wave function of the 
fluctuation operator is the one of the zero mode wave-function 
\begin{align}
\Psi_G (t) = \Psi_0 (t),  \qquad e_G= e_0=0\, . 
\end{align}
 This is an exact zero mode. In fact,  the fluctuation 
operator is factorizable and hence, a partner in a supersymmetric pair. 

 The parity-odd anti-symmetric combination is a parametrically small 
eigenmode of the fluctuation operator, and requires special care. Let us 
call it $\Psi_1(t)$.  The energy $e_1$ of the $\Psi_1(t)$ must be 
parametrically small with respect to the rest of the spectrum of ${\bf M}$. 
A way to see this is to realize that the parametric smallness of $e_1$ 
itself is an instanton effect. But this instanton is extremely exotic, it 
actually lives in the fluctuation operator of the real bion solution.  So, we 
call this instanton an  ``f-instanton''. The non-perturbative splitting  
$e_1-e_0=e_1$  is proportional to $e^{-S_{fI}}$.    
  The f-instanton is a configuration which interpolates between one maximum 
of  $-V''(z)|_{z=z_{\text{cl}}}$ and the other maximum, but with a flat potential 
in between. Here, we do not discuss the details of the f-instanton, except 
for pointing out that it accounts for the parametric smallness of the 
quasi-zero mode eigenvalue of fluctuation operator.  
  
\noindent
 {\bf Bounce:} For the real bounce the zero mode wave function is 
\begin{align}
 \Psi_0 (t) = \dot z_{\text {bn}}(t)\, .  
\end{align}
 But now, the zero mode  with energy $e_0=0$ is parity odd, $\Psi_0 (-t)
= - \Psi_0 (t)$ and has a single node.  Since $[{\bf M}, {\bf P}]= 0$, 
the ground state must be parity even and nodeless. Thus, the zero mode 
cannot be the ground state,  $\Psi_G (t)$: 
\begin{align}
 \Psi_G (t) \neq  \Psi_0 (t), \qquad e_G < e_0=0\, . 
\end{align}
Since $\Psi_0 (t)$ has only one node, there can  only be one negative 
eigenmode, denoted as  $\Psi_G (t)$,  with a negative eigenvalue. $e_G$ 
is a parametrically small, negative eigenmode. Similar to the previous case, 
the parametric smallness of the eigenvalue $e_G$ is due to f-instanton 
effect, the instanton living in the fluctuation operator.  A more detailed 
study of the fluctuation operators, including the complex fluctuation operators,  is deferred to future work. 

\section{Complex Saddles and Resurgence}
\label{sec:csr}

In both the  tilted DW and double-SG  quantum mechanics  example, we have shown that inclusion of the complex saddles prevents a potential 
discrepancy between semi-classical analysis and the constraints of supersymmetry.   In this sense, the necessity of the inclusion of these saddles in 
supersymmetric theory is not in doubt.  However, one may still ask a more 
general question:   In a non-supersymmetric theory,  what is the 
 guiding principle which instructs us that complex solutions with multi-fold ambiguous actions must be included?    In particular, why does multi-valuedness of these saddles 
not lead to multi-valuedness in the energy spectrum (which must be unambiguous) of the theory?
   (This second question and its various incarnations in QFT are  prime arguments against the physical significance of multi-valued saddles.) 
  In this brief section,  we show that the multi-valuedness associated with the amplitudes of the complex bion cancels with the ambiguity in the left/right Borel resummation of the perturbation series.   This is a natural implication of resurgence theory, and a realization of Borel-Ecalle (BE)-summability \cite{Ecalle-book,Costin-book,2007arXiv0706.0137S}, providing a mechanism for the reality of the resurgent transseries in the physical domain where it indeed has to be real \cite{Aniceto:2013fka, Basar:2013eka, Marino:2012zq}.    A general 
discussion of all-orders cancellations in real  trans-series  is elegantly 
described in   \cite{Aniceto:2013fka}, also see other examples in  
\cite{Marino:2007te, Aniceto:2011nu}.
  Some  aspects of resurgence in  supersymmetric quantum mechanics  from  a complex WKB point of view is also discussed in \cite{Getmanenko-1,Getmanenko-2}.

 The contribution of the complex bions  to the vacuum energy  for the 
tilted-DW system is  
\begin{align}
 {\cal E}_0 &= + \frac{1}{2\pi} \left(\frac{-g}{16a^3}\right)^{p-1}  \; \Gamma(p)  \; e^{-8a^3/3g} \cr
 &= + \frac{1}{2\pi} \left(\frac{g}{16a^3}\right)^{p-1}  \left( -\cos (p \pi) \Gamma(p)  \pm i \frac{\pi}{\Gamma(1-p)} \right)  e^{-8a^3/3g}\, ,
 \label{en-gr}
\end{align}
 which has both real and imaginary  (ambiguous) parts.  In the second step, we assumed that $p$ is real.  Of course, as it stands, this expression is  not acceptable on physical grounds. 
 

Consider perturbation theory for the energy spectrum, in particular the 
ground state energy,  for the tilted DW system. Let ${\cal E}_n( g)$  be 
the energy of the state with quantum number $n$  ($n$ may be a collective 
quantum number) in units of natural frequency (boson mass) $m_b$ of the 
system.    Let
\begin{equation}
 {\cal E}_n( g) = a_{n,0} +  a_{n,1} g   +  a_{n,2} g^2 + \ldots   
= \sum_{q=0}^{\infty} a_{n,q} g^{q}\, . 
 \end{equation}
The large-order behavior for the ground and   first few states are derived 
 using the methods of 
Bender-Wu \cite{Bender:1969si,Bender:1973rz} in  Ref.\cite{Verbaarschot:1990ga},   and alternatively,  using instantons and dispersion relations  in \cite{Verbaarschot:1990fa}. 
The Bender-Wu recursive method 
is based on a Hamiltonian approach,  and does not rely on instanton calculus.  
The leading order factorial divergence of the perturbative series are  
\cite{Verbaarschot:1990fa,Verbaarschot:1990ga}:
 \begin{align}
  a_{0,q }  \sim -\frac{6^{-p+1}}{2\pi}   \frac{\Gamma(q-p +1) }{\Gamma(1-p )} 
  \frac{1}{(2S_I)^q} \, ,  \cr
  a_{1,q }  \sim -\frac{6^{-p+3}}{2\pi}   \frac{\Gamma(q-p +3) }{\Gamma(2-p )} 
  \frac{1}{(2S_I)^q} \, .  
    \label{large-order}
 \end{align}
The perturbative series is a formal asymptotic  non-alternating series  of 
the form $q!/(2S_I)^q$. It is not Borel summable, due to singularities on 
the positive real axis on the Borel plane,  but it is left and right Borel 
summable, giving a two-fold ambiguous result  ${\cal S}_{\pm }  {\cal E} (g^2)$.

Remarkably, the ambiguity inherent to the complex saddle   and the ambiguity 
in the lateral Borel resummation of perturbation theory  cancel each other 
{\it exactly}:
\begin{align} 
\Im \Big[   {\cal S}_{\pm }  {\cal E}  (g^2)   + I_{\rm cb}^{\pm}   \Big]= 0  \, ,
\end{align}
and the non-perturbative contribution to the ground state energy is 
\begin{align}
 {\cal E}_0^{\rm n.p.} = - \frac{1}{2\pi} \left(\frac{g}{16a^3}\right)^{p-1}  \Gamma(p)  e^{-8a^3/3g} \cos (p \pi) \, . 
 \label{np-gr}
 \end{align}
This presents a mechanism in which the discontinuities  in saddle amplitudes 
can cancel in the sum over different saddle points,   despite the fact that 
different saddles are hierarchical, providing a resolution of a puzzle 
addressed in  \cite{Harlow:2011ny}.  

The crucial (and almost paradoxical)  point is that the reality 
of  resurgent trans-series  associated with a physical observable   for real 
physical values of the couplings  demands complexification of the path integral, and contribution of complex multi-valued saddles.    

\section{Comments on the literature} 
\label{sec:comments}
\subsection{Witten and Harlow-Maltz-Witten on analytic continuation 
in Chern-Simons and Liouville theory}
\label{sec:HMW}
  
 Prior to this work, the most serious study of complex, multi-valued, 
and singular saddle points of the path integral is described in the 
recent work of Harlow, Maltz, and Witten (HMW) on Liouville theory  
\cite{Harlow:2011ny}.  First, we would like to emphasize that there 
are two closely related ideas: 
\begin{itemize}
\item{The analytic continuation of Euclidean path integrals as coupling 
constants are continued to the complex domain.}
\item{The complexification of Euclidean path integrals even if the coupling 
constants are real. }
\end{itemize} 
In the following we will try to explain the relationship  between 
HMW and our work:

\begin{itemize}
\item[\bf 1)]  
The first idea is the starting point of the recent work of Witten 
on Chern Simons theory \cite{Witten:2010cx}, and of Harlow et al.~on
Liouville theory \cite{Harlow:2011ny}. These papers emphasize the need 
to complexify the path integral once one performs analytic continuation
of a coupling parameter into an  {\it unphysical region of parameter space}.
In the case of Chern-Simons theory Witten considers analytic continuation 
to non-integer values of the coupling $k$.

 In the present work we show that even for {\it physical values of the 
parameters}, that means for values of the coupling for which a Hilbert 
space interpretation exists, the path integral still has to be complexified 
in order to obtain the correct semi-classical expansion.
 
\item[\bf 2)]  
 A more complicated issue raised by HMW is the role of singular
and multi-valued solutions. HMW find solutions of this type in 
analytically continued Liouville theory. In the introduction
to their paper HMW state

\begin{quote}
``Rather surprisingly, we have found that allowing ourselves to use the 
multi-valued  ``solutions" just mentioned in the semiclassical expansion 
enables us to account for the asymptotics of the DOZZ formula throughout 
the full analytic continuation in the $\eta_i$.  $\ldots$   
We do not have a clear rationale for why this is allowed."
\end{quote}

 In the present paper we found a simple quantum-mechanical example 
in which we provide evidence that multi-valued discontinuous solutions
with finite action must be included in the semi-classical expansion. In 
particular, in the SUSY QM case, we showed that the inclusion of 
singular solutions is needed to achieve consistency between the 
supersymmetry analysis and the semi-classical expansion.  
 
 \item[\bf 3)]  HMW raise a more specific concern related to multi-valued 
actions. In footnote 33,  HMW state that perhaps discontinuities can cancel 
in the sum over different saddle points. However, they reject this possibility
based on the fact that for generic values of the parameters there will 
be a complex saddle that is parametrically larger than the rest.   
 
 We note that a hierarchy of saddle points also occurs in our examples. 
However, in the context of resurgence theory this is an expected feature
of the semi-classical expansion. Indeed, consider a dominant and subdominant 
saddle, and assume the action associated with the sub-dominant saddle is 
multi-valued. What happens in resurgent expansions is that the ambiguity 
associated with perturbation theory around the dominant saddle, is of 
the same order as the subdominant saddle. Consequently, there may in 
fact be cancellations of the discontinuities between hierarchical saddles.  
In Sec.~\ref{sec:csr}, we provided a concrete realization of such 
cancellations. 

\end{itemize}
 
 Other studies on complexification of path integration, with different 
physical motivations,  are discussed in 
\cite{Witten:2010zr,Guralnik:2007rx,Ferrante:2013hg,Jaffe:2014pea,Jaffe:2013yia}.

\subsection{Brezin, Le Guillou, and Zinn-Justin on complex instantons}

 Almost four decades ago Brezin, Le Guillou, and Zinn-Justin (BGZ)
\cite{Brezin:1976wa} considered the potential
\begin{align}
V(x)=  \half x^2 - \gamma x^3 + \half x^4\, ,
\end{align}
which, for $\gamma>1$, is the same as our tilted double-well example, and 
for $\gamma<1$ corresponds to a potential with a unique minimum. $\gamma=0$ 
is the focus of the Bender-Wu analysis of the large order behavior of
perturbation theory.
 
 In the case $\gamma>1$, the analysis of BGZ corresponds to ``half'' 
of our real  bounce.  BGZ find the real solution interpolating from 
the local maximum of the inverted potential $-V(x)$ to a real turning 
point. However, the complex (and sometimes multi-valued) configuration that 
we call the complex bion,  which dictates  ground state properties, is 
not considered in \cite{Brezin:1976wa}.   We note that the complex bion 
corresponds  
to either a singularity in the Borel plane on $\R^{+}$  $ (p\neq \mathbb N^{+})$,
where the complex bion amplitude has a genuine ambiguity with amplitude  
$e^{-\Re S_{\rm cb} \pm i p \pi} $, or to a hidden topological angle  $(p \in  
\mathbb N^{+}) $, where the complex bion amplitude is unambiguous. In the 
second case, the  leading singularity in the Borel plane fades away 
as $p$ approaches a positive integer. The Stokes multiplier of this 
configuration is non-zero, and contribute crucially to ground state properties.

 
In the case $\gamma<1$ BGZ  construct complex instantons, which interpolate 
between $x=0$ and a complex turning point $z_T, z_T^{*}$.  This type of 
complex instanton are argued to lead to a Borel summable series, similar to  and are 
associated with  singularities in the $\mathbb C \backslash \mathbb R^{+} $ 
portion of the Borel plane. For example, in the case $\gamma=0$, the 
associated singularity is on $\mathbb \R^{-}$ and leads to the famous alternating 
series, found by Bender and Wu \cite{Bender:1969si,Bender:1973rz}. In this 
case, although complex instantons dictate the perturbative result, their 
Stokes multiplier is zero. If that were not the case,  complex instantons
would give a pathological, exponentially increasing, contribution to the path integral of 
the form $e^{A/g}, A >0$, similar to a particular type of complex instantons in QFT \cite{Lipatov:1976ny}.   
 In the present work, the complex saddles have nonzero Stokes multipliers, their weight is  exponentially suppressed, 
 $e^{-A/g}, A >0$.   It is important to realize that generically the complex bion 
 lead to a Borel non-summable series, unlike the complex instantons,  and  they are crucial  for the correct Euclidean description 
of the ground state physics. 

\subsection{Balitsky and Yung on complex quasi-solutions}
\label{sec:BY}

  Some of the ideas in our  work  are inspired  from  the work  
of Balitsky and Yung (BY) \cite{Balitsky:1985in}. In a beautiful, and 
under-appreciated, paper on semi-classical aspects of supersymmetric 
quantum mechanics BY considered both supersymmetric QM and the deformation 
of the Yukawa term in \eqref{lagsusy} away from $p=1$, which is the
supersymmetric point. The BY deformation corresponds to the two $SU(N_f)$ 
singlet sub-sectors in our construction with $k=0$ and $k=N_f$.

  BY constructed, following Zinn-Justin, an approximate real 
instanton-anti-instanton configuration. This is not an exact solution, 
but a good approximation to the exact bounce solution in the weak coupling 
regime. They also constructed the approximate complex solution by using 
analytic continuation from the approximate bounce solution, following 
the idea from \cite{Brezin:1976wa}. BY correctly interpret this 
solution as governing the ground state properties. 
   However, BY do not obtain the exact complex solutions found here. 
 
\subsection{Other related works}
 
 There are earlier important works in  quantum mechanics in which 
complexification plays some role.  The main distinction between our 
present study  and these works is the following.  In our work, we have 
shown that the ground state properties of a generic quantum mechanical 
system are governed by complex (sometimes even multi-valued) saddles. In 
the quantum mechanical papers mentioned below, the ground state properties 
are always governed by real saddles, such as instantons, but complex 
classical solutions become important when considering the entire spectrum, 
for example the spectral resolvent.
Gutzwiller \cite{Gutzwiller:1971fy,MuratoreGinanneschi:2002tm}  pioneered 
the idea of summing over all classical solutions in semiclassical expansions 
of Green's functions and spectral problems. This was further developed by many 
authors, from a general formulation  in  \cite{Balian:1974ah,Lapedes:1981tz}, 
to specific analyses of the symmetric double-well 
\cite{Carlitz:1984ab,Richard:1980ei,Richard:1981gn} and 
periodic potentials \cite{Millard:1984qt}.  A complex version of WKB  analysis 
of the pure quartic oscillator also requires inclusion of complex semiclassical 
configurations in addition to the familiar real Bohr-Sommerfeld ones 
\cite{PhysRevLett.41.1141,Balian:1978et,Voros1983}.

\section{Conclusion} 

 In this work we studied the semiclassical expansion in two classes
of quantum mechanical models with fermions, as well as closely related purely 
bosonic systems with non-degenerate harmonic minima. We showed that correct 
semi-classical results can only be obtained if the path integral
is complexified, and finite action complex solutions are taken into
account. This includes solutions that are singular and multi-valued. 
Our main conclusions are:

\begin{itemize}
\item{We argued that in quantum mechanical models with multiple Grassmann 
fields,  the natural setting for finding the semi-classical expansion is a 
graded formulation in which fermions are integrated out. This induces a 
quantum modified potential for the bosonic fields. }
  
\item{The real Euclidean equations of motion in the  inverted potential   
may or may not have real  finite action solutions.   But once the path 
integral is complexified, the critical points are described by the solutions 
of the holomorphic Newton equation in the inverted quantum modified potential: 
\begin{align} 
 \frac{d^2 z}{dt^2} =  +\frac{ d V}{d z} 
  =      W'(z)  W'' (z) + \frac{p g}{2} W'''(z)  \, . 
\label{bion-eq}
\end{align}
This equation generically admits complex, multi-valued finite action 
solutions.}  

\item{The characteristic size of the real and complex bion saddle solutions  
is parametrically larger than  the natural size of the instanton solution 
in the original formulation. The instanton size is parametrically  
$\omega^{-1}$  where $\omega$ is natural frequency, and bion size is 
$ \tau^* \sim \omega^{-1} \ln \frac{Aa^3}{gp}$.  The  non-BPS  bion  
solutions  are exact  for any value of the coupling. In the weak coupling 
regime, it can be described as a correlated two-instanton event  (which 
is only approximate quasi-solution). }

\item{The action of the exact non-BPS solution is in general multi-valued, 
and weight of these saddles is of the form  $\sim e^{ -S_{\rm r}  \pm i p \pi} $. 
For $N_f=1,2, \ldots$, where $p=N_f-2k$, the multi-valuedness of the action 
disappears since the imaginary part of the action is defined modulo $2\pi$. 
This gives the $\mathbb Z_2$ valued hidden topological angle 
\cite{Behtash:2015kna}.

\item For non-integer values of $p$, the multi-valuedness of the fugacity 
is canceled by Borel resummation of perturbation theory, an explicit 
realization of resurgence. In this sense, neither divergence of a  series 
(or its ambiguous Borel resummation), nor the multi-valued saddles  are 
meaningless or a nuisance. Rather, they are essential in order to give 
proper meaning to the path integral.} 

\item{In generic quantum mechanical systems, the Euclidean description 
of the vacuum is a dilute gas of complex and real bions. In particular, the 
much better known  instantons, and the  dilute  instanton gas picture, are 
relatively rare cases relevant for potentials with symmetry. } 
\end{itemize}

\subsection{Prospects in quantum field theories}

In this paper we pursued two strategies for constructing saddles in path 
integration, one of which is exact, and the other approximate. 
\begin{itemize}
\item [\bf (A)]  Exact solution of {\it complexified} Newton equations 
after integrating out fermions exactly. 
\item  [\bf (B)]  Construct BPS solutions, and glue them together by 
integrating over the {\it complexified} quasi-zero mode thimble, 
$\Gamma^{\rm qzm}$,  using the finite dimensional steepest descent method.  
\end{itemize}
In weak coupling we have shown that all the essential features of the 
exact solutions, their real or complex nature, multi-valuedness, singular 
or smooth behavior, hidden topological angle,  characteristic size, 
and monodromy properties, are reproduced by the second (approximate) method.
This makes us confident that the second method, {\bf (B)}, is indeed a 
systematic approximation to exact saddle point calculations.\footnote{Proving 
this  statement starting with the complex gradient flow equations by using 
techniques from partial differential equations  (PDEs) would represent major 
progress in Picard-Lefschetz theory.} 
The physical reason for this conjecture is based on the following argument:
Consider a two-instanton event, and fluctuations around it. The center 
position of the two instantons is an exact zero mode, the relative position 
is a quasi-zero mode, and the remaining modes are Gaussian.   The quasi-zero 
mode is special in this sense. An exact integration over the 
$\Gamma^{\rm qzm}$-cycle, which passes through the saddle of the interaction 
potential of the two instantons is capable of producing the features of the 
exact solution. 


The first method, {\bf (A)}, is  hard to generalize to QFT. In studying the
quantum mechanics of a single particle with spin degrees of freedom we are
in a somewhat fortunate situation in the sense that we can write the 
notorious fermion determinant $N_f \log \det (D) $ (where $D$ is the 
Dirac operator) as a sum of local modifications of the bosonic potential. 
This is important for constructing of exact solutions.  In QFT, this is 
a formidable task, although in certain background field problems it has 
been shown, using the world-line representation of the fermion determinant, 
that complex saddles govern the physics \cite{Dumlu:2011cc}.

 
 Another consideration is that in bosonic quantum field theories and quantum 
mechanics,  when the topological theta angle is set to zero, the contribution 
of instantons  (as well as other  real saddles) to the ground state energy 
is universally negative. In theories with fermions (multiple Grassmann 
valued fields), instantons do not contribute to the ground state energy 
because of the fermionic  zero modes. At second order,  there can be either 
a negative or  positive contribution.    In fact, for real saddles (at second 
order or otherwise), the contribution to the 
ground state energy is  universally negative. For complex saddles, whether 
the contribution is negative or positive depends on the hidden topological 
angle. For example, in the present work, we showed that complex saddles 
contribute to the ground state energy as  $-\cos( {p \pi})\, e^{-S_{\rm r}}$, 
so it is positive for $p=1$,  and negative for $p=2$. 
 
If we accept the idea that the second method, {\bf (B)}, is indeed a 
systematic approximation to exact saddle points in QFT, we find ourselves 
in an interesting situation. In particular, such real and complex saddles 
must exist  in four dimensional gauge theories,  QCD and SQCD  on $\R^4$, 
$\N=1$  SYM and QCD(adj) on $\R^3 \times S^1$,  $\N=2$  SYM  on $\R^3$,  
and sigma models such as $\mathbb{CP}^{N-1}$ and $O(N) $ models   with 
fermions on $\R^2$   as well as $\R^1 \times S^1$, and more. 

 
Consider for example ${\cal N}=1$ SYM. It is a   vector-like (QCD-like)  
theory, without an elementary scalar.   It is also a minimal supersymmetric 
gauge theory, and an integral part of all SQCD theories. In this sense, it 
is  a useful model to describe phenomena that may take place both in  QCD 
and SQCD.   On $\R^3\times S_1$,  at leading order in the semi-classical 
expansion,  this theory has monopole-instantons which induce a superpotential 
\cite{Seiberg:1996nz,Davies:1999uw}.   At second order in the semi-classical 
expansion, the microscopic origin of the bosonic potential can be understood 
in terms of magnetic bions  (in one-to-one correspondence with the positive 
entries of the extended Cartan matrix) and neutral bions (in one-to-one 
correspondence with the negative  entries of extended Cartan matrix) 
\cite{Unsal:2007jx,Argyres:2012ka,Poppitz:2012sw}. For example, 
for $SU(2)$ gauge theory, the bosonic potential is  
\begin{align}
V(\phi, \sigma)  \sim - e^{ -2S_m } \cos( 2 \sigma)  
  -e^{-2S_m \pm   i \pi}   \cosh(2\phi) \, ,
\end{align} 
where $\sigma$ is the dual photon and $\phi$ is the fluctuation of the 
holonomy relative to the center symmetric point.  Magnetic bions contribute 
negatively to the ground state energy (found by setting $\sigma=0, \phi=0$), 
and neutral bions contribute positively: 
 \begin{align}
 \label{np-vanish-2}
E_{\rm gr} \propto   - e^{ -2S_m} - e^{ -2S_m \pm  i   \pi} =0\, . 
\end{align}
Note that this formula is identical to the one we found in supersymmetric 
quantum mechanics with periodic superpotential \eqref{np-vanish}. There 
is a relative $e^{i \pi}$ hidden topological angle  difference between the 
two amplitudes. According to our criterion, it seems  plausible that the 
magnetic bion is associated with an exact real saddle while the neutral  
bion is  associated with an exact complex saddle. All statements concerning 
two-events (second order semi-classics) in ${\cal N}=1$ SYM are also valid 
for QCD(adj) on  $\R^3 \times S^1 $,   as well as non-linear sigma models 
on $\R^1 \times S^1$ \cite{Dunne:2012ae,Cherman:2013yfa,Cherman:2014ofa}.

Similarly,  in three-dimensional ${\cal N}=2$ SUSY  gauge theory 
\cite{Affleck:1982as}, the bosonic potential can either be derived from 
the superpotential or by performing a quasi-zero mode integration, and 
provides a positive definite run-away potential,  
\begin{align}
V(\phi)  \sim  -e^{-2S_m \pm   i \pi}   e^{-2\phi}  \, . 
\end{align}
 The origin of the positive definiteness of the run-away potential is  
a complex phase that arises from the  QZM integration.   

Finally, in SQCD with $N_f=N_c-1$ in $\R^4$, the semi-classical analysis  
is reliable, because   the IR-divergence due to the  instanton size modulus 
is cut-off by the scalar meson vacuum expectation value. The  theory has a  
bosonic run-away potential, which can either be derived from the 
instanton-induced ADS-superpotential  \cite{Affleck:1983mk}, or alternatively, 
it can be derived by computing the $[\I\bar \I]$ two-event amplitude  
\cite{Yung:1987zp}. In order for the potential to be positive semi-definite, 
the contribution  of the  $[\I\bar \I]$ saddle must be $-e^{-2S_I + i \pi}$. 
This also suggest that the $[\I\bar \I]$  event in SQCD may in fact be  be 
an exact  complex saddle.  

Our results suggest that  the natural  formalism  for the semi-classical 
analysis of  quantum field theory,  and quantum mechanics, is the complexified 
version of the path integral.  This perspective provides a physical 
interpretation for non-BPS solutions 
\cite{Din:1980jg,Taubes:1982ie,Taubes:1985qt,sibner}, and possibly also 
for complex instantons \cite{Dallagnol:2008zz}. 

\subsection{Other Related Directions}

{\bf Lattice constructions and sign problem: } The Lefschetz thimble 
decomposition is applied to QFTs with sign problems in various lattice 
theories, see, e.g. \cite{Cristoforetti:2012su, Cristoforetti:2013wha, Fujii:2013sra, Aarts:2014nxa,Tanizaki:2015pua}.
Very often, these simulations are run in the vicinity of one saddle.  In the 
present work, we have seen that the effect of different saddles generically 
have different phases, leading to a (Euclidean version of)  interference in 
the sum over saddles.  For example, vanishing of the vacuum energy in 
supersymmetric theory is of the form $E_{\rm gr} = 0 e^{-S_{\rm pert}} -e^{-S_{\rm rb}} 
- e^{i \pi}   e^{-S_{\rm rb}}$, where $e^{i \pi}$ is the $\Z_2$ valued  hidden 
topological angle (HTA), leading to destructive interference among saddles  
\cite{Behtash:2015kna,Behtash:2015kva}. This important point is also 
emphasized in the context of sign problem in 
\cite{Tanizaki:2015rda,Fujii:2015bua}.  Clearly, future lattice (as well 
as continuum) studies must take into account the sum over all contributing 
saddles due to interference effects induced by the HTA. 

 
{\bf Complex gradient flow equations and  ${\cal N}=2$ instantons:} Readers 
familiar with extended  (${\cal N}=2$) supersymmetric quantum mechanics  
will realize that the complex gradient flow equation  \eqref{cgf-inst} in 
a $d=0$ dimensional "field theory" is identical to  the  instanton equation 
for the ${\cal N}=2$ system with superpotential $W(z)=f(z)$. 
\begin{align}
 \frac{dz}{du} = e^{i \theta} \frac{\partial \bar W}{\partial \bar z} \, , 
 \label{cgf-inst-2}
 \end{align} 
where $\theta$ in QM may be identified with the phase of the central charge, 
and $u$ is the gradient flow time (which becomes Euclidean time in the 
second interpretation). Namely, thimbles in the simple zero dimensional 
integrals are related to integration cycles in path integrals of   
(${\cal N}=2$) supersymmetric QM related to instantons. This also 
generalizes to certain QFTs, see  \cite{Witten:2010cx} and the following
item. We find this connection intriguing. 


{\bf Parabolic vs. elliptic gradient flow:} The first path integral analyses 
of the Lefschetz thimble  decomposition were performed for Chern-Simons
theory and the phase space path integral (where the Hamiltonian is set to 
zero) \cite{Witten:2010cx, Witten:2010zr}. In both cases, the complex 
gradient flow equations are elliptic PDEs, and possess a higher 
dimensional (Euclidean version of Lorentz) symmetry,  e.g. for Chern-Simons, 
these equations are  
\begin{align} 
\frac{\partial {\cal A}_i}{\partial u} 
 = - e^{i \theta}  \frac{ \delta \bar {\cal S}_{\rm CS} }{\delta \bar {\cal A}_i} 
 = \half e^{i \theta}  \epsilon_{ijk} \bar {\cal F}_{jk} \, , 
\label{cgf-inst-3}
\end{align}
which is the complex generalization of the instanton equation in $A_4=0$ 
gauge: 
\begin{align} 
{\cal F}^{(2)}    =  e^{i \theta} * \bar {\cal F}^{(2)} \qquad      
   {\rm or } \qquad 
{\cal F}_{\mu \nu} =   e^{i \theta} \half  \epsilon_{\mu\nu \rho \sigma} 
     \bar {\cal F}_{\rho \sigma} \, .
\label{cgf-inst-4}
\end{align}
The reason for the $3d-4d$ connection is actually not an accident: Thimbles 
in $d=3$ dimensional  Chern-Simons theory \cite{Witten:2010cx} are related 
to the complexified instanton equations in the  {\it twisted} version of 
the ${\cal N} = 4 $ super Yang-Mills theory in four dimensions. Indeed, 
\eqref{cgf-inst-4} appeared first  in the context of  lattice supersymmetry 
\cite{Unsal:2006qp}, and in the Geometric Langlands program 
\cite{Kapustin:2006pk}, again, not accidentally, because both of these are  
applications of the same  twist. Witten argues that the nice properties of 
the finite dimensional Lefschetz thimble  decomposition carry over to the 
infinite dimensional functional integral for the elliptic flow equations 
\cite{Witten:2010cx}.  On the other hand, for almost all interesting QFTs 
and configuration space path integrals, the complex gradient flow equations 
are parabolic. It will be interesting to understand more precisely the 
relation between the nature of the flow equation  (parabolic vs. elliptic) 
and the nature of the saddles (smooth/singular/multivalued), and the 
implications for the the thimble construction. 

  Clearly, it is of interest to study general field theories  using 
complexified path integrals. We expect other interesting applications in 
cosmology, string theory, and condensed matter; indeed in any problem in 
which the path integral is a useful tool.

\acknowledgments
We thank  P.~Argyres,  G.~Ba\c sar, A.~Cherman, O.~Costin,  D.~Dorigoni,  
D.~Harlow, C.~Howls,  M.~Kaminski,  K.~Konishi, J.~Maltz,  D.~O' Connor, 
R.~Pisarski, E.~Poppitz, R.~Riwar,  M.~Shifman,  Y.~Tanizaki, B.~Tekin, 
E.~Witten, S.~T.~Yau and A.~Yung  for useful comments and discussions.~   
M.\"U.~thanks especially E.~Witten for encouragement and  probing questions. 
M.\"U.'s work  was  partially supported by the Center for Mathematical 
Sciences and Applications  (CMSA) at Harvard University. Part of this
work was completed while T.S.~was a visitor at the Aspen Center for 
Physics, which is supported by National Science Foundation grant 
PHY-1066293. We acknowledge support from DOE grants DE-FG02-03ER41260 
(T.S.),  DE-SC0013036 (M.U.) and DE-SC0010339 (G.D.).

\appendix

\section{Review of some relevant  elliptic integrals}
\label{app:elliptic}

In this appendix, we summarize some basic result concerning elliptic functions 
and elliptic integrals \cite{WhWa27}. These result are used in constructing
the exact bounce and bion solutions. Let 
\begin{equation}
t=\int_{z_{a}}^{z}\frac{d w}{\sqrt{F(w)}},
\end{equation}
where $F(w)$ is a quartic polynomial with no repeated factors
\begin{align}
F(w):=a_{4}w^{4}+4a_{3}w^{3}+6a_{2}w^{2}+4a_{1}w+a_{0} \, . 
\label{curve}
\end{align} 
Then, the  general solution to 
\begin{eqnarray}
\ddot{z}= \frac{\partial F}{\partial z}
\label{eq:eom}
\end{eqnarray}
is given by
\begin{equation}
z(t) =  z_a + 
  \frac{\sqrt{F(z_a)}\wp'(t;g_{2},g_{3})
        +\tfrac{1}{2}F'(z_a)[\wp(t;g_{2},g_{3})
        -\tfrac{1}{24}F''(z_a)]
        +\tfrac{1}{24}F(z_a)F'''(z_a)}
       {2[\wp(t;g_{2},g_{3})-\tfrac{1}{24}F''(z_a)]^{2}
           -\tfrac{1}{48}F(z_a)F''''(z_a)}\, , 
\label{eq:general}
\end{equation}
where $\wp(t; g_2, g_3)$ is the Weierstrass elliptic function with algebraic 
quadratic and cubic lattice invariants $g_2$ and $g_3$ related to the 
coefficients of the quartic polynomial $F$ in \eqref{curve} by
\begin{align}
g_{2}&:=a_{4}a_{0}-4a_{3}a_{1}+3a_{2}^{2}, \cr 
g_{3}&:=a_{4}a_{2}a_{0}-4a_{3}a_{2}a_{1}-a_{2}^{3}-a_{4}a_{1}^{2}-a_{3}^{2}a_{0}.
\label{lat-in}
\end{align}
The function $\wp(t; g_2, g_3)$ is an even, doubly-periodic, elliptic 
function with periods $T_{1}$ and $T_{2}$ expressed in terms of the 
invariants $g_{2}, g_{3}$ of the associated  algebraic curve of genus $g=1$:
\begin{align}
g_2= 60\sum_{(m, n) \neq (0,0)}(m   T_1 + n   T_2)^{-4}, \qquad 
g_3= 140\sum_{(m, n) \neq (0,0)}(m   T_1 + n  T_2)^{-6}.
\label{ALI-2}
\end{align}
The complex  periods $T_{1}$ and $T_{2}$ define a 2D lattice, and  $\wp$ 
parametrizes the algebraic curve $F(z)$ corresponding to a complex Riemann
surface with genus $g=1$; i.e.,  a torus with two periods corresponding to
each cycle $S^{1}.$ More technically, this torus is obtained by ramifying
a Riemann sphere at 4 points, i.e., zeros of $F(z),$ whose branched
double covering is the function $\wp.$ In the generalization of our 
solutions to higher degree potentials, the energy conservation relation 
defines a higher genus complex algebraic curve, and the role of the 
Weierstrass function is played by hyperelliptic and automorphic functions.  

If $z_a$ is a zero of $F(z)$, which corresponds to the relevant  situation 
of $z_a$ being a turning point, then the general solution \eqref{eq:general} 
simplifies significantly to:
\begin{align}
\label{master-3}
z(t) =
  z_a + \frac{\tfrac{1}{4}F'(z_a)}{\wp(t)-\tfrac{1}{24}F''(z_a)}\, . 
\end{align} 
This is our ``master formula'', which gives the general solution to the 
Euclidean equations of motion. To find bounce or bion solutions which 
pass through certain physically important turning points and critical 
points, we must specify the point $z_a$ and determine the appropriate 
invariants $g_2$ and $g_3$. Note that the ``Euclidean energy'' $E$ in 
\eqref{eq:curve1} enters into the quartic polynomial $F(z)\equiv 2(E+V(z))$, 
so the invariants $g_2$ and $g_3$ depend on $E$, and take especially simple 
values when $z_a$ is a critical point of the potential.

Note the interesting but elementary algebraic fact that for a quartic 
polynomial $F(z)$, as in \eqref{curve}, if $z_1$ and $z_2$ are two 
points such that $F(z_1)=F(z_2)$, then 
\begin{eqnarray}
F^\prime(z_1)+F^\prime(z_2)=\frac{1}{6}(z_1-z_2)\left(F^{\prime\prime}(z_1)
  - F^{\prime\prime}(z_2)\right)\, . 
\label{eq:id}
\end{eqnarray}

\section{Absence of  a Hidden Topological Angle  in one-dimensional  real 
integrals with real parameters}
\label{app:hta}


Consider a one dimensional  integral
\begin{equation}\label{eq:ordint}
I=\int_{-\infty}^\infty dx\; e^{- \frac{1}{\hbar} S(x)}\;.
\end{equation}
with $S(x)$ real for any $x\in \mathbb R$, and $\hbar >0$.   Consider decomposing the real integration cycle ${\mathbb R}$  into a sum over Lefschetz thimbles: 
${\mathbb R}=  \sum_\sigma  n_\sigma {\cal J_\sigma}$. Here, we prove that if $z_\sigma$ is a contributing saddle to the integration, i.e, $n_\sigma \neq 0$, then the phase associated with this cycle must be zero. This proves that the HTA phenomenon  that we found in  the semi-classical analysis of the path integral over real fields and real values of the parameters does not have a counterpart in one-dimensional exponential type integrals, and does not provide an accurate intuition for the problem.  
This may be one reason why the possibility of complex saddles  {\it contributing} to the semi-classical analysis of path integrals  was not taken sufficiently seriously in the past. 

\begin{figure}[htbp] 
   \centering
   \includegraphics[width=0.6\textwidth]{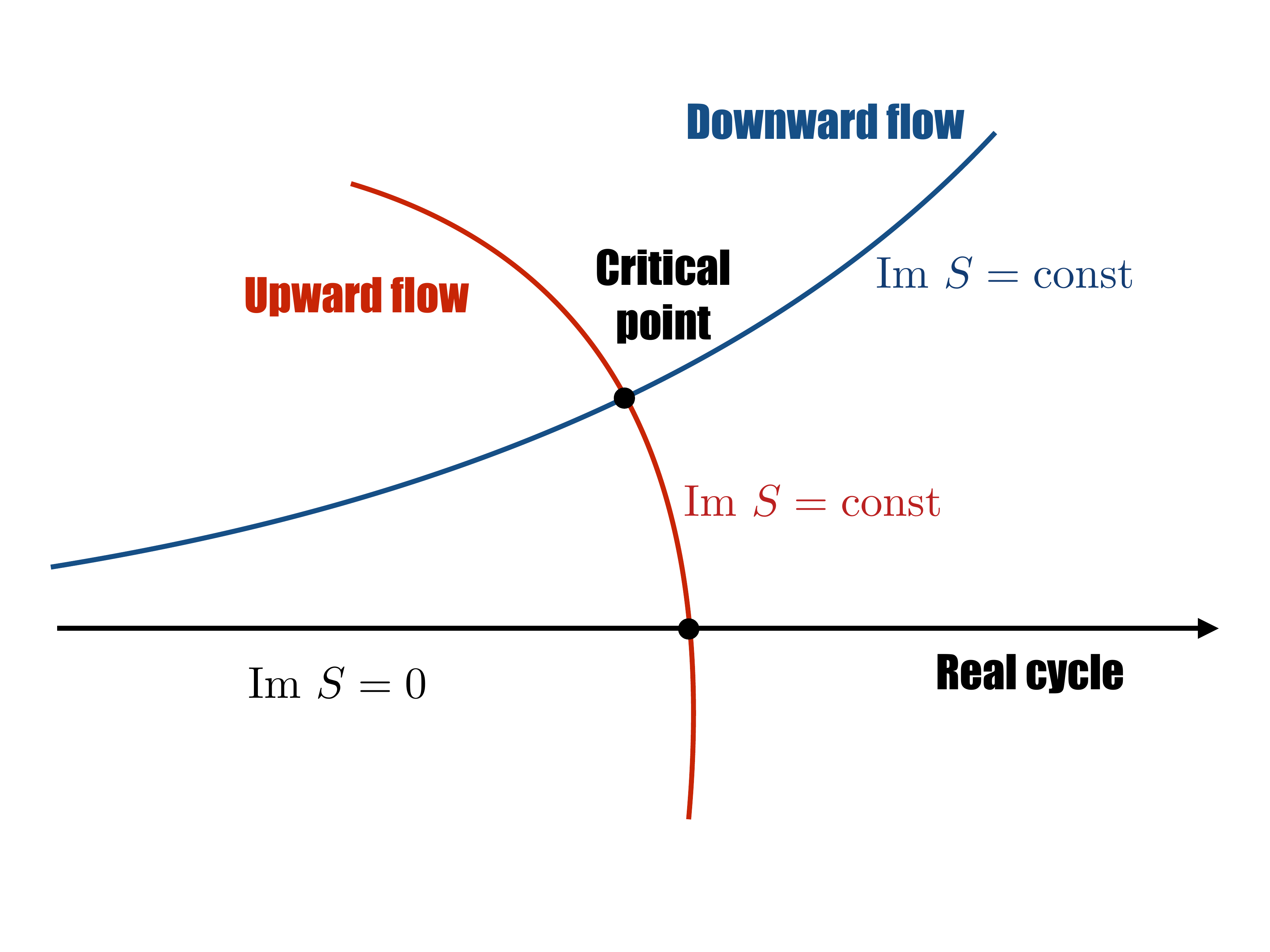} 
   \caption{Sketch of upward and downward flows from a critical point.}
   \label{fig:flows}
\end{figure}
 If the integral has a critical point at $z_\sigma$, then the steepest 
descent equation $\text{Im } S(z)=\text{Im }S(z_\sigma)$ has two solutions: 
``the upward flow'' $\mathcal J_\sigma$, and the ``downward flow'' 
$\mathcal K_\sigma$ (see Fig. \ref{fig:flows}). The ``upward flow'' 
(``downward flow'') is the solution under which $\text{Re }( -S(z))$ 
increases (decreases) away from the critical point. For the cycle 
$\mathcal J_\sigma$ to contribute to the cycle decomposition of the real 
line $\mathbb R$, its upward flow $\mathcal K_\sigma$ must intersect the 
real line. However since $\text{Im }S(x)=0, \forall x\in \mathbb R$, and 
since the phase is stationary over the ascent thimble,  $\text{Im }S(x)=
\text{const},\forall z\in \mathcal K_0$, it follows that $\text{Im }S(z_\sigma)
=0$. This shows that HTA of any critical point which contributes to the 
integral  \eqref{eq:ordint} is always identically zero.

\bibliography{bibliography}
\bibliographystyle{JHEP}

\end{document}